\documentclass[letterpaper]{article}

\usepackage{arxiv}
\usepackage[utf8]{inputenc}
\usepackage[english]{babel}
\usepackage{blindtext}
\usepackage{xr} % corss reference different .tex files

\usepackage{times}

\usepackage[round]{natbib} % natbib citation style
\usepackage[dvipsnames]{xcolor} % text color
\usepackage{amsmath} % basic maths symbols 
\usepackage{amsthm,amssymb} % define proof environment
\usepackage{amsfonts} % extended maths symbols and fonts 
\usepackage{mathtools}
\usepackage{nicefrac} % compact symbols for 1/2, etc.
\usepackage{enumitem} % customized numbering style
\usepackage{setspace} % double space
\usepackage{algorithm2e} % algorithm package 
\usepackage[hidelinks]{hyperref} % clickable links
\usepackage{graphicx}
\usepackage[most]{tcolorbox}
\usepackage{marginnote}
\usepackage{stackengine}
\usepackage{cleveref}
\usepackage{cancel} %to cancel math terms in equation
\usepackage{imakeidx}
\usepackage{listings} %to format code snippets
\usepackage{tikz} % drawing graph
\usetikzlibrary{arrows.meta} % drawing graph
\usetikzlibrary{decorations.markings}
\usetikzlibrary{backgrounds}
\usetikzlibrary{positioning,chains,fit,shapes,calc}
\usetikzlibrary{angles,patterns}
\usepackage{subcaption}
\usetikzlibrary{matrix,quotes}
\usetikzlibrary{decorations.pathmorphing, calc}
\usetikzlibrary{mindmap}

\newtheorem{theorem}{Theorem}[subsection]
\newtheorem{proposition}[theorem]{Proposition}
\newtheorem{lemma}[theorem]{Lemma}
\newtheorem{remark}[theorem]{Remark} % [section]
\newtheorem{corollary}[theorem]{Corollary}
\newtheorem{definition}[theorem]{Definition} % [section] % definition numbers are dependent on theorem numbers
\newtheorem{example}[theorem]{Example} % [section] % same for example numbers

\newcommand{\Z}{\mathbb{Z}}
\newcommand{\Q}{\mathbb{Q}}
\newcommand{\R}{\mathbb{R}}
\newcommand{\C}{\mathbb{C}}
\newcommand{\F}{\mathbb{F}}
\newcommand{\N}{\mathbb{N}}

\newcommand{\T}{\mathbb{T}}
\newcommand{\OO}{\mathcal{O}}
\newcommand{\DD}{\mathcal{D}}

\newcommand{\calA}{\mathcal{A}}
\newcommand{\vc}[1]{\mathbf{#1}}
\newcommand{\tO}{\Tilde{O}}
\newcommand{\dual}[1]{#1^{\vee}}
\newcommand{\inv}[1]{#1^{-1}}
\newcommand{\mfq}{\mathfrak{q}}
\newcommand{\mfp}{\mathfrak{p}}
\newcommand{\ok}{\mathcal{O}_K}

\newcommand{\floor}[1]{\left\lfloor{#1}\right\rfloor} %floor
\newcommand{\round}[1]{\left\lfloor{#1}\right\rceil} %round
\newcommand{\innerprod}[1]{\langle{#1}\rangle} %inner product
\newcommand{\sqbracket}[1]{\left[{#1}\right]} %square bracket
\newcommand{\bracket}[1]{\left({#1}\right)} %bracket

\newcommand{\he}{\text{HE}}
\newcommand{\keygen}{\text{Keygen}}
\newcommand{\seckeygen}{\text{SecretKeygen}}
\newcommand{\pubkeygen}{\text{PublicKeygen}}
\newcommand{\evalkeygen}{\text{EvalKeygen}}
\newcommand{\enc}{\text{Enc}}
\newcommand{\dec}{\text{Dec}}
\newcommand{\eval}{\text{Eval}}
\newcommand{\setup}{\text{Setup}}
\newcommand{\add}{\text{Add}}
\newcommand{\mult}{\text{Mult}}

\newcommand{\pk}{\text{pk}}
\newcommand{\sk}{\text{sk}}
\newcommand{\evk}{\text{evk}}
\newcommand{\rlk}{\text{rlk}}

\newcommand{\reg}{\text{Regev}}
\newcommand{\bvv}{\text{BV}^*}
\newcommand{\bv}{\text{BV}}
\newcommand{\para}{\text{params}}
\newcommand{\bgv}{\text{BGV}}
\newcommand{\brak}{\text{B}}
\newcommand{\bfv}{\text{BFV}}

\newcommand{\powersoftwo}{\text{PowersOfTwo}}
\newcommand{\bitdecomp}{\text{BitDecomp}}
\newcommand{\switchkey}{\text{SwitchKey}}
\newcommand{\switchkeygen}{\text{SwitchKeyGen}}

\newcommand{\kl}[1]{{\color{orange}#1}}

\newcommand*\diff{\mathop{}\!\mathrm{d}}

\title{A Tutorial Introduction to Lattice-based Cryptography and Homomorphic Encryption}

\author{
  Yang Li \\
  School of Computing\\
  Australian National University\\
  Canberra, ACT, 2600 \\
  \texttt{kelvin.li@anu.edu.au} \\
   \And
 Kee Siong Ng \\
  School of Computing\\
  Australian National University\\
  Canberra, ACT, 2600 \\
  \texttt{keesiong.ng@anu.edu.au} \\
  \And
  Michael Purcell \\
  School of Computing \\
  Australian National University \\
  Canberra, ACT, 2600 \\
  \texttt{michael.purcell1@anu.edu.au}\\
}

\usepackage{subfiles} % Best loaded last in the preamble

\makeindex

\begin{document}
\maketitle

\tableofcontents

%\newpage
%\section{Course Design}
%\subfile{sections/course design}

\newpage
\section{Introduction}

\subsection{Motivations}
\paragraph{Why study Lattice-based Cryptography?}
There are a few ways to answer this question.
\begin{enumerate}\itemsep1mm\parskip0mm
 \item It is useful to have cryptosystems that are based on a variety of hard computational problems so the different cryptosystems are not all vulnerable in the same way. 
 \item The computational aspects of lattice-based cryptosystem are usually simple to understand and fairly easy to implement in practice.
 \item Lattice-based cryptosystems have lower encryption/decryption computational complexities compared to popular cryptosystems that are based on the integer factorisation or the discrete logarithm problems. 
 \item Lattice-based cryptosystems enjoy strong worst-case hardness security proofs based on approximate versions of known NP-hard lattice problems. % that are conjectured to be hard to approximate
% are (mostly) based on known NP-hard problems % like the Shortest Integer Problem, 
 \item Lattice-based cryptosystems are believed to be good candidates for post-quantum cryptography, since there are currently no known quantum algorithms for solving lattice problems that perform significantly better than the best-known classical (non-quantum) algorithms, unlike for integer factorisation and (elliptic curve) discrete logarithm problems.
 \item Last but not least, interesting structures in lattice problems have led to significant advances in Homomorphic Encryption, a new research area with wide-ranging applications.
\end{enumerate}

Let’s look at that fourth point in more detail.

Note first that the discrete logarithm and integer factorisation problem classes, which underlie several well-known cryptosystems, are only known to be in NP, they are not known to be NP-complete or NP-hard. The way we understand their complexity is by looking at the average run-time complexity of the current best-known (non-polynomial) algorithms for those two problem classes on randomly generated problem instances. Using that heuristic complexity measure, we can show that 
\begin{enumerate}\itemsep1mm\parskip0mm
    \item  there are special instances of those problems that can be solved in polynomial time but, in general, both problems can be solved only in sub-exponential time; and
    \item on average, most of the discrete logarithm and integer factorisation problem instances are as hard as each other.
\end{enumerate} 
So we believe these two problems to be average-case hard problem classes, but we cannot yet prove that. 
Interestingly, we know there are quantum algorithms that can solve these two problems efficiently \citep{bernstein09}.

The above then begs the question of whether we can design cryptosystems based on known NP-hard or worst-case hard problem classes. %The challenge, of course, is that most NP-hard problem classes are only hard in the worst case but are efficiently solvable on many if not most cases. 
In constructing a (public-key) cryptosystem using a problem class $\mathit{ACH}$ with average-case hardness like Integer Factorisation or Discrete Logarithm, it is sufficient to show that the generation of a key pair (at random) and the solution of the private key corresponds to a problem instance $I \in \mathit{ACH}$, and we rely on average hardness to say $I$ is hard to solve with good probability. But in constructing a (public-key) cryptosystem using a problem class $\mathit{WCH}$ with only known worst-case complexity, we need to do a bit more work, in that it is not sufficient to generate a key pair (at random) and show the solution of the private key is a problem instance $I \in \mathit{WCH}$, we need to actually show that $I$ is one of the hard or worst cases in $\mathit{WCH}$.

In other words, to build a cryptosystem based on a worst-case hard problem class, we do not just need to know that hard instances exist, but we need a way to explicitly generate the hard problem instances. And that is an issue because we do not know how to do that for most worst-case hard problem classes. But this is what makes lattice problems interesting: we know how to generate, through reductions, the worst-case problem instances of approximation versions of NP-hard lattice problems and build efficient cryptosystems based on them. In practice, this means breaking these cryptosystems, even with some small non-negligible probability, is provably as hard as solving the underlying lattice problem approximately to within a polynomial factor in polynomial time.

How hard are these approximation lattice problems?
In most cases, the underlying lattice problem is the Shortest Vector Problem (SVP), and the approximation version is called the GapSVP$_{\lambda}$ problem for an approximation factor $\lambda$.
These gap lattice problems are known to be NP-hard only for small approximation factors like $n^{O(1/\log^2 n)}$.
We also know that these gap lattice problems are not NP-hard for approximation factors above $\sqrt{n/\log n}$, unless the polynomial time hierarchy collapses. See \citet{micciancio-goldwasser02,khot05,khot10} for surveys of these results.
The best-known algorithm for solving these gap lattice problems to within poly(n) factor has time complexity $2^{O(n)}$ \citep{ajtaiKS01}, which leads us to the following conjecture that underlies the security of lattice-based cryptography:
\begin{quote}
    Conjecture: There is no polynomial time algorithm that approximates lattice problems to within polynomial factors.
\end{quote}

\paragraph{Why another paper on Lattice Cryptography and Homomorphic Encryption?}
It is important to state early that this tutorial is a compilation of known results in the literature and we do not claim any research originality.
% To make life easier for readers, we have tried to make it as self-contained as possible, with all relevant background given either in the body of the tutorial or in the appendix.
In contrast with some existing works \citet{peikert16decade, halevi2017homomorphic, chi15}, this tutorial 
\begin{enumerate}\itemsep1mm\parskip0mm
 \item is written primarily with pedagogical considerations in mind;
 \item is as self-contained as possible, with essentially all required background given either in the body of the tutorial or in the appendix;
 \item focuses mostly on the narrow development path from the Learning With Errors (LWE)\index{LWE} problem to Ring LWE and homomorphic encryption schemes built on top of them; we do not cover other lattice cryptographic systems like NTRU, Ring SIS-based systems, and homomorphic signatures. %;
 % \item provides practical programming examples to help illustrate key concepts throughout the paper. 
\end{enumerate}
The target audiences are students, practitioners and researchers who want to learn the ``core curriculum" of lattice-based cryptography and homomorphic encryption from a single source. %, but not the research frontiers.

In writing the tutorial, we have benefited from peer-reviewed published papers as well as many less-formal explanatory material in the form of lecture notes and blog articles.
We are not always careful and comprehensive in citing the latter class of material, and we apologise in advance for errors of omission.

\subsection{Tutorial organisation}

The tutorial can be divided into three parts in pedagogical order as follows. Each part will be presented with definitions, examples, discussions around the intuitions of abstract concepts and more importantly corresponding computer code to help develop the understanding.

After brief introductions to the basics of Computational Complexity Theory in Section~\ref{sec:computational complexity} and Cryptography in Section~\ref{sec:crypto}, the first part of the tutorial focuses on the LWE problem, a foundational hard lattice problem. This part begins with some Lattice Theory in Section~\ref{section:lattice theory}, followed by material on Discrete Gaussian Distributions in Section~\ref{sec:discrete gaussian}.
The LWE problem is then described in some detail in Section~\ref{section:lwe}, including hardness proofs.

% , both are essential in later lattice-based cryptosystems.

The second part discusses the Ring LWE (RLWE) problem, which is a generalization of LWE from the integer domain to an algebraic number field domain that allows more computationally efficient cryptosystems to be built. %in order to improve reduce LWE's public key size from roughly quadratic to linear. 
As LWE does not straightforwardly generalize to its ring version, some required background knowledge will be presented with intuition, examples and computer code, including cyclotomic polynomials and their Galois groups in \Cref{sec:cyclotomic} and algebraic number theory in Section~\ref{sec:ant short}.
(For readers that require a more extensive background, the appendix covers Abstract Algebra, Galois Theory and Algebraic Number Theory in significantly more details.)
The RLWE problem is described in some detail in Section~\ref{section:rlwe}, including hardness proofs.
(A mindmap is given in \Cref{sec:mind maps} to help readers navigate and remember the many components of RLWE proofs.)

Having introduced the LWE and RLWE problems, the final part of the tutorial (Section~\ref{sec:he}) shows how efficient homomorphic encryption (HE) schemes can be developed based on the LWE and RLWE problems.
% aims to develop an intuition of how to design efficient HE schemes based on these hard problems. In particular, the course focuses on a series of works that are considered as the second generation of HE developments. 
These schemes are both similar and different to Gentry's  original fully HE scheme. The similarity is in designing a somewhat HE scheme first, then using bootstrapping to achieve fully HE. The difference is that they avoided using Gentry's ``squashing'' technique, but used the algebraic properties of (R)LWE instances to make the somewhat HE schemes bootstrappable.

% \subsection{Background}

\subsection{A simple lattice-based encryption scheme}
\label{subsec:regev scheme}

Before diving into the technical details of lattice-based cryptosystems and homomorphic encryption schemes, we describe a simple public-key encryption scheme introduced by  \citet{regev2009lattices} to illustrate the connection between the scheme's security and lattice problems. This scheme is based on the learning with errors (LWE) problem, see Section \ref{section:lwe} for details. Its simplicity inspired subsequent developments in homomorphic encryption schemes that are based on lattices, and is a fundamental building block in many such schemes. 

Note that in this example $\Z_q$ is the collection of integers in the range $[-q/2, q/2)$ rather than its standard usage for representing the ring $\Z/\Z_q$, and $[x]_q$ is the reduction of $x$ into $\Z_q$ such that $[x]_q=x \bmod q$.
We use boldface to denote vectors and matrices. When working with matrices, all vectors are by default considered as column vectors. Vector multiplications are denoted by $\vc{a} \cdot \vc{b}$, whilst matrix and scalar multiplications are denoted without the ``dot'' in the middle. For simplicity, we use $[\vc{b} \mid -\vc{A}]$ to denote the action of appending the column vector $\vc{b}$ to the front of the matrix $-\vc{A}$.
The parameters  $n,q,N,\chi$ correspond to the vector dimension, the plaintext modulus, the number of LWE samples, and the noise distribution over $\Z_q$, respectively. In particular, $\chi$ is chosen such that $\text{Pr}(|\vc{e} \cdot \vc{r}| < \floor{\frac{q}{2}}/2) > 1 - \text{negl}(n)$ for a random binary vector $\vc{r} =\{0,1\}^N$.
The scheme is summarized as follows, but in an alternative format to be consistent with later homomorphic encryption schemes that will be presented in \Cref{sec:he}. 
% Note that all additions in this example are modular additions within the domain $\Z_q = [-q/2,q/2)$. \textcolor{red}{This isn't quite true. We bounce back and forth between values mod $q$, values mod $2$, and integers. It isn't really a problem since we have obvious embedding maps between the various spaces, but we should probably try to be precise about what is really going on.} \kl{You are right, Regev used $Z_q=[0,\dots,q-1]$, Brakerski's schemes used the symmetric range $[-q/2,q/2)$. I saw it in somewhere it's for computational purpose.}
% The scheme is summarized as follows. 

\begin{tcolorbox}
\noindent
\textbf{Private key}: Sample a private key $\vc{s} = (1,\vc{t})$, where $\vc{t} \leftarrow \Z_q^n$.\\

\textbf{Public key}: Sample a random matrix $\vc{A} = \left[\begin{array}{@{}c@{}}
    \vc{a}_1 \\
    \vdots \\
    \vc{a}_N 
    \end{array} \right] \leftarrow \Z_q^{N \times n}$ and compute $\vc{b} = \vc{A}  \vc{t} + \vc{e}$ for a random noise vector $\vc{e} \leftarrow \chi^N$. Output the public key $\vc{P}=\sqbracket{\vc{b} \mid -\vc{A}} \in \Z_q^{N \times (n+1)}$.\\

\textbf{Encryption:} Encrypt the message $m \in \{0,1\}$ by computing 
\begin{align*}
    \vc{c}=\sqbracket{\vc{P}^T  \vc{r} + \floor{\frac{q}{2}}  \vc{m}}_q \in \Z_q^{n+1},
\end{align*}
where $\vc{m}=(m, 0, \dots, 0)$ has length $n+1$.\\

\textbf{Decryption:} Decrypt the ciphertext $\vc{c}$ using the secret key by computing 
\begin{align*}
    m = \sqbracket{\round{\frac{2}{q}\sqbracket{\vc{c} \cdot \vc{s}}_q}}_2.
\end{align*}
\end{tcolorbox}

The purpose of the binary vector $\vc{r}$ is to randomize the use of the public key so that it is impossible to derive $\vc{m}$ from the ciphertext $\vc{c}$. To demonstrate how decryption works, the ciphertext can be re-written as 
\begin{align*}
    \vc{c}=\sqbracket{\vc{b}^T  \vc{r} + \floor{\frac{q}{2}}  m \mid -\vc{A}^T  \vc{r}}_q,
\end{align*}
%it produces   
%\begin{align*}
 %   m = \sqbracket{\round{\frac{2}{q}\sqbracket{\vc{c} \cdot \vc{s}}_q}}_2. %\leftarrow \dec(\vc{s},\vc{c},q).
%\end{align*}
which implies 
%\noindent{}Notice that
\begin{align*}
    \sqbracket{\vc{c} \cdot \vc{s}}_q
    &= \sqbracket{\vc{b}^T  \vc{r} + \floor{\frac{q}{2}} m - \vc{t}^T \vc{A}^T  \vc{r}}_q
    = \sqbracket{(\vc{t}^T  \vc{A}^T + \vc{e}^T) \vc{r} + \floor{\frac{q}{2}} m - \vc{t}^T  \vc{A}^T  \vc{r}}_q\\
    &=\sqbracket{\vc{e}^T \vc{r} + \floor{\frac{q}{2}} m}_q.
    % &= \sqbracket{\vc{P}^T \cdot \vc{r} + \floor{\frac{q}{2}} \cdot \vc{m}}_q \cdot \vc{s}\\
    % &= \sqbracket{\begin{bmatrix}\vc{b}^T \\ -\vc{A}\end{bmatrix} \cdot \vc{r} + \floor{\frac{q}{2}} \cdot \vc{m}}_q \cdot \vc{s}\\
    %  &= \sqbracket{\begin{bmatrix}\vc{t}^T\vc{A} + \vc{e}^T \\ -\vc{A}\end{bmatrix} \cdot \vc{r} + \floor{\frac{q}{2}} \cdot \vc{m}}_q \cdot \vc{s} \\
     %&= \sqbracket{\begin{bmatrix} \vc{A}^T \cdot \vc{t} \cdot \vc{r} + \vc{e} \cdot \vc{r} \\ -\vc{A}\cdot \vc{r} \end{bmatrix} + \floor{\frac{q}{2}} \cdot \vc{m}}_q \cdot \vc{s},
\end{align*}
%which implies 
%\begin{align*}
%     \sqbracket{\vc{c} \cdot \vc{s}}_q =
    %  \sqbracket{\begin{bmatrix} \vc{t}^T\vc{A}\vc{r} + \vc{e} \cdot \vc{r} \\ -\vc{A}\cdot \vc{r} \end{bmatrix} \cdot \vc{s} + \floor{\frac{q}{2}} \cdot \vc{m} \cdot \vc{s}}_q \\
    %  &= \sqbracket{\begin{bmatrix} \vc{t}^T\vc{A}\vc{r} + \vc{e} \cdot \vc{r} \\ -\vc{A}\cdot \vc{r} \end{bmatrix} \cdot \begin{bmatrix} 1 \\ \vc{t}\end{bmatrix} + \floor{\frac{q}{2}} \cdot \vc{m} \cdot \begin{bmatrix} 1 \\ \vc{t}\end{bmatrix}}_q \\
%     \sqbracket{\vc{e} \cdot \vc{r} + \floor{\frac{q}{2}} \cdot m}_q.
%\end{align*}

\noindent{}Because $\text{Pr}(|\vc{e}^T \vc{r}| < \floor{\frac{q}{2}}/2) > 1 - \text{negl}(n)$, we have (with overwhelming probability)
% \begin{equation*}
% \sqbracket{\vc{c} \cdot \vc{s}}_q \in
% \begin{cases}
% (-q/4,q/4) & \text{if $m=0$;} \\
% [-q/2, -q/4) \cup (q/4,q/2) & \text{if $m = 1$}.
% \end{cases}
% \end{equation*}
% Similarly, we have
\begin{equation*}
\frac{2}{q}\sqbracket{\vc{c} \cdot \vc{s}}_q \in
\begin{cases}
(-1/2,1/2) & \text{if $m=0$;} \\
[-1, -1/2) \cup (1/2,1) & \text{if $m = 1$}.
\end{cases}
\end{equation*}
%and
%\begin{equation*}
%    \text{Dec}(\vc{s}, \text{Enc}(\vc{P}, m, n, q, N), q) = \text{Dec}(\vc{s}, \vc{c}, q) = \sqbracket{\left\lfloor\frac{2}{q}\sqbracket{\vc{c} \cdot \vc{s}}_q\right\rceil}_2 = m.
%\end{equation*}

Notice that if $\vc{b}^{\prime} = \vc{A} \vc{t}$ then an attacker who knows $\vc{A}$ and $\vc{b}^{\prime}$ could recover the secret $\vc{t}$ by solving a system of linear equations. The security of the system therefore depends on the presence of the noise vector $\vc{e}$.

If an attacker knows $\vc{b}$ instead of $\vc{b}^{\prime}$, then the attack described above will not work. If, however, such an attacker could recover the noise vector $\vc{e}$, then they could use that information to compute $\vc{b}^{\prime}$. They could then recover $\vc{t}$ as described above. Recovering $\vc{e}$ is an instance of a well-known lattice problem called the bounded distance decoding (BDD) problem. So, an attacker that can solve the BDD problem could recover the secret $\vc{t}$. In other words, recovering $\vc{t}$ is ``no harder'' than solving the BDD problem.

Conversely, \citeauthor{regev2009lattices} showed that the BDD problem is ``no harder'' than recovering $\vc{t}$. That is, an attacker who could recover $\vc{t}$ given $\vc{A}$ and $\vc{b}$ could solve the BDD problem as well. This result implies that if the BDD problem is hard, then attacking the cryptosystem is hard as well. This kind of result is called a \emph{reduction}.
Crucially, the BDD problem is believed to be hard. So, Regev's result constitutes a proof of security for the LWE-based cryptosystem described above.

\begin{figure}
\centering
\caption{A Sage implementation of the simple lattice-based encryption system described above.\\ \textbf{Note:} This implementation is not suitable for use in real-world applications.}
\begin{tcolorbox}
\begin{verbatim}
#!/usr/bin/env sage

from sage.misc.prandom import randrange
import sage.stats.distributions.discrete_gaussian_integer as dgi

# Define parameters
def sample_noise(N, R):
    D = dgi.DiscreteGaussianDistributionIntegerSampler(sigma=1.0)
    return vector([R(D()) for i in range(N)])

q = 655360001
n = 1000
N = 500

R = Integers(q)
Q = Rationals()
Z2 = Integers(2)

# Generate keys
t = vector([R.random_element() for i in range(n)])
secret_key = vector([R(1)] + t.list())

A = matrix(R, [[R.random_element() for i in range(N)]
                                   for i in range(n)])
e = sample_noise(N, R)
b = A.T * t + e

public_key = block_matrix([matrix(b).T, -A.T],
                          ncols=2)

# Encrypt Message
message = R(randrange(2))
m_vec = vector([message] + [R(0) for i in range(n)])
r = vector(R, [randrange(2) for i in range(N)])

ciphertext = public_key.transpose() * r + (q//2) * m_vec

# Decrypt Message
temp = (2/q) * Q(ciphertext*secret_key)
decrypted_message = R(Z2(temp.round()))

# Verification
print(decrypted_message == message)
\end{verbatim}
\end{tcolorbox}
\end{figure}

\newpage
\section{Computational Complexity Theory}

\label{sec:computational complexity}

Computational complexity theory is the foundation of computational security of modern cryptography by allowing one to emphasize the security of a cryptosystem by drawing an efficient reduction from a computationally hard problem (that either has been proved or is believed with high confidence to be unsolvable in a reasonable time, e.g., polynomial time). That being said, a cryptosystem that is provably secure is still vulnerable to real-world attacks, depending on what threat model was considered, how close to reality the underlying security definitions and assumptions are and so on. 

In this section,
%\footnote{This section is part of the work \textit{A Tutorial Introduction to Lattice-based Cryptography and Homomorphic Encryption} by the authors Yang Li, Kee Siong Ng, Michael Purcell from the School of Computing, Australian National University @2022.}
we start by introducing some basic definitions in computational complexity theory, then go on to talk about inapproximability, which are variants of the standard decision and optimization problems and commonly used to prove the computational security of cryptosystems. We then introduce gap problems, which are generalization of decision problems and proving their hardness is a useful technique of proving inapproximability. We finish the chapter by briefly introducing \citet{ajtai1996generating}'s worst-case to average-case reduction. Ajtai's work is considered as the first published average-case problem whose hardness is based on the worst-case hardness of some well-known lattice problems.

\subsection{Basic time complexity classes}

The following concepts are introduced under the assumption that a general purpose computer is of the form of a \textit{Turing machine}. The primary reference of this subsection is \citet{sipser2013introduction}'s book \textit{Introduction to the Theory of Computation, Third Edition}. 

A \textit{language (or decision problem)} 
\reversemarginpar
\marginnote{\textit{Decision problem}}
is a set of strings that are decidable by a Turing machine. We use $\Sigma$ to denote the alphabet and $\Sigma^*$ to denote the set of all strings over the alphabet $\Sigma$ of all lengths. A special case is when $\Sigma=\{0,1\}$ and $\Sigma^*=\{0,1\}^*$ is the set of all strings of $0$s and $1$s of all lengths. In this case, a language $A=\{x \in \{0,1\}^* \mid f(x)=1\}$, where $f:\{0,1\}^* \rightarrow \{0,1\}$ is a \textit{Boolean function}.

Let $M$ be a deterministic Turing machine that halts on all inputs. We measure the 
\reversemarginpar
\marginnote{\textit{Time complexity}}
\textbf{time complexity} or \textbf{running time} of $M$ by the function $t:\N \rightarrow \N$, where $t(n)$ is the maximum number of steps that $M$ takes on any input of length $n$. Generally speaking, $t(n)$ can be any function of $n$ and the exact number of steps may be difficult to calculate, so we often just analyse $t(n)$'s \textbf{asymptotic behaviour} by taking its leading term, denoted by $O(t(n))$. We also relax its codomain by letting $t:\N \rightarrow \R^+$ be a non-negative real valued function. 

It is worth mentioning that when analysing the time complexity of a function, we often consider its time complexity in the worst case, i.e., the longest running time of all inputs of a particular length $n$. At the end of this chapter, we will emphasize the importance of the worst-case complexity in the proof of security of modern cryptosystems. We will give a clue of how this was achieved by \citeauthor{ajtai1996generating} through an average-case to worst-case reduction.

\begin{definition}
The \textbf{time complexity class},
\reversemarginpar
\marginnote{\textit{Time complexity class}}
$\mathbf{TIME(t(n))}$, is defined as the set of all languages that are decidable by a Turning machine in time $O(t(n))$.
\end{definition}

Obviously, $t$ can be any function, e.g., logarithm, polynomial, exponential, etc. In practice, polynomial differences in running time are considered to be much better than exponential differences due to the super fast growth rate of the latter. For this reason, we separate languages into different classes according to their worst case running time on a deterministic single-tape Turing machine. 

\begin{definition}
\textbf{P} \index{time complexity class!P}
\reversemarginpar
\marginnote{\textit{P}}
is the class of languages that are decidable in polynomial time by a deterministic single-tape Turing machine, i.e., % for a constant $k$, 
\begin{equation*}
    P = \bigcup_{k \in \N} \it{TIME}(n^k).
\end{equation*}
\end{definition}

Some problems are computationally hard, so cannot be decided by a deterministic single-tape Turing machine in polynomial time. But given a possible solution, sometimes we can efficiently \textit{verify} whether or not the solution is genuine. The length of the solution has to be polynomial in the length of the input string length, for otherwise the verification process cannot be done efficiently. Based on the ability to efficiently verify, we can define the complexity class NP. 

\begin{definition}
\textbf{NP} \index{time complexity class!NP}
\reversemarginpar
\marginnote{\textit{NP}}
is the class of languages that can be verified in polynomial time.
\end{definition}

\iffalse % KS: we don't seem to need this definition in the rest of the paper
An alternative definition of NP is based on a \textbf{non-deterministic Turing machine}.
\reversemarginpar
\marginnote{\textit{NDTM}}
We can define the time complexity class $NTIME(t(n))$ for a non-deterministic Turing machine as the set of all languages that are decidable by a non-deterministic Turing machine. 

\begin{definition}
\textbf{NP} is the class of languages that are decidable in polynomial time by a non-deterministic Turing machine, i.e., % for a constant $k$
\begin{equation*}
    NP=\bigcup_{k \in \N} \it{NTIME}(n^k).
\end{equation*}
\end{definition}
\fi

Sometimes, a problem can be solved by reducing it to another problem, whose solution can be found relatively easier, provided the reduction between the two problems is efficient. For example, a polynomial time reduction is often acceptable.   
%The efficient reduction can be formally defined as the next. 

\begin{definition}
A language $A$ is \textbf{polynomial time reducible} 
\reversemarginpar
\marginnote{\textit{PT reduction}}
to another language $B$, written as $A \le_P B$, if a polynomial time computable function $f:\Sigma^* \rightarrow \Sigma^*$ exists, where for every $w$, 
\begin{equation*}
    w \in A \iff f(w) \in B.
\end{equation*}
\end{definition}

A polynomial time reduction $A \le_P B$ implies $A$ is no harder than $B$, so if $B \in \text{P}$ then $A \in \text{P}$. Based on this reduction, we can define another complexity class. 

\begin{definition}
A language $B$ is \textbf{NP-complete} \index{time complexity class!NP-complete}
\reversemarginpar
\marginnote{\textit{NP-complete}}
if it is in NP and every problem in NP is polynomial time reducible to $B$. 
\end{definition}
Essentially, we are saying that NP-complete is the set of the hardest problems in NP. There are, however, hard problems that are not in NP such as an \textbf{optimization problem}. Given a solution of an optimization problem, it is often not trivial to verify the solution is optimal among all the answers, so this type of problems are not polynomial time verifiable and hence not in NP. For these problems, we can define a similar complexity class as NP-complete but without requiring their solutions to be polynomially checkable. 

\begin{definition}
A language is \textbf{NP-hard} \index{time complexity class!NP-hard}
\reversemarginpar
\marginnote{\textit{NP-hard}}
if every problem in NP is polynomial time reducible to it. 
\end{definition}
 
The two terms NP-complete and NP-hard are sometimes used  interchangeably because an optimization problem can also be formed as a decision problem. For example, instead of asking for the shortest route from the \textit{travelling salesman problem}, we can ask whether there exists a route that is shorter than a threshold. 

Many optimization problems are NP-hard, which means there is no polynomial time solution under the assumption $\text{P}\neq \text{NP}$. Hence, when an answer for an NP-hard problem is needed, the fallback is to use an approximation algorithm to compute a near-optimal solution that is within an acceptable range. 
% However, giving the best possible answers is not in the most interest of cryptographers. Instead, 
For a NP-hard problem, it is sometimes easier to build a cryptosystem based on its approximated version rather than the NP-hard problem itself. For this reason, cryptographers are concerned about whether or not an optimization problem is hard to be approximated within a certain range. % for an approximated answer.
This brings us to the study of the hardness of approximation or inapproximability in the next subsection. 

\subsection{Hardness of approximation}
\label{subsection:gapProb}

An optimization problem aims at finding the optimum result of a computational problem. This optimum result can either be the maximum or minimum of some value. Throughout this section, we focus on minimization problems only. The same results also hold for maximization problems. \footnote{Lecture 18: \textit{Gap Inapproximability}, 6.892 \textit{Algorithmic Lower Bounds: Fun with Hardness Proofs} (Spring 2019), Erik Demaine, available at
 \url{http://courses.csail.mit.edu/6.892/spring19/lectures/L18.html}} 
In the previous section, we said an optimization problem can be made into a decision problem by comparing the solution with a threshold. More formally, it is defined as the next. 

\begin{definition}
An \textbf{NP-optimization} (NPO) problem
\reversemarginpar
\marginnote{\textit{NPO}}
is an optimization problem such that 
\begin{itemize}
    \item all instances and solutions can be recognized in polynomial time,
    \item all solutions have polynomial length in the length of the instance,
    \item all solution's costs can be computed in polynomial time.
\end{itemize}
\end{definition}
% NPO is the class of all NP-optimization problems. 
For a minimization problem in NPO, its decision version asks ``Is $OPT(x) \le q$?'', where $OPT(x)$ is the unknown optimal solution (or its cost, we use interchangeably) to the instance $x$. For example, in the \textit{maximum clique} problem, an instance is a graph, an optimal solution is the maximum clique in the given graph and its cost is the clique size. Given an NPO problem, its decision version is an NP problem, so NPO is an analogy of NP but for optimization problems. On the other hand, PO (P-optimization) problem is the set of optimization problems whose decision versions are in P, such as finding the shortest path. 

\begin{definition}
An algorithm $ALG$ for a minimization problem is called \textbf{$c$-approximation algorithm} \index{$c$-approximation} for $c \ge 1$
\reversemarginpar
\marginnote{\textit{$c$-approx}}
if for all instances $x$, it satisfies 
\begin{equation}
    \frac{cost(ALG(x))}{cost(OPT(x))} \le c.
\end{equation}
\end{definition}
The ratio $c$ is not necessarily a constant, it can be any function of the input size, i.e., $c=f(n)$ for an arbitrary function $f(\cdot)$. Practically, we prefer a near optimal solution $ALG(x)$ such that the ratio $c$ is as small as possible or at least does not grow quickly in the input size. This, however, may not be possible for some problems such as the maximum clique problem, whose best possible ratio is $O(n^{1-\epsilon})$ for small $\epsilon > 0$. 
From a provable security's perspective, the smaller the ratio $c$ is, the harder the c-approximation problem is. This leads to a cryptosystem with higher security because it requires more time and computational resources for an attacker to break the system. 

For a given $c=f(n)$, there are different ways of proving $c$-approximating a problem is hard. One way is by proving a c-gap problem is hard, which is in direct analogy to the c-approximation problem in hand. This way, if the gap problem is hard, then the $c$-approximation problem is also hard.  

\begin{definition}
For a minimization problem, a \textbf{$c$-gap problem} \index{$c$-gap problem}
\reversemarginpar
\marginnote{\textit{$c$-gap}}
(where $c > 1$) distinguishes two cases for the optimal solution $OPT(x)$ of an instance $x$ and a given $k$ as follows:
\begin{itemize}
    \item $x$ is an YES instance if $OPT(x) \le k$,
    \item $x$ is an NO instance if $OPT(x) > c\cdot k$.
\end{itemize}
\end{definition}

The value $k$ is a given input. For example, in the $c$-gap version of the shortest vector problem, we can set $k=\lambda_1(L)$ to be the shortest vector in a given lattice $L$. Intuitively, a $c$-gap problem is a decision problem where the unknown optimal solution $OPT$ of the corresponding optimization problem is mapped to the opposite side of a gap. It is, however, different from a decision problem in the sense that there is a gap between $k$ and $c\cdot k$.

\iffalse % KS: Promise Problems don't seem to be needed in the rest of the paper
To be more precise, a $c$-gap problem is a promise problem. A \textbf{promise problem}
is a generalization of a decision problem, where only a subset of the entire input space is considered. Alternatively, we can think of the algorithm is only defined for a subset of the entire input space. For example, instead of taking the set of all graphs as inputs, a promise problem may only consider connected graphs. Hence, it is the union of two disjoint subsets, one contains all the YES instances and the other contains all the NO instances. These two classes of instances make the entire set of connected graphs. But if a given graph is not connected, the algorithm for solving the promise problem can return any output or may not even halt. It is named ``promise'' because all inputs are promised to be on either side of the gap, nothing falls in between.\footnote{See Erik Demaine's MIT lecture notes for more details. \url{http://courses.csail.mit.edu/6.892/spring19/lectures/L18.html}}
\fi

The connection between c-gap and c-approximation problems is that if a c-gap problem is proved to be hard, then the corresponding c-approximation problem is also hard. 
\reversemarginpar
\marginnote{\textit{$c$-gap implies inapprox}}
In other words, there is a reduction from a c-gap problem to a c-approximation problem. The proof is straightforward. 
Assuming the problem can be c-approximated in polynomial time by an algorithm $A$, so for an input $x$ we have $OPT(x) \le A(x) \le c\cdot OPT(x)$.
If $x$ is a YES instance of the gap problem, then 
\begin{equation*}
    OPT(x) \le k \implies A(x) \le c \cdot OPT(x) \le c\cdot k.
\end{equation*}
If $x$ is a NO instance, then 
\begin{equation*}
    OPT(x) > c \cdot k \implies A(x) > c \cdot k.
\end{equation*}
Either way the instance $x$ can be distinguished easily using the decision procedure $A(x) \leq c\cdot k$. %, so it solves the $c$-gap problem. 

Gap and approximation lattice problems are the foundation of provable security for latticed-based cryptosystems. We will see more of these problems in \Cref{section:lattice theory} and some of their cryptographical applications in the hardness proofs of the short integer solution problem, learning with error problem and ring learning with error problem. 

\iffalse
\subsection{One way function}

We now introduce one-way functions which will be mentioned in the next section for worst-case to average-case reduction. 

One-way functions are essential for the security of cryptographic primitives. Intuitively, a one-way function is a function $f: \{0,1\}^n \rightarrow \{0,1\}^n$ that can be computed in polynomial time, but hard to be inverted.

In particular, if $f$ is a one-way function and injective (or one-to-one), then $f$ is a \textbf{one-way permutation}.

Formal definition will be filled in, see Chapter 7.1 in \citep{katz2014introduction} or Chapter 10.6 in \citep{sipser1996introduction}, may need to introduce the concept of probabilistic polynomial-time algorithm, see Chapter 10.2 in \citep{sipser1996introduction}.

The existence of one-way functions is an open problem. That is, no one can prove the existence of one-way functions unconditionally. 

Some examples of one-way functions are prime factorization (complexity class is known), subset sum problem (known to be NP-complete), discrete logarithm problem and permutations. 
\fi

\subsection{Average-case hardness}
\label{subsec:averagecasehard}

So far, we have introduced the time complexity classes P and NP in the worst case scenario. That is, the longest running time over all inputs at a given input length. A problem that is hard to be solved in polynomial time in the worst case is known as worst-case hard. There is another related concept called average-case hardness, \index{average-case hardness} which is stronger than worst-case hardness, in the sense that the former implies the latter but not vice versa. To finish section, we briefly discuss the critical role of average-case hard problems for cryptography and how they can be constructed by a worst-case to average-case reduction\index{worst-case to average-case} that was achieved by \citet{ajtai1996generating}. 

%Average-case hardness is a stronger assumption than worst-case hardness, as if a problem is average-case hard then it must be worst-case hard, but not vice versa. For that matter, it is preferable to build a cryptosystem, whose security is based on a weaker assumption (i.e., worst-case hardness) rather than a stronger assumption (i.e., average-case hardness). Intuitively, this also makes sense. Because if an attacker is able to solve a decent number of instances of an average-case hard problem, he may claim that the problem is not hard and hence any cryptosystem that is build upon it is not safe. On the other hand, the attacker cannot make such a claim against a worst-case hard problem unless he has solved every instance of the problem. 

Without going into the details, we state some remarks of average-case problems to help the reader to get an intuitive understanding of these problems. More discussions of these problems can be found in Chapter 18 of \citet{arora2009computational}. First, an average-case problem consists of a decision problem and a probability distribution, from which inputs can be sampled in polynomial time. Such a problem is called a \textbf{distributional problem}. This is different from a worst-case decision problem, where all inputs are considered when determining its hardness. Second, the first remark entails that average-case complexity is defined with respect to a specific distribution over the inputs. This suggests that a problem may be difficult with one distribution but easy with another distribution. For example, integer factorization may be difficult for large prime numbers, but easy for small integers. Hence, which probability distribution is used is crucial for the hardness of the integer factorization problem. Finally, average-case complexity has its own complexity classes \textbf{distP} and \textbf{distNP}, which are the average-case analogs of P and NP, respectively. 

%One can prove a problem is NP-complete by producing a polynomial time reduction from an existing NP-complete problem to the problem at hand. This implies that there must be some instances or inputs of the problem that are hard to be solved efficiently. 
%Next, we discuss why an average-case problem is a better candidate for the provable security of modern cryptosystems. 
To prove a cryptosystem is computationally secure, one could build an efficient reduction from a known worst-case problem to it, so that if the cryptosystem can be attacked successfully, such an attack model provides a solution to the worst-case problem. However, knowing alone the underlying problem is worst-case hard is not sufficient to build a secure cryptosystem in real-world, because many of the system's instances may correspond to easy instances of the worst-case problem, which can be solved efficiently. 

\begin{figure}[h!]
    \centering
    \includegraphics[page=1]{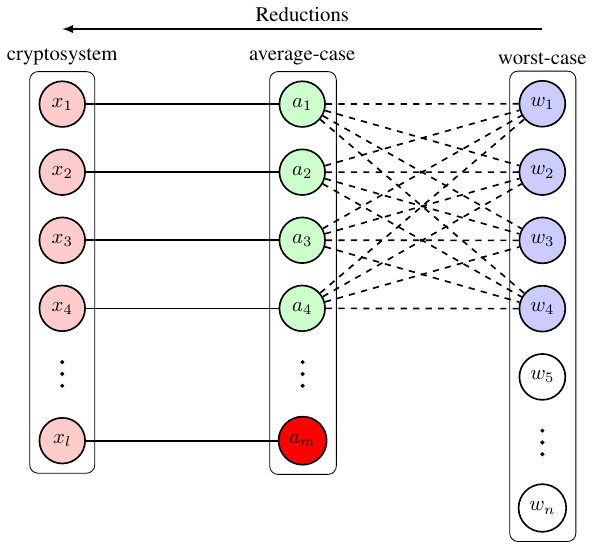}    
    \caption{A demonstration of a cryptosystem's computational security is based on an average-case problem. Each cryptographic instance $x_i$ corresponds to a random average-case instance $a_j$. Almost all random instances in the average-case problem can be mapped with the hard instances in a worst-case problem. There may be a fraction of average-case instances (colored in red) that can be solved easily, so their solutions entail solutions of the worst-case problem. But the fraction of such instances is negligible. The hard and easy instances in the worst-case problem are colored blue and white, respectively. The dashed lines indicate the worst-case to average-case reduction is random.}
    \label{fig:average2worst}
\end{figure}

For this reason, an ideal situation is when a cryptosystem's security is based on an average-case problem and the exact distribution to sample hard instances is known. But this is hard to achieve. It is more difficult to prove that a certain distribution generates only hard instances, because this would imply the problem is also worst-case hard. An alternative is to construct an average-case problem, such that its instances correspond to the hard instances in a worst-case problem. This is known as the worst-case to average-case reduction. A visual representation of this type of constructions is illustrated in \Cref{fig:average2worst}. In this figure, a random cryptographic instance corresponds to an average-case instance. By construction, it is almost always true that an average-case instance links to a hard instance of some worst-case problem. This reduction implies that if the worst-case problem is known or believed (in high confident) to be hard, then the cryptosystem is guaranteed to be secure with high probability.  
% In other words, to solve an average-case instance, the attacker needs to be able to solve all instances of the worst-case problem, which is essentially solving the worst-case problem that is either proved or believed to be hard with high confidence. 

The work by \citeauthor{ajtai1996generating} served exactly this purpose by introducing the \textit{short integer solution} (SIS) problem and proving that SIS is an average-case problem with polynomial time reductions from three worst-case lattice problems to it. This work is knowable the first worst-case to average-case reduction. The significant implication of Ajtai's work in cryptography is the fact that it laid the foundation for the security of modern cryptosystems to be based on worst-case problems (via average-case problems). 
More importantly, this work sparked a number of important following up works including the learning with error and ring learning with error problems that advanced lattice-based cryptography to a new era.

%\newpage
%\bibliography{references}
%\bibliographystyle{abbrvnat}

\newpage
\section{Cryptography Basics}

\label{sec:crypto}

The history of cryptography dates back to the pre-computer era, but with the same goal as today's, that is, securely sharing secret information between parties on public communication channels. A simple but motivating example is shown next, which is a \textit{shift cipher} encryption technique used by Julius Caesar (during 81-45BC) to securely communicate with his troops on battlefields \citep{hoffstein2008introduction}.
\begin{align*}
	\frac{\text{j\;s\;j\;r\;d\;k\;f\;q\;q\;n\;s\;l\;g\;f\;h\;p\;g\;w\;j\;f\;p\;y\;m\;w\;t\;z\;l\;m\;n\;r\;r\;n\;s\;j\;s\;y\;q\;z\;h\;n\;z\;x}}{\text{e\;n\;e\;m\;y\;f\;a\;l\;l\;i\;n\;g\;b\;a\;c\;k\;b\;r\;e\;a\;k\;t\;h\;r\;o\;u\;g\;h\;i\;m\;m\;i\;n\;e\;n\;t\;l\;u\;c\;i\;u\;s}}
\end{align*}  	
As the name of the technique suggests, each letter in the plaintext (below the horizontal line) was shifted by a pre-determined number of places along a fixed direction in the alphabet. This transforms it into a ciphertext (above the horizontal line) that do not hold the original information any more.

\subsection{Computational security}
Back then, Caesar's method was still able to effectively protect his secret messages to the troops from eavesdroppers. But with the help of nowadays multi-core GHz processor-computers that handle billions of instructions per second, this encryption method will fail within seconds. The example motivates the need to design more complex ciphertexts that are hard to decrypt, where the hardness should both be measurable and tunable by some parameters in order to cope with the increasing computing resources of potential attackers.

With the help of mathematics and computer science, in particular probability theory and computational complexity theory, the safety of modern encryption methods can be captured by
\reversemarginpar
\marginnote{\textit{computational security}}
\textit{computational security} \index{computational security}, a security notion, which allows an attacker to succeed in guessing the secret message with a measurable chance and computational effort such as running time. A frequently used approach to realize this security notion is to parameterize the probability of success and algorithmic running time of an attack by an integer-valued security parameter. This was named ``asymptotic approach'' and discussed in more details in Chapter 3 of \citet{katz2014introduction}. Some of the following results are also taken from that chapter, but presented in different orders and notations to ensure consistency of this tutorial paper.  

Under the notion of computational security, one can draw the connection between an encryption scheme and a computational problem that has been proved (or believed with high confidence) to be hard to solve within a practical time. A famous example is that the security of the RSA encryption scheme relies on the large integer factorization problem, which is presumed (without an actual proof) hard to solve by an efficient non-quantum algorithm. The RSA problem is to solve the unknown $x$ in the equation $x^e = c \bmod N$.\footnote{Throughout the paper, we use $=$ instead of $\equiv$ to denote the \textit{congruent modulo} relation in order to be consistent with most others in the field. This is also noted in the Notation table in \Cref{sec:notation}.} The problem is easy when $N$ is prime, so it comes down to primality test of $N$.    

The 
\reversemarginpar
\marginnote{\textit{security parameter}}
\textit{security parameter} \index{security parameter} described above, sometimes denoted by $n$ (or $\lambda$ or $\kappa$), reflects the input size of the underlying hard computational problem. The larger the security parameter, the larger the input size, so the problem is more difficult to be solved in a practical time frame, which ensures the encryption scheme is less likely to be attacked with success. In the RSA scheme, the security parameter is the bit length $n$ of the modulus $N$. The larger $n$ is, the more difficult it is to prime factor $N$ to efficiently solve the RSA problem. By convention, the security parameter $n$ is often supplied to a scheme in the unary format $1^n$ by repeating the number 1 $n$ times.

\subsection{Private and public encryptions}
Now that we discussed the security parameter, we formally introduce two types of encryption schemes, that is, the private (or symmetric) and public (asymmetric) key encryption schemes. The two types are similar in the sense that they both consists of three sub-steps for key generation, encryption and decryption. The main difference is that private key encryption uses only one key for both encryption and decryption (hence the name symmetric), whilst public key encryption uses one key for each purpose.

\begin{definition}
	Define the following three polynomial time algorithms: 
	\begin{itemize}
		\item Key generation: A probabilistic algorithm that generates a key $k \leftarrow \keygen(1^n)$ for encryption and decryption, where $|k| > n$.   
		\item Encryption: A probabilistic algorithm that encrypts the plaintext $m \in \{0, 1\}^*$ to a ciphertext $c \leftarrow \enc(k, m)$ using the key.
		\item Decryption: A deterministic algorithm that decrypts the ciphertext with the key to get the plaintext  $m \leftarrow \dec(k,c)$.
	\end{itemize} 
	The collection $(\keygen, \enc, \dec)$ forms a \textbf{private key encryption scheme} \index{private key encryption scheme} if for all $n, k, m$, it satisfies $m \leftarrow \dec(k,\enc(k,m))$.
\end{definition}

\begin{definition}
	Define the following three polynomial time algorithms: 
	\begin{itemize}
		\item Key generation: A probabilistic algorithm that generates a pair of keys $(\pk,\sk) \leftarrow \keygen(1^n)$, where $\pk$ is the public key for encryption and $\sk$ is the secret key for decryption and both have sizes larger than $n$.  
		\item Encryption: A probabilistic algorithm that encrypts the plaintext $m \in \{0, 1\}^*$ to a ciphertext $c \leftarrow \enc(\pk, m)$ using the public key.
		\item Decryption: A deterministic algorithm that decrypts the ciphertext using the secret key to get $m \leftarrow \dec(\sk,c)$.
	\end{itemize} 
	The collection $(\keygen, \enc, \dec)$ forms a \textbf{public key encryption scheme} \index{public key encryption scheme} if for all $n, (\pk,\sk), m$, it satisfies $m \leftarrow \dec(\sk,\enc(\pk,m))$. 
\end{definition}

\subsection{Security definitions}
Generally speaking, public key encryption uses longer keys due to the fact that one key is public. This in return makes it slower than private key encryption. It is, however, more convenient when under private key encryption, no secure channel is available for sharing the key or the key needs to be changed constantly for different parties. Regardless, the requirement for the keys (in both private and public key encryptions) to be larger than $n$ is to ensure the keys are at least of certain sizes in order to indicate the lower bound of an encryption scheme.  

As $n$ directly reflects the security of an encryption scheme, it is convenient to parameterize an attacker's running time and probability of success by $n$. More specifically, the running time is defined as the time taken to attack the scheme by a randomized algorithm. For practical purpose, this is often preferred to be polynomial in $n$, denoted by $poly(n)$. From the designer's point of view, an encryption scheme is only considered secure if both the probability of success is significantly small and such a probability decreases as $n$ gets larger. A frequently used function that captures these two characteristics is called a \textit{negligible function}. 

\begin{definition}
	A function $\mu: \N \rightarrow \R$ is \textbf{negligible} \index{negligible}, if for every positive integer $c$, there exists an integer $N_c$ such that for all $n > N_c$, we have $|\mu(n)| < n ^{-c}$.
\end{definition}
An example is the negative exponential function $\mu(n)=2^{-n}$. For $c=6$, the threshold to satisfy the above condition is $N_c=30$. 

When a function is not defined explicitly, we use $negl(n)$ to indicate it is negligible. Another characteristic that makes negligible function a suitable candidate for measuring an attacker's probability of success is due to the fact that it is still negligible even after multiplied by a polynomial function of $n$, that is, $|poly(n)| \cdot negl(n)$ is also negligible (Proposition 3.6 \citep{katz2014introduction}). This assures that if an attacker has a negligible probability of success, his chance stays extremely small even if the same attack is repeated a polynomial number of times (in $n$). 

An example (Example 3.2 \citep{katz2014introduction}) to illustrate this negligible probability and the running time is when an adversary's probability of success is $2^{40} \cdot 2^{-n}$ by running an attacking algorithm for $n^3$ minutes. If the security parameter is set to $n=40$, the adversary only needs to run the attack for roughly $40^3 \approx 44$ days to break the system with a probability 1. But if the security parameter is set large $n=500$, the adversary's chance of breaking the system is $2^{-460}$ that is almost 0 even if the attack runs for 237 years. 

\begin{definition}
	An encryption scheme is \textbf{secure} \index{secure!} if any probabilistic polynomial time (PPT) \index{PPT} adversary has only a negligible probability of success to break the scheme. 
\end{definition}
Here, probabilistic refers to the attack being a randomized algorithm, which typically runs faster than deterministic algorithms.

So far, we have implicitly discussed the notion of security (or breaking an encryption scheme) without formally defining the meaning of it. The concrete security definition that is most relevant to this tutorial paper is semantic security.
Below we give a formal definition of it and an equivalent definition, called indistinguishability which is easier to work with in practice. Both definitions can be defined for either private or public key encryptions, with the difference being a public key is also given for the public key encryption case. 

At a high level, semantic security means given a ciphertext that encrypts one of two messages, a PPT adversary has no better chance than random guessing that the ciphertext is an encryption of one message or the other. 

\begin{definition}
	\label{def:semSec}
	\reversemarginpar
	\marginnote{\textit{Semantic security}}
	An (public or private key) encryption scheme $\Pi$ is \textbf{semantically secure} \index{secure! semantically} if for every PPT adversary $\calA$, there is another PPT adversary $\calA'$ such that their chances of guessing the plaintext $m$ are almost identical, regardless $\calA'$ is only given the length of $m$. That is, let $c \leftarrow \enc(k,m)$, then
	\begin{equation*}
	\left|Pr[\calA(1^n, c)=m] - Pr[\calA'(1^n, |m|)=m]\right| \le negl(n).
	\end{equation*}
\end{definition}

It is convenient to consider the attack model as a distinguisher (i.e., a PPT algorithm) that tries to exhibit the non-randomness from the ciphertexts in order to associate a ciphertext with a particular plaintext. If the adversary's chance of success is better than random, then the encryption scheme is vulnerable to attacks. The process of guessing the source of a given ciphertext can be formalized as an \textbf{adversarial indistinguishability experiment}  (Section 3.2.1 \citep{katz2014introduction}). Given a PPT adversary $\calA$ and a (public or private) encryption scheme $\Pi$, the experiment outputs $\text{IndisExp}_{\calA, \Pi}(n)=1$ for a successful guess of the source plaintext. 

\begin{definition}
	An (private or public key) encryption scheme $\Pi$ is 
	\reversemarginpar
	\marginnote{\textit{Indistinguishable}}
	\textbf{indistinguishable} \index{secure! indistinguishable} if it satisfies 
	\begin{equation*}
	Pr\left[\text{IndisExp}_{\calA, \Pi}(n)=1\right] \le \frac{1}{2} + negl(n)
	\end{equation*}
	for all PPT adversary $\calA$ and security parameter $n$.
\end{definition}

The following theorem states the equivalent relationship between semantic security and indistinguishability. The same equivalent relation can also be proved under the public key encryption setting.\footnote{See a proof in Lecture 9: \textit{Public Key Encryption of} the course CS 276 – \textit{Cryptography} (Oct 1, 2014) at UC Berkeley by the instructor Sanjam Garg.}

\begin{theorem}[Theorem 3.13 \citep{katz2014introduction}]
	A private key encryption scheme is indistinguishable in the presence of an eavesdropper if and only if it is semantically secure in the presence of an eavesdropper. 
\end{theorem}

Both semantic security and indistinguishability discussed above are in the presence of an eavesdropper, who passively receives/intercepts a plaintext and tries to guess the corresponding plaintext. In the case of public-key encryption, the adversary has access to the public key and the encryption method, so it is possible for the adversary to compare the intercepted ciphertext with a self-encrypted ciphertext, and use this piece of information to increase the probability of successfully guessing the plaintext. By assuming the adversary has an oracle access to the encryption scheme which allows repeated interactions, this attack model is valid for both public and private key encryptions (Section 3.4.2 \citep{katz2014introduction}). The security notion defined under such a \textit{chosen-plaintext attack} (CPA) \index{secure! CPA} model is called CPA security and is a stronger security definition than the previous one which is defined in the presence of an eavesdropper. Similarly, semantic security and indistinguishability can also be defined under chosen plaintext attack, and a similar equivalent relations can be established between semantic security under CPA and IND-CPA \index{secure! IND-CPA}. This stronger level of security is useful when introducing homomorphic encryption.

%mention level of security, brute force attack, 

%cpa, semantic secure, public and private key enc,

\newpage
\section{Lattice Theory}

\label{section:lattice theory}
%\section{Lattice theory (light)}

\subsection{Lattice basics}

%Start with a review of some linear algebra concepts. 

%\textbf{Keywords: vector space, basis, span, linearly independent, matrix determinant, etc}

Lattices are useful mathematical tools for connecting different areas of mathematics, computer science and cryptography. They are widely used for cryptoanalysis and building secure cryptosystems. In this section,
%\footnote{This section is part of the work \textit{A Tutorial Introduction to Lattice-based Cryptography and Homomorphic Encryption} by the authors Yang Li, Kee Siong Ng, Michael Purcell from the School of Computing, Australian National University @2022.}
we will introduce the basics of lattices in the general setting $\R^n$. In addition, we introduce dual lattices and some computational lattice problems that are commonly used to achieve provable security of lattice-based hard problems and cryptosystems. At the end of this section, we will sketch \citet{ajtai1996generating}'s polynomial time worst-case-to-average-case reduction to reinforce our understanding of lattices as well as appreciate the great breakthrough in provable security of lattice-based cryptography, even against quantum computing in some cases. Although we introduce lattices in the most general setting, their results also hold for special lattices such as ideal lattices in the ring learning with error problem. 

\begin{figure}[ht]
  \centering
  \includegraphics[page=2]{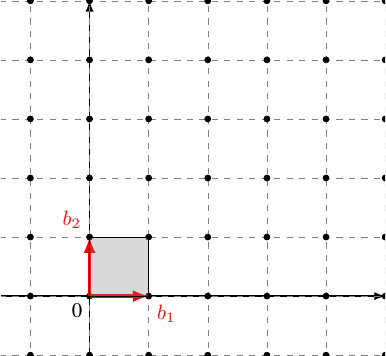}
  \caption{A lattice $L$ with a basis $B=\{b_1, b_2\}$ and its fundamental domain $F$.}
  \label{fig:lattice1}
\end{figure}

Intuitively, a lattice is similar to a vector space except that it consists of discrete vectors only, that is, elements in lattice vectors have discrete values as opposed to real-valued vectors in a vector space. For example, Figure \ref{fig:lattice1} is a lattice in $\R^2$. More formally, we have the following definition. 

\begin{definition}
Let $\mathbf{v_1}, \dots, \mathbf{v_n} \in \R^m$ be a set of linearly independent vectors. The \textbf{lattice} \index{lattice} $L$
\reversemarginpar
\marginnote{\textit{Lattice}}
generated by $\mathbf{v_1}, \dots, \mathbf{v_n}$ is the set of integer linear combinations of $\mathbf{v_1}, \dots, \mathbf{v_n}$. That is, 
\begin{equation*}
    L = \{a_1 \mathbf{v_1} + \cdots + a_n \mathbf{v_n} \mid a_1, \dots, a_n \in \Z\}.
\end{equation*}
\end{definition}
Here, the difference with vector spaces is that the coefficients in the linear combination are integers. The integers $m$ and $n$ are the \textbf{dimension} and \textbf{rank} 
\reversemarginpar
\marginnote{\textit{Dimension, rank}}
of the lattice respectively. If $m=n$, then $L$ is a \textbf{full-rank} lattice. In most cases, we work with full-rank lattices. 

It follows from the definition that a lattice is closed under addition. Hence, we can say that an n-dimensional lattice is a discrete additive subgroup of $\mathbb{R}^n$. It is isomorphic to the additive group of $\mathbb{Z}^n$. That is, 
\begin{equation*}
    (L, +) \cong (\Z^n, +) \subsetneq (\R^n, +).
\end{equation*}

It is often convenient to work with lattices whose coordinates are integers. These are called \textbf{integer lattices} or \textbf{integral lattices}. 
For example, the set of even integers forms an integer lattice, but not the set of odd integers because it is not closed under addition. 

A \textbf{basis} \index{lattice! basis}
\reversemarginpar
\marginnote{\textit{Basis}}
of a lattice $L$ is a set of linearly independent vectors $B = \{b_1, \dots, b_n\}$ that spans the lattice, that is,
\begin{equation*}
    L(B)= \{z_1 b_1 + \dots + z_n b_n \mid z_i \in \mathbb{Z}\}.
\end{equation*}
For example, the vectors $\{b_1,b_2\}$ form a basis of the lattice in Figure \ref{fig:lattice1}.
\begin{figure}[ht]
  \centering
  \includegraphics[page=3]{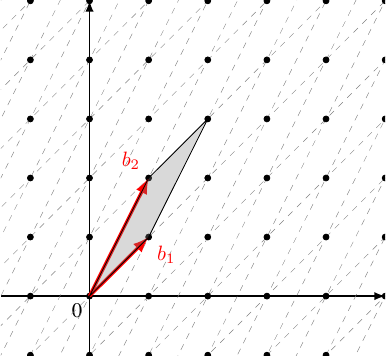}
  \caption{The same lattice $L$ with a different basis $B'=\{b_1', b_2'\}$ and its fundamental domain $F'$, where $B'=AB$ for a unimodular change of basis matrix $A=\big(\begin{smallmatrix}
  1 & 1\\
  1 & 2
\end{smallmatrix}\big)$.}
  \label{fig:lattice2}
\end{figure}

In what follows, we will frequently appeal to properties of a class of matrices known as \emph{unimodular matrices}. Unimodular matrices can be used to translate between different lattice bases. They are also used, sometimes implicitly, when performing important lattice operations such as lattice basis reduction.
\begin{definition}
\reversemarginpar
\marginnote{\textit{Unimodular matrix}}
A matrix $A \in \Z^{n\times n}$ is \textbf{unimodular} if it has a multiplicative inverse in $\Z^{n\times n}$. That is, $A \in \Z^{n\times n}$ is unimodular if and only if $A^{-1} \in \Z^{n\times n}$. Equivalently, a matrix $A \in \Z^{n\times n}$ is unimodular if and only if $|\det(A)| = 1$.
\end{definition}

Similar to a vector space, a lattice does not need to have a unique basis. The following proposition establishes the fact that one basis can be transformed to another via multiplication by the matrix $A$ provided that $A$ is a unimodular matrix.
\begin{proposition}
If $B$ and $B^{\prime}$ be two basis matrices, then $L(B) = L(B^{\prime})$ if and only if $B^{\prime} = AB$ for some unimodular matrix $A$.
\end{proposition}
\begin{proof}
Suppose that $B^{\prime} = AB$ for some unimodular matrix $A$. Then, by definition both $A$ and $A^{-1}$ have integer entries.  Therefore we have $L(B^{\prime}) \subset L(A^{-1}B^{\prime}) = L(B)$ and $L(B) \subset L(AB) = L(B^{\prime})$.

Now suppose that $L(B) = L(B^{\prime})$. Then there exist integer square matrices $A, A^{\prime} \in \Z^{n \times n}$ such that $B^{\prime} = AB$ and $B = A^{\prime}B^{\prime}$. Therefore we have $B = A^{\prime}AB$ or equivalently $(I - A^{\prime}A)B = 0$. Because $B$ is non-singular, we have $A^{\prime} = A^{-1}$ and $A$ is unimodular.
\end{proof}

For example, the vectors $\{b_1',b_2'\}$ in Figure \ref{fig:lattice2} form a different basis for the lattice in Figure \ref{fig:lattice1}, with the relation $B'=AB$ where the change of basis matrix 
$A=\big(\begin{smallmatrix}
  1 & 1\\
  1 & 2
\end{smallmatrix}\big)$
is unimodular.  

%For example, let $\{v_1, \dots, v_n\}$ be a basis for $L$ and $\{w_1, \dots, w_n\}$ be a different set of vectors that can be written in terms of $v_1, \dots, v_n$ as 
%\begin{align*}
%w_i = \sum_j A_{ij}v_j.
%\end{align*}

An important concept of a lattice is the fundamental domain. It is closely related to the sparsity of a lattice as can be seen from the following definition.

\begin{definition}
\reversemarginpar
\marginnote{\textit{Fundamental domain}}
Let $L$ be an $n$-dimensional lattice with a basis $\{v_1, \dots, v_n\}$. The \textbf{fundamental domain} \index{fundamental domain (parallelepiped)} or (\textbf{fundamental parallelepiped}) of $L$ is a region defined as 
\begin{equation*}
    F(v_1, \dots, v_n) = \{t_1 v_1 + \cdots + t_n v_n \mid t_i \in [0, 1)\}.
\end{equation*}
\end{definition}

The lattice $L$ and the given basis in Figure \ref{fig:lattice1} has the fundamental domain coloured in grey. It is the convex region that is surrounded by the given basis vectors and the nearby lattice points. 

\begin{definition}
\reversemarginpar
\marginnote{\textit{Determinant}}
Let $L$ be an $n$-dimensional lattice with a fundamental domain $F$. Then the $n$-dimensional volume of $F$ is called the \textbf{determinant} \index{lattice! determinant} of $L$, denoted by $\det(L)$. 
\end{definition}

Given a basis $\{v_1, \dots, v_n\}$ of an $n$-dimensional lattice $L$, we can write each basis vector $v_i = (v_{i1}, \dots, v_{in})$ as a vector of its coordinates. Then we have a \textbf{basis matrix} 
\begin{equation}
\label{equation:basis matrix}
B = 
\begin{pmatrix}
v_{11} & \cdots & v_{1n} \\ 
\vdots & \ddots & \vdots \\
v_{n1} & \cdots & v_{nn}
\end{pmatrix}.
\end{equation}
In cryptography, we are interested in full-rank lattices, whose determinant can be easily calculated using a basis matrix as stated in the next proposition. 

\begin{proposition}
If $L$ is an $n$-dimensional full-rank lattice with a basis $\{v_1, \dots, v_n\}$ and an associated fundamental domain $F = F(v_1, \dots, v_n)$, then the volume of $F$ (or determinant of $L$) is equal to the absolute value of the determinant of the basis matrix $B$, that is,
\begin{equation*}
    \det(L)=Vol(F) = |\det B|.
\end{equation*}
\end{proposition}

Although the fundamental domain may have a different shape under another choice of a basis, it can be proved that area (or volume) stays unchanged. This gives rise to the determinant of a lattice which is an invariant quantity under the choice of a fundamental domain. 

\begin{corollary}
\reversemarginpar
\marginnote{\textit{Invariant determinant}}
The determinant of a lattice is an invariant \index{lattice! invariant determinant} quantity under the choice of a basis for $L$.
\end{corollary}
\begin{proof}
Let $L$ be a lattice and let $B$ and $B^\prime$ be the basis matrices for two different bases for $L$. There exists a unimodular matrix $A$ such that $B^{\prime} = AB$. Consequently, we have
\begin{equation*}
|\det(B^{\prime})| = |\det (AB)|=|\det(A)|\cdot |\det(B)|=|\det(B)|.
\end{equation*}
So, we have $|\det(L)| = |\det(B^{\prime})| = |\det(B)|$.
\end{proof}

\begin{example}
Let $L$ be a 3-dimensional lattice with a basis \[ \{v_1=(2,1,3), v_2=(1,2,0), v_3(2,-3,-5)\}. \] Then a basis matrix is 
\begin{equation}
B = 
\begin{pmatrix}
2&1&3\\
1&2&0\\
2&-3&-5
\end{pmatrix}.
\end{equation}
The determinant of the lattice is $\det(L)=|\det(B)|=36$.
\end{example}

Geometrically, this also makes sense. 
By definition, each fundamental domain contains exactly one lattice vector (in Figure \ref{fig:lattice1} and \ref{fig:lattice2} the origin). 
% \marginnote{KS: This doesn't sound right} \textcolor{red}{Micciancio gave the following nice argument for this in this video: https://youtu.be/21IzHN9-CjE?t=866.} 
Consider fundamental domains that are centered on lattice points rather than having lattice points at one corner. That is, consider
\begin{equation*}
\tilde{F}(v_1, v_2, \ldots, v_n) = \{t_1v_1 + t_2v_2 + \ldots + t_nv_n \mid t_i \in [-1/2, 1/2)\}.
\end{equation*}
Take a large ball centered at the origin and notice that, because each fundamental domain contains exactly one lattice point, the volume of the ball is approximately equal to the number of lattice points in the ball multiplied by the volume of the fundamental domain. More precisely, we have
\begin{equation*}
    \lim_{r \rightarrow \infty} \frac{\text{Vol}\left(B_r(\vc{0})\right)}{\left|B_r(\vc{0}) \cap L\right|} = \text{Vol}\left(\tilde{F}(v_1, v_2, \ldots, v_n)\right) = \det(L).
\end{equation*}
By definition, choosing a different basis doesn't change the lattice. So, the volume of the fundamental domain, and therefore the determinant of the lattice, is a property of the lattice and does not depend on the basis used to represent that lattice.
%After changing basis, the number of lattice point is unchanged, so is the number of fundamental domains. The total area of the lattice is also unchanged, so the volume of a fundamental domain or the determinant of the lattice must be invariant.
%The concepts of fundamental region and lattice determinant are widely used in lattice-based cryptography as we will see in \citet{ajtai1996generating}'s proof. 

Two remarks. First, a lattice $L$ can be partitioned into disjoint fundamental domains, the union of which covers the entire $L$. Second, since the choice of a fundamental domain is arbitrary and it covers real vectors that are not in $L$, each real vector can be uniquely identified by a lattice vector and a real vector in a fundamental domain. These are captured in the following proposition. For the proof, see Proposition 6.18 of \citet{hoffstein2008introduction}.  

\begin{proposition}
Let $L$ be an $n$-dimensional lattice in $\R^n$ with a fundamental domain $F$. Then every vector $w \in \R^n$ can be written as 
\begin{equation}
\label{equation:unique lattice vector}
    w = v + t
\end{equation}
for a unique lattice vector $v \in L$ and a unique real vector $t \in F$. 

Equivalently, the union of the translated fundamental domains cover the span of the lattice basis vectors, i.e., 
\begin{equation*}
    \text{span}(L) = \{F + v \mid v \in L\}.
\end{equation*}
\end{proposition}

Another useful interpretation of Equation \ref{equation:unique lattice vector} is that for any vector $w \in \R^n$, there is a unique real vector $t \in F$ in the fundamental domain such that $w-t \in L(B)$ is a lattice vector. In other words, given an arbitrary vector $w \in \R^n$ in the span, we can efficiently reduce it to a vector $t \in F$ in the fundamental domain 
\reversemarginpar
\marginnote{\textit{Modulo basis}}
by taking $w$ modulo the basis (or modulo the fundamental domain as used by some authors). More precisely, for a basis $\{\vc{v}_1, \dots, \vc{v}_n\}$ of $L \in \R^n$, it is obvious that the basis is also a basis of the span $\R^n$, so we have $\vc{w}=\alpha_1 \vc{v}_1 + \cdots + \alpha_n \vc{v}_n$ for coefficients $\alpha_1, \dots, \alpha_n \in \R$. The coefficients can also be written as $\alpha_i = a_i + t_i$ for $a_i \in \Z$ and $t_i \in (0,1)$. This implies the real vector can be re-written as $\vc{w}=(a_1 \vc{v}_1 + \cdots + a_n \vc{v}_n) + (t_1 \vc{v}_1 + \cdots + t_n \vc{v}_n)=\vc{v}+\vc{t}$, where in the first pair of parentheses is a lattice vector $\vc{v}$ and in the second pair is a real vector $\vc{t}$ within the fundamental domain. From this, we can compute $\vc{t}=\vc{w}-\vc{v}$. This also gives an alternative formula for computing the modulo basis operation by
\begin{equation}\label{eq:modulo lattice}
    \vc{w} \bmod \vc{B} = \vc{w} - \vc{B} \cdot \lfloor \vc{B}^{-1} \cdot \vc{w} \rfloor.
\end{equation}
For example, given a 2-dimensional lattice $L \in R^2$ with a basis 
$\vc{B}=\big(\begin{smallmatrix}
  3 & 0\\
  0 & 2
\end{smallmatrix}\big)$ and a real vector $\vc{w}=(2,3)$. By reducing $\vc{w}$ modulo the fundamental domain we get $\vc{w} \bmod \vc{B} = (2,1)$.

Similar to a real vector, the length a lattice vector can also be measured by a norm function $||\cdot||$. However, unlike in a vector space where there is no shortest non-zero vector, it is possible to define shortest non-zero vector in a lattice because of the discreteness, although this shortest vector may not be unique.  
\begin{definition}
\reversemarginpar
\marginnote{\textit{Shortest vector}}
Given a lattice $L$, \textbf{the length of a shortest non-zero vector} \index{shortest lattice vector} in $L$ which is also a \textbf{minimum distance} between two lattice vectors is defined as 
\begin{align*}
    \lambda_1(L) &= \min\{||\mathbf{v}|| \mid \mathbf{v} \in L \setminus \{\mathbf{0}\} \} \\
    &= \min \{||\mathbf{x} - \mathbf{y}|| \mid \mathbf{x}, \mathbf{y} \in L, \mathbf{x} \neq \mathbf{y} \}.
\end{align*}
\end{definition}

The shortest vector problem (formally defined in \Cref{subsec:lattice problem}) is to find the shortest non-zero vector in a given lattice. For a lattice $L$, notice that $\lambda_1(L)$ is the solution to the shortest vector problem for that lattice.

The shortest vector problem  can be generalized to the problem of finding the $i^{th}$ successive minima. The $i$th successive minima is the minimum length $r$ such that the lattice contains $i$ linearly independent vectors of length at most $r$. This can also be defined in relation to the dimension of the space spanned by the intersection between $L$ and a zero-centered closed ball $\Bar{B}(0,r)$ with radius $r$.

\begin{definition}
\reversemarginpar
\marginnote{\textit{Successive minima}}
Given a lattice $L$, the $i^{th}$ \textbf{successive minima} \index{successive minima} of $L$ is defined as 
\begin{equation*}
    \lambda_i(L) = \min\{r \mid \dim(span(L \cap \Bar{B}(0,r)))\ge i\},
\end{equation*}
where $\Bar{B}(0,r) = \{ x \in \mathbb{R}^n \mid ||x||\leq r \}$ is the closed ball of radius $r$ around 0.
\end{definition}

For example, if the lattice $L=\Z^n$, then the 1st to the $n^{th}$ successive minima $\lambda_1= \cdots = \lambda_n=1$ are equal to 1. The length of a shortest vector is a special case of the successive minima when $i=1$. We will see the successive minima again when introducing shortest independent vector problem as a generalization of the shortest independent problem in \ref{subsec:lattice problem}. 

Notice that a set of vectors that achieves the successive minima of a lattice is not necessarily a basis for that lattice. Consider the following example which is derived from the work by \citet{korkine1873} and was presented its current form in \cite{nguyen2010}. Let
\begin{equation*}
    \vc{B} = \begin{pmatrix}
        2 & 0 & 0 & 0 & 1 \\
        0 & 2 & 0 & 0 & 1 \\
        0 & 0 & 2 & 0 & 1 \\
        0 & 0 & 0 & 2 & 1 \\
        0 & 0 & 0 & 0 & 1 
    \end{pmatrix}.
\end{equation*}
Notice that $2\vc{e}_5 \in L(\vc{B})$ and that $\lVert v \rVert \geq 2$ for all $\vc{v} \in L(\vc{B}) \setminus \{\vc{0}\}$.  So, $\lambda_i(L(\vc{B})) = 2$  for $1 \leq i \leq 5$. If we let
\begin{equation*}
    \tilde{\vc{B}} = \begin{pmatrix}
        2 & 0 & 0 & 0 & 0 \\
        0 & 2 & 0 & 0 & 0 \\
        0 & 0 & 2 & 0 & 0 \\
        0 & 0 & 0 & 2 & 0 \\
        0 & 0 & 0 & 0 & 2 
    \end{pmatrix}.
\end{equation*}
then we have $L(\tilde{\vc{B}}) \subset L(\vc{B})$ and $\det(\tilde{\vc{B}}) = 32$. On the other hand, we see that $\det(\vc{B}) = 16$. Therefore, $\tilde{\vc{B}}$ cannot be a basis for $L(\vc{B})$. In fact, it can be shown that no basis of $L(\vc{B})$ realizes all of the successive minima of $L(\vc{B})$.

%smoothing parameter is relatd to discrete gaussian, if discrete gaussian variance is larger than smoothing parameter, then adding discrete gaussian noise to lattice vectors results in near uniform distribution, also if discrete gaussian variance is large enough, it behaves like the continuous gaussian. see \cite{micciancio07worst} section 4. 

\subsection{Dual lattice}

In this subsection, we introduce dual lattices. This is a useful concept that will be used at several different places, such as defining smoothing parameter for discrete Gaussian distribution and in the hardness proof of the ring learning with error problem. It is important to develop a geometric intuition of the relationship between a lattice and its dual. %We will see in a later section that how the dual lattice is closed related to the smoothing parameter of a lattice. 

The dual (sometimes also called reciprocal) of a lattice is the set of vectors in the span of the lattice (e.g., the span is $\R^n$ if the lattice is $\Z^n$) whose inner product with the lattice vectors are integers.

\begin{figure}[hbt!]
	\centering
	\begin{subfigure}[b]{0.95\textwidth}
		\centering
    	\includegraphics[page=5]{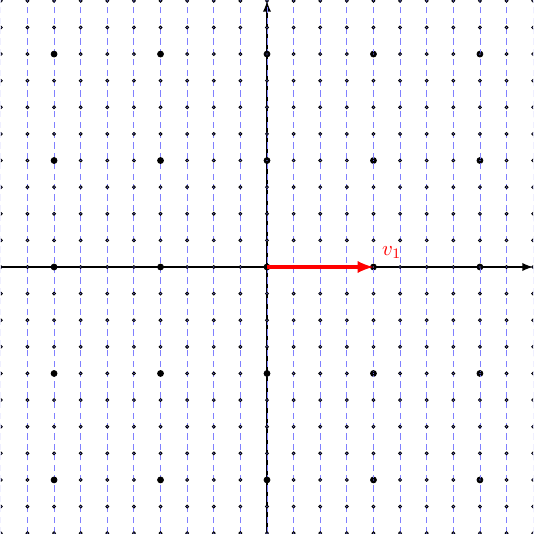}
		\caption{The dual lattice is partitioned into hyperplanes according to the given lattice vector $v=(2,0)$.}
		\label{subfig:dualLatHypExp1}
	\end{subfigure}
	
	\begin{subfigure}[b]{0.95\textwidth}
		\centering
		\includegraphics[page=6]{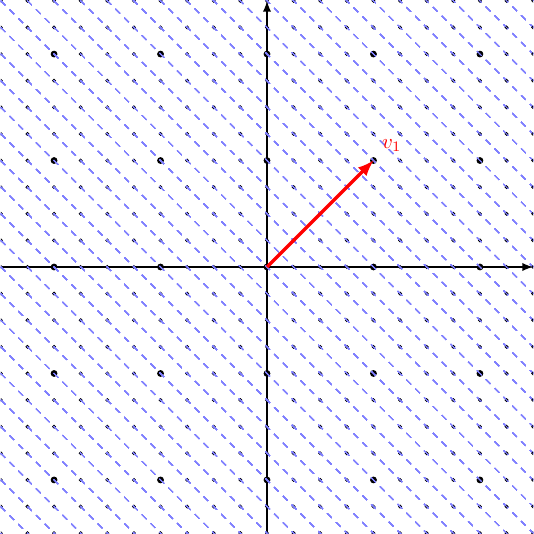}
		\caption{The dual lattice is partitioned into hyperplanes according to the given lattice vector $v=(2,2)$.}
		\label{subfig:dualLatHypExp2}
	\end{subfigure}
	\caption{For a given lattice vector $v \in L=2\Z^2$, the dual lattice $L^*=\frac{1}{2}\Z^2$ can be partitioned into parallel non-overlapping hyperplanes (vertical lines) that are perpendicular to $v$. Elements in the same hyperplane have the same dot product with $v$, so they form an equivalence class.}
	\label{fig:dualLatHyp}
\end{figure}

\begin{definition}
\reversemarginpar
\marginnote{\textit{Dual lattice}}
Given a full-rank lattice $L$, its \textbf{dual lattice} \index{dual lattice} is defined as 
\begin{equation*}
    L^* = \{\mathbf{y} \in span(L) \mid \forall \mathbf{x} \in L, \mathbf{x} \cdot \mathbf{y} \in \Z\}.
\end{equation*}
\end{definition}
For example, the dual lattice of $\Z^n$ is $\Z^n$ and the dual lattice of $2\Z^n$ is $\frac{1}{2}\Z^n$ as shown in Figure \ref{fig:dualLatHyp}. An important observation is that the more vectors a lattice has, the less vectors its dual has and vice versa, because there are more (or less) constraints. Most importantly, it can be verified that the dual of a lattice is also a lattice.

\begin{proposition}
If $L$ is a lattice then $L^*$ is a lattice.
\end{proposition}
\begin{proof}
It suffices to show that $L^*$ is closed under subtraction.  That is, to show that if $x,y \in L^*$ then $x-y \in L^*$. This follows from the linearity of the inner product. More explicitly, for every $\vc{z} \in L$ we have $(\vc{x}-\vc{y})\cdot \vc{z} = \vc{x}\cdot \vc{z} - \vc{y} \cdot \vc{z}$. Because $\vc{x}\cdot \vc{z} \in \Z$ and $\vc{y} \cdot \vc{z} \in \Z$, we have $(\vc{x}-\vc{y})\cdot \vc{z} \in \Z$. The result then follows from the definition of $L^*$.
\end{proof}
%\marginnote{KS: Use green colour in the diagram}
\begin{figure}[hbt!]
	\centering
	\includegraphics[page=4]{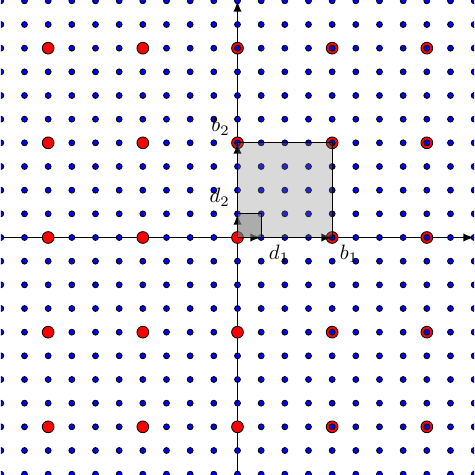}
	\caption{A lattice $L=2\Z^2$ (black points) and its dual $L^*=\frac{1}{2}\Z^2$ (blue points). The basis of $L$ is $B=\{b_1=(2,0),b_2=(0,2)\}$ and the dual basis of $L^*$ is $D=\{d_1=(\frac{1}{2},0),d_2=(0,\frac{1}{2})\}$.}
	\label{fig:dualLat}
\end{figure}

Given a lattice $L$, it is natural to ask if we can find a basis for $L^*$. This leads us to define the dual basis of a lattice.

\begin{definition}
\reversemarginpar
\marginnote{\textit{Dual basis}}
For a lattice $L$ and a basis $B=(b_1, \dots, b_n) \in \R^{m \times n}$, the \textbf{dual basis} \index{dual lattice! dual basis} $D=(d_1, \dots, d_n) \in \R^{m \times n}$ is defined as the unique basis that satisfies 
\begin{itemize}
    \item $span(B)=span(D)$ and 
    \item $B^T D = I$.
\end{itemize}
\end{definition}
The first condition says both bases span the same vector space. The second condition implies that $b_i \cdot d_j = \delta_{ij}=1$ if $i=j$ and 0 otherwise. Abusing notation, we use $B$ to denote both the basis of a lattice and the basis matrix. If $L$ is a full-rank lattice (i.e., $m=n$), then the basis matrix $B$ is invertible, so the dual basis matrix can be expressed as $D=(B^T)^{-1} = (B^{-1})^T$.

\begin{proposition}
If $L$ is a lattice with basis $B$, then the dual basis is a basis for $L^*$.
\end{proposition}
\begin{proof}
This follows immediately from the definition of the dual lattice and the linearity of the inner product.
\end{proof}

Having established that the dual of a lattice is itself a lattice, we can ask what we get if repeat the process and compute the dual of a dual lattice.
\begin{proposition}
For any lattice $L$, we have $(L^*)^*=L$.
\end{proposition}
\begin{proof}
If $B$ is a basis for a full-rank lattice $L$, then a dual basis is $D=(B^T)^{-1}$. Then the dual basis of $D$ is $(D^T)^{-1}$ that is equal to $B$. The same argument works for rank-deficient lattices, but with slight variation because their bases are non-square matrices. 
\end{proof}

\begin{proposition}
For any lattice $L$, we have $\det(L^*) = \frac{1}{\det(L)}$.
\end{proposition}
\begin{proof}
Again, we give a proof for full-rank lattices. If $L$ is full-rank, then 
\begin{align*}
    \det(L^*) = |\det (D)| = |\det ((B^T)^{-1})| = \frac{1}{|\det (B^T)|} = \frac{1}{|\det (B)|}=\frac{1}{\det (L)}.
\end{align*}
\end{proof}

Although a lattice and its dual are both lattices, they are fundamentally different objects. The dual of a lattice can be thought as functions that are applied to the lattice such that the inner products of the lattice vectors and each dual vector are integers. 

Here is a geometric interpretation of a lattice and its dual. For each lattice vector $\mathbf{v}$, its inner products with the dual vectors produce integers of different values. So $\mathbf{v}$ partitions the dual lattice into parallel non-overlapping hyperplanes that are perpendicular to $\vc{v}$
\reversemarginpar
\marginnote{\textit{Hyperplanes}}
according to its inner product values with the dual vectors. Elements in the same hyperplane have the same inner product with the lattice vector $\vc{v}$, so they form an equivalence class. Alternatively, we can say $\vc{v}$ partitions the dual lattice into a set of equivalence classes. Figure. \ref{fig:dualLatHyp} gives two examples of how a lattice vector $\vc{v}\in L=2\Z^2$ partitions the dual lattice $L^*=\frac{1}{2}\Z^2$. In addition, the distance between two neighbouring hyperplanes is the inverse of the vector length (i.e., $1/||\mathbf{v}||$).

\begin{example}
When $L = 2\Z$ and $L^*= \frac{1}{2}\Z$, the vector $\mathbf{v}=\frac{1}{2}$ partitions $L$ to $|2\Z|$ hyperplanes, each contains exactly one integer from $L$ and the neighbouring hyperplanes are distance 2 apart. 

When $L=2\Z^2$ and $L^*=\frac{1}{2}\Z^2$, the vector $\vc{v}=(2,0)$ partitions the dual lattice into hyperplanes as shown in Figure \ref{subfig:dualLatHypExp1}, where the hyperplanes are the vertical lines that are perpendicular to the lattice vector $\vc{v}$. The distance between the neighbouring hyperplanes is $\frac{1}{||\vc{v}||} = \frac{1}{2}$. So the dual is denser than $L$. 
If $\vc{v}=(2,2)$, the dual is partitioned into hyperplanes as shown in Figure \ref{subfig:dualLatHypExp2}. The distance between the neighbouring hyperplanes is $\frac{1}{||\vc{v}||}=\frac{1}{2\sqrt{2}}$.
\end{example}

%This is related to the covering radius of the lattice. Hence, the shorter the shortest vector in the dual, the larger the covering radius of the lattice because the more distance two neighbouring lattice layers are.  

%see K. Conrad on "The Different Ideal" for more details about lattice dual. 

%Other relations between a lattice and its dual, such as fundamental region, 

%Why do we introduce dual lattice? Because CVP can be formulated (using dual lattice) as SVP in a shifted lattice. 

\subsection{Some lattice problems}
\label{subsec:lattice problem}
%see Peikert's end of sec 2.2 (A decade of Lattice...)
Having briefly introduced lattices and some related concepts, we are ready to define some computational lattice problems in this subsection. The most well known two are the shortest vector problem and closest vector problem. %are the two central lattice problems that have been studied in computational complexity theory. 
These two are search problems because the aims are to find a shortest or closest lattice vector.  
Few cryptosystems, however, are based on these two problems directly. Instead, most cryptosystems are based on their decision versions or relaxed approximation variants. Below, we state the two well known lattice problems and some variants. 
%of these problems such as the short integer solution problem and bounded distance decoding problem. Some of these problems are known to be hard, while others were conjectured to be hard with high confidence. 

%svp
\begin{tcolorbox}
\noindent
\textbf{The Shortest Vector Problem (SVP) \index{lattice problems! SVP}}\\
Given a lattice basis $B$, find a shortest non-zero vector in the lattice $L(B)$, i.e., find a non-zero vector $\mathbf{v} \in L(B)$ such that $||\mathbf{v}|| = \lambda_1(L(B))$. 
\end{tcolorbox}

SVP is hard to solve in high-dimensional lattices. An important variant of SVP is finding a set of short linearly independent lattice vectors as stated below.  
%sivp
\begin{tcolorbox}
\noindent
\textbf{The Shortest Independent Vectors Problem (SIVP)}\\
Given a lattice basis $B$ of an $n$-dimensional lattice $L(B)$, find $n$ linearly independent vectors $\mathbf{v_1}, \dots, \mathbf{v_n} \in L(B)$ such that $\max_{i \in [1,n]} ||\mathbf{v_i}|| = \lambda_n(L(B))$.
\end{tcolorbox}

%cvp
\begin{tcolorbox}
\noindent
\textbf{The Closest Vector Problem (CVP) \index{lattice problems! CVP}}\\
Given a lattice basis $B$ and a target vector $\mathbf{t}$ that is not in the lattice $L(B)$, find a vector in $L(B)$ that is closest to $\mathbf{t}$, i.e., find a vector $\mathbf{v} \in L(B)$ such that for all $w \in L(B)$ it satisfies $||\mathbf{v} - \mathbf{t}|| \le ||\mathbf{w} - \mathbf{t}||$. 
\end{tcolorbox}

A special case of CVP is the bounded distance decoding problem, which is used in the learning with error problem's hardness proof \citep{regev2009lattices}. The name reflects that the problem is to ``decode'' a given $\R^n$ vector. The extra condition makes it a special case of CVP is that the given non-lattice vector is within a bounded distance to the lattice. 
\begin{tcolorbox}
\noindent
\textbf{The $\alpha$-Bounded Distance Decoding Problem (BDD$_{\alpha}$) \index{lattice problems! BDD$_{\alpha}$}}\\
Given a lattice basis $B$ of an $n$-dimensional lattice $L$ and a target vector $\mathbf{t} \in \R^n$ satisfies $dist(\vc{t},B) \le \alpha \lambda_1(L)$, find a lattice vector $\vc{v} \in L$ that is closest to $\mathbf{t}$, i.e., for all $\vc{w} \in L$ it satisfies $||\mathbf{v} - \mathbf{t}|| \le ||\mathbf{w} - \mathbf{t}||$. 
\end{tcolorbox}
An alternative way of defining BDD is to find the lattice vector $\vc{x} \in L$ given the instance $\vc{y} = \vc{x} + \vc{e} \in \R^n$, where $\vc{e}$ is often interpreted as a noise with norm $||e|| \le \alpha \lambda_1(L)$.

As discussed in \Cref{subsection:gapProb}, knowing c-gap problems are hard implies the corresponding c-approximate  problems are also hard. But c-approximations are often used to prove some problems are hard to solve (e.g., SIS) because it is relatively easier to build reductions from them. Below we state the gap/approximate variants of the standard lattice problems. Let $\gamma(n): \N \rightarrow \N$ be a gap function in the input size such that $\gamma(n) \ge 1$, for example $\gamma(n)$ is a polynomial of $n$.
%gapsvp
\begin{tcolorbox}
\noindent
\textbf{The $\gamma$-GAP Shortest Vector Problem (GAPSVP$_{\gamma}$) }\\
INSTANCE: For a function $\gamma(n) \ge 1$, given a real number $d > 0$ and a lattice basis $B$, the instance $(B, d)$ is 
\begin{itemize}  
    \item either a YES instance if $\lambda_1(L(B)) \le d$
    \item or a NO instance if $\lambda_1(L(B)) \ge \gamma(n) d$.
\end{itemize}
QUESTION: Is $(B,d)$ a YES or NO instance? 
\end{tcolorbox}

%gapsvp
\begin{tcolorbox}
\noindent
\textbf{The $(\zeta,\gamma$)-GAP Shortest Vector Problem (GAPSVP$_{\zeta,\gamma}$)}\\
INSTANCE: For functions $\zeta(n) \ge \gamma(n) \ge 1$, given a real number $d > 0$ and a lattice basis $B$ of an $n$-dimensional lattice $L(B)$ such that
\begin{itemize}
    \item $\lambda_1(L(B)) \le \zeta(n)$,
    \item $\min_{i \in [1,n]} ||\Tilde{b}_i|| \ge 1$,
    \item $1 \le d \le \zeta(n) / \gamma(n)$, 
\end{itemize}
the instance $(B, d)$ is 
\begin{itemize}
    \item either a YES instance if $\lambda_1(L(B)) \le d$
    \item or a NO instance if $\lambda_1(L(B)) \ge \gamma(n) d$.
\end{itemize}
QUESTION: Is $(B,d)$ a YES or NO instance? 
\end{tcolorbox}

%The following two problems are not presented in their gap formats. They are c-approximation of SIVP and SBP, which make the reduction to the SIS problem easier. 
\begin{tcolorbox}
\noindent
\textbf{The $\gamma$-Shortest Independent Vectors Problem (SIVP$_{\gamma}$)}\\
Given a lattice basis $B$ of an $n$-dimensional lattice $L(B)$, find $n$ linearly independent vectors $\mathbf{v_1}, \dots, \mathbf{v_n} \in L(B)$ such that $\max_{i \in [1,n]} ||\mathbf{v_i}|| \le \gamma(n) \lambda_n(L(B))$.
\end{tcolorbox}

% \kl{More lattice problems, e.g. covering radius problem (CRP), etc. See \citep{micciancio07worst}?}

\subsection{Ajtai's worst-case to average-case reduction}

\begin{figure}[hbt!]
    \centering
    \includegraphics[page=15,width=10em]{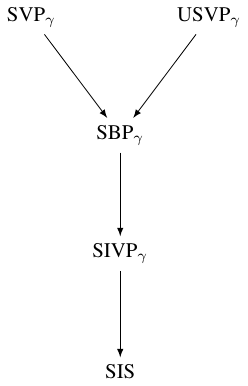}
    \caption{Reductions to the SIS problem from hard lattice problems (SVP$_{\gamma}$, USVP$_{\gamma}$ and SBP$_{\gamma}$). The intermediate lattice problem in the reductions is the $\gamma$-approximation of the shortest independent vector problem (SIVP$_\gamma$).}
    \label{fig:sisReduction}
\end{figure}
To finish off this section, we present a high level overview of \citeauthor{ajtai1996generating}'s worst-case to average-case reduction.\index{worst-case to average-case} As briefly explained in  \Cref{subsec:averagecasehard}, such a reduction allows one to build cryptosystems based on an average-case hardness problem, so that users can rest assured that their random encryption instances are guaranteed to be secure with high confidence. %Ajtai's work shows that breaking a random SIS-based cryptographic instance is as hard as solving some hard lattice problems in the worst case. %,  that either proved or conjectured to be difficult. 

Ajtai's proof is based on three well-studied lattice problems, SVP$_{\gamma}$, USVP$_{\gamma}$\index{lattice problems! USVP} and SBP$_{\gamma}$.\index{lattice problems! SBP}
The second problem is a variant of SVP that finds the unique shortest non-zero vector in the lattice $L(B)$, i.e., find the non-zero vector $\mathbf{v} \in L(B)$ such that $||\mathbf{v}|| = \lambda_1(L(B))$ and if $\vc{w} \in L(B)$ such that $||\vc{w}|| \le n^c ||\vc{v}||$ then $\vc{w}$ is parallel to $\vc{v}$. 
The third problem is to find a shortest basis $\{\vc{b}_1, \dots, \vc{b}_n\}$ of a given lattice, where the basis length is defined as $\max_{i=1}^n||\vc{b}_i||$. All three problems are used in their gap (or approximation) versions. 

The average-case hard problem constructed by \citeauthor{ajtai1996generating} is known as the \textbf{short integer solution (SIS)} problem.\index{lattice problems! SIS}
\reversemarginpar
\marginnote{\textit{SIS}}
Let $\mathbf{a_i} \in \Z_q^n$ be a length $n$ vector with entries taken uniformly from $\Z_q$. Let $A = [\mathbf{a}_1 \mid \cdots \mid \mathbf{a}_m]$ be an $n \times m$ matrix whose columns are $m$ linearly independent $\mathbf{a_i}$s. The SIS problem is to find a non-zero vector $\mathbf{x} \in \Z^m$ such that
\begin{itemize}
    \item $||\mathbf{x}|| \le \beta$ and 
    \item $A \mathbf{x} = \mathbf{0} \in \Z_q^n$, i.e., $\mathbf{x}_1 \mathbf{a}_1 + \cdots + \mathbf{x}_m \mathbf{a}_m = \mathbf{0} \bmod q$.
\end{itemize}
Notice that the norm bound exists to ensure the problem is not easily solvable by for example Gaussian elimination. It must satisfy $\beta < q$ to avoid the trivial solution $\vc{x}=(q, 0, \dots, 0)$. Moreover, $\beta$ and $m$ must be large enough to allow a solution to exist. A sufficient condition of guaranteeing a solution is given in a subsequent work \cite{micciancio07worst}. See Section 4 of \citet{peikert16decade} for more detailed insights. 

\begin{lemma}[Lemma 5.2 \cite{micciancio07worst}]
\label{lm:sufficientCondSIS}
For any $q, A, \beta \ge \sqrt{m}q^{n/m}$, the SIS instance $(q,A,\beta)$ admits a solution. 
\end{lemma}

\begin{proof}
The proof is done using the \textit{pigeonhole principle} by constructing $\vc{x}=(x_1, \dots, x_m)$ where each $x_i \in \{0, \dots, 0, q^{n/m}\}$, so that there are $(q^{n/m})^m=q^n$ this type of vectors, more than the size of the codomain $A\vc{x} \in \Z_q^n$. Hence, there must exist two distinct vectors $\vc{x}_1$ and $\vc{x}_2$ of this form such that $A \vc{x}_1=A \vc{x}_1 \bmod q$. This entails $A\vc{x}' = 0 \bmod q$ for $\vc{x}'=\vc{x}_1 - \vc{x}_2$. The norm of this vector satisfies $||\vc{x}'||\le \sqrt{m q^{2n/m}}=\sqrt{m}q^{n/m}$ because each of its coordinate is at most $q^{n/m}$. Hence, there always exist a solution with such maximum norm.
\end{proof}

% The average-case problem Ajtai proposed is the SIS problem that has been defined above. 
% Given a randomly generated matrix $A$, denote the set of solutions of the SIS instance by 
% \begin{equation*}
%    L_q^{\perp}(A) = \{\mathbf{x} \mid A\mathbf{x} = \mathbf{0} \bmod q\}.
% \end{equation*}
% It is not hard to check that this is a lattice (closed under addition). So finding a solution of an SIS instance is equivalent to finding a short lattice vector in $L_q^{\perp}(A)$. 

The structure of the reduction is shown in Figure \ref{fig:sisReduction}. The essential part of the proof is a polynomial-time reduction from the lattice problem SBP$_{\gamma}$ to SIS. The other two lattice problems can be reduced to SBP$_{\gamma}$ (See \citet{ajtai1996generating} Appendix).

To simplify the reduction, note $\text{SBP}_{\gamma}$ is related to $\text{SIVP}_{\gamma}$ because given a set of linearly independent lattice vectors $\mathbf{r_1}, \dots, \mathbf{r_n} \in L$, a basis $\{\mathbf{s_1}, \dots, \mathbf{s_n}\}$ of $L$ can be constructed in polynomial time such that $\max_{i=1}^n ||\vc{s_i}|| \le n \max_{i=1}^n ||\vc{r_i}||$. Hence, the task becomes reducing the lattice problem SIVP$_{\gamma}$ to SIS, where the approximation factor $\gamma=n^{c_3-1}$ is polynomial in $n$. This is also a well accepted hard lattice problem \cite{micciancio2009lattice}. 

%finding linearly independent lattice vectors $\mathbf{r_1}, \dots, \mathbf{r_n} \in L$ such that $\max_{i=1}^n ||\vc{r_i}|| \le n^{c_3-1}bl(L)$, where $bl(L)$ is the length of the shortest basis of $L$. In other

The reduction starts by assuming 
\reversemarginpar
\marginnote{\textit{SIVP$_{\gamma}$ to SIS}}
there is a probabilistic polynomial time (PPT) algorithm $\mathcal{A}$ that solves SIS with a non-negligible probability.\footnote{Ajtai related SIS with finding a short vector in a \textit{q-ary lattice} $L_q^{\perp}(A) = \{\mathbf{x} \mid A\mathbf{x} = \mathbf{0} \bmod q\}$. His reduction starts with assuming $\mathcal{A}$ is a PPT algorithm to find a short lattice vector in a given $L_q^{\perp}(A)$. For the purpose of sketching the main steps of the proof, it is not necessary to relate SIS with the q-ary lattice problem.}
%outputs a vector $\mathbf{x} \in L(\lambda_{n,c_1,c_2}, \lfloor n^{c_2}\rceil)$ such that $||\mathbf{x}||\le n$.
The next step is to transform a hard SIVP$_{\gamma}$ instance to a random SIS instance and show that if such an SIS solution $\mathcal{A}$ exists, it gives rise to a PPT algorithm $\mathcal{B}$ that solves $\text{SIVP}_{\gamma}$ for a polynomial factor. %by finding a set of linearly independent vectors $\{\vc{a}_1, \dots, \vc{a}_n\}$ such that the maximum length $\max_i ||a_i|| =M \le n^{c_3-1}bl(L)$  is bounded . 
This solution then transforms into a solution for SBP$_{\gamma}$, as well as SVP$_{\gamma}$ and USVP$_{\gamma}$.    

% , which follows a near uniform distribution over $\Z_q^n$, so the PPT algorithm $\calA$ for SIS can be used as a subroutine to compute a solution to the SBP$_{\gamma}$ problem instance. % can output a solution of SIS. 
% Figure \ref{fig:average2worstDuplicate} provides a visual aid to understand this reduction.  

%    \item For certain parameter settings, assume the SIS problem can be solved in polynomial time with  non-negligible probability; i.e. there is a probabilistic polynomial time (PPT) algorithm $\mathcal{A}$ that outputs a vector $\mathbf{x} \in L(\lambda_{n,c_1,c_2}, \lfloor n^{c_2}\rceil)$ such that $||\mathbf{x}||\le n$. We want to show that there is a PPT algorithm $\mathcal{B}$ that solves the $\text{SBP}_{\gamma}$ problem, i.e., finds a basis $d_1, \dots, d_n$ such that $\max_i ||d_i|| \le n^{c_3}bl(L)$.  
    
%    \item $\text{SBP}_{\gamma}$ is related to $\text{SIVP}_{\gamma}$, because given a set of linearly independent lattice vectors $\mathbf{r_1}, \dots, \mathbf{r_n} \in L$, it can be constructed in polynomial time a basis $\{\mathbf{s_1}, \dots, \mathbf{s_n}\}$ of $L$ such that $\max_{i=1}^n ||\vc{s_i}|| \le n \max_{i=1}^n ||\vc{r_i}||$. So the task becomes finding linearly independent lattice vectors $\mathbf{r_1}, \dots, \mathbf{r_n} \in L$ such that $\max_{i=1}^n ||\vc{r_i}|| \le n^{c_3-1}bl(L)$.
    
\begin{figure}[hbt!]
    \centering
    \includegraphics[page=16]{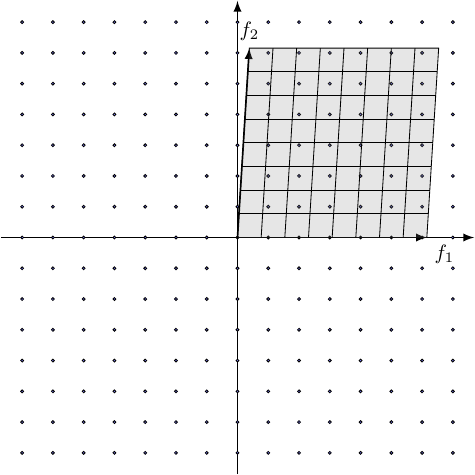}
    \caption{In a lattice $L=\Z^2$, the near cubic parallelepiped $W$ formed by the large independent vectors $\{f_1,f_2\}$. It is divided into $q^2$ smaller pieces, each of which is hit with equal probability by random lattice vectors reduced within $W$.}
    \label{fig:sisIteStep}
\end{figure}

For simplicity, denote $M = \max_i ||a_i||$ and $bl(L)$ the length of the shortest basis. The key to guarantee $M < n^{c_3-1}bl(L)$ is to iteratively shorten the longer vectors by half to achieve $\frac{M}{2}$. Repeating this steps at most $\log_2 M$ steps we get vectors of the desired length. 
Each iteration of this process is as follows:
    \begin{enumerate}
        \item \textbf{Construct near cubical parallelepiped:} Starting from the lattice vectors $\mathbf{a_1}, \dots, \mathbf{a_n}$, construct other lattice vectors $\mathbf{f_1}, \dots, \mathbf{f_n}$ such that they are nearly pairwise orthogonal and have similar length, but constraint the maximum length $\max_{i=1}^n ||\vc{f_i}|| \le n^3 M$. The reason is to form a parallelepiped $W=P(\mathbf{f_1}, \dots, \mathbf{f_n})$ that is almost a hypercube, as shown in a 2-dimensional lattice in Figure \ref{fig:sisIteStep}. This step was proved in Lemma 3 of \citet{ajtai1996generating}.
                
        \item \textbf{Induce near uniform SIS instance:} We then evenly cut $W$ into $q^n$ small non-overlapping parallelepipeds which have the form $w_j=(\sum_{i=1}^n \frac{t_i^j}{q}\mathbf{f_i}) +\frac{1}{q}W$, where $t_i^j \in [0,q)$ is an integer.  
        Now sample $m$ random lattice vectors from $L$, then reduce them modulo $W$ to ensure they are within the bigger parallelepiped. Denote these reduced vectors by $\mathbf{\xi_1}, \dots, \mathbf{\xi_m}$. If $\mathbf{\xi_k}$ is in a smaller parallelepiped $w_j=(\sum_{i=1}^n \frac{t_i^j}{q}\mathbf{f_i})+\frac{1}{q}W$, then take  $(t_1^j, \dots, t_n^j)$ and put it as a column of a matrix $A$. The claim is that each of the $w_j$'s is selected with almost equal chance, so we have a random $n \times m$ matrix $A$.
        The key intuition is that for a short basis of $L$, if $W$ intersects with a translation of the fundamental domain formed by the short basis, then $W$ will contain a large proportion of the translated fundamental domain. This property remains true for an arbitrary translation and scaling of $W$ using $\vc{u} + \frac{1}{q}W$ for a vector $\vc{u} \in \R^n$. With this property, if $W$ is cut into small non-overlapping regions evenly, then random lattice vectors within $W$ will induce a near uniform distribution over the pieces $w_j$'s. This implies that the matrix $A$ is a random instance of SIS. This step was proved in Lemma 8 of \citet{ajtai1996generating}. %is important for generating uniformly random column vectors in $A$ for the SIS algorithm $\calA$. 

        \item \textbf{Halve vector length:} Now give the matrix $A$ to the PPT algorithm $\mathcal{A}$ to output an SIS solution $(h_1, \dots, h_m) \in \Z^m$. It remains to prove that the vector $\vc{u}=\sum_{i=1}^n h_i \vc{\xi_i}$ is only half of size of the starting vectors, i.e., $||\vc{u}|| \le \frac{M}{2}$ and they are non-zero. This step was proved in Lemma 13 of \citet{ajtai1996generating}. %The shorter length of $\vc{u}$ is because each $w_j$ is a smaller parallelepiped (detail skipped, see page 7 \& 8 in \cite{ajtai1996generating}). 
    \end{enumerate}

% To recap, the key is to construct a large parallelepiped $W$ that has a nice hypercube shape, then cut it into smaller parallelepipeds correspond to the ring elements in $\Z_q^n$, which are neither too big in order to produce shorter vectors, nor too small to be sampled uniformly. 

\iffalse
\begin{figure}[hbt!]
    \centering
    \includegraphics[page=1]{images/Lattice_crypto_tikz_folder.pdf}
    \caption{A demonstration of a cryptosystem's computational security that is based on an average-case problem. Each cryptographic instance $x_i$ corresponds to a random average-case instance $a_j$. All random instances in the average-case problem can be mapped with the hard instances in the worst-case problem. There may be a fraction of average-case instances (colored in red) that can be solved easily, so their solutions entail solutions of the worst-case problem. But the fraction of such instances is negligible. The hard and easy instances in the worst-case problem are colored blue and white, respectively. The dashed lines indicate the worst-case-to-average-case reduction is random.}
    \label{fig:average2worstDuplicate}
\end{figure}
%https://www.youtube.com/watch?v=qZIjVX61NFc&t=433s&ab_channel=SimonsInstitute, at 39mins
\fi

In order to motivate subsequent works inspired by SIS, we make two remarks about the above reduction. First, the polynomial approximation factor in the lattice problems are large enough to raise a minor security concern of SIS-based encryption schemes, because the larger the factor is the easier the problems could be. As analysed in \citet{cai1997improved}, a typical factor size is larger than $n^8$. In a following section, we will introduce the discrete Gaussian technique to reduce these factors down to $\tO(n)$ in SIS hardness proof. 
%The proof strategy will follow Ajtai's technique of constructing, from a set of linearly independent vectors of a lattice, a random set of lattice vectors that are well spread out and as short as possible.
Second, the public key size required by an SIS-based cryptosystem is $\tO(n^4)$ that is quite inefficient for practical purposes. This will be dramatically improved by developing different average-case problems as we will see in the learning with error and ring learning with error problems. 
%Read Section 1.1 in \citep{micciancio07worst} for a more detailed interpretation of earlier techniques on worst-case-to-average-case reductions.

%\kl{what's the gap factor? mention it here, because we will introduce an improvement using discrete Gaussian.}

\subsection{An application of SIS: Collision resistant hash functions}
% it's better to replace with sis-based cryptosystem, such as ajtai and dwork's 
SIS has been used as the foundation of one-way functions and hash functions \citep{lyubashevsky2010ideal}.

A hash function maps inputs of arbitrary length and compresses them into short fixed-length outputs known as \emph{digests}.

\begin{definition}
A \textbf{(keyed) hash function} \index{hash function}
\reversemarginpar
\marginnote{\textit{Hash function}}
with output length $l$ is a pair of probabilistic polynomial-time algorithms $(\text{Gen},H)$ satisfying the following: 
\begin{itemize}
    \item The algorithm $\text{Gen}(1^n) \rightarrow s$ generates a key $s$ from the security parameter $1^n$.
    \item For a string $x\in \{0,1\}^*$ of arbitrary length, the algorithm $H$ outputs a string $H^s(x) \in \{0,1\}^{l(n)}$.
\end{itemize}
\end{definition}

The general interest in hash functions is the case when the outputs are shorter than the inputs for both computational and storage efficiency. In such a case,  a hash function's domain is larger than its range, which implies the possibility of having two distinct inputs being mapped to the same output. We often say the two distinct inputs \textit{collide} and the scenario is called a \textit{collision}. 

For a hash function $\Pi=(\text{Gen},H)$, an adversary $\mathcal{A}$ and the security parameter $n$, we can define the \textbf{collision-finding experiment} \textbf{Hash-coll}$_{\mathcal{A},\Pi}(n)$ as:
\reversemarginpar
\marginnote{\textit{\textbf{Hash-coll}$_{\mathcal{A},\Pi}(n)$}}
\begin{enumerate}
    \item Run the algorithm $\text{Gen}(1^n) \rightarrow s$. 
    \item The adversary $\mathcal{A}$ is given the key $s$.
    \item The adversary produces two strings $x$, and $x'$.
    \item \textbf{Hash-coll}$_{\mathcal{A},\Pi}(n)=1$ if $x \neq x'$ and $H^s(x)=H^s(x')$ and 0 otherwise.  
\end{enumerate}

A cryptographic hash function requires the chance of finding a collision is negligible, which is defined more formally as follows. 
\begin{definition}
A hash function $\Pi=(\text{Gen},H)$ is \textbf{collision resistant} \index{hash function! collision resistant}
\reversemarginpar
\marginnote{\textit{Collision resistant}}
if for any probabilistic polynomial time adversary $\mathcal{A}$, it satisfies 
\begin{equation*}
    Pr[\text{Hash-coll}_{\mathcal{A},\Pi}(n)=1] \le \text{negl}(n).
\end{equation*}
\end{definition}

From Ajtai's SIS problem and the worst-case-to-average-case reduction, one can easily build a collision resistant hash function where the key is the matrix $A \in \Z_q^{n \times m}$ and the hash function is given by 
\begin{align*}
    &f_A: \{0, \dots, d-1\}^m \rightarrow \Z_q^n \\
    &f_A(\mathbf{x})=A\mathbf{x} \bmod q.
\end{align*}
If there is a collision $f_A(\vc{x})=f_A(\vc{x'})$ between distinct inputs $\vc{x}$ and $\vc{x'}$, then $A(\vc{x}-\vc{x}^{\prime}) = 0$ and $\vc{x} - \vc{x}^{\prime} \in L^{\perp}_q(A)$. Furthermore, because each element of $\vc{x} - \vc{x}^{\prime}$ is in the set $\{-1,0,1\}$, we see that $\vc{x} - \vc{x}^{\prime}$ is a short vector. Hence, an efficient algorithm that produces collisions for this hash function could be used to solve SIS in the lattice $L^{\perp}_q(A)$.

\newpage
\section{Discrete Gaussian Distribution}

\label{sec:discrete gaussian}

%\section{Discrete Gaussian distribution}

%The discrete Gaussian distribution... see section 5.2 of \cite{mukherjee2016cyclotomic}.

Discrete Gaussian distribution is an important ingredient in the provable security of lattice-based cryptosystems. The distribution behaves in a similar fashion as the continuous Gaussian distribution, but with a discrete lattice support. The technique was first employed in \citet{micciancio07worst} to improve the hardness proof certain lattice-based problems. More precisely, it was used to reduce the approximation factors to nearly linear in $n$ (i.e., $\Tilde{O}(n))$ of the lattice problems in \citeauthor{ajtai1996generating}'s SIS hardness proof. After being proved as a useful and efficient standalone mathematical tool, this sampling technique was then widely adopted by subsequent works to demonstrate the hardness of certain lattice-based problems, including the popular learning with error (LWE) and ring learning with error (RLWE) problems. This section is primarily based on \cite{micciancio07worst}. 
%\footnote{This section is part of the work \textit{A Tutorial Introduction to Lattice-based Cryptography and Homomorphic Encryption} by the authors Yang Li, Kee Siong Ng, Michael Purcell from the School of Computing, Australian National University @2022.}
We will discuss some essential properties of the discrete Gaussian distribution and how such a distribution can be used to simplify and strengthen the hardness proof of SIS in the preceding section. 

\subsection{Discrete Gaussian distribution}

We start by reviewing some terms and intuitions about the better-understood continuous Gaussian distribution. A \textbf{Gaussian function} is a continuous function of the form 
\begin{equation*}
    f(x) = a \cdot \exp \left(-\frac{(x-c)^2}{2\sigma^2} \right).
\end{equation*}
The mostly common  Gaussian function  is the probability density function (PDF) of the Gaussian distribution. For simplicity, we work with the case when $a=1$, so we can define the \textbf{Gaussian measure} in $\R$ as 
\reversemarginpar
\marginnote{\textit{Gaussian measure}}
\begin{equation*}
    \rho_{\sigma,c}(x) = \exp\left( -\frac{(x-c)^2}{2\sigma^2}\right). 
\end{equation*}
Another algebraic expression of the Gaussian measure is by using a \textbf{scale} parameter $s = \sqrt{2\pi} \sigma$. Substitute $\sigma$ in the above equation and generalize the Gaussian measure to the higher dimensional space $\R^n$, we get 
\begin{equation}
\label{equ:gauMeasure}
    \rho_{s,\vc{c}}(\vc{x}) = \exp\left( -\frac{-\pi ||\vc{x}-\vc{c}||^2}{s^2}\right).
\end{equation}
Integrating the measure over $\R^n$, the total measure is\footnote{The total measure is not 1 because the coefficient $a$ in the Gaussian function is ignored.} % See appendix for a proof.}
\begin{equation*}
    \int_{\vc{x} \in \R^n} \rho_{s,\vc{c}}(\vc{x}) \diff \vc{x} = s^n,
\end{equation*}
hence we can define the $n$-dimensional (continuous) Gaussian 
\reversemarginpar
\marginnote{\textit{Gaussian PDF}}
probability density function as 
\begin{equation}
\label{equ:ctsGauPDF}
    D_{s,\vc{c}}(\vc{x}) = \frac{\rho_{s,\vc{c}}(\vc{x})}{s^n}.
\end{equation}
This is the $n$-dimensional Gaussian PDF that we know from probability theory, but presented in a non-standard way.  
%The expected squared distance from the Gaussian variable $\vc{x} \in \R^n$ to its distribution center $\vc{c}$ (i.e., ) is 
%\begin{equation*}
%    E\left[||\vc{x}-\vc{c}||^2\right] = \sigma^2 = \frac{n s^2}{2\pi}.
%\end{equation*}
%Geometrically, most of the Gaussian samples lie in the $n$-dimensional ball centered at $\vc{c}$ with radius $s\sqrt{n/2\pi}$. %Obviously, the larger the scale $s$ is, the wider the Gaussian distribution is, so the larger the radius of the ball. 

\Cref{equ:gauMeasure} and \Cref{equ:ctsGauPDF} would still make sense if $\vc{x}$ is a non-continuous lattice vector. Since a lattice $L$ is a countable set, the total Gaussian measure over $L$ and the ``discretized'' density function are
\begin{align*}
    \rho_{s,\vc{c}}(L) &= \sum_{\vc{x} \in L} \rho_{s,\vc{c}}(\vc{x})\\
    D_{s,\vc{c}}(L) &= \frac{\rho_{s,\vc{c}}(L)}{s^n}.
\end{align*}
Hence, we can define the \textbf{discrete Gaussian distribution} \index{discrete Gaussian distribution} 
\reversemarginpar
\marginnote{\textit{Discrete Gaussian}}
over the lattice $L$ for all lattice vectors $\vc{x} \in L$ as 
\begin{equation*}
    D_{L,s,\vc{c}}(\vc{x}) = \frac{D_{s,\vc{c}}(\vc{x})}{D_{s,\vc{c}}(L)} = \frac{\rho_{s,\vc{c}}(\vc{x})}{\rho_{s,\vc{c}}(L)}.
\end{equation*}
%This can also be interpreted as the probability of $\vc{x}$ conditioning on the fact that it is a lattice vector. The numerator is the probability of $\vc{x}$ being a lattice vector and follows a Gaussian distribution. The denominator is the probability of an arbitrary Gaussian random variable in $\R^n$ takes on a value of a lattice vector in $L$. 

The discrete Gaussian distribution is commonly used nowadays to introduce randomness in the proof of lattice problems and lattice-based cryptosystems. Unlike a uniform distribution over a space (e.g., the way uniformity was proved in Ajtai's SIVP$_{\gamma}$ to SIS problem), Gaussian distribution does not have sharp boundaries, which is useful when smoothing a distribution over a space. More precisely, given a Gaussian distribution $\rho_{s,\vc{c}}(\vc{s})$ whose center is a lattice point (i.e., $\vc{c} \in L$), if random samples from this distribution are taken modulo the lattice fundamental domain, the resulting samples will induce a distribution within the fundamental domain. Whether or not such a distribution is close to the uniform distribution depends on the scale $s$ of the Gaussian distribution. Obviously, the larger $s$ is, the closer the induced distribution is to uniform. 

% \kl{Work out the statistical distance b/w discrete gaussian and uniform, which motivates the definition of smoothing parameter, see \url{https://cims.nyu.edu/~regev/teaching/lattices_fall_2004/ln/averagecase.pdf}}
To give a quantitative threshold on how large $s$ needs to be, \citeauthor{micciancio07worst} introduced the smoothing parameter. As the name suggests, the purpose of this parameter is to measure the minimum Gaussian noise magnitude, so that if the noise is added to a lattice $\Z^n$, the lattice is ``blured'' to almost a uniform distribution over $\R^n$ (formally stated in \Cref{lm:nearUniform}). 
For the rest of this section, we assume $\epsilon(n)>0$ (or just $\epsilon>0$ if the context is clear) is a negligible function of the space dimension $n$.   

\begin{definition}
\label{def:smthPara}
The \textbf{smoothing parameter} \index{smoothing parameter} 
\reversemarginpar
\marginnote{\textit{Smoothing parameter}}
of an $n$-dimensional lattice $L$, denoted $\eta_{\epsilon}(L)$, is the smallest scale $s$ such that the Gaussian measure gives almost all weights to the origin in the dual lattice, that is, $\rho_{1/s}(L^* \setminus \{0\}) \le \epsilon$.
\end{definition}

The parameter is defined in terms of the dual lattice. A possible reason is that the dual lattice also appears in the \textit{Poisson summation formula} (Lemma 2.8 \cite{micciancio07worst}) that is key tool to prove some properties of the discrete Gaussian distribution, for example, \Cref{lm:nearUniform}.

%In other words, the Gaussian measure gives almost all weights to the origin in the dual lattice. 
% It is not clear why the smoothing parameter is expressed in terms of the dual lattice rather than the lattice itself. Perhaps the motivation is to relate this concept to the dual so that some results can be proved later. 
%Note $\rho_{1/s}(L^* \setminus \{0\})$ is a well-defined decreasing function of $s$ with the range $(0, \infty)$. More precisely, $\lim_{s \rightarrow \infty} \rho_{1/s}(L^* \setminus \{0\}) = 0$ because when $s \rightarrow \infty$ its inverse $\frac{1}{s} \rightarrow 0$, which implies the Gaussian measure puts almost all weights on $0$. Conversely, $\lim_{s \rightarrow 0} \rho_{1/s}(L^* \setminus \{0\}) = \infty$.

Next, we relate the smoothing parameter to two standard lattice quantities. These relations tight the smoothing parameter hence discrete Gaussian, with lattice problems and lattice-based cryptosystems. The proofs of these lemmas can be found in the reference paper. 

\begin{lemma}[Lemma 3.2 \cite{micciancio07worst}]
\reversemarginpar
\marginnote{\textit{relate to $\lambda_1(L^*)$}}
The smoothing parameter of an $n$-dimensional lattice $L$ satisfies $\eta_{\epsilon}(L) \le \frac{\sqrt{n}}{\lambda_1(L^*)}$, where $\epsilon = 2^{-n}$.
\end{lemma}

The key to prove this lemma is to assume the discrete Gaussian scale satisfies $s > \sqrt{n}/\lambda_1(L^*)$, so removing a closed ball of radius $\sqrt{n}/s$ from the dual lattice is the same as removing only the zero vector, that is, $L^* \setminus (\sqrt{n}/s) \mathcal{B} = L^* \setminus \{\vc{0}\}$. This assumption of the scale also inversely relates the smoothing parameter to the shortest vector in the dual lattice as stated in the lemma. The factor $\sqrt{n}$ comes from Equation (5) in Lemma 2.10 \cite{micciancio07worst}.

To intuitively understand the inverse relation between $\eta_{\epsilon}(L)$ and $\lambda_1(L^*)$, the definition of smoothing parameter suggests that the parameter is to give almost all weights to the lattice origin, so the longer the dual's shortest vector is the smaller $\eta_{\epsilon}(L)$ needs to be. This also connects $\eta_{\epsilon}(L)$ with the shortest vector in the original lattice $L$. Given $\lambda_1(L)$ is in an inverse relation with $\lambda_1(L^*)$, hence the smoothing parameter is related to $\lambda_1(L)$.

\begin{lemma}[Lemma 3.3 \cite{micciancio07worst}]
\label{lm:smthParUpperBd}
\reversemarginpar
\marginnote{\textit{relate to $\lambda_n(L)$}}
The smoothing parameter of an $n$-dimensional lattice $L$ satisfies 
\begin{equation*}
    \eta_{\epsilon}(L) \le \sqrt{\frac{\ln (2n(1+1/\epsilon))}{\pi}} \cdot \lambda_n(L).
\end{equation*}
\end{lemma}

We finish this subsection by stating two key properties of the discrete Gaussian distribution. These properties make discrete Gaussian extremely useful when proving the hardness of lattice-based problems and building lattice-based cryptosystems. 

Recall that any vector $t \in \R^n$ in the span of a lattice $L$ is uniquely identifiable by a lattice vector $v$ and a (translation of) vector $w \in F$ in the lattice fundamental domain $F$. This gives rise to a way of reducing an arbitrary vector in $\R^n$ to a vector within $F$ by taking $w = t \bmod F$ the vector modulo the fundamental domain. The next lemma addresses the near uniformity of the distribution over $F$ induced by applying this modulo operation. 

\begin{lemma}[Lemma 4.1 \cite{micciancio07worst}]
\label{lm:nearUniform}
\reversemarginpar
\marginnote{\textit{Near uniformity}}
Let $L$ be an $n$-dimensional lattice and $D_{s,\vc{c}}$ be a Gaussian distribution with arbitrary scale $s \ge \eta_{\epsilon}(L)$ and center $\vc{c} \in \R^n$, the statistical distance between $D_{s,\vc{c}} \bmod F$ and a uniform distribution $U(F)$ over the fundamental domain $F$ is 
\begin{equation*}
    \Delta(D_{s,\vc{c}} \bmod F, U(F)) \le \frac{\epsilon}{2} .
\end{equation*}
\end{lemma}

The uniform distribution over $F$ has a PDF $U(F)=1/\text{vol}(F)=\det(L^*)$, so the proof in \cite{micciancio07worst} employed Poisson summation formula to rewrite the discrete Gaussian in terms of $\det(L^*)$ too, so that this term can be cancelled when computing the statistical distance. As discussed before, this Lemma motivates the definition of smoothing parameter, which is a useful criterion when sampling uniform samples in the fundamental domain from a discrete Gaussian distribution.  
%It can be proved that for $s>0$ the statistical distance $\Delta \le \rho_{1/s}(L^* \setminus \{\vc{0}\})$. The lemma then follows since the Gaussian's scale is at least as large as the smoothing parameter. The proof can be found in Lemma 4.1 of \citet{micciancio07worst}. It uses the Fourier transform of the Gaussian function and the properties of the discrete Gaussian described before.  

The next lemma proves that the discrete and continuous Gaussian distributions share similar characteristics when the scale of the discrete Gaussian is sufficiently large. 

\begin{lemma}[Lemma 4.3 \cite{micciancio07worst}]
\reversemarginpar
\marginnote{Similar to continuous Gaussian}
Let $D_{L, s,\vc{c}}$ be a discrete Gaussian distribution over an $n$-dimensional lattice $L$ with arbitrary scale $s \ge 2\eta_{\epsilon}(L)$ and center $\vc{c} \in \R^n$. For $0 < \epsilon < 1$, the following are satisfied
\begin{align*}
    \left|\left|E_{\vc{x} \sim D_{L, s, \vc{c}}}\left[\vc{x}-\vc{c}\right] \right|\right|^2 &\le \left(\frac{\epsilon}{1-\epsilon}\right)^2 s^2 n,  \\
    E_{\vc{x} \sim D_{L, s, \vc{c}}}\left[\left|\left|\vc{x}-\vc{c}\right|\right|^2\right] &\le \left(\frac{1}{2\pi}+\frac{\epsilon}{1-\epsilon}\right)^2 s^2 n.
\end{align*}
\end{lemma}
The first inequality suggests that on expectation the random samples from $D_{L,s,\vc{c}}$ are close to the distribution center, with the distance at most $s\sqrt{n}$. So if the discrete Gaussian is centered at the origin, the sampled lattice vectors will have norms at most $s\sqrt{n}$. The second inequality suggests the discrete version has almost the same variance as the continuous Gaussian whose variance is $\frac{ns^2}{2\pi}$).

\subsection{Discrete Gaussian for provable security}

In this subsection, we revisit the hardness proof of Ajtai's short integer solution (SIS) problem, but use the discrete Gaussian tool as an important technique to reduce the gaps of the hard lattice problems. \index{discrete Gaussian distribution! SIS security proof}
Recall that SIS is parameterized by a modulus $q$, the number of linearly independent vectors $m$ and a norm bound $\beta$. These parameters are often considered as functions of the security parameter $n$. The purpose of SIS is to find a short integer vector $\vc{x} \in \Z^m$ such that 
\begin{itemize}
    \item $||\vc{x}|| \le \beta$ and 
    \item $A \vc{x} =\vc{0} \in \Z_q^n$ for an arbitrary integer matrix $A \in \Z_q^{n \times m}$.
\end{itemize}
As stated in \Cref{lm:sufficientCondSIS} and Lemma 5.2 in \citet{micciancio07worst}, the norm bound of $\vc{x}$ needs to satisfy $\beta(n) \ge \sqrt{m} q^{n/m}$ in order to guarantee an SIS solution. 

The overal proof strategy in \citeauthor{micciancio07worst} is similar to Ajtai's by introducing an intermediate lattice problem - \textbf{incremental guaranteed distance decoding} - for a simple reduction to SIS. The standard lattice problems can be reduced to this intermediate problem, but are not covered in this section because the focus is the discrete Gaussian sampling technique. %In \citet{ajtai1996generating}'s proof, the intermediate problem is the GAPSBP$_{\gamma}$ or GAPSIVP$_{\gamma}$. 
This intermediate problem is different to the bounded distance decoding (BDD) problem (\Cref{section:lattice theory}), in the sense that it finds a lattice vector within a bounded distance to the target, not necessarily the closest to the target which is given close to the lattice in BDD.

\begin{definition}
\label{def:incgdd}
Given a basis $B$ of an $n$-dimensional lattice $L$, a set of linearly independent lattice vectors $S \subseteq L$, a target vector $\vc{t} \in \R^n$ and a real $r > \gamma(n) \lambda_n(B)$, the \textbf{incremental guaranteed distance decoding (INCGDD)} problem\index{lattice problems! INCGDD} outputs a lattice vector $\vc{v} \in L$ such that $||\vc{v}-\vc{t}|| \le (||S||/g)+r$.
\end{definition}

The norm $||S||$ of the set is the length of the longest lattice vector in $S$. The additional parameter $r$ is needed to guarantee a solution exists for certain settings of $S$ and $g$, as illustrated by the example in \cite{micciancio07worst}. If $S$ is the basis of $\Z^n$ and $g=4$, there is no solution to the target $\vc{t} = (1/2, \dots, 1/2)$ satisfies $||\vc{v}-\vc{t}|| \le ||S||/g=1/4$, since the closest lattice vector is at distance $\sqrt{n}/2$. Hence, if $\gamma(n) = \sqrt{n}/2$ and $\phi(B) = \lambda_n(B)$, then $r > \sqrt{n}/2 \cdot \lambda_n(B) = \sqrt{n}/2$ and it guarantees a solution $\vc{v}$ where the distance bound $1/4 + \sqrt{n}/2$ is met. Unless otherwise mentioned, the rest of this section assumes $\phi(B)=\lambda_n(B)$.

%For a set of vectors $S=\{\vc{s}_1, \dots, \vc{s}_n\}$, denote $P(S)=\{\sum_{i=1}^n x_i \vc{s}_i \mid x_i \in [0,1)\}$ the half-opened parallelepiped generated by $S$. If $S$ is a lattice basis, $P(S)$ is the fundamental domain. The discrete Gaussian tool is used to prove one of the key steps of the reduction. More precisely, the discrete Gaussian distributions gives rise to a sampling mechanism (as shown next), whose output is a pair $(\vc{c}, \vc{y})$ where $\vc{c}$ is almost uniform in $P(B)$ and $\vc{y}$ is a sample from a discrete Gaussian distribution over the lattice $L(B)$.

Recall $P(B)$ is the fundamental domain (or parallelepiped) of the lattice $L(B)$. This is generalized to the half-opened parallelepiped $P(S)=\{\sum_{i=1}^n x_i \vc{s}_i \mid x_i \in [0,1)\}$ generated by the set of linearly independent vectors $S=\{\vc{s}_1, \dots, \vc{s}_n\}$. 

The next lemma presents a sampling technique to produce uniformly random vectors within a lattice's fundamental domain as well as Gaussian lattice vectors. This sampling procedure is the core technique to reduce INCGDD to SIS as shall be seen later. The intuition of this sampling technique is really simple. It is based on the observation that every vector in $\R^n$ can be uniquely identified by a lattice vector plus a small ``noise'' vector in the shifted fundamental domain. Hence, we generate a Gaussian sample in $\R^n$, then split it into the ``noise'' vector and the lattice vector. The former is almost uniformly distributed in the fundamental domain and the latter follows a discrete Gaussian with a shifted center by the ``noise'' magnitude.   

\begin{lemma}[Lemma 5.7 \cite{micciancio07worst}]
\label{lm:disGauSampling}
Given an $n$-dimensional lattice $L(B)$, a vector $\vc{t} \in \R^n$ and a scale $s \ge \eta_{\epsilon}(L)$ for some $\epsilon>0$, there is a PPT sampling algorithm $\mathcal{S}(B,\vc{t},s)$ to output a pair $(\vc{c},\vc{y}) \in P(B) \times L(B)$ such that 
\begin{itemize}
    \item $\vc{c}$ is nearly (with statistical distance at most $\epsilon/2$) uniformly distributed over $P(B)$,
    \item for any vector $\vc{\hat{c}} \in P(B)$, given $\vc{c}=\vc{\hat{c}}$ it entails $\vc{y} \sim D_{L,s,\vc{t}+\vc{\hat{c}}}$.
\end{itemize}
\end{lemma}

\begin{proof}
The sampling procedure $\mathcal{S}$ simply generates a continuous Gaussian sample $\vc{r} \leftarrow D_{s,\vc{t}}$. This sample is then reduced to within the fundamental domain by $\vc{c} = -\vc{r} \bmod P(B)$. Since the Gaussian scale is at least as large as the smoothing parameter, it implies that this sample is nearly uniformly random by \Cref{lm:nearUniform}.

Let $\vc{y} = \vc{r}+\vc{c}$. Since $\vc{c} = -\vc{r} \bmod P(B)$, it implies $\vc{r} = \vc{v} - \vc{c}$, where $\vc{v} \in L(B)$ is a lattice vector. Hence, $\vc{y}$ is a lattice vector. For any $\hat{\vc{c}} \in P(B)$, the new sample $\vc{r}+\hat{\vc{c}} \sim D_{s,\vc{t}+\hat{\vc{c}}}$ is still Gaussian with a shifted center. Since $\vc{y} = \vc{r}+\vc{c}$, the condition $\vc{c} = \hat{\vc{c}}$ is the same as saying $\vc{y} = \vc{r}+\hat{\vc{c}}$ is a lattice vector. Therefore, the distribution of $\vc{y}$ conditioning on $\vc{y}$ being a lattice vector (equivalently $\vc{c} = \hat{\vc{c}}$) is just the discrete Gaussian distribution $D_{L,s,\vc{t}+\vc{\hat{c}}}$.

\end{proof}

From the outputs of the sampling procedure, one is able to build a random matrix $A$ to call the SIS oracle to produce a short non-zero integer vector $\vc{x}$ that is an SIS solution. More importantly, $\vc{x}$ is used to produce a lattice vector $\vc{s}$ that is the solution of the INCGDD problem. Let the $n$ by $m$ matrix $C = [\vc{c}_1, \dots, \vc{c}_m] \in P(B)^m$ be the output by running the sampling procedure $m$ times, where each $\vc{c}_i$ is one part of the pair $(\vc{c}_i, \vc{y}_i) \leftarrow S(B, \vc{t}, s)$.

\begin{lemma}[Lemma 5.8 \cite{micciancio07worst}]
	Given an $n$-dimensional lattice $L(B)$, a full-rank sublattice $S \subseteq L(B)$, the sampling output $C=[\vc{c}_1, \dots, \vc{c}_m]$ and an integer $q$, there is a PPT algorithm $\mathcal{A}^{\mathcal{F}}(B, S, C, q)$ that makes a single call to the SIS oracle $\vc{z} \leftarrow \mathcal{F}(A)$ to produce a vector $\vc{x} \in \R^n$ such that 
	\begin{itemize}
		\item $A$ is uniformly random,
		\item $\vc{x} \in L(B)$ is a lattice vector, 
		\item $||\vc{x}-C\vc{z}|| \le \sqrt{m}n||S|| ||\vc{z}||/q$.
	\end{itemize}
\end{lemma}

Recall that a strong motivation to study discrete Gaussian distribution is to simply Ajtai's SIS reduction. The following proof indeed states a simpler way of building a random matrix $A$ for the SIS oracle.  

\begin{proof}
The PPT procedure is as follows:
\begin{enumerate}
	\item Generate uniformly random lattice vectors $\vc{v}_1, \dots, \vc{v}_m \in L(B) \bmod P(S)$.
	\item Build the matrix $W=[\vc{w}_1, \dots, \vc{w}_m]$ where $\vc{w}_i=\vc{v}_i + \vc{c}_i \bmod P(S)$.
	\item Build the matrix $A=\floor{qS^{-1}W} \in \Z_q^{n\times m}$.
	\item Invoke the SIS oracle $\vc{z} \leftarrow \mathcal{F}(A)$.
	\item Output the vector $\vc{x}=(C-W+SA/q)\vc{z}$.
\end{enumerate}
Since $\vc{v}_i$ and $\vc{c}_i$ are all uniformly random, so is their modulo sum $\vc{w}_i$. The first two steps create uniformly distributed samples within the parallelepiped $P(S)$. They are much simpler than the procedure in Ajtai's reduction, which has to start with a larger parallelepiped to ensure near orthogonal which is a key step to generate uniform samples from the smaller parallelepiped. From here, it is not hard to see $A$ is uniform too. 

Step 2 suggests that $W=V+C$, so $C-W=-V$ contains only lattice vectors. Given $\vc{z}$ is an SIS solution, $SA\vc{z}/q=\vc{k}S$ for an integer vector $\vc{k}$. Hence, $\vc{x}=-V\vc{z}+\vc{k}S$ is also a lattice vector in $L(B)$. We skip the last part of the proof which can be found in \cite{micciancio07worst}.

\end{proof}

We finish this section by stating the final reduction theorem without proving it. The proof of this theorem is nothing but calling the two procedures above to produce an INCGDD solution, and a justification that the change of producing a solution is non-negligible. %All the parameters $g, m, \beta, \epsilon, q, \gamma$ are functions of $n$, so this functional relationship is omitted for simplicity. 

\begin{theorem}
	For any $g(n)>0$, polynomially bounded functions $m(n), \beta(n)=n^{O(1)}$, negligible function $\epsilon(n)=n^{-\omega(1)}$, and $q(n) > g(n) n \sqrt{m(n)} \beta(n)$, there is a PPT reduction from INCGDD$_{\gamma,g}^{\eta_{\epsilon}}$ for $\gamma(n) = \beta(n)\sqrt{n}$ to SIS$_{q,m,\beta}$, so that if there is a solution to a random SIS instance then it solves INCGDD in the worst case with a non-negaligible probability.
\end{theorem}

%The parameter $\gamma(n)$ is the approximation factor...

%The output of the sampling mechanism is then used by a combining procedure to solve the INCGDD problem together with the hypothetical SIS oracle. We skip that part of the reduction and refer the reader to Lemma 5.8 and Theorem 5.9 of \citet{micciancio07worst} for the detailed proofs.  

%should come back to this subsection to enrich the content, in particular, emphasize the simplicity of using gaussian sampling to produce inputs for the SIS oracle, 1) no need to construct a large hypercube whose volumn, surface area and min height are bounded below and above, 2) no need to ensure smaller hypercubes are near orthogonal, 

%Comparing with the strategy of generating random ring elements in $\Z_p^n$ in Ajtai's reduction, that is, constructing a large fundamental domain then cutting it down gradually, the discrete Gaussian sampling mechanism is much simpler. This is also the key of reducing the gap factors in the hard lattice problems. We will not go through the rest of the proof of a reduction from INCGDD to SIS, the readers and refer to Section 5.2 of \citep{micciancio07worst}. 

%%%%%%%%%%%%%%%%%%%%%%%%%%%%%%%%%%%%%%%%%%%%%%%%%%%%%%%%%%%%%%%%%%%%%%%%%%%%%%%%%%%%%%%%%%%%%%%%%%%

%\newpage
%\bibliography{references}
%\bibliographystyle{abbrvnat}

\newpage
\section{Learning with Errors}

%\section{Learning with errors (LWE)}
\label{section:lwe}

% \kl{lwe has been used as the security base for many applications including public key encryption, HE, etc, see the 4th paragraph in \citet{lyubashevsky2010ideal}.}

In Section \ref{section:lattice theory}, we have introduced the SIS\index{SIS} problem, which is an average-case problem whose difficulty is based on the worst-case hardness of three lattice problems. The main drawback of SIS-based cryptosystems is the impractical public key size and ciphertext size. Typically, the key size is $\Tilde{O}(n^4)$ and the plaintext size is $\Tilde{O}(n^2)$, where $n$ is a security parameter with typical values in the hundreds. \footnote{$\Tilde{O}(\cdot)$ is a variation of the $O(\cdot)$ notation that ignores logarithmic terms: $\Tilde{O}(g(n))=O(g(n) \log^k n)$ for some $k$. This time complexity class is known as \textbf{quasilinear time} and sometimes expressed as $O(n^{1+\epsilon})$ for an $\epsilon >0$.}
% \kl{Remain of the SIS problem and give an example of a specific SIS-based cryptosysmte, so the readers can see the computational inefficiency.}

The \textbf{learning with error (LWE)}\index{LWE} problem was introduced by \citet{regev05} as another foundational problem for building lattice-based cryptosystems with provable security but smaller key and ciphertext size. 
% It allows one to build a cryptosystem with smaller key size and ciphertext size, while still retaining the same provable security from some worst-case lattice problems. 
In particular, LWE-based cryptosystems' public key size is $\Tilde{O}(n^2)$, which is a considerable improvement from SIS-based ones, although still not practical for large $n$. In addition, the plaintext size  is increased by only $\Tilde{O}(1)$ times once encrypted.

Intuitively, the LWE problem tries to recover a secret key from a system of noisy linear equations. To draw an analogy, if the linear equations are not noisy, the problem can be solved efficiently using Gaussian elimination as shown in the following example. 

\begin{example}
Given three linear equations of the form $Ax=B$, where $A$ is a 3 by 3 matrix, $B$ is a 3 by 1 matrix and $x$ is a 1 by 3 matrix, we can use Gaussian elimination (a.k.a. row reduction) to turn $A$ into an upper triangular matrix, hence solving for the solution $x$. 
\begin{small}
\[
\left[
\begin{array}{ccc|c}
1 & 3 & 1 & 9  \\
1 & 1 & -1 & 1  \\
3 & 11 & 5 & 35  \\
\end{array}
\right]
\]    

\[
\left[
\begin{array}{ccc|c}
1 & 3 & 1 & 9  \\
0 & -2 & -2 & -8  \\
0 & 2 & 2 & 8  \\
\end{array}
\right]
\]    

\[
\left[
\begin{array}{ccc|c}
1 & 3 & 1 & 9  \\
0 & -2 & -2 & -8  \\
0 & 0 & 0 & 0  \\
\end{array}
\right]
\]

\[
\left[
\begin{array}{ccc|c}
1 & 0 & -2 & 3  \\
0 & 1 & 1 & 4  \\
0 & 0 & 0 & 0  \\
\end{array}
\right]
\]
\end{small}
\end{example}
\noindent The LWE problem, however, introduces noises (or errors) into the linear equations, making the above problem significantly harder. More precisely, Gaussian elimination involves linear combinations of rows. This process may amplify the noises so that the resulting rows are unable to maintain the original information that is embedded in the equations. 

\iffalse 
In LWE setting, $\chi$ is often assumed to concentrate on ``small integers''. That is, for small $\alpha \ll 1$,
\begin{equation*}
    Pr\left(\mathbf{x} \leftarrow \chi \mid ||\mathbf{x}|| > \alpha q\right) < negl(n),
\end{equation*}
where $negl(n)$ is a
\reversemarginpar
\marginnote{\textit{Negligible function}}
\textbf{negligible function}.\footnote{A function $\mu: \N \rightarrow \R$ is \textbf{negligible} if for every positive integer $c$, there exists an integer $N_c$ such that for all $x > N_c$, we have $|\mu(x)| < x ^{-c}$.} 
We use the \textbf{big-Theta} notation $f(n) = \Theta(g(n))$ to denote the asymptotic behaviour of the function $f(n)$ if there exists constants $c_1$, $c_2$ and $n_0$ such that for all $n > n_0$, the function is bounded between by $c_1 g(n) \le f(n) \le c_2 g(n)$. 
\fi

%%%%%%%%%%%%%%%%%%%%%%%%%%%%%%%%%%%%%%%%%%%%%%%%%%%%%%%%%%%%%%%%%%%%%%%%%%%%%%%%%%%%%%%%%%%%%%%%%%%

\subsection{LWE distribution}
\label{subsec:lweDist}

We introduce and recall some notations before going into the main content of this section. Denote $\Z / q\Z$ by $\Z_q$ and let $\Z_q^n = \{( a_1, \dots, a_n) \mid a_i \in \Z_q\}$ be its $n$-dimensional generalization. 
The notation $\mathbf{x} \leftarrow \Z_q^n$ indicates $\mathbf{x}$ is uniformly sampled from $\Z_q^n$. 
Let $\T=\R/\Z=[0,1)$ be $\R \bmod 1$.

In regards to errors in the LWE samples, we use $\phi$ and $\chi$ to denote the error distributions over $\T$ and $\Z_q$, respectively. In the hardness proof, \cite{regev2009lattices} set the error distribution $\phi=\Psi_{\alpha}$ which can be obtained by sampling from a continuous Gaussian with mean 0 and standard deviation $\frac{\alpha}{\sqrt{2\pi}}$ (or scale $\alpha$) and reducing the outputs modulo 1. But in practice, these errors are discretized for convenience by multiplying samples from $\Psi_{\alpha}$ by $q$ and rounding to the nearest integer modulo $q$. This gives rise to the discretized error distribution $\Bar{\Psi}_{\alpha}$ over $\Z_q$. 

Throughout his work, \citeauthor{regev2009lattices} proved the hardness result of LWE based on the continuous error distribution $\Psi_{\alpha}$ and only used the discretized error $\Bar{\Psi}_{\alpha}$ when presenting a secure LWE-based cryptosystem. In fact, both error distributions entail the same hardness of the LWE problem as emphasized by Lemma 4.3 of \citet{regev2009lattices}. For simplicity, we present the LWE problem and its hardness proof based on the discretized error distribution $\chi=\Bar{\Psi}_{\alpha}$ over $\Z_q$, the reader should keep in mind the original proofs were based on the continuous error distribution $\phi=\Psi_{\alpha}$ over $\T=\R/\Z=[0,1)$.

\begin{definition}
\label{def:lweDist}
Given the following parameters 
\begin{itemize}\itemsep1mm\parskip0mm
    \item $n$ - the security parameter (usually $n=2^k$ for an integer $k \ge 0$),
    \item $q$ - an integer (not necessarily prime) that is a function of $n$, i.e., $q=q(n)$,
    %\item $m$ - a dimension parameter satisfies $m = \Theta(n \log q)$,
\end{itemize}
a fixed $\mathbf{s} \in \Z_q^n$ and an error distribution $\chi$ over $\Z_q$, %that is concentrated on small integers,
the 
\reversemarginpar
\marginnote{LWE distribution}
\textbf{LWE distribution} \index{LWE! distribution} $A_{\vc{s},\chi}$ over $\Z_q^n \times \Z_q$ is obtained by these steps
\begin{itemize}\itemsep1mm\parskip0mm
    \item sample a vector $\vc{a} \leftarrow \Z_q^n$,
    \item sample a noise element $\mathbf{\epsilon} \leftarrow \chi$ over $\Z_q$, 
    \item compute $b = \mathbf{s} \cdot \mathbf{a} + \epsilon \bmod q$,
    \item output $(\mathbf{a}, b)$.
\end{itemize}
\end{definition}

% The security parameter $n$ is often implicitly mentioned. 
The integer $q$ which controls the size of the ring $\Z_q$ is often a large integer and a function of $n$, but it does not need to be a prime number for the hardness proof of the LWE search problem. It is only required to be a prime when reducing the search to decision LWE, in which the ring $\Z_q$ needs to be a field to build the connection between the two problems as we will see next. 

It has been demonstrated that solving a system of exact linear equations can be done efficiently with Gaussian elimination, but solving a system of noisy linear equations is conjectured to be hard.\footnote{Another way of seeing the hardness of this problems is that LWE is a generalization of the \textit{Learning Parity with Noise} problem \citep{pietrzak12}, in which $q=2$ and the error distribution $\chi$ is a Bernoulli distribution with $p(1)=\epsilon$ and $p(0)=1-\epsilon$. This problem is believed to be hard too.} This motivates the search version of the LWE problem stated next. For simplicity, we denote by $(\vc{A},\vc{b}) \subseteq \Z_q^{n \times N} \times \Z_q^N$ the $N$ samples generated from a LWE distribution. 

\begin{definition}
Given the parameter $q$ and the error distribution $\chi$ over $\Z_q$, the \textbf{search version of the LWE} (or just \textbf{LWE)} problem \index{LWE! search}, denoted by LWE$_{q,\chi}$, is to compute the secret key $\mathbf{s}$ given samples  $(\vc{A},\vc{b})$ from the LWE distribution $A_{\vc{s},\chi}$. 
\end{definition}

Although all hardness proofs were done on search LWE, the decision version is what is often used to build secure cryptosystems upon. 

\begin{definition}
Given the parameter $q$ and the error distribution $\chi$ over $\Z_q$, the \textbf{decision version of the LWE} (or \textbf{DLWE)} problem \index{LWE! decision (DLWE)}, denoted by DLWE$_{q,\chi}$, is to distinguish between the LWE samples $(\mathbf{A}, \mathbf{b})$ and uniformly random samples $(\mathbf{A}, \mathbf{u})$ over $\Z_q^{n \times N} \times \Z_q^N$. 
\end{definition}

An efficient reduction \index{LWE! search to decision}
\reversemarginpar
\marginnote{\textit{Search to decision}}
from LWE to DLWE can be constructed so that if there is a solution for DLWE, there is a solution for LWE. The reduction is by applying the same procedure to guess (at most $poly(n)$ times) each element $s_i$ of the secret key $\vc{s}$. To guess the first element $s_1$, we generate a random $r \in \Z_q$ and add it to the first element of each column vector $\vc{a}_i \in \Z_q^n$, so we get the new random column vectors
\begin{equation*}
    \Tilde{\vc{a_i}} = \vc{a_i}+ (r, 0, \dots, 0) \in \Z_q^n.
\end{equation*}
To utilize the DLWE oracle, we output the pair
\begin{align}
\label{equ:dlweToSlwe}
    (\Tilde{\vc{a_i}}, b+r \cdot k \bmod q)
\end{align}
for each $k \in \Z_q$. If $k$ is the correct guess of the first secret vector component, i.e., $k=s_1$, then $b+r\cdot k = \Tilde{\vc{a_i}} \cdot \vc{s} + \epsilon_i \,(\bmod \,q)$, so the corresponding pair in \Cref{equ:dlweToSlwe} looks like $(\Tilde{\vc{a_i}}, \Tilde{\vc{a_i}} \cdot \vc{s} + \epsilon_i)$ which follows the LWE distribution. If $k \neq s_1$, then the corresponding pair is uniform in the domain $\Z_q^n \times \Z_q$, provided $q$ is prime to make $\Z_q$ a field so the product $r\cdot k$ can map to each field element with equal chance. Apply the DLWE oracle to distinguish the LWE pair from the uniform pair to obtain the correct guess of $s_1$. We have a simple reduction from LWE to DLWE. 

Before going forward, it should be made clear that there are different variants of LWE from three different perspectives, which are decision or search, discrete or continuous error distribution, average-case or worst-case. We have explicitly discussed the first two perspectives above. The last one suggests that the LWE distribution and LWE problem can be defined either for all secret $\vc{s}$ or for a uniform random $\vc{s}$. The next lemma shows a reduction from the search, continuous error, worst-case LWE to decision, discrete error, average-case LWE.  

\begin{lemma}
Let $q=poly(n)$ be a prime integer, $\phi$ be an error distribution over $\T$ and $\Bar{\phi}$ be its discretization over $\Z_q$. Assume there is a DLWE$_{q,\Bar{\phi}}$ oracle that distinguishes the LWE distribution $A_{\vc{s},\Bar{\phi}}$ from the uniform distribution for a non-negligible fraction of $\vc{s}$, then there is an efficient algorithm that solves LWE$_{q,\vc{\phi}}$ for all $\vc{s}$. 
\end{lemma}

To keep things simple in this paper, we illustrate the hardness proof in terms of the search, discrete error, worst-case LWE problem. The only difference from the original proof is the discretized error distribution rather than continuous.

%%%%%%%%%%%%%%%%%%%%%%%%%%%%%%%%%%%%%%%%%%%%%%%%%%%%%%%%%%%%%%%%%%%%%%%%%%%%%%%%%%%%%%%%%%%%%%%%%%%

\subsection{LWE hardness proof}
\label{subsec:lweHardnessProof}

%Below, we state the main hardness result of the LWE problem.

\begin{theorem}[Theorem 1.1 \citep{regev2009lattices}]
Let $n, p$ be integers and $\alpha \in (0, 1)$ be such that $\alpha p > 2 n$. If there exists an efficient algorithm that solves $LWE_{p,\Bar{\Psi}_{\alpha}}$ then there exists an efficient quantum algorithm that approximates the decision version of the shortest vector problem (GAPSVP) and the shortest independent vectors problem
(SIVP) to within $\Tilde{O}(n/\alpha)$ in the worst case.   
\end{theorem}

\begin{figure}[hbt!]
    \centering
    \includegraphics[page=14]{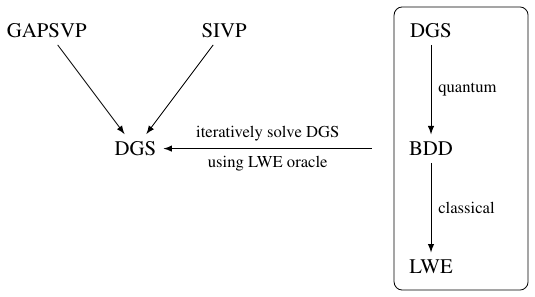}
    \caption{Reductions to the LWE decision problem. If DGS can be solved for a small scale $r$ close to its lower bound $\sqrt{2n} \eta_{\epsilon}(L)/\alpha$, then both lattice problems can be solved with close to optimal solutions. The key to solve DGS for small $r$ is to iteratively apply a subroutine to gradually reduce the scale. The subroutine supplies discrete Gaussian samples to an LWE oracle to classically solve BDD, the result of which is then used by a quantum algorithm to produce shorter discrete Gaussian samples.}
    \label{fig:lweReduction}
\end{figure}

The major steps of the hardness proof of the LWE problem, as outlined by \citeauthor{regev2009lattices}, is sketched in \Cref{fig:lweReduction}. In the box, there is a classical (i.e., non-quantum) reduction from BDD to LWE, which suggests LWE is hard. The more preferable reduction is from the more standard (and well studied) lattice problem GAPSVP, but involves both quantum and classical reductions.\index{LWE! hardness proof} The focus of this subsection is the classical reduction in the box. For details of the others steps, the read is referred to the original paper \citep{regev2009lattices}.

As it is often convenient to build a cryptosystem based on DLWE and there is an efficient reduction from LWE to DLWE, if there is a solution to the cryptosystem, such a solution can be used to solve LWE. This in return can solve the worst-case GAPSVP (and SIVP) using a quantum algorithm, which is conjectured to be difficult with high confidence.
Note that the assumption that these lattice problems are hard to be solved using quantum algorithms is a stronger assumption than using classical algorithms, which obviously are more difficult to be achieved. \citet{peikert2009public} proposed a classical reduction that can replace the quantum step in this proof, but compromising the hardness to be based on non-standard (variant) of lattice problems, or a large modulus $q$ that weakens a cryptosystem's security that is inverse proportional to the size of $q$. 

\begin{theorem}[Theorem 3.1 \citep{regev2009lattices}]
Let $\epsilon=\epsilon(n)$ be some negligible function of $n$. Also, let $p = p(n)$ be some integer and $\alpha=\alpha(n)\in (0, 1)$ be such that $\alpha p > 2 n$. Assume that we have access to an oracle $W$ that solves $LWE_{p,\Psi_{\alpha}}$ given a polynomial number of samples. Then there exists an efficient quantum algorithm for $DGS_{\sqrt{2n} \eta_{\epsilon}(L)/\alpha}$.
\end{theorem}

The \textit{Discrete Gaussian Sampling} (DGS) problem 
\reversemarginpar
\marginnote{\textit{DGS problem}}
is defined as generating a lattice vector in $L$ according to a discrete Gaussian distribution $D_{L,r}$ over $L$ with the scale $r \ge \sqrt{2n} \eta_{\epsilon}(L)/\alpha$ that is larger than the lattice's smoothing parameter $\eta_{\epsilon}(L)$. 
%It is with high confidence that both GAPSVP$_{\gamma}$ and SIVP$_{\gamma}$ are hard to approximate within polynomial factors.
For the sake of explaining only the BDD to LWE reduction, we accept (without proving) that GAPSVP$_{\gamma}$ and SIVP$_{\gamma}$ are more likely to be solved if DGS can be performed with as small scale $r$ as possible. Hence, it is sufficient to show that one can run DGS with a small $r$.
It turns out that this can be achieved by using an LWE oracle and an iterative step which involves the use of classical and quantum algorithms (in the box of Figure \ref{fig:lweReduction}) in order to produce samples from a discrete Gaussian distribution with small $r$. More specifically, starting from $n^c$ samples of a discrete Gaussian distribution $D_{L, r}$ where $r$ is large, the iterative step is able to produce $n^c$ samples from a narrower Gaussian distribution $D_{L, r'}$ where $r' < r/2$. Repeating this step a polynomial number of times so that the last step produces samples from a Gaussian $D_{L, r_0}$ where the width $r_0 \ge \sqrt{2n} \eta_{\epsilon}(L)/\alpha$ reaches its lower bound. One part of the iterative step requires an LWE oracle and an efficient DGS algorithm for $r > 2^{2n} \lambda_n(L)$ to solve the intermediate problem using a classical algorithm. The intermediate problem is CVP for a given vector that has bounded norm, which is also known as the \textit{Bounded Distance Decoding} (BDD) problem \index{lattice problems! BDD}. The efficient DGS algorithm for large scale is proved plausible by the Bootstrapping Lemma 3.2 of \citet{regev2009lattices}.  The other part of the iterative step is a quantum algorithm that uses the solution of the intermediate problem to solve DGS for a narrower distribution that is at most half of the previous scale. The quantum part is out of the scope of this material, hence is not included. 

The classical step was demonstrated using the special lattice $L=\Z^n$ in a follow up paper (Proposition 2.1 \citep{regev2010learning}). Although the original reduction in \citet{regev2009lattices} involves working in the dual lattice $L^*$, the lattice and its dual are identical when $L=\Z^n$. Note as BDD can be solved easily in $\Z^n$ (without the LWE oracle), so this restricted context is for demonstration purpose only and does not guarantee LWE hardness. 

\begin{proposition}
\label{prop:bddToLWE}
Let 
\reversemarginpar
\marginnote{\textit{BDD to LWE}}
$q \ge 2$ be an integer and $\alpha \in (0,1)$ be a real number. Assume there is an LWE oracle for the modulus $q$ and error distribution $\Psi_{\alpha}$. Then, given as input 
an $n$-dimensional lattice $L$, 
a sufficient polynomial number of samples from the discrete Gaussian distribution $D_{L^*, r}$ %for some (not too small) $r$, 
and a BDD instance $\vc{x} =\vc{v}+\vc{e} \in \R^n$ such that $||\vc{e}|| \le \alpha q/\sqrt{2}r$, there is a polynomial time algorithm finds the (unique) closest lattice vector $\vc{v} \in L$.
\end{proposition}

It is worth mentioning that the scale $\alpha$ of the error distribution $\Psi_{\alpha}$ for LWE is restricted to $(0,1)$ in order to ensure the Gaussian error distribution is still distinguishable from the uniform distribution once reduced to within a smaller region. In fact, as long as $\alpha < \eta_{\epsilon}(L)$, the Gaussian error is still distinguishable. This implies that it is sufficient to have $\alpha \in (0, O(\sqrt{\log n}))$, because the smoothing parameter $\eta_{\epsilon}(L) \le O(\sqrt{\log n}) \cdot \lambda_n(L)$ by Lemma \ref{lm:smthParUpperBd} and the $n$th successive minima $\lambda_n(\Z^n)=1$.

\begin{proof}[Sketch of proof]
To utilize the LWE oracle, we wish to construct random LWE samples from the given BDD instance $\vc{x}$ such that its closest lattice vector $\vc{v}\in L$ is the secret vector $\vc{s} \in \Z_q^n$ for the LWE distribution. Hence, the problem becomes producing from the given BDD instance sufficient LWE samples in the domain $\Z_q^n \times \Z_q$. %This corresponds to the classical reduction as shown in Figure \ref{fig:lweReduction}.

To do so, we need help from the given discrete Gaussian samples. The rational is that such a discrete Gaussian sample behaves like a random element in a smaller domain after modulo reduction. Furthermore, it still distributes normally after multiplying with a random continuous element. So by manipulating this discrete Gaussian element, it outputs an LWE sample that can be used by the oracle. More precisely, sample $\vc{y}$ according to the discrete Gaussian distribution $D_{\Z^n, r}$ over $\Z^n$ with a relatively large scale $r$, then output the pair
\begin{align}
\label{equ:bdd2lwe}
    (\vc{a}=\vc{y} \bmod q, b=\lfloor \langle \vc{y}, \vc{x}\rangle \rceil \bmod q) \in \Z_q^n \times \Z_q.
\end{align}
To see why the pair is in the LWE domain, we notice first $r$ being large ensures that $\vc{y}$ is almost uniformly distributed in $\Z_q^n$. This is consistent with LWE's first component distribution.

Expressing $\vc{y}$ in terms of $\vc{a}$ and $q$, we get $\vc{y}=q \Z^n + \vc{a}$. Substitute $\vc{y}$ and $\vc{x}$ into Equation \ref{equ:bdd2lwe}, we get 
\begin{align*}
    b &= \lfloor \langle q\Z^n, \vc{v} \rangle + \lfloor \langle \vc{a}, \vc{v} \rangle + \langle \vc{y}, \vc{e}\rangle \rceil \bmod q\\
    &=\lfloor \langle \vc{a}, \vc{v} \rangle + \langle \vc{y}, \vc{e}\rangle \rceil \bmod q.
\end{align*}
%Both the first and second terms are in $\Z_q$, because of the rounding to the nearest integer and modulo $q$. 
The first term is an integer, so rounding is ignored. For the second term, since $\vc{y} \in D_{\Z^n, r}$ its expected norm is roughly $||\vc{y}|| \le \sqrt{n}r$. In addition, given $||\vc{e}||\le \alpha q / \sqrt{2}r$, then by Corollary 3.10 of \citet{regev2009lattices}, the second term is almost normally distributed with norm approximately at most $\alpha q \sqrt{n/2}$ and then reduced to roughly $\alpha \sqrt{n/2}$, which is consistent with the error distribution $\Psi_{\alpha}$ for the LWE oracle. Therefore, the pair $(\vc{a},b)$ follows the LWE distribution and hence can be used by the oracle to recover the secret key $\vc{s}$.

Since $\vc{s}=\vc{v} \bmod q$, the LWE oracle and the modulo operation reveal the least significant digits of $\vc{v}$ in base $q$. Next, we update the non-lattice vector from $\vc{x}$ to $(\vc{x} - \vc{s})/q \in \R^n$ which gets rid of the least significant digits of $\vc{x}$, and employ the above BDD to LWE process to search for the next set of least significant digits in base $q$ in the new secret vector $(\vc{v} - \vc{s})/q  \bmod q \in L$. % , which is essentially the second least significant digit in the original vector $\vc{v} \in L$. 
Iterating this process enough times, we will recover the entire closest lattice vector $\vc{v} \in L$ to the given BDD instance $\vc{x}$. 
\end{proof}

Two remarks about the proof. First, to completely hide the discreetness of $\vc{y}$ by additive noise, additional Gaussian noise is needed to add to $b$ as shown in Equation 12 of \citet{regev2009lattices}. Second, the assumed LWE oracle may only work for a noise distribution of a certain magnitude. However, the noise magnitude $\langle \vc{y},\vc{e} \rangle$ is strongly related to the distance $\vc{e} = \vc{x} -\vc{v}$ from the given vector to the lattice. The way to address this potential issue is by adding to the second element $b$ in equation \ref{equ:bdd2lwe} an extra noise, whose magnitude can be varied to ensure the LWE oracle works (Lemma 3.7 \citep{regev2009lattices}). We will see in Section~\ref{section:rlwe} that this becomes a challenge in the ring-LWE problem, in which a vector of Gaussian noises is added rather than a single noise whose effect on the result is much easier to be controlled.  

The last paragraph of the above proof is formalized in the next lemma for general lattices. It gives rise to reduction from CVP$_{L,d}$ to CVP$_{L,d}^{(q)}$. The latter problem is to find the closest lattice vector reduced modulo $q$. 
That is, for a given vector $\vc{x}=\vc{v}+\vc{e} \in \R^n$ with $||\vc{e}||\le d$, finds the coefficient vector $L^{-1} \vc{v} \bmod q \in \Z_q^n$. Here, the notation $L$ is used in a non-standard way to denote the basis matrix, where the columns of $L$ are the basis vectors $\vc{v}_1, \dots, \vc{v}_n$, so $L^-1$ is the inverse of the basis matrix. 
 
\begin{lemma}[Lemma 3.5 \citep{regev2009lattices}]
\label{lm:lweCoeffModQ}
Given a lattice $L$, an integer $p \ge 2$ and a CVP$_{L,d}^{(p)}$ oracle for $d < \lambda_1(L)/2$, there is an efficient algorithm that solves CVP$_{L,d}$.   
\end{lemma}

\begin{proof}
The lemma can be proved using the same bit-by-bit iterating strategy as in the special case $L=\Z^n$ in the above proof. Let $\vc{x} = \vc{v}+\vc{e} \in \R^n$ be a BDD instance. Create a sequence of vectors $\vc{x}_1 = \vc{x}, \vc{x}_2, \dots$. Start from $\vc{x}_1$, use the CVP$_{L,d}^{(p)}$ oracle to find the coefficient vector $\vc{a}_1 = L^{-1} \vc{v}_1 \bmod q$ of $\vc{x}_1$'s, and update the vector by 
\begin{align*}
    \vc{x}_{i+1} = (\vc{x}_i - L(\vc{a}_i \bmod q))/p,
\end{align*}
where $L(\vc{a}_i \bmod q)$ denote the lattice vector corresponds to $\vc{a}_i \bmod q$, the least significant bit of the coefficient vector in base $q$. Substitute $\vc{x}_i= \vc{v}_i + \vc{e}_i$ into the above equation, we get 
\begin{align*}
    \vc{x}_{i+1} = (\vc{v}_i - L(\vc{a}_i \bmod q))/q + \vc{e}_i/q,
\end{align*}
where the error is reduced by a factor of $q$ in the updated instance. Repeat this process $n$ times, we get a BDD instance $\vc{x}_{n+1}$ with much smaller error $||\vc{e}_{n+1}||\le d/p^n$.
Unlike in the special case where the process is repeated to solve all bits of the vector, it is sufficient to get down to $\vc{x}_{n+1}$ that is very close to the lattice, then use an algorithm (e.g., the nearest plane algorithm \citep{babai1986lovasz}) to solve for its closest lattice vector $\vc{a}_{n+1}$. Work backwards to add the solved bits to $\vc{a}_{n+1}$, we obtain a solution $\vc{a}_1$ for the given BDD instance $\vc{x}_1$.
\end{proof}

%%%%%%%%%%%%%%%%%%%%%%%%%%%%%%%%%%%%%%%%%%%%%%%%%%%%%%%%%%%%%%%%%%%%%%%%%%%%%%%%%%%%%%%%%%%%%%%%%%%

\subsection{An LWE-based encryption scheme}
\label{subsec:lweSecurity}

To finish off this section, we state the LWE-based encryption scheme that was proposed by \citeauthor{regev2009lattices}\index{Regev's LWE-based encryption scheme}. Later, this scheme became a popular building block for LWE-based homomorphic encryption schemes as we will see in \Cref{sec:he} (especially in the second generation of homomorphic encryption schemes). 

The scheme is parameterized by $n$, $N$, $q$ and $\chi$ that correspond to the dimension (or security parameter), sample size, modulus and the noise distribution over $\Z_q$ of, same as the setting for the LWE distribution. The parameters need to be set to appropriate values to ensure the system is correct, secure and efficiently computable. An example setting in \cite{regev2009lattices} is taking a prime number $q \in [n^2, 2n^2]$, $N=(1+\epsilon)(n+1)\log q$ for an arbitrary constant $\epsilon > 0$, and $\chi=\Bar{\Psi}_{\alpha(n)}$, where the scale $\alpha(n)=1/(\sqrt{n}\log^2 n)$

For 
\reversemarginpar
\marginnote{\textit{Correctness}}
the correct choices of the parameters, it can be proved (Lemma 5.1 and Claim 5.2  \citep{regev2009lattices}) that there is only a negligible chance that the norm of an error sampled from the distribution $\chi$ is greater than $\lfloor \frac{q}{2}\rfloor/2$. Hence, when decrypting the ciphertext of 0, the scheme gives $c_2-\vc{s}\cdot \vc{c}_1=\sum_{i \in S} \vc{\epsilon_i}$, whose norm $|\sum_{i \in S} \vc{\epsilon_i}| < \lfloor \frac{q}{2}\rfloor/2$, which implies the result is closer to 0 than to $\lfloor \frac{q}{2}\rfloor$. Use the same argument, the decryption of the ciphertext of 1 is also correct. 

The
\reversemarginpar
\marginnote{\textit{security}}%the security proof can be enriched 
semantic security of the cryptosystem is based on the hardness of the DLWE problem. If there is a PPT distinguisher that can tell apart the encryptions of 0 and 1, then we can build another distinguisher that tells apart the LWE distribution from the uniform distribution for a non-negligible fraction of all secret keys $\vc{s}$ (Lemma 5.4 \citep{regev2009lattices}). More specifically, assuming $W$ is a distinguisher between the encryptions of 0 and 1, that is, $|p_0(W)-p_1(W)| \ge \frac{1}{n^c}$ for some constant $c > 0$, then it is possible to build another distinguisher $W'$ such that $|p_0(W')-p_u(W')| \ge \frac{1}{2n^c}$. By the above remark, it is sufficient to prove a DLWE distinguisher for a non-negligible fraction of $\vc{s}$. Define a set $Y=\{\vc{s} \mid |p_0(\vc{s})-p_u(\vc{s})| \ge \frac{1}{4n^c}\}$. Construct a distinguisher $Z$ that estimates $p_0((\vc{A},\vc{b}))$ and $p_u((\vc{A},\vc{b}))$ up to an additive error $\frac{1}{64n^c}$ by applying $W'$ a polynomial number of times. Then $Z$ accepts if the two estimates differ by more than $\frac{1}{16n^c}$, otherwise it rejects. 
% It remains to show that $Z$ behaves noticeably different for ciphertext encrypted by public keys from the LWE distribution or from the uniform distribution over the same domain. The details are skipped.  

\begin{tcolorbox}
\noindent
\textbf{Private key:} choose a private key $\vc{s} \leftarrow \Z_q^n$.\\
\textbf{Public key:} choose a public key $(\vc{A},\vc{b})$, where $\vc{A} = \left[\vc{a_1}, \dots, \vc{a_N}\right] \leftarrow \Z_q^{n \times N}$ and $\vc{b} = \vc{s} \cdot \vc{A} + \vc{\epsilon}$ for random $\vc{\epsilon} \leftarrow \chi^N$.\\
\textbf{Encryption:} to encrypt a message $m \in \{0, 1\}$, choose a random subset $S \subseteq [N]$, then
\begin{align*}
    Enc(0)&=(\vc{c}_1, c_2) = \left(\sum_{i \in S} \vc{a_i}, \sum_{i \in S} b_i\right), \\
    Enc(1) &= (\vc{c}_1, c_2) =\left(\sum_{i \in S} \vc{a_i}, \lfloor\frac{q}{2}\rfloor + \sum_{i \in S} b_i\right).
\end{align*}
\textbf{Decryption:} given a ciphertext $(\vc{c}_1, c_2)$, then
\begin{align*}
    Dec((\vc{c}_1, c_2)) &=0 \text{ if } c_2-\vc{s} \cdot \vc{c}_1 \text{ is close to 0}\\
    Dec((\vc{c}_1, c_2)) &=1 \text{ if } c_2-\vc{s} \cdot \vc{c}_1 \text{ is close to } \lfloor \frac{q}{2}\rfloor.
\end{align*}
\end{tcolorbox}

\newpage
\section{Cyclotomic Polynomials and Cyclotomic Extensions}

\label{sec:cyclotomic}

Cyclotomic polynomials are frequently used in the construction of homomorphic encryption schemes that are based on the ring learning with error (RLWE) problem \index{ring LWE} as we will see later in this tutorial. The motivations of using cyclotomic polynomials are the fact that cyclotomic fields have additional algebraic properties to reduce encryption scheme's time complexity and also make security proofs feasible by following the LWE proof paradigm. 
In this section, we will introduce the cyclotomic polynomials and the Galois groups of cyclotomic extensions.\index{cyclotomic extension} 
We have tried to make this section as self-contained as possible.
The appendix contains a more general treatment of field extensions and the Galois groups of field extensions for interested readers.
Some useful references for material covered in this section include \citet{mukherjee2016cyclotomic}, \citet{conradcyclotomic} and \citet{porter15cyclotomic}.

\subsection{Cyclotomic polynomials}
Cyclotomic polynomials\index{cyclotomic polynomial} are polynomials whose roots are the primitive roots of unity. To understand what it means, we define next. 

\begin{definition}
For any positive integer $n$, the $n$-th roots of unity\reversemarginpar\marginnote{\textit{Roots of unity}}\index{roots of unity} are the (complex) solutions to the equation $x^n = 1$, and there are $n$ solutions to the equation. 
% 
% A number $r$ is an \textbf{nth root of unity} for a positive integer $n$ if it satisfies $r^n=1$.
\end{definition}

\begin{theorem}\label{thm:roots of unity}
Let $n$ be a positive integer and define $\zeta_n = e^{2\pi i/n}$.
Then the set of all $n$-th roots of unity is given by 
\begin{equation}
    \{ \zeta_n^k \,|\, k = 0,1,\ldots, n-1 \},
\end{equation}
\end{theorem}
\begin{proof}
By Euler's formula\index{Euler's formula}, we have
\[ e^{2\pi i} = \cos(2\pi) + i \sin(2\pi) = 1 \]
and that $(e^{2\pi i})^k = e^{2k\pi i} = 1$ for all $k \in \{0,1,\ldots, n-1\}$.
To solve for $x^n = 1$, note that
\[ x^n = 1 = e^0 = e^{2\pi i} = e^{4\pi i} = e^{6\pi i} = \cdots = e^{2k\pi i}. \]
Raising each term to the power of $1/n$ yields
\[ x = (x^n)^{1/n} = 1 = e^{2\pi i/n} = e^{4\pi i/n} = e^{6\pi i/n} = \cdots = e^{2k\pi i/n}. \]
Therefore, there are $n$ distinct solutions to $x^n = 1$, each given by $\zeta_n^k$, for $k=0,1,\ldots,n-1$
\end{proof}

\iffalse
In the complex field $\C$, for example, both $\pm 1$ are the 2nd roots of unity. In general, the root $r$ can either be a real or complex number. So we can re-write the root $r$ as a complex number in the polar form as 
\begin{align*}
    r^n = 1 &= \cos 2\pi + i \sin 2\pi \\
    &= \cos 2k\pi + i \sin 2k\pi \\
    &= e^{2k \pi i}
\end{align*}
for $k \in [0, \infty)$. Hence, the nth root of unity in this representation is 
\begin{align*}
    r &= \cos 2k\pi/n + i \sin 2k\pi/n \\
    &= e^{2k\pi i/n} 
\end{align*}
for $k \in [0, n-1]$. The nth roots of unity is often denoted by the set $\{\zeta_n^0, \zeta_n^1, \dots, \zeta_n^{n-1}\}$, where $\zeta_n^0=1$ is always a root of itself. The subscript can be dropped if the context is clear. 
\fi

\begin{example}
The 1st root of unity is 1. The 2nd roots of unity are $\zeta_2^0=1$ and $\zeta_2^1=-1$. The 3rd roots of unity are $\zeta_3^0=1$, $\zeta_3^1=-\frac{1}{2}+i\frac{\sqrt{3}}{2}$ and $\zeta_3^2=-\frac{1}{2}-i\frac{\sqrt{3}}{2}$.
\end{example}

\begin{figure}[h]
    \centering
    \includegraphics[page=7]{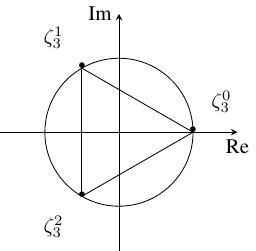}
    \caption{The 3rd roots of unity $\zeta^0=1$, $\zeta^1=-\frac{1}{2}+i\frac{\sqrt{3}}{2}$ and $\zeta^2=-\frac{1}{2}-i\frac{\sqrt{3}}{2}$. We sometimes drop the subscript to simplify the notation to $\zeta^k$ if the context is clear.}
    \label{fig:roots of unity}
\end{figure}

Geometrically, we can interpret the nth roots of unity as the points that are evenly spread on the unit circle in the complex plane, starting from 1 on the real axis. (The word ``cyclotomic'' means ''circle-dividing''.) Equivalently, they are the vertices of a regular n-gon that lies on the unit circle, with the real value 1 as one of the $n$ vertices.  Figure \ref{fig:roots of unity} illustrates the 3rd roots of unity. 

In general, the equation $x^n = 1$ can be defined over different fields.
In the real field $\R$,  the only possible roots of unity are $\pm 1$. In the complex field $\C$, the nth roots of unity form a cyclic group under multiplication. The generator is $e^{2\pi i / n}$ and the group order is $n$, as shown in Theorem~\ref{thm:roots of unity}. 
%
% Both $\R$ and $\C$ are fields of characteristic zero. 
In a finite field, for example $\F_7 =\Z/7\Z= \{0, 1, 2, 3, 4, 5, 6\}$, the 3rd roots of unity are $\{1,2,4\}$, because these are the only numbers equal to 1 modulo 7 when raising to the third power. 

\iffalse
% KS: This requires the definition of the order of a root of unity, which is probably unnecessary here. The Primitive Root definition below is understandable without this next paragraph.
The order of an nth root of unity may not be $n$. For example, the 4th root of unity $-1$ has order 2, but the order of the complex root $i$ is 4. This sets the distinction between the nth roots of unity that are primitive and non-primitive.
\fi

\begin{definition}
An $n$-th root of unity $r$ is called \textbf{primitive}\reversemarginpar
\marginnote{Primitive root}\index{primitive roots of unity}
 if it is not a $d$-th root of unity for any integer $d$ smaller than $n$; i.e. $r^n=1$ and $r^d \neq 1$ for $d < n$. % d, n \in \N$ and $d < n$.
\end{definition}

% Algebraically, $r$ is a primitive $n$-th root of unity if $n$ is the smallest positive integer such that $r^n=1$. 
Geometrically, $r$ is primitive if it is a vertex of a regular polygon that lies on the unit circle, but not a vertex of a smaller regular polygon that lies on the unit circle.  

\begin{figure}[h]
    \centering
    \includegraphics[page=8]{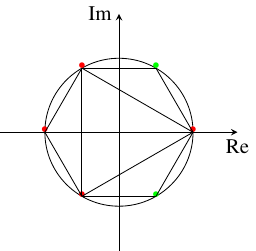}
    \caption{The 6th roots of unity $\zeta^0=1,\zeta^1=\frac{1}{2}+i\frac{\sqrt{3}}{2},\zeta^2=-\frac{1}{2}+i\frac{\sqrt{3}}{2},\zeta^3=-1, \zeta^4=-\frac{1}{2}-i\frac{\sqrt{3}}{2},\zeta^5=\frac{1}{2}-i\frac{\sqrt{3}}{2}$. The primitive roots are $\zeta^1,\zeta^5$ that are coloured in green. $\zeta^0,\zeta^2,\zeta^4$ are not primitive because they are also the 3rd roots of unity. $\zeta^0,\zeta^3$ are not primitive because they are also the 2nd roots of unity.}
    \label{fig:primitive roots}
\end{figure}

\begin{example}
1 is not primitive. The two real roots $\pm 1$ of the 4th roots of unity are not primitive, because they are also the 2nd roots of unity. Both complex roots of the 3rd roots of unity are primitive. The primitive 6th roots of unity are shown in Figure \ref{fig:primitive roots}.
\end{example}

The following theorem provides an easy way to find the $n$-th primitive roots of unity.\index{primitive roots of unity} % $\zeta_n^k$ by looking at its power $k$.

\begin{theorem}
The $n$-th primitive roots of unity are $\{\zeta_n^k \mid 1 \le k \le n-1 \text{ and } \gcd(k, n) = 1 \}$. 
\label{thm:primitive roots of unity}
\end{theorem}
If $n$ is prime, then all the $n$-th roots of unity except 1 are primitive. It follows from Theorem~\ref{thm:primitive roots of unity} that the number of $n$-th primitive roots of unity is equal to the number of natural numbers smaller than $n$ that is coprime with $n$, which is also known as the \textbf{Euler's totient function}\index{Euler's totient function} 
\[ \varphi(n)=|\{k \mid 1 \le k \le n-1 \text{ and } \gcd(k,n)=1\}|. \] % of $n$. 
For example, there are four 12th primitive roots of unity $\{\zeta, \zeta^5, \zeta^7, \zeta^{11}\}$.

% At the beginning of this section, we said that the nth cyclotomic polynomial is the polynomial whose roots are the nth primitive roots of unity. Hence, it can be defined formally as the following. 
We now have the necessary components to formally define cyclotomic polynomials.

\begin{definition}
The \textbf{$n$-th cyclotomic polynomial} $\Phi_n(x)$ \reversemarginpar
\marginnote{\textit{Cyclotomic polynomial}}\index{cyclotomic polynomial}
is the polynomial whose roots are the $n$-th primitive roots of unity. That is,
\begin{equation*}
    \Phi_n(x) = \prod_{\substack{1 \le k < n \\ \gcd(k,n)=1}} (x-\zeta_n^k),
\end{equation*}
where $\zeta_n^k=e^{2k\pi i/n}$ is an nth root of unity (as before in Theorem~\ref{thm:roots of unity}). 
\end{definition}

\begin{example}
The first few cyclotomic polynomials and their roots are listed in Table~\ref{tab:first few cyclotomics}.
\begin{table}[!htbp]
    \centering
    \begin{tabular}{|c|l|l|}
        \hline 
    $n$ & $\Phi_n(x)$ & roots \\
        \hline \hline
        1 & $x - 1$ & 1 \\
        \hline
        2 & $x + 1$ & $\zeta^1 = -1$ \\
        \hline
        3 & $x^2 + x + 1$ & $\zeta^1, \zeta^2$ \\
        \hline
        4 & $x^2 + 1$ & $\zeta^1 = i, \zeta^3 = -i$ \\
        \hline
        5 & $x^4 + x^3 + x^2 + x + 1$ & $\zeta^1, \zeta^2, \zeta^3, \zeta^4$ \\
        \hline
        6 & $x^2 - x +1$ & $\zeta^1, \zeta^5$ \\
        \hline
        7 & $x^6 + x^5 + x^4 + x^3 + x^2 + x + 1$ & $\zeta^1, \zeta^2, \zeta^3, \zeta^4, \zeta^5, \zeta^6$ \\
        \hline
        8 & $x^4$ + 1 & $\zeta^1, \zeta^3, \zeta^5, \zeta^7$ \\
        \hline 
    \end{tabular}
    \caption{First few cylotomic polynomials}
    \label{tab:first few cyclotomics}
\end{table}
For $n = 4$, the 4th cyclotomic polynomial is $\Phi_4(x) = (x-i)(x+i)=x^2+1$, because the 4th roots of unity are $\{\pm 1, \pm i\}$ and the primitive roots are $\pm i$. 
\end{example}

In lattice-based cryptography, we are only interested in some special forms of cyclotomic polynomials as they make certain proofs feasible and computations easier. Next, we introduce two special cases. % \footnote{More special cyclotomic polynomials can be found at  \url{https://brilliant.org/wiki/cyclotomic-polynomials/}.}

\begin{remark}
\label{rmk:speCycPoly}
If $n$ is prime, then the $n$-th cyclotomic polynomial is given by
\begin{equation*}
    \Phi_n(x) = x^{n-1} + x^{n-2} + \cdots + 1 = \sum_{t=0}^{n-1} x^{t}.
\end{equation*}
%If $n=2p$ where $p$ is an odd prime, then the nth cyclotomic polynomial 
%\begin{equation*}
%    \Phi_n(x) = x^{p-1} - x^{p-2} + \cdots - x + 1.
%\end{equation*}
If $n=p^k$ is a prime power, then the $n$-th cyclotomic polynomial is given by
\begin{equation*}
    \Phi_n(x) = \Phi_p(x^{n/p}) = \Phi_p(x^{p^{k-1}}) = \sum_{t=0}^{p-1} x^{tp^{k-1}}.
\end{equation*}
As a special case, when $p=2$ we have $n=2^k$ or $n= 2m \ge 2$ where $m=2^{k-1}$, the $n$-th cyclotomic polynomial is
\begin{equation*}
    \Phi_n(x) = x^{m}+1.
\end{equation*}
This directly relates to the underlying ring in the RLWE \index{ring LWE} problem as we shall see in Section \ref{section:rlwe}.
\end{remark}

The definition of cyclotomic polynomial implies it is monic (i.e., the leading coefficient is equal to 1) and has $\varphi(n)$ linear factors. In addition, $\Phi_n(x)$ divides $x^n - 1$ because the roots of the former are also roots of the latter, but not vice versa. This implies an important relationship: % \marginpar{KS: is this used anywhere?}
\begin{equation}
\label{equation:xn and cyclotomic polynomial}
    x^n-1 = \prod_{d \mid n} \Phi_d(x).
\end{equation}
Here are some special cases of Equation~(\ref{equation:xn and cyclotomic polynomial}),
\begin{align*}
& x^2 - 1 = (x-1)(x+1) \\
& x^3 - 1 = (x-1)(x^2 + x + 1) \\
& x^4 - 1 = (x^2-1)(x^2+1) = (x-1)(x+1)(x^2+1) \\
& x^5 - 1 = (x-1)(x^4 + x^3 + x^2 + 1)\\
& x^6 - 1 = (x^2 - 1)(x^2+x+1)(x^2 - x + 1) = (x^3-1)(x+1)(x^2 - x + 1).
\end{align*}
Note the pattern that if $d$ divides $n$, then $x^d-1$ divides $x^n -1$:
\[ x^n - 1 = (x^d - 1)(x^{n-d} + x^{n-2d} + \cdots + x^d + 1).    
\]
% To show Equation~(\ref{equation:xn and cyclotomic polynomial} 
More formally, note that
\begin{align*}
% \prod_{d \mid n} \Phi_d(x)
x^n - 1
&= \prod_{1 \le k \le n} (x-\zeta_n^k) \\
&= \prod_{d: d \mid n} \prod_{\substack{1 \le k \le n \\ \gcd(k, n)=d}} (x - \zeta_n^k) \\ 
&=  \prod_{d: d \mid n} \Phi_{\frac{n}{d}}(x) \\
 &=\prod_{d: d \mid n} \Phi_{d}(x).
\end{align*}
The second equality is because $d\mid n$ splits $[1,n]$ into $\frac{n}{d}$ mutually exclusive subsets. The third equality uses the definition of cyclotomic polynomial. The last equality is because the subset of integers $\frac{n}{d}$ and $d$ are identical. 

Equation~(\ref{equation:xn and cyclotomic polynomial}) says that a number is an $n$-th root of unity if and only if it is a $d$-th primitive root\index{primitive roots of unity} of unity for some natural number $d$ that divides $n$. 

\begin{example}
The 6th roots of unity are shown in Figure \ref{fig:primitive roots}. $\zeta^0=1$ is the 1st primitive root. $\zeta^3$ is the 2nd primitive root. $\zeta^2$ and $\zeta^4$ are the 3rd primitive roots. $\zeta^1$ and $\zeta^5$ are the 6th primitive roots. Hence, the product of these four cyclotomic polynomials is a polynomial whose roots are the 6th roots of unity, i.e., $\Phi_1(x)\Phi_2(x)\Phi_3(x)\Phi_6(x)=x^6-1$.
\end{example}

% We state below two important theorems about cyclotomic polynomials without proving them.
Here are some important properties of cyclotomic polynomials.\index{cyclotomic polynomial}

\begin{theorem}
The $n$-th cyclotomic polynomial $\Phi_n(x)$ is a degree $\varphi(n)$ monic polynomial with integer coefficients.
\end{theorem}
% It is clear that the degree of $\Phi_n(x)$ is $\varphi(n)$ the Euler's totient function. The integer coefficient part can be proved by induction. 

\begin{theorem}\label{thm:cyclotomics are minimal polynomials}
The $n$-th cyclotomic polynomial is the minimal polynomial \reversemarginpar
\marginnote{\textit{Minimal polynomial}}\index{minimal polynomial}
 of an $n$-th primitive root of unity. 
\end{theorem}
This theorem implies that cyclotomic polynomials are irreducible\index{irreducible} over the field of rationals $\Q$. 
%
\iffalse
\begin{remark}
The coefficient for the term $x^{\varphi(n)-1}$ is the negative sum of the n-th primitive roots of unity, which is also known as the \textbf{Mobius function} $-\mu(n)$. For $n < 105$, the coefficients of $\Phi_n(x)$ are $\{-1, 0, 1\}$. 
\end{remark}
\fi 
%
As we will see in \Cref{section:rlwe}, ring LWE\index{ring LWE} is defined with respect to the quotient ring of polynomials $\Z[x]$ by the ideal generated by a cyclotomic polynomial.
Theorem~\ref{thm:cyclotomics are minimal polynomials}, together with the First Isomorphism Theorem (Theorem~\ref{thm:first isomorphism theorem}),\index{First Isomorphism Theorem} gives the following characterisation of these quotient rings. \begin{theorem}\label{thm:ring LWE isomoprhic 1}
For all $m\in \N$, we have
\[ \Z[x]/(\Phi_m(x)) \cong \Z[\zeta_m] \]
\end{theorem}
\begin{proof}
This is a direct consequence of Theorems~\ref{thm:cyclotomics are minimal polynomials} and \ref{thm:fieldExtEquiv}.
\end{proof}

\subsection{Galois Group of Cyclotomic Polynomials}
\label{subsec:galois group cyclotomics}

 Galois theory associates to every polynomial a group, called the Galois group of the polynomial, that holds useful algebraic information about the roots of the polynomial that can be used to answer important questions about the polynomial. 
 % By studying this group, we can translate this algebraic information back to the world of polynomials.
In this subsection, we use Galois theory to study the roots of cyclotomic polynomials and the symmetric structure in their permutations that will turn out to be useful in the RLWE\index{ring LWE} hardness proof.
We will start with a simple example to motivate the discussion. 

\begin{example}\label{ex:quadratic formula}
Consider a quadratic polynomial with roots $r$ and $s$:
\begin{equation}
f(x) = x^2 + bx + c \label{eq:a polynomial}    
\end{equation}
The polynomial can be written in the alternative form of $(x-r)(x-s)$, which expands out to 
\[ x^2 - (r+s)x + rs. \] 
Equating coefficients with (\ref{eq:a polynomial}), we get
\begin{align}
-b = r + s \label{eq:brs} \\
c = rs. \label{eq:crs}
\end{align}
To express $r$ and $s$ in terms of $b$ and $c$, we can first square (\ref{eq:brs}) to obtain
\[ b^2 = (r+s)^2 = r^2 + 2rs + s^2. \]
Subtracting both sides by $4c$ then yields
\[ b^2 - 4c = r^2 -2rs + s^2 = (r - s)^2. \]
Taking square roots, we now get
\begin{align}
    r - s = \sqrt{b^2 - 4c} \label{eq:r-s}\\
    s - r = -\sqrt{b^2 - 4c}. \label{eq:s-r}
\end{align}
Adding (\ref{eq:brs}) to (\ref{eq:r-s}) and (\ref{eq:s-r}) now gives the familiar quadratic formula.
\[ r = \frac{-b + \sqrt{b^2 - 4c}}{2}  \;\;\text{and}\;\; s = \frac{-b + \sqrt{b^2 - 4c}}{2}. \]
% the familiar quadratic formula.
\end{example}

Equations (\ref{eq:brs}) and (\ref{eq:crs}) and their equivalents for arbitrary higher-degree polynomials are called the elementary symmetric polynomials (of the roots).\index{elementary symmetric polynomials}
For another example, a cubic polynomial $x^3 + bx^2 + cx + d$ with roots $r,s,t$ have the following elementary symmetric polynomials:
\begin{gather*}
-b = r + s + t \\
    c = rs + rt + st \\
    -d = rst.    
\end{gather*}
% Historically, Galois theory tells us
The high-level steps outlined briefly in Example~\ref{ex:quadratic formula}, codified properly in Galois theory, can be used to answer the question of whether the roots of an arbitrary polynomial $f$ can be expressed in terms of its coefficients: start with the elementary symmetric polynomials of $f$ and then systematically simplify the formulas by breaking the symmetries in them.
We are thus led to the following definition of the splitting field of a polynomial, which contains the elementary symmetric polynomials and other polynomials of (subsets of) the roots that can be obtained from them.
% and the use of group theory to study the classification of symmetric polynomials in splitting fields.

% \begin{theorem}\label{thm:symmetric polynomial of roots}
% If $P(x)$ is a polynomial of degree $n$ with leading coefficient 1, then any symmetric polynomial in the roots of $P(x)$ can be written as a polynomial in the coefficients of $P(x)$.
% \end{theorem}

\iffalse
\begin{example}
Consider a cubic polynomial with roots $r,s,t$:
\begin{align*} 
P(x) &= x^3 + bx^2 + cx + d \\
     &= (x - r)(x - s)(x - t) \label{eq:cubic polynomial roots}
\end{align*}
Expanding out the second line, we get 
\[ x^3 - (r+s+t)x^2 + (rs + rt + st)x - rst. \]
Equating coefficients, we get the so-called elementary symmetric polynomials\index{symmetric polynomial}
\begin{gather*}
    -b = r + s + t \\
    c = rs + rt + st \\
    -d = rst
\end{gather*}
Let $Q(r,s,t) = r^3 + s^3 + t^3$ be a symmetric polynomial\index{symmetric polynomial}, which means switching any pair of variables results in the same polynomial. Then one can show that $Q(r,s,t) = -b^3 - 3bc - 9d$. 
\end{example}
\fi

\begin{definition}
Let $f$ be a polynomial with rational coefficients. The splitting field\reversemarginpar
\marginnote{\textit{Splitting field}}\index{splitting field} $K$ of $f$ is the smallest field that contains the roots of $f$. 
($K$ is called the splitting field because we can split $f$ into linear factors in $K$. Also, by the properties of a field, $K$ can be understood as the set of multi-variate polynomial expressions in the roots of $f$ with rational coefficients.)
\end{definition}

The symmetric polynomials in the splitting field for a polynomial $f$ are exactly those that are invariant under permutations of the roots of $f$, and 
% We want to understand the symmetricity of the elements of splitting field $K$ of $P(x)$ with respect to permutations of the roots of $P(x)$. 
these permutations can be obtained via automorphisms.

\begin{definition}
An automorphism\reversemarginpar\marginnote{\textit{Automorphism}}\index{automorphism} $\alpha$ of the splitting field $K$ of a polynomial $f$ is a bijection from $K$ to $K$ such that
\begin{gather*}
    \alpha(a + b) = \alpha(a) + \alpha(b) \\
    \alpha(ab) = \alpha(a) \alpha(b).
\end{gather*}
\end{definition}
Note that for all $a \in K$ that is a rational number, $\alpha(a) = a$ by the property of $\alpha$.
It then follows that for all polynomials $Q(r_1,\ldots,r_n) \in K$, where each $r_i$ is a root of $f$, we have
\[ \alpha(Q(r_1,\ldots,r_n)) = Q(\alpha(r_1),\ldots,\alpha(r_n)). \]
Now consider $f(r_i)$, which is in $K$ because it is a polynomial in a root of $f$.
Since 
\[ f(\alpha(r_i)) = \alpha(f(r_i)) = \alpha(0) = 0, \]
we can see that an automorphism always send a root of $f$ to another root of $f$; further, given automorphisms are bijections, each automorphism can be identified with a permutation of the roots of $f$.

A collection of permutations is a group if it is closed under composition of permutations. Since automorphisms compose, the set of permutations of the roots of a polynomial $f$ that correspond to an automorphism is a group, called the Galois Group of the polynomial $f$\index{Galois group of a polynomial}, or equivalently the Galois Group $Gal(K(\zeta)/K)$ of the field extension $K(\zeta)/K$, where the cyclotomic extension $K(\zeta)$ is the splitting field of $f$.\index{cyclotomic extension}\index{Galois group of a field extension}

For most polynomials $f$, every permutation of the roots induces an automorphism so the Galois Group of $f$ is the set of all permutations of the roots. But for some polynomials, the Galois Group is a strict subset of the permutations of the roots because some permutations do not induce an automorphism. This is the case for cyclotomic polynomials.

Let $G$ be the Galois group of the $n$-th cyclotomic polynomial, where $n$ is prime. The roots of the polynomial are $\{\zeta, \zeta^2, \ldots, \zeta^{n-1} \}$.
Each $\alpha \in G$ maps $\zeta$ by $\alpha(\zeta) = \zeta^a$ for some $a \in \{1,\ldots,n-1\}$.
Since
\[ \alpha(\zeta^k) = \alpha(\zeta)^k = \zeta^{ak}, \]
the number $a$ completely determines where all the other roots go.
In general, the Galois group of a polynomial can permute the roots arbitrarily, but the Galois group of cyclotomic polynomials only allow permutations of the form
\[ (\zeta,\zeta^2,\ldots,\zeta^{n-1}) \mapsto (\zeta^a, \zeta^{2a \bmod n}, \ldots, \zeta^{(n-1)a \bmod n}) \]
for all $a \in \{1,\ldots,n-1\}$.

\begin{example}
For $n=5$, these are the only permutations induced by automorphisms:
\begin{gather*}
    (\zeta^1,\zeta^2,\zeta^3,\zeta^4) \;\textnormal{ for}\; a=1 \\
    (\zeta^2,\zeta^4,\zeta^1,\zeta^3) \;\textnormal{ for}\; a=2 \\
    (\zeta^3,\zeta^1,\zeta^4,\zeta^2) \;\textnormal{ for}\; a=3 \\
    (\zeta^4,\zeta^3,\zeta^2,\zeta^1) \;\textnormal{ for}\; a=4 
\end{gather*}
\end{example}

The above chain of reasoning can be more formally stated in the following theorem, where $(\Z/n\Z)^*$ is the multiplicative integer modulo $n$ group.
% The above lemma implies an injective homomorphism from the Galois group to $(\Z/n\Z)^*$, the multiplicative integer modulo $n$ group as stated in the following theorem. 
\begin{theorem}
\label{theorem:injective homomorphism from galois group}
The mapping 
\begin{gather*}
    \omega: Gal(K(\zeta_n)/K) \rightarrow (\Z/n\Z)^* \\
     \omega(\sigma) = a_{\sigma} \bmod n 
\end{gather*}
that is given by $\sigma(\zeta) = \zeta^{a_{\sigma}}$ for all $n$-th roots of unity $\zeta$ is an injective group homomorphism.\reversemarginpar
\marginnote{\textit{Injective homomorphism}}\index{injective homomorphism}
\end{theorem}

\begin{proof}
% It is not hard to see that the map is a homomorphism. 
For any automorphisms $\sigma, \tau \in Gal(K(\zeta_n)/K)$, a primitive root $\zeta_n \in \mu_n$ satisfies $\sigma \tau(\zeta_n) = \sigma(\zeta_n^{a_{\tau}}) = \zeta_n^{a_{\sigma}a_{\tau}}$ by applying the automorphism one after the other. In addition, the two automorphisms gives another automorphism in the Galois group by composition, so $\sigma \tau(\zeta_n) = \zeta_n^{a_{\sigma \tau}}$. Hence, we have $\zeta_n^{a_{\sigma}a_{\tau}}=\zeta_n^{a_{\sigma \tau}}$. This implies $a_{\sigma}a_{\tau}= a_{\sigma \tau} \bmod n$, because $\zeta_n$ has order $n$. Therefore, we have $\omega(\sigma \tau) = a_{\sigma \tau} = a_{\sigma}a_{\tau} \bmod n = \omega(\sigma) \omega(\tau)$ which entails $\omega$ is a homomorphism. The injectivity is not difficult to see either. 
\end{proof}

We know the group $(\Z/n\Z)^*$ is abelian. The map $\omega$ embeds the Galois groups of cyclotomic extensions to this abelian group, so the Galois group is also abelian. For a general base field $K$, the group homomorphism need not be surjective. There are two special cases, $K=\Q$ and $K=\F_p$, for a prime $p$, that are of most interest for building lattice cryptosystems. We will look at the property of the map $\omega$ in each special case one by one. 
%In the first case, the above homomorphism is surjective, hence it is an isomorphism. In the second case, it is not necessary surjective, but we can still describe the image of the embedding. 

\begin{theorem}
\label{thm:galGrpCycField}
The Galois group of the cyclotomic extension $\mathbb{Q}(\zeta_n)$ is isomorphic\index{isomorphic} to the multiplicative integer modulo $n$ group.\reversemarginpar
\marginnote{\textit{Isomorphism when $K=\Q$}}
That is, 
\begin{equation*}
    Gal(\mathbb{Q}(\zeta_n) / \mathbb{Q}) \cong (\mathbb{Z}/ n\mathbb{Z})^*.
\end{equation*}
For each automorphism $\sigma \in Gal(\mathbb{Q}(\zeta_n) / \mathbb{Q})$, there is an integer $i \in (\mathbb{Z}/ n\mathbb{Z})^*$ such that the automorphism $\sigma \mapsto [i]$ is mapped to the equivalent class of $i$ if and only if $\sigma(\zeta_n) = \zeta_n^i$.
\end{theorem}
The automorphisms in the Galois group are functions on the roots of unity. We can think of the equivalent class $[i]$ as a function too given by $[i]:\zeta \mapsto \zeta^i$ for all roots $\zeta \in \mu_n$. The theorem says each automorphism in the Galois group is uniquely mapped to an integer in the multiplicative group  (or a function). 
Theorem~\ref{thm:galGrpCycField} is useful for proving the pseudorandomness of the ring LWE\index{ring LWE} distribution as we will see in a later section.

% \iffalse
% Without proving the theorem, we can at least verify the two groups have the same order. 
Observe that the order of the Galois group is equal to the degree of the Galois extension over $\Q$, which is equal to the degree $\varphi(n)$ of the $n$-th cyclotomic polynomial. The order of the multiplicative group is equal to the number of integers in $[0,n-1]$ that are coprime with $n$. 
The two numbers are obviously equal. 
% \fi

%\kl{introduce Kronecker and Weber's result?} no
When $K$ is a field with non-zero prime characteristic $char(K)=p$ (e.g., $K=\F_p$), as is often the case in cryptography, the homomorphism $\omega$ is not necessarily surjective. 
Theorem~\ref{theorem:galois group finite field} caters for this case.
For our purpose, we are primarily interested in the cyclotomic polynomials $\Phi_d(x)$ where $\gcd(d, p)=1$. 

\begin{theorem}
\label{theorem:galois group finite field}
Let $\mathbb{F}_q$ be a finite field with a prime power order $q$ and $\gcd(q,n)=1$, the Galois group of a cyclotomic extension $\mathbb{F}_q(\zeta_n)$ of the finite field \reversemarginpar
\marginnote{Image of Galois group when $K=\F_p$}
is mapped by the homomorphism $\omega$ to the cyclic group $\langle q \bmod n \rangle$ in $(\mathbb{Z}/ n\mathbb{Z})^*$. That is, 
\begin{equation*}
    \omega(Gal(\mathbb{F}_q(\zeta_n) / \F_q))= \langle q \bmod n \rangle \subseteq (\mathbb{Z}/ n\mathbb{Z})^*.
\end{equation*}
In particular, the dimension of the cyclotomic extension is the order of $q$ modulo $n$.
\end{theorem}

% Before stating a similar theorem for $K$ being a finite field, we give some useful theorems. %The next theorem explicitly states the image of $\omega$ in the finite field $F_q$. 

\iffalse
\begin{theorem}
\label{thm:quoRngIsField}
Let $p$ be a prime and $f(x) \in \F_p[x]$ be a monic irreducible polynomial of degree $n$. The quotient ring $\F_p[x]/f(x)$ is a field of order $p^n$. 
\end{theorem}
When interpreting the ring of polynomials $\F_p[x]/f(x)$, it means each polynomial in this quotient ring has coefficients taken from the field $\F_p$ and the polynomial degree is at most $n-1$. We will see this notation  more often when introducing \textit{Algebraic Number Theory}, which has a different interpretation. 
Theorem~\ref{thm:quoRngIsField} is needed when defining the underlying ring of the ring LWE problem\index{ring LWE}.

\begin{proof}
Each coset in the quotient ring $\F_p[x]/f(x)$ has the form $a_0 + a_1 x + \cdots + a_{n-1} x^{n-1}$, where $a_i \in \F_p$. So there are $p^n$ different cosets. The polynomial $f(x)$ is irreducible implies the quotient ring is also a field. 
\end{proof}
\fi

% I am skipping a lot of details and important results because they are not quite relevant to our topic, but for short 
% Without going into details, we can think of $\F_{p^n}=\F_p(\zeta_n)$.

To prove Theorem~\ref{theorem:galois group finite field}, we need this next result.
% With the help of this theorem, we can prove the next theorem when $K$ is a finite field of a prime power order. 

\begin{theorem}
\label{theorem:finite field pth power map}
For a prime $p$ and prime power $q=p^n$, the pth power map\reversemarginpar
\marginnote{\textit{Power map}}
 $\omega_p:x \mapsto x^p$ on $\F_q$ generates the Galois group $Gal(\F_q(\zeta_n)/\F_q)$. 
\end{theorem}
% see theorem 4.1 in https://kconrad.math.uconn.edu/blurbs/galoistheory/finitefields.pdf

% We prove the theorem for the special case when $q=p$ for a prime $p$. % To do so, we need another theorem from finite fields. 

\begin{proof} (of Theorem~\ref{theorem:galois group finite field} for the special case when $q = p$ for a prime $p$)
Theorem \ref{theorem:finite field pth power map} implies that the Galois group $Gal(\F_q(\zeta_n)/\F_q)$ is generated by the pth power map $\omega_p: x \mapsto x^p$ for all $x \in \F_q(\zeta_n)$. In addition, by Theorem \ref{theorem:injective homomorphism from galois group} the group homomorphism $\omega$ associates to $\omega_p$ an non-negative integer $a \bmod n$ such that $\omega_p(\zeta) = \zeta^a$ for all nth roots of unity $\zeta \in \mu_n$. This entails $\zeta^p = \zeta^a$, which is true if $a = p \bmod n$. Hence, the homomorphism $\omega$ maps the pth power map $\omega_p$ in the Galois group to $p \bmod n$ in the group $(\Z/n\Z)^*$. Since $Gal(\mathbb{F}_q(\zeta_n) / \F_q) = \langle \omega_p \rangle$, its image is the cyclic group $\langle p \bmod n \rangle \in (\Z/n\Z)^*$.

The assumption $char(\F_q)=p$ implies the polynomial $x^n-1$ is separable in $\F_q[x]$, so $\F_q(\zeta_n)$ is an Galois extension given that it is also the splitting field of $x^n-1$. Hence, we have $[\F_q(\zeta_n): \F_q]=|Gal(\mathbb{F}_q(\zeta_n) / \F_q)|=|\langle p \bmod n \rangle|$, which is the order of $p$ modulo $n$.
\end{proof}

Knowing cyclotomic polynomials are irreducible over $\Q$, we would like to know whether they are also irreducible in a finite field $\F_q$ of prime power order $q$. This brings out the following theorem and corollary. Denote $\bar{\Phi}_n(x)$ as reducing the coefficients of $\Phi_n(x)$ modulo $q$. 

\begin{theorem}
\label{thm:splitCycPoly}
\reversemarginpar
\marginnote{\textit{Factor $\Phi_n(x)$ in $F_p$}}
Let $q$ be prime power and $\gcd(q,n)=1$, the monic irreducible factors of the polynomial $\bar{\Phi}_n(x) \in \F_p[x]$ are distinct and each has a degree equal to the order of $q$ modulo $n$. 
\end{theorem}

\begin{corollary}
The polynomial $\bar{\Phi}_n(x)$ is irreducible in $\F_q[x]$ if $\gcd(q,n)=1$ and $\langle q \bmod n \rangle = (\Z / n\Z)^*$. That is, $q \bmod n$ is a generator of the group $(\Z / n\Z)^*$. 
\end{corollary}

\begin{example}
For $n=5$, the polynomial 
\begin{equation*}
    \bar{\Phi}_5(x) = x^4+x^3+x^2+x+1
\end{equation*}
can be factored in $\F_{11}$ as 
\begin{align*}
    (x-3)(x-4)(x-5)(x-9)
\end{align*}
because the order of 11 modulo 5 is 1. Similarly, it can be factored in $\F_{19}$ as 
\begin{align*}
    (x^2+5x+1)(x^2+15x+1) 
\end{align*}
because the order of 19 modulo 5 is 2. Similarly, it can be factored in $\F_3$ as 
\begin{align*}
    x^4+x^3+x^2+x+1
\end{align*}
because the order of 3 modulo 5 is 4. The last case is an example of the corollary where the cyclic group $\langle3 \bmod 5\rangle$ is a generator of the group $(\Z / 5\Z)^*$.
More details on the derivation of these factorizations can be found in Example~\ref{ex:q ideal factorisation}.
\end{example}

%Efficiency of using cyclotomic number fields are covered in Section 6.2 of 

%\kl{Up to this point, all the topics introduced in the previous sections should make sense now, because they mainly serve under the aim of introducing number fields, in particular cyclotomic number fields.}

%%%%%%%%%%%%%%%%%%%%%%%%%%%%%%%%%%%%%%%%%%%%%%%%%%%%%%%%%%%%%%%%%%%%%%%%%%%%%%%%%%%%%%%%%%%%%%%%%%%
%%%%%%%%%%%%%%%%%%%%%%%%%%%%%%%%%%%%%%%%%%%%%%%%%%%%%%%%%%%%%%%%%%%%%%%%%%%%%%%%%%%%%%%%%%%%%%%%%%%

%%%%%%%%%%%%%%%%%%%%%%%%%%%%%%%%%%%%%%%%%%%%%%%%%%%%%%%%%%%%%%%%%%%%%%%%%%%%%%%%%%%%%%%%%%%%%%%%%%%

% \subsection*{Useful References}
% \cite{mukherjee2016cyclotomic}, \cite{conradcyclotomic}, \cite{porter15cyclotomic}. 

%\newpage
%\bibliography{references}
%\bibliographystyle{abbrvnat}

\newpage
\section{Algebraic Number Theory}

\label{sec:ant short}
%\section{Algebraic Number Theory (medium+)}

This section
%\footnote{This section is part of the work \textit{A Tutorial Introduction to Lattice-based Cryptography and Homomorphic Encryption} by the authors Yang Li, Kee Siong Ng, Michael Purcell from the School of Computing, Australian National University @2022.}
introduces some of the results in \textit{Algebraic Number Theory} that will be needed in the hardness proof of the ring LWE\index{ring LWE} (RLWE) problem. In RLWE, proofs and computations are conducted in number fields and rings of integers, which are generalizations of the rational field $\Q$ and integers $\Z$. However, unlike elements in $\Z$ that can be uniquely factorized, which is an essential property that guarantees the validity of some hard computational problems such as integer factorization, elements of rings of integers are not necessarily uniquely factorizable in general. Instead we need to work with sets of elements that possess such unique factorization. As we will see in this section, the ideals of these rings of integers are natural candidates for this purpose and we will state some useful properties of the ideals. 
In particular, the connection with lattice theory comes from a natural mapping between these ideals of a ring of integers to full-ranked lattices that we call ideal lattices.

\textit{Algebraic Number Theory} is a deep and interesting area and we do not attempt to cover all important results in this compact section. Instead, we cover only those mathematical results that are directly relevant to the future sections. Additional results that may assist the reader to better understand the main content are kept in the appendix. This section is organized as follows:
\begin{enumerate}
	\item First, we familiarize the reader with algebraic number field, its ring of integers and ideals of the ring of integers including the generalized fractional ideals. The most important observation is that a fractional ideal can be uniquely factorized into prime ideals. This plays a significant part when employing the \textit{Chinese Remainder Theorem} (CRT) for number fields. 
	 
	\item Second, to build the geometric interpretation of these algebraic objects, we introduce canonical embedding, which maps fractional ideals to special lattices called \textit{ideal lattices}. The embedding allows us to talk about geometric quantities of algebraic objects and enables certain features of ideal lattices that are convenient for the RLWE's proof and computations. 
	
	\item Finally, we go through dual lattices in number fields and relate them with fractional ideals.
	
	% \item Last, to finish this section, we state the Chinese Remainder Theorem in the ring of integers and connect some of the important results that will be used in the proof of RLWE.
	
\end{enumerate}

It's worth noting that many of the concepts covered in this section are used primarily for analysis of the hardness results of the RLWE problem. As such, some readers may find it useful to first skim this section quickly to identify key concepts, and only come back for details as they work through Section~\ref{section:rlwe}.
The only computations that are explicitly needed in RLWE-based cryptosystems are Fast Fourier Transform operations to transform polynomials between their natural and canonical embeddings. 

%-----------------------------------------------------------------------------------------------

\subsection{Ring of integers and its ideal}
\label{subsection:number field}

%\label{subsection:algebraic number field}
%Some of the concepts in this section have been discussed in Section \ref{subsection:field extension}. 

We have seen the LWE problem, which was defined in the integer domain $\Z$ and proved to be hard by reductions from hard lattice problems in the domain in $\R^n$. The drawback of LWE is the large public key that is a matrix of $m$ independent length $n$ column vectors. The RLWE problem (as will be introduced in \Cref{section:rlwe}) is defined in a more general domain, called \textit{the ring of integers}. It greatly reduces the public key size by defining the problem in domain with additional algebraic structures.      

Recall that an algebraic number (integer) is a complex number that is a root of a non-zero polynomial with rational (integer) coefficients. For example, $\sqrt{1/2}$ and $\sqrt{2}$ are roots of the polynomials $x^2 - 1/2$ and $x^2-2$ respectively, so the former is an algebraic number and the latter is an algebraic integer. Algebraic numbers and algebraic integers generalize rational numbers and rational integers by forming the notions of number field and ring of integers, just like the rational field $\Q$ and the integer ring $\Z$. 

\begin{definition}
\reversemarginpar
\marginnote{\textit{Number field}}
An \textbf{algebraic number field} (or simply \textbf{number field}) is a finite extension of the field of rationals by algebraic numbers, i.e., $\mathbb{Q}(r_1, \dots, r_n)$, where $r_1, \dots, r_n$ are algebraic numbers.
\end{definition}

In a special case when the element $\zeta_n$ adjoins to $\Q$ is an nth root of unity, which is also an algebraic number, the number field $\Q(\zeta_n)$ is also known as the  
\reversemarginpar
\marginnote{\textit{Cyclotomic field}}
\textbf{nth cyclotomic (number) field}.\index{cyclotomic field} This is the working domain for reducing the RLWE search to decision problem. 
In a number field $K$, the set of all algebraic integers forms a ring under the usual addition and multiplication operations in $K$. These elements form a ring and is the generalization of the ring of rational integers. 

\begin{definition}
\reversemarginpar
\marginnote{\textit{Ring of integers}}
The \textbf{ring of integers} of an algebraic number field $K$, denoted by $\ok$, is the set of all algebraic integers that lie in the field $K$. 
\end{definition}

Some examples of a number field and its ring of integers are the basic $\Q$ and $\Z$, the quadratic field $\Q(\sqrt{2})$ and $\Z[\sqrt{2}]$, the nth cyclotomic field $\Q(\zeta_n)$ and $\Z[\zeta_n]$. In general, determining the ring of integers is a difficult problem, unless for special cases, see Theorem \ref{app thm:roiQuadField} in Appendix \ref{appen:ant}. 
% It is worth mentioning that a number field $K$ can always be constructed by adjoining a single element $r$, i.e., $K=\Q(r)$, which is known as the primitive element (Theorem \ref{app thm:primEleThm}). But this is not true for its ring of integers. In other words, $\ok$ cannot always be written in the form $\ok=\Z[r]$. 

%A similarity between $\Z$ and $\ok$ is being an integral domain (ID). Recall that an ID is a non-zero commutative ring in which the product of two non-zero elements is non-zero. This is the consequence of being a subset of a number field $K$, which is a field hence always an ID. 

%Recall that a module is a generalization of a vector space where scalar multiplications are defined in a ring rather than a field. 
Since $\Z$ is contained in $\ok$, it is easy to see $\ok$ is also $\Z$-module. In addition,  
\reversemarginpar
\marginnote{\textit{$\ok$ is a free $Z$-module}}\index{$Z$-module}
$\ok$ is a free $Z$-module, as there always exists a $\Z$-basis $B=\{b_1, \dots, b_n\} \subseteq \ok$ such that every element $r \in \ok$ can be written as $r=\sum_{i=1}^n a_i b_i$, where $a_i \in \Z$. 
% The basis is called a \textbf{$\Z$-basis} of $\ok$. 
% It is also a \textbf{$\Q$-basis} of $K$, because $r$ can also be written in terms of rational coefficients $a_i \in \Q$. 
The basis $B$ is called an \textbf{integral basis} 
\reversemarginpar
\marginnote{\textit{Basis}}\index{integral basis}
of the number field $K$ and its ring of integers $\ok$.
If the basis can be written as $\{1,r,\dots,r^{n-1}\}$ the powers of an element $r \in K$, then it is called a \textbf{power basis}.\index{power basis} A field $K$ always has a power basis by the Primitive Element Theorem (Appendix \ref{appen:ant} Theorem \ref{app thm:primEleThm}). % , but not for $\ok$, unless 
If $K=\Q(\zeta_m)$ is a cyclotomic field, the power basis $\{1, \zeta_m, \dots, \zeta_m^{\varphi(m)-1}\}$ is also an integral basis of $\ok$.

\subsubsection{Integral ideal}
In the applications of this tutorial, we do not work with individual elements in $\ok$ because they lack the unique factorization property; instead, we work with ideals of $\ok$ (\Cref{eq:ideal unique factorization}).
Ideals of a ring are useful for constructing a field, for the same reason they are important in the ring of integers. 
% Recall that an ideal of a ring is an additive subgroup of the ring that is closed under multiplication by the ring elements. 
To distinguish ideals of $\ok$ from fractional ideals that will be introduced later, we sometimes refer the former as integral ideals. 
	
\begin{definition}
Given a number field $K$ and its ring of integers $\ok$, an 
\reversemarginpar
\marginnote{\textit{Integral ideal}}
(\textbf{integral}) \textbf{ideal} $I$ of $\ok$ is a non-empty (i.e., $I \neq \emptyset$) and non-trivial (i.e., $I \neq \{0\}$) additive subgroup of $\ok$ that is closed under multiplication by the elements in $\ok$, i.e., for any $r \in \ok$ and any $x \in I$, their product $rx \in I$. 
\end{definition}

As $\ok$ is commutative, we do not differentiate left and right ideals. 
The definition intentionally excluded the zero ideal $\{0\}$ in order to simplify the work of defining ideal division later.
Since $\ok$ has a $\Z$-basis, each of its ideals has a $\Z$-basis too, which entails the ideal is a free $\Z$-module too. As we will see later, this basis will be mapped to a basis of an ideal lattice by canonical embeddings. 
	
We now define ideal multiplication and division which lead to the definition of prime ideals.

Recall that if $I$ and $J$ are ideals then the set sum $I+J = \{x + y \mid x\in I, y\in J\}$ is also an ideal.  The set product $S = \{xy \mid x \in I, y \in J\}$, however, may not be an ideal because it is not necessarily closed under addition. For this reason, the \textbf{product of two ideals} $I$ and $J$ is defined as the set of all finite sums of products of two ideal elements:
\reversemarginpar
\marginnote{\textit{Ideal product}}\index{product of ideals}
\begin{align*}
    IJ := \left\{\sum_{i = 1}^n a_i b_i \mid a_i \in I \text{ and } b_i \in J, n \in \mathbb{N} \right\},
\end{align*}
% which consists of all finite sums of the products of two ideal elements.
%\footnote{Again, it can be proved that $IJ$ and $(IJ)$ are equivalent.} 
By grouping all finite sums of products, the set is closed under addition. Furthermore, it is closed under multiplication by $\ok$, so the  above definition of product is also an ideal. Since $\ok$ is commutative, ideal multiplication is commutative too. 
	
\begin{example}
Given the ring of integers $\ok = \Z$ and two ideals $I = 2\Z = \{2, 4,6,8,\dots,\}$ and $J = 3\Z=\{3,6,9,12,\dots,\}$, their product is $IJ=\{2\cdot 3, 2\cdot 6,2\cdot 3 + 2\cdot 6,\dots\}$.  
\end{example}
	 
Since the zero ideal is excluded from the ideal definition, it is convenient to define ideal division. The intuition is the same as non-zero integer division. 

\begin{definition}
Let $I$ and $J$ be two ideals of $\ok$. We say 
\reversemarginpar
\marginnote{\textit{Ideal division}}
$J$ \textbf{divides} $I$, denoted $J \mid I$, if there is another ideal $M \subseteq \ok$ such that $I = JM$.
\end{definition}

% As ideals are finite sets, their sizes matter when dividing one by another. 
The following theorem gives a more intuitive way of thinking about ideal division by relating division with containment. 

\begin{theorem}
\label{thm:divCont}
%\reversemarginpar
%\marginnote{\textit{Divisibility $\iff$ containment}}
Let $I$ and $J$ be two ideals of $\ok$. Then $J \mid I$ if and only if $I \subseteq J$. 
\end{theorem}
The intuition of divisibility implies containment is that if $J\mid I$ then $I=JM\subseteq J$, so $I\subseteq J$. The converse may not be true in general, but is certainly true in the context of $\ok$. % (\kl{Proposition 5.0.2 Ben Green}). 

The standard definition of a prime ideal $I \subseteq \ok$ is that it is a proper ideal such that if $xy \in I$, then either $x \in I$ or $y \in I$. The next lemma gives an alternative definition in terms of ideal containment. 
\begin{lemma}
An ideal $I$ of $\ok$ is prime if and only if for ideals $J$ and $K$ of $\ok$, whenever $JK \subseteq I$, either $J \subseteq I$ or $K \subseteq I$. 
\end{lemma}

By this lemma and Theorem \ref{thm:divCont}, we can define a prime ideal in analogy to a prime number. 

\begin{definition}
A proper ideal $I \subsetneq \ok$ is \textbf{prime} 
\reversemarginpar
\marginnote{\textit{Prime ideal}}\index{prime ideal}
if whenever $I \mid JK$, either $I \mid J$ or $I \mid K$.
\end{definition}

Principal ideals and maximal ideals are defined in the same way as that in general rings. An important observation is that in $\ok$, prime ideals are also maximal. 

\begin{lemma}
%\reversemarginpar
%\marginnote{\textit{Prime is maximal}}
All prime ideals in $\ok$ are maximal. 
\end{lemma}
The proof relies on the results that the quotient of a commutative ring by a prime ideal gives an integral domain, and the quotient by a maximal ideal gives a field. See Lemma \ref{app lm:primeIsMax} in Appendix \ref{appen:ant}.
The importance of this lemma is that when working in $\ok/I$, the quotient ring by a prime ideal $I$ is a field, as implied by Proposition \ref{prop:quotRngIsField} in Appendix \ref{appen:abstract algebrac}.

The most important result of this subsection, which is also one of the main theorems in \textit{Algebraic Number Theory}, is that ideals of $\ok$ can be uniquely factorized into prime ideals. Alternatively, we say the ideals of $\ok$ form a unique factorization domain.
%Note that it is not always true that $\ok$ is a unique factorization domain. As we have seen, an counter example is when $K=\Q(\sqrt{-5})$ and $\ok=\Z(\sqrt{-5})$, in which $6 = 2 * 3=(1+\sqrt{-5})*(1-\sqrt{-5})$.\footnote{It is also necessary to check that 2, 3, $1+\sqrt{-5}$ and $1-\sqrt{-5}$ are irreducible and are not associates of each other. For more details, see the example on Page 30 of Ben Green's notes on algebraic number theory.}

\begin{definition}
An integral domain $D$ is a 
%\reversemarginpar
%\marginnote{\textit{UFD}}
\textbf{unique factorization domain (UFD)} if every non-zero non-unit element $x \in D$ can be written as a product 
\begin{equation*}
    x= p_1 \cdots p_n
\end{equation*}
of finitely many irreducible elements $p_i \in D$ uniquely up to reordering of the irreducible elements.
\end{definition}

We know $\Z$ is a UFD, because every integer can be uniquely factored into a prouct of prime numbers. But the extension $\Z(\sqrt{5})$ is not a UFD, because not every element has a unique factorization, for example $6=2 \cdot 3 = (1+\sqrt{-5})(1-\sqrt{-5})$, which can be factored in two ways. 
% The issue of such non-unique factorization is that when working with the principal ideal $(6)$ in $\ok$, we cannot expect it to be uniquely factored into the prime factors of 6 as we do in $\Z$, in which $(6)=(2)(3)$. 
To avoid such issues, 
% For this reason, 
% when factoring principal ideals $(x)$ in $\ok$, 
we do not work with the individual elements in $\ok$, but study the ideals of $\ok$, which do form a UFD because $\ok$ is a Dedekind domain. 
% The general context of proving such a property and some other properties of ideals of $\ok$ is in a Dedekind domain. But for simplicity, we restrict our domain to $\ok$ which is a Dedekind domain. 
(See Appendix \ref{appen:ant} for more detail about Dedekind domain.) 

\begin{theorem}
\label{thm:idealsOKUFD}
\reversemarginpar
\marginnote{UFD}
For an algebraic number field $K$, every proper ideal $I$ of $\ok$ admits a unique factorization
\begin{equation}
\label{eq:ideal unique factorization}
    I = \mfq_1 \cdots \mfq_k,
\end{equation}
into prime ideals $\mfq_i$ of $\ok$. 
\end{theorem}

\begin{example}
When working in the 5th cyclotomic field $K=\F_{11}(\zeta_5)$ and $\ok=\Z_{11}[\zeta_5]$, the ideal $I=(11)$ of $\ok$ can be uniquely factorized into the product of these four prime ideals:
\begin{align*}
%    I&=\mfq_1 \mfq_2 \mfq_3 \mfq_4 \\
    (11)&=(11,\zeta_5-3)(11,\zeta_5-9)(11,\zeta_5-5)(11,\zeta_5-4).
\end{align*}
The detailed derivation is given in Example~\ref{ex:q ideal factorisation}.
\end{example}

% Needless to say, there are many benefits of being a UFD. 
The usefulness of UFD in our context is that it gives a unique isomorphism between a quotient ring $\ok/I$ and its Chinese Remainder Theorem (CRT) representation\index{CRT}\index{Chinese Remainder Theorem}. 
To generalize CRT to the ring of integers $\ok$, we first define coprime ideals in $\ok$. Since ideals in $\ok$ can be uniquely factorized, it makes sense to talk about coprimality. The standard definition is similar to coprime integers, which do not share a common divisor.  

\begin{definition}
\reversemarginpar
\marginnote{\textit{Ideal GCD}}\index{ideal GCD}
Let $I$ and $J$ be integral ideals of $\ok$, their \textbf{greatest common divisor (GCD)} $\gcd(I, J) = I+J$. 
\end{definition}

\begin{definition}
\label{def:coprimeIdeal}
\reversemarginpar
\marginnote{\textit{Coprime}}\index{coprime}
Two ideals $I$ and $J$ in $\ok$ are \textbf{coprime} if $I+J=\ok$.
\end{definition}
In other words, two integral ideals are coprime if their sum is the entire ring of integers.
For example, the integral ideals $(2)$ and $(3)$ in $\Z$ are coprime because $(2)+(3)=(1)=\Z$. But the integral ideals $(2)$ and $(4)$ are not coprime because $(2)+(4)=(2) \neq \Z$. 

\begin{theorem}
\label{thm:crtInOK}
\reversemarginpar
\marginnote{\textit{CRT in $\ok$}}
   Let $I_1, \dots, I_k$ be pairwise coprime ideals in a ring of integers $\ok$ and $I = \prod_{i=1}^k I_i$. Then the map 
   \begin{equation*}
       \ok \rightarrow (\ok / I_1, \dots, \ok/I_k)
   \end{equation*}
   induces an isomorphism 
   \begin{equation*}
       \ok / I \cong \ok / I_1 \times \cdots \times \ok / I_k.
   \end{equation*}
\end{theorem}

The core element of the proof of CRT in $\ok$ % , first prove the map is surjective. Then prove 
is to show that the kernel of the map is $I_1 \cap \cdots \cap I_k$, which is identical to $\prod_{i=1}^k I_i$ under the assumption that the ideals are pairwise coprime. The result then  follows from the First Isomorphism Theorem.\index{First Isomorphism Theorem} 

By CRT in $\ok$, the factorization (\ref{eq:ideal unique factorization}) yields the isomorphism 
\begin{align}\label{eq:ok prime ideal crt}
    \ok/I \cong \ok / \mfq_1 \times \cdots \times \ok / \mfq_k. 
\end{align}
This isomorphism is essential for the hardness proof of RLWE\index{ring LWE}. % , as well as renders  polynomial multiplications simpler. 
If the factorization is not unique, the same proof will not follow through. We will discuss more detail of the proof in Section~\ref{section:rlwe}.

\subsubsection{Fractional ideal}\index{fractional ideal}
As briefly mentioned earlier, fractional ideals are generalizations of integral ideals and they are one of the main ingredients in the hardness proof of RLWE\index{ring LWE}.
% although some parts of the proof also work for integral ideals. 
On the one hand, fractional ideals share some common properties with integral ideals including the important unique factorization characteristic. On the other hand, they are neither ideals of the ring of integers $\ok$ nor ideals of the number field $K$ as we will see soon. 

\begin{definition}
\label{def:fracIdeal2}
Let $K$ be a number field and $\ok$ be its ring of integers. A \textbf{fractional ideal}\reversemarginpar
\marginnote{\textit{Fractional ideal}}\index{fractional ideal}
 $I$ of $\ok$ is a set such that $dI \subseteq \ok$ is an integral ideal for a non-zero $d \in \ok$.  
\end{definition}

% Fractional ideals can also be defined in an integral domain, which is a more general context than number field as we shall see from Definition \ref{app def:fracIdeal} in Appendix \ref{appen:ant}. 
Given an integral ideal $J \subseteq \ok$ and an invertible element $x \in K$, the corresponding fractional ideal $I$ can be expressed as 
\begin{equation*}
I = x^{-1} J := \{x^{-1} a \mid a \in J\} \subseteq K.
\end{equation*}
From this expression, it is clearer that the non-zero element $d \in K$ in the above definitions is for cancelling the denominator $x$ of elements in the fractional ideal. When $x=1$, it entails the integral ideals of $\ok$ including $\ok$ itself are all fractional ideals. This is also why fractional ideals are generalizations of them.  
%Note $x$ is in $K$ but not $\ok$ because it needs to be invertible. 
Since an integral ideal is a free $\Z$-module and a fractional ideal is related to an integral ideal by an invertible element, it follows that a fractional ideal is a free $\Z$-module too with a $\Z$-basis.   

It can be seen that a fractional ideal is closed under addition and multiplication by the elements in $\ok$, but it is NOT an ideal of $\ok$, because it is not necessarily a subset of $\ok$. Neither it is an ideal of the number field $K$, because a field has only zero and itself as ideals. % Throughout, we refer to a fractional ideal as in a number field. 

\begin{example}
Let $K= \Q$ and $\ok= \Z$. Given the integral ideal $5\Z$ and $x=4 \in \Q$, whose inverse is $\frac{1}{4}$, the corresponding fractional ideal in $\Q$ is $\frac{5}{4}\Z$. 
% Alternatively, given the fractional ideal, there exists an element $4 \in \Z$ that clears the denominator of elements in the fractional ideal so that the resulting set $5\Z$ is an integral ideal of $\Z$.
\end{example}

% \begin{example}
% A counter example is when $I=\Z[\frac{1}{2}]$, because there does not exist a denominator $d \in \ok$ to make $dI$ is an integral ideal of $\ok$. 
% \end{example}

The product of two fractional ideals can be defined analogous to the product of two 
\reversemarginpar
\marginnote{\textit{Frac ideal product}}
integral ideals. That is, for fractional ideals $I$ and $J$, 
\begin{align*}
    IJ := \left\{\sum_{i=1}^n a_i b_i \mid a_i \in I \text{ and } b_i \in J, n \in \N \right\}.
\end{align*}
It is also easy to check that the product of two fractional ideals is still a fractional ideal. 

The fractional ideals in a number field $K$ form a multiplicative group. To see this, we have demonstrated that they are closed under multiplication and the unit ideal $(1)=\ok$ is the multiplicative identity in the group. It remains to show that every fractional ideal has an inverse in the group. This is done via the following two lemmas. The first lemma states that every prime ideal of $\ok$ has an inverse. The second lemma states that every non-zero integral ideal of $\ok$ has an inverse, which uses the result of the first lemma and the fact that every prime ideal in $\ok$ is also maximal. See Appendix \ref{appen:ant} for the proofs of these two lemmas. 

\begin{lemma}
%\reversemarginpar
%\marginnote{\textit{Prime ideal inverse}}
If $P$ is a prime ideal in $\ok$, then $P$ has an inverse $P^{-1} = \{a \in K \mid a P \subseteq \ok\}$ that is a fractional ideal.
\end{lemma}

\begin{lemma}
%\reversemarginpar
%\marginnote{Integral ideal inverse}
Every non-zero integral ideal of $\ok$ has an inverse. 
\end{lemma}

The two lemmas combined prove that a fractional ideal has an inverse. For more detail of the proof, see Theorem 3.1.8 of \citet{stein2012algebraic}. To be more precise, the inverse 
\reversemarginpar
\marginnote{\textit{Frac ideal inverse}}
of a fractional ideal $I$ has the form 
\begin{equation}
\label{equ:fracIdInv}
    I^{-1} = \{x \in K \mid xI \subseteq \ok\}.
\end{equation}
In the special case when the product of two fractional ideals is a principal fractional ideal $IJ=(x)$, the inverse has the form $I^{-1}=\frac{1}{x}J$.
%It can be proved that this inverse is also a fractional ideal and it is unique for the given fractional ideal $I$. See Conrad's lecture notes on ``Ideal Factorization'' (Definition 2.5, Theorem 2.7 and Theorem 4.1). 

\begin{theorem}
\label{thm:fracIdealGroup}
\reversemarginpar
\marginnote{\textit{Multiplicative group}}
The set of fractional ideals in a number field $K$ is an abelian group under multiplication with the identity element $\ok$. 
\end{theorem}

A key result of this subsection is that a fractional ideal can also be uniquely factorized into a product of prime ideals. 

\begin{theorem}
\reversemarginpar
\marginnote{\textit{UFD}}
Let $K$ be a number field. If $I$ is a fractional ideal in $K$, then there exist prime ideals $\mfp_1, \dots, \mfp_n$ and $\mfq_1, \dots, \mfq_m$ in $\ok$, unique up to ordering, such that 
\begin{equation*}
    I = (\mfp_1 \cdots \mfp_n)(\mfq_1 \cdots \mfq_m)^{-1}.
\end{equation*}
\end{theorem}
The theorem follows from the fact that a fractional ideal has the form $I=\frac{1}{a}J$, where $J$ is an integral ideal and $a \in \ok$. Since both $J$ and $(a)$ are integral ideals of $\ok$, Theorem \ref{thm:idealsOKUFD} implies they have unique prime ideal factorization. %, so the theorem holds.

\subsubsection{Applications in Ring LWE}
\label{subsec:appInRLWE}
As we will see in \Cref{section:rlwe}, when working on the hardness proof of the ring LWE problem, it is easier to view the underlying ring $\Z[x]/(\Phi_m(x))$ as a ring of integers in a cyclotomic number field, as opposed to the (more direct) interpretation of a ring of polynomials. 
This perspective change in interpretation is supported by the following two results.

\begin{theorem}\label{thm:ring of integers of Q(zeta)}
The ring of integers in $\Q(\zeta_m)$ is generated by $\zeta_m$: 
\[ \OO_{\Q(\zeta_m)} = \Z[\zeta_m]. \]
\end{theorem}
% \kl{I think this theorem is obvious, perhaps we shouldn't state it as an theorem?}
% The proof of the theorem is obvious but it has great importance in understanding the content of this paper. Happy to "downgrade" this to a Corollary if that makes you more comfortable?

\begin{theorem}\label{thm:ring LWE isomoprhic 2}
For all $m\in \N$, we have
\[ \Z[x]/(\Phi_m(x)) \cong \OO_{\Q(\zeta_m)} \]
\end{theorem}
\begin{proof}
This is a direct consequence of \Cref{thm:ring of integers of Q(zeta)} and \Cref{thm:ring LWE isomoprhic 1}.
\end{proof}

We state here two technical lemmas that will be needed in the RLWE result. The first lemma shows that given two ideals $I, J \subseteq R$ of a Dedekind domain $R$ (e.g., a ring of integers $\ok$ of a number field $K$ is a Dedekind domain), it is possible to construct another ideal that is coprime with either one of them. 

\begin{lemma}[Lemma 5.2.2 \citep{stein2012algebraic}, Lemma 2.1.4 \citep{lyubashevsky2010ideal}]
\label{lm:coprimeIdeals}
If $I$ and $J$ are non-zero integral ideals of a Dedekind domain $R$, then there exists an element $t \in I$ such that $(t)I^{-1} \subseteq R$ is an integral ideal coprime to $J$. 
\end{lemma}
\begin{proof}
Let $\mathfrak{p}_1, \dots, \mathfrak{p}_r$ be the prime factors of the ideal $J$. We  create a coprime ideal of $J$ as follows. Let $n_i$ be the largest power of $\mathfrak{p}_i$ such that $\mathfrak{p}_i ^{n_i} | I$ for all $i \in [1,r]$. 
As $\mathfrak{p}_i$ is a prime ideal, $\mathfrak{p}_i ^{n_i+1} \subsetneq \mathfrak{p}_i ^{n_i}$.%IDEAL FACTORIZATION KEITH CONRAD corollary 3.5
So there exits an element $t_i \in \mathfrak{p}_i^{e_i}$ such that it is not in $\mathfrak{p}_i ^{n_i+1}$. By construction, we know the ideals $\mathfrak{p}_1^{e_1+1}, \dots, \mathfrak{p}_r^{e_r+1}, I/\prod_{i=1}^r \mathfrak{p}_i^{e_i}$ are pairwise coprime, so by the Chinese Remainder Theorem, there is an element $t \in R$ such that $t \equiv t_i \bmod \mathfrak{p}_i^{e_i+1}$ and $t \equiv 0 \bmod I/\prod_{i=1}^r \mathfrak{p}_i^{e_i}$. Since $t_i \in \mathfrak{p}_i^{e_i}$, it entails $t \equiv 0 \bmod \mathfrak{p}_i^{e_i}$ for all $i \in [1,n]$, so $t \in I$ as in the lemma. 

To prove $(t)I^{-1}$ is coprime to $J$, it sufficient to show none of $J$'s prime divisor can divide it. Suppose $\mathfrak{p}_i | (t)\inv{I}$, then $\mathfrak{p}_i I | (t)$. The assumption $\mathfrak{p}_i^{e_i} | I$ implies that $\mathfrak{p}_i^{e_i+1} | (t)$, so $(t) \subseteq \mathfrak{p}_i^{e_i+1}$. This contradicts with the above that $t \equiv a_i \bmod \mathfrak{p}_i^{e_i+1}$. So the two are coprime. 
\end{proof}

The element $t \in I$ can be efficiently computable using CRT in $\ok$. Hence, given two ideals in $R$, we can efficiently construct another one that is coprime with either one of them. The next lemma is essential in the reduction from K-BDD problem to RLWE. 

% state the lemma according to lemma 2.15 lyubashevsky2010ideal, which is a special case of prop 5.2.4 in stein2012algebraic.
\begin{lemma}[Lemma 5.2.4 \citep{stein2012algebraic}, Lemma 2.1.5 \citep{lyubashevsky2010ideal}]
\label{lemma:MJM}
% to show homomorphism, need to prove the map is well-defined, from the way theta is defined, it's easy to see it's well-defined. 
% to prove isomorphism, first show the matp from M to IM/IJM has kernel JM, then we know a homomorphism quotient by its kernel induces an injective homomorphism, see JS Milne's group theory, theorem 1.45. 
Let $I$ and $J$ be ideals in a Dedekind domain $R$ and $M$ be a fractional ideal in the number field $K$. Then there is an isomorphism 
\begin{align*}
    M/JM \cong IM/IJM.
\end{align*}
\end{lemma}
\begin{proof}
Given ideals $I,J\subseteq R$, by Lemma \ref{lm:coprimeIdeals} we have $(t)I^{-1} \subseteq R$ is coprime to $J$ for an element $t \in I$. Then we can define a map 
\begin{align*}
    \theta_t: K &\rightarrow K \\
    u &\mapsto tu.
\end{align*}
This map induces a homomorphism 
\begin{align*}
    \theta_t: M \rightarrow IM/IJM.
\end{align*}
First, show $ker(\theta_t)=JM$. Since $\theta_t(JM)=tJM \subseteq IJM$, then $\theta_t(JM)=0$. Next, show any other element $u \in M$ that maps to 0 is in $JM$. To see this, if $\theta_t(u)=tu=0$, then $tu \in IJM$. To use Lemma \ref{lm:coprimeIdeals}, we re-write it as $(tI^{-1}) (uM^{-1})\subseteq J$. Since $tI^{-1}$ and $M$ are coprime, we have $uM^{-1}\subseteq J$, which implies $u\subseteq JM$. Therefore, $ker(\theta_t)=JM$ and
\begin{align*}
    \theta_t: M/JM \rightarrow IM/IJM
\end{align*}
is injective. 

Second, show the map is surjective. That is, for any $v \in IM$, its reduction $v \mod IJM$ has a preimage in $M/JM$. Since $tI^{-1}$ and $J$ are coprime, by CRT we can compute an element $c \in tI^{-1}$ such that $c = 1 \bmod J$. Let $a = cv \in tM$, then $a-v=cv-v=v(c-1) \in IJM$. Let $w=a/t \in M$, then $\theta_t(w)=t (a/t)=a = v \bmod IJM$. Hence, any arbitrary element $v \in IM$ satisfies the preimage of $v \bmod IJM$ is $w \bmod IM$. 
\end{proof}

In the hardness proof of RLWE as will be shown in Section~\ref{section:rlwe}, we can use Lemma~\ref{lemma:MJM} to show that for $R = \Z[x]/(\Phi_m(x))$, an ideal $I$ and a prime integer $q$,
% we let $M=R$ or $M=\dual{I}=I^{-1}\dual{R}$ and $J=(q)$ for a prime integer $q$, then the isomorphism becomes 
\begin{align*}
    R/(q)R &\cong I/(q)I  \\
    \dual{I} / (q)\dual{I} &\cong \dual{R} / (q)\dual{R},
\end{align*}
where $\dual{R}$ denotes the dual of $R$ that we will define later in Section~\ref{subsec:dualLatInNumField}.

We end this subsection by looking at the (unique) factorisation of the ideal $(q)$ in the ring of integers $R_q = \Z_q[x]/(\Phi_m(x))$.
Since $q$ is prime, the principal ideal generated by it can be split into prime ideals $\mathfrak{q}_i$ as follows: % in $R_q$
% \kl{(can q be a non-prime for the reduction to work? such as a prime power $q=p^k$? Intuitively it seems possible, if $q=ab$ is non-prime with prime factors $a$ and $b$, then we can use the same strategy to split $(a)$ and $(b)$ then combine the results.)}
\begin{equation*}
    (q)=\prod_{i=1}^{n/(ef)} \mfq_i^e = \prod_{i=1}^{n/(ef)} (q, F_i(\zeta_m))^e,
\end{equation*}
where $n = \varphi(m)$, $e=\varphi(q')$ is the Euler totient function of $q'$, the largest power of $q$ that divides $m$, $f$ is the multiplicative order of $q$ modulo $m/q'$, i.e., $q^f = 1 \bmod (m/q')$,
and
each $\mfq_i$ is generated by two elements, the prime number $q$ and the monic irreducible factor $F_i(x)$ of the cyclotomic polynomial $\Phi_m(x)= \prod_i (F_i(x))^e$ when splitting over $\Z_q[x]$ (see Theorem \ref{thm:splitCycPoly}). 
% Given $q$ and $m$, we can compute $q'$ that is the largest power of $q$ and divides $m$. 
% The exponent $e=\varphi(q')$ is the Euler totient function of $q'$, the largest power of $q$ that divides $m$, and $f$ is the multiplicative order of $q$ modulo $m/q'$, i.e., $q^f = 1 \bmod (m/q')$. 
% This gives a way to split an ideal $(q)$ into prime ideals in the ring of integers $R$ of an $n$-dimensional number field $K$. 
For details, see Chapter 4 of \citet{stein2012algebraic}.

\begin{example}\label{ex:q ideal factorisation}
For $m=5$, the 5th cyclotomic polynomial is
\begin{equation*}
    \Phi_5(x)=x^4+x^3+x^2+x+1,
\end{equation*}
so $n=4$ and $K=\Q(\zeta_5)$ the 4-dimensional cyclotomic field. Let $q=19$, then we have $q'=19^0=1$ to be the largest power of $q$ that divides $5$. So $e=\varphi(1)=1$ and the multiplicative order of $19 \bmod (4/1)$ is $f=2$. Assuming we are given how the cyclotomic polynomial splits in $\Z_{19}[x]$, i.e., 
\begin{equation*}
    \Phi_5(x)=x^4+x^3+x^2+x+1=(x^2+5x+1)(x^2+15x+1),
\end{equation*}
then we can split the ideal into prime ideals in the ring of integers $R=\Z[\zeta_5]$ as 
\begin{align*}
    (q)&=\mfq_1 \mfq_2 \\
    \implies (19)&=(19,(\zeta_5)^2+5\zeta_5+1)(19,(\zeta_5)^2+15\zeta_5+1).
\end{align*}
\end{example}

If we further restrict $q = 1 \bmod m$, it follows that $f=1$. In addition, it also entails that $q'=1$ and $e=1$. In addition, the cyclotomic polynomial $\Phi_m(x)=x^n+1$ can be split into $n$ linear factors $(x-\omega^i)$, where $\omega^i$ is a primitive $m$th root of unity in $\Z_q$. This satisfies the condition of Theorem \ref{thm:splitCycPoly} for $q$ and $m$ being coprime.\footnote{Note this also works if $q=p^k$ is a prime power coprime with $m$.} Hence, the ideal can be factored as 
\begin{align*}
    (q) % =\prod_{i=1}^{m} \mfq_i 
    &= \prod_{\substack{i=1,\ldots,m \\ \gcd(i,m)=1}} (q, \zeta_m - \omega^i) \\%, \text{ where } \gcd(i,m)=1 \\
    &=\prod_{i \in \Z_m^*} (q, \zeta_m - \omega^i).
\end{align*}
Note the index $i$ is not any integer between 1 and $m$, but those coprime with $m$. So for the above example, when $q=11 \cong 1 \bmod 5$, the polynomial splits in $\Z_{11}[x]$ as 
\begin{align*}
    \Phi_5(x)=(x-3)(x-9)(x-5)(x-4),
\end{align*}
where each 3, 9, 5, 4 is a primitive 5th root of unity in $\Z_{11}$, generated by the 1st, 2nd, 3rd and 4th power of 3 in $\bmod\, 11$. So the ideal splits as % in $R$ as 
\begin{align*}
    (q)&=\mfq_1 \mfq_2 \mfq_3 \mfq_4 \\
    \implies (11)&=(11,\zeta_5-3)(11,\zeta_5-9)(11,\zeta_5-5)(11,\zeta_5-4).
\end{align*}

%-----------------------------------------------------------------------------------------------

\subsection{Number field embedding}

Similar to LWE, the RLWE\index{ring LWE} problem's hardness is also based on hard lattice problems, except these are special lattices called \textit{ideal lattices}. In this subsection, we will study how algebraic objects such as ring of integers and its ideals are mapped to full-ranked lattices via embeddings. The embedding we will build is from a number field $K$ to the $n$-dimensional Euclidean space $\R^n$ or a space $H$ that is isomorphic to $\R^n$. 
% By the Primitive Element Theorem (Theorem \ref{app thm:primEleThm} in Appendix \ref{appen:ant}), there is an element $r \in K$ such that $K=\Q(r)$. Let $f(x) \in \Q[x]$ be the minimal polynomial of $r$ of degree $n$, by Theorem \ref{thm:fieldExtEquiv} in Appendix \ref{appen:galois theory} the number field can also be written as $K=\Q[x]/(f)$.
As $\ok$ and its ideals are additive groups, our embedding must preserves the additive group structure of these objects.

As a degree $n$ polynomial can be uniquely identified by its coefficients, our naive choice of embedding is by sending a polynomial $f=a_0 + a_1 x + \cdots a_{n-1} x^{n-1}$ to a coefficient vector $(a_0, a_1, \cdots, a_{n-1}) \in \R^n$. This coefficient embedding\index{coefficient embedding} is clearly an additive ring homomorphism and hence satisfies our basic requirements. Furthermore, it is related by a linear transformation to the canonical embedding that will be introduced next. However, the RLWE's proof and computations do not use the coefficient embedding. We list some reasons here and leave the details to Section~\ref{section:rlwe}.  
\begin{itemize}\itemsep1mm\parskip0mm
    \item Firstly, when working with cyclotomic fields, the canonical embedding makes both polynomial addition and multiplication efficient component-wise operations (under the point-value representation). These operations have simple geometric interpretations that lead to tight bounds.
    
    \item Secondly, in the coefficient embedding, specifying the error distribution in RLWE, which is an $n$-dimensional Gaussian, requires an $n$-by-$n$ covariance matrix in general. With the canonical embedding, the error distribution in RLWE takes the simple form of a product of one-dimensional Gaussians. This dramatically decreases the number of parameters that need to be taken care of when working with RLWE.
    
    \item Finally, the canonical embedding makes the Galois automorphisms simply permutations of the embedded vector components. This is important for the reduction from decision to search RLWE, and is not possible with the coefficient embedding.
    
\end{itemize}

\subsubsection{Canonical embedding}
\label{subsec:canonical embedding}\index{canonical embedding}

Let $K = \Q(\alpha) = \Q[x]/(f)$ be an extension field with degree $n$. Let $\alpha$ be a primitive element of $K$ (whose existence is proved by \Cref{app thm:primEleThm}) and $f \in \Q[x]$ be its minimal polynomial\index{minimal polynomial}.
Apart from the coefficient embedding, we will study an alternative embedding of $K$ into $\C^n$. 
Since $f$ is monic and irreducible in $\Q[x]$, and $\Q$ has characteristic 0, by \Cref{thm:minPolyIsSepInChar0Field} $f$ is separable, so it has $n$ distinct roots $\{\alpha_1, \dots,\alpha_n\}$ where the primitive element $\alpha$ is one of them. 
%So we can define $n$ different such embeddings.
For each root $\alpha_i$, we define a map 
% can define an embedding from the number field $K$ to the complex plane $\C$ by fixing $\Q$ and mapping $r$ to another root of $f$. That is, 
\begin{align*}
\sigma_i: K &\rightarrow \Q(\alpha_i) \subseteq \C \\
\alpha &\mapsto \alpha_i 
\end{align*}
sending $\alpha$ to $\alpha_i$ by 
\begin{align*}
    \sigma_i(a_0 + a_1\alpha + a_2 \alpha^2 + \cdots + a_{n-1} \alpha^{n-1}) = 
  a_0 + a_1\alpha_i + a_2 \alpha_i^2 + \cdots + a_{n-1}\alpha_i^{n-1},
\end{align*}
where $a_i \in \Q$. The map fixes $\Q$ in the sense that $\sigma_i(x)= x$ for all $x \in \Q$, so it is an automorphism \index{automorphism} (of the extension field \Cref{def:automorphismGroup}). 
One can show that %$\{ \sigma_i \}_{i=1}^n$ are the only embeddings of $K$ into $\C$, which implies 
these embeddings are independent of the choice of the primitive element.

Since the roots of $f$ consist of real and complex numbers, we can distinguish these embeddings as real and complex embeddings. If $\sigma_i(\alpha) \in \R$, then it is a \textbf{real embedding}, otherwise it is a \textbf{complex embedding}. 
By the Complex Conjugate Root Theorem, which states that the complex roots of real coefficient polynomials are in conjugate pairs, we know the images of the complex embeddings are in conjugate pairs. Let $s_1$ be the number of real embeddings and $s_2$ be the number of conjugate pairs of complex embeddings, then the total number of embeddings is $n=s_1 + 2s_2$. Let $\{\sigma_i\}_{i = 1}^{s_1}$ be the real and $\{\sigma_j\}_{j = s_l+1}^{n}$ be the complex embeddings, where $\sigma_{s_1 + j} = \overline{\sigma_{s_1 + s_2 + j}}$ are in the same conjugate pair for each $j \in [1,\ldots,s_2]$, then we have the following definition of a canonical embedding. 

\begin{definition}
	\label{def:canEmbd}
	\reversemarginpar
	\marginnote{\textit{Canonical embedding}}
	A \textbf{canonical embedding} $\sigma$ of an $n$-dimensional number field $K$ is defined as 
	\begin{align}
	\label{eq:canEmbed1}
	&\sigma: K \rightarrow \R^{s_1} \times \C^{2s_2} \subseteq \C^{s_1}\times \C^{2s_2} \cong \C^n \nonumber \\
	%\sigma(r) &\mapsto (\sigma_1(r), \dots, \sigma_{s_1}(r), \sigma_{s_1+1}(r),\dots, \sigma_{s_1+s_2}(r)),
	&\sigma(r) \mapsto (\sigma_1(r), \dots, \sigma_{s_1}(r), \sigma_{s_1+1}(r),\dots, \sigma_{s_1+2s_2}(r)).
	\end{align}
\end{definition}
By this definition, the canonical embedding maps a number field to an $n$-dimensional space, 
\reversemarginpar
\marginnote{\textit{Canonical space}}
named \textbf{canonical space}, which is expressed as  
\begin{equation*}
H = \left\{(x_1, \dots, x_n) \in \R^{s_1} \times \C^{2s_2} \mid x_{s_1 + j} = \overline{x_{s_1 + s_2 + j}}, \text{ for all } j \in [s_2]\right\}.   
\end{equation*}
Intuitively, one can think of the canonical embedding as sending each element $r \in K$ (i.e., a polynomial) to a coordinate (i.e., length $n$ vector) in the canonical space, where the coordinates are where $r$ sends the roots of $f$ to. 
%This intuition comes from the fact that canonical embedding $\sigma: K \rightarrow \C^n$ maps every element $r \in K$ to the vector $(r(\alpha_i), \dots, r(\alpha_n))$, where $\alpha_1, \dots \alpha_n$ are the $n$ (complex) roots of the polynomial $f(x)$ in $\C$.

%The \textbf{Minkowski space} $H_{\R}$ in \citep{mukherjee2016cyclotomic} is isomorphic to this canonical space $H$ with a minor difference. In $H_{\R}$, only one element from each complex conjugate pair is kept and it is written in two parts as $(Re, Im)$.
The canonical space $H$ can be shown to be isomorphic to $\R^n$ by establishing a one-to-one correspondence between the standard basis of $\R^n$ and a basis of $H$ as the row vectors in the following matrix
\begin{equation*}
\label{eq:basisMtxForH}
B = \left(
    \begin{matrix}
    I_{s_1\times s_1} & 0 & 0 \\
    0 & I_{s_2 \times s_2} & iI_{s_2 \times s_2} \\
    0 & I_{s_2 \times s_2} & -iI_{s_2 \times s_2}
    \end{matrix}
\right).
\end{equation*}
The matrix $I_{s_1 \times s_1}$ is the $s_1$ by $s_1$ identity matrix.\footnote{Note in \citet{lyubashevsky2010ideal}, the row vectors are multiplied by $\frac{1}{\sqrt{2}}$ to make them an orthonormal basis, so $B$ is a unitary matrix (i.e., $BB^*=I$, where $B^*$ is $B$'s conjugate transpose).} The image $\sigma(r) \in H$ can then be written in terms of this basis as a real vector 
\begin{align}
\label{eq:canEmbed2}
\tau(r) = (&\sigma_1(r), \dots, \sigma_{s_1}(r), \nonumber \\  
&Re(\sigma_{s_1+1}(r)),\dots,Re(\sigma_{s_1+s_2}(r)),Im(\sigma_{s_1+1}(r)),\dots, Im(\sigma_{s_1+s_2}(r)))
\end{align}
by taking the real and complex parts from two conjugate complex embeddings respectively. Taking the dot product of each row vector in $B$ with $\tau(r)$, we get back to $\sigma(r)$ in Equation \ref{eq:canEmbed1}, that is, 
\begin{align*}
    \sigma(r) = B \cdot (\tau(r))^T.
\end{align*}
\iffalse 
To be consistent with the reference books we used, we re-order entries of Equation \ref{eq:canEmbed3} so that 
\begin{align}
\label{eq:canEmbed2}
\sigma(r) = (&\sigma_1(r), \dots, \sigma_{s_1}(r), \nonumber \\  
&Re(\sigma_{s_1+1}(r)),Im(\sigma_{s_1+1}(r)), \dots,
Re(\sigma_{s_1+s_2}(r)),Im(\sigma_{s_1+s_2}(r)))
\end{align}
\fi 
%We suggest the reader to visualize the canonical embedding as in Equation \ref{eq:canEmbed1}, but use Equation \ref{eq:canEmbed2} when computing things related to the canonical embedding. 

%For computational purpose, we treat $\C\cong \R^2$ and it is not necessary to keep both complex embeddings in each conjugate pair, so a practical way of explicitly writing out Equation \ref{eq:canEmbed1} is  

Here are some examples to illustrate canonical embedding, canonical space and its basis. 
\begin{example}
When $K=\Q(\sqrt{2})$ is a quadratic field. The minimal polynomial of $\sqrt{2}$ is $x^2-2$, which has two roots $\pm \sqrt{2}$. The canonical embedding consists two real embeddings only and is defined as 
\begin{align*}
    \sigma(\sqrt{2}) = (\sqrt{2},-\sqrt{2}).
\end{align*}
The basis of the canonical space $H$ is 
\begin{equation*}
B = \left(
    \begin{matrix}
    1 & 0 \\
    0 & 1
    \end{matrix}.
\right)
\end{equation*}
%and
%\begin{align*}
%    \tau(\sqrt{2}) = (\sqrt{2},-\sqrt{2}).
%\end{align*}
Given the integral basis $\{1,\sqrt{2}\}$ of $K$, the basis vectors are mapped to the canonical space $H$ and can be written in terms of the basis of $H$ as real vectors    
\begin{align*}
    \tau(1) &= (1,1) \\
    \tau(\sqrt{2}) &= (\sqrt{2},-\sqrt{2}),
\end{align*}
which form a $\Z$-basis of the image $\sigma(\ok)$, that is, $\sigma(\ok) = \{a(1,1)+b(\sqrt{2},-\sqrt{2}) \mid a, b \in Z\}$.
\end{example}

\begin{example}
When $K=\Q(\zeta_8)$ is the 8th cyclotomic field. The 8th primitive root of unity $\zeta_8 = \frac{\sqrt{2}}{2}+ i \frac{\sqrt{2}}{2}$ and its minimal polynomial is the 8th cyclotomic polynomial $\Phi_8(x) = x^4+1$. The roots of $\Phi_8(x)$ are
\begin{align*}
\zeta_8 &= \frac{\sqrt{2}}{2}+ i \frac{\sqrt{2}}{2}, \text{ }
\zeta_8^3 = -\frac{\sqrt{2}}{2}+ i \frac{\sqrt{2}}{2}, \\
\zeta_8^5 &= -\frac{\sqrt{2}}{2}- i \frac{\sqrt{2}}{2}, \text{ }
\zeta_8^7 = \frac{\sqrt{2}}{2}- i \frac{\sqrt{2}}{2}.
\end{align*}
The canonical embedding consists of exactly four complex embeddings, i.e., $\sigma=(\sigma_1,\sigma_2,\sigma_3,\sigma_4)$,  
\begin{align*}
    \sigma_1\left(\frac{\sqrt{2}}{2}+ i \frac{\sqrt{2}}{2}\right) &= \frac{\sqrt{2}}{2}+ i \frac{\sqrt{2}}{2}, \text{ }
    \sigma_2\left(\frac{\sqrt{2}}{2}+ i \frac{\sqrt{2}}{2}\right) = -\frac{\sqrt{2}}{2}+ i \frac{\sqrt{2}}{2}, \\
    \sigma_3\left(\frac{\sqrt{2}}{2}+ i \frac{\sqrt{2}}{2}\right) &= \frac{\sqrt{2}}{2}- i \frac{\sqrt{2}}{2}, \text{ }
    \sigma_4\left(\frac{\sqrt{2}}{2}+ i \frac{\sqrt{2}}{2}\right) = -\frac{\sqrt{2}}{2}- i \frac{\sqrt{2}}{2}, 
\end{align*}
where $\sigma_1=\overline{\sigma_3}$ and $\sigma_2=\overline{\sigma_4}$ are in conjugate pairs. 
The basis of the canonical space $H$ is 
\begin{equation*}
B = \left(
    \begin{matrix}
    1 & 0 & i & 0 \\
    0 & 1 & 0 & i \\
    1 & 0 & -i & 0 \\
    0 & 1 & 0 & -i
    \end{matrix}
\right).
\end{equation*}
By Equation \ref{eq:canEmbed2}, the canonical embedding of the primitive element $\zeta_8$ can be written in terms of this basis as the real vector 
\begin{equation*}
    \tau\left(\frac{\sqrt{2}}{2}+ i \frac{\sqrt{2}}{2}\right)=\left(Re(\sigma_1),Re(\sigma_2),Im(\sigma_1),Im(\sigma_2)\right) =\left(\frac{\sqrt{2}}{2}, -\frac{\sqrt{2}}{2}, \frac{\sqrt{2}}{2}, \frac{\sqrt{2}}{2} \right).
\end{equation*}
By multiplying each row of $B$ with this expression, we get back to the canonical embedding $\sigma=(\sigma_1,\sigma_2,\sigma_3,\sigma_4)$. 
\end{example}

Given the canonical embedding, it allows us to talk about the geometric norm of an algebraic element $x \in K$. More precisely, we can define the \textbf{$L_p$-norm} 
\reversemarginpar
\marginnote{\textit{$L_p$-norm}}
of $x$ by looking at the $L_p$-norm of its image $\sigma(x)$ that is embedded into the real space $\R^n$
\begin{equation}
\label{eq:lpNorm}
||x||_p = ||\sigma(x)||_p = 
\begin{cases}
\left( \sum_{i \in [n]} |\sigma_i(x)|^p \right)^{1/p} & \text{ if $p < \infty$},  \\
\max_{i \in [n]} |\sigma_i(x)| & \text{ if $p = \infty$}.
\end{cases}
\end{equation}

In the next example, we illustrate the $L_p$-norm of a root of unity in a cyclotomic field. 
\begin{example}
	
	Let $K=\Q(\zeta_n)$ be the nth cyclotomic field and $\sigma: K \rightarrow H$ be its canonical embedding. The cyclotomic polynomial $\Phi_n(x)$ is the minimal polynomial of $\zeta_n$ and it has only complex roots for $n \ge 3$, as the two real roots are non-primitive. Since the Galois group $Gal(K/\Q) \cong (\Z/n\Z)^*$ is isomorphic to the multiplicative group (Theorem \ref{thm:galGrpCycField}), the complex embeddings are given by $\sigma_i(\zeta_n) = \zeta_n^i$ for $i \in (\Z/n\Z)^*$ and $n = 2s_2 = |(\Z/n\Z)^*|$.
	Since the primitive roots of unity are closed under $\sigma_i$, the magnitude $|\sigma_i(\zeta_n^j)|=1$. So the $L_P$-norm of an nth root of unity is  $||\zeta_n^j||_p = n^{1/p}$ for $p < \infty$ or $||\zeta_m^j||_{\infty} = 1$.
\end{example}

We have shown that the canonical embedding $\sigma$ sends a number field to a space isomorphic to $\R^n$. When restricted to the ring of integers $\ok$ that is closed under addition, we would like to see what $\sigma$ does to preserve the discreteness and the additive group structure of $\ok$. The following theorem states that the canonical embedding maps $\ok$ to a full-rank lattice. 

%Towards the end of this section, we will discuss the minimum distance (or the shortest vector) of this lattice and how the determinant of this lattice $\sigma(\ok)$ is related to a quantity of the number field, called the discriminant.

\begin{theorem}
	\label{thm:rngIntLat}
	\reversemarginpar
	\marginnote{\textit{$\tau(\ok)$ is lattice}}
	Let $K$ be an $n$-dimensional number field, then $\sigma(\ok)$ is a full-rank lattice in $\R^n$. 
\end{theorem}

\begin{proof}	
	Let $\{e_1,\dots,e_n\}$ be an integral basis of $\ok$, then every element $x \in \ok$ can be written as $x=\sum_{i=1}^n z_i e_i$, where $z_i \in Z$. The embedding of $x$ can then be written as $\sigma(x)=\sum_{i=1}^n z_i \sigma(e_i)$, where the coefficients are fixed because $\sigma$ fixes $\Q$. Hence, $\sigma(\ok)$ is also a $\Z$-module generated by $\{\sigma(e_1),\dots,\sigma(e_n)\}$. 
	
	By definition, a lattice is a free $\Z$-module. If we can show $\{\sigma(e_1),\dots,\sigma(e_n)\}$ is a basis of $\sigma(\ok)$, then $\sigma(\ok)$ is a free $\Z$-module. To do so, write each $\sigma(e_i)$ in terms of the canonical space basis according to Equation \ref{eq:canEmbed2} as a real vector, so we have the following basis matrix for $\sigma(\ok)$
	\begin{equation*}
	N^T = \left(
	\begin{smallmatrix}
	\sigma_1(e_1) & \cdots & \sigma_{s_1}(e_1) & Re(\sigma_{s_1+1}(e_1)) & \cdots & Re(\sigma_{s_1+s_2}(e_1)) & Im(\sigma_{s_1+1}(e_1)) & \dots & Im(\sigma_{s_1+s_2}(e_1)) \\
	\vdots & & \vdots & \vdots & \vdots & & \vdots & \vdots \\
	\sigma_1(e_n) & \cdots & \sigma_{s_1}(e_n) & Re(\sigma_{s_1+1}(e_n)) & \cdots & Re(\sigma_{s_1+s_2}(e_n)) & Im(\sigma_{s_1+1}(e_n)) & \dots & Im(\sigma_{s_1+s_2}(e_n)) \\
	\end{smallmatrix}
	\right).
	\end{equation*}
	Then show that the matrix has a non-zero determinant, and consequently the rows are independent. By Equation \ref{eq:canEmbed1} of canonical embedding, we can write the images of the integral basis $\{e_1, \dots, e_n\}$ under the canonical embedding as the matrix
	\begin{equation*}
	M^T = \left(
	\begin{smallmatrix}
	\sigma_1(e_1) & \cdots & \sigma_{s_1}(e_1) & \sigma_{s_1+1}(e_1) & \overline{\sigma_{s_1+1}}(e_1) & \cdots & \sigma_{s_1+s_2}(e_1) & \overline{\sigma_{s_1+s_2}}(e_1) \\
	\vdots & & \vdots & \vdots & \vdots & & \vdots & \vdots \\
	\sigma_1(e_n) & \cdots & \sigma_{s_1}(e_n) & \sigma_{s_1+1}(e_n) & \overline{\sigma_{s_1+1}}(e_n) & \cdots & \sigma_{s_1+s_2}(e_n) & \overline{\sigma_{s_1+s_2}}(e_n)
	\end{smallmatrix}
	\right).
	\end{equation*}
	The two matrices are of the same dimension and their determinants are related by 
	\begin{align}
	\label{eq:detN}
		\det N = \frac{1}{2^{s_2}}\det M,
	\end{align}
	so it remains to show $\det M \neq 0$. If a rational matrix $A$ changes a basis of $K$ to another basis by 
	\begin{align*}
		e_j'=\sum_{k} A_{kj} e_k,
	\end{align*}
	then the above matrix $M$ is also changed to a new matrix $M'=M A$. % (Lemma 1.7.1 Ben Green). 
	We know $K$ always has a power basis $\{1, r, \dots, r^{n-1}\}$ (Theorem~\ref{app thm:primEleThm}) 
	and the matrix $M^T$ in terms of the power basis is a \textit{Vandermonde matrix}\index{Vandermonde matrix} with a non-zero determinant as the powers of $r$ are all distinct. Then we can conclude that the above matrix $M$ has non-zero determinant and so does the matrix $N$. 
	%See the proof of Lemma 10.6.1 on page 65 of Ben Green's book or the proof of Proposition 4.26 on page 80 of Milne's book. 
\end{proof}

An important corollary of Theorem~\ref{thm:rngIntLat} is that every fractional ideal of $K$ is also mapped to a full-rank ideal. 

\begin{corollary}
If $I$ is a fractional ideal in an $n$-dimensional number field $K$, then $\sigma(I)$ is a full-rank lattice in $\R^n$.
\end{corollary}
\begin{proof}
Given $I$ is a fractional ideal in $K$, for a non-zero integer $m \in K$ we have $m \ok \subseteq I \subseteq \frac{1}{m}\ok$, and both the subset and superset of $I$ are full-rank lattices in $\R^n$, so is $I$. See Lemma 7.1.8 of \citet{stein2012algebraic} for more detail. 
\end{proof}

As mentioned earlier, the canonical embedding allows polynomial addition and multiplication to be done component-wise efficiently, which is a convenient feature for both the deduction from search to decision RLWE and polynomial computations. We explain next why such a nice feature comes with the canonical embedding. We know a polynomial can be uniquely represented by both the coefficient and point-value representations, and the latter allows us to multiply two polynomials component-wise \citep{cormen01introduction}.
To allow efficient transformation $O(n\log n)$ between the two representations, we should evaluate a degree $n$ polynomial at the n-th roots of unity, which is essentially what \textit{fast Fourier transform} (FFT) does. We know both the n-th cyclotomic field $K$ and its ring of integers $\ok$ have a power basis $B=\{1, \zeta_n, \dots, \zeta_n^{\varphi(n)-1}\}$, which consists of the n-th roots of unity just as we need. 
We can use the power basis to build a Vandermonde matrix $M^T$.\index{Vandermonde matrix} 
Since $K$ can also be interpreted as a polynomial ring quotient by the ideal $(f)$, an element $a \in K$ can be viewed as $a(x) = \sum_{i=0}^{n-1} a_i x^i$ and its image under the embedding is $\sigma_i(a(x))=a(\sigma_i(x))$. Hence, each embedding $\sigma_i(a(x))$ is equivalent to evaluate $a(x)$ at $\sigma_i(x)$. Therefore, we have 
\begin{align*}
    M^T \cdot (a_0, \dots, a_{n-1})^T = \sigma(a) = B \cdot (\tau(a))^T. 
\end{align*}
Therefore, for a polynomial $a \in \ok$, its image $\sigma(a)$ (or $\tau(a)$ in terms of the basis $B$) is precisely its point-value representation evaluated at the n-th roots of unity.   

% \kl{This is a great interpretation. I think 
In short, when using the canonical embedding, the image of $K$ is a lattice with a power basis consisting of the primitive roots of unity. Since each element in $K$ is also a polynomial, when converting to the point-value representation, the primitive roots of unity are the precise points that are needed. So adding or multiplying two polynomials in the point-value representation is equivalent to adding or multiplying two elements $\sigma(K)$ w.r.t. the power basis.
% }

%is mapped to a basis $\sigma(B)=\{\sigma(1), \sigma(\zeta_n), \dots, \sigma(\zeta_n^{n-1})\}$ for the ideal lattice $\sigma(\ok)$. Furthermore, the canonical embedding for cyclotomic field are simply permutations of the roots of unity, so the basis $\sigma(B)$ also consists of the nth roots of unity. 

\subsubsection{Geometric quantities of ideal lattice}\index{ideal lattice}

We know from the previous subsection that a fractional ideal $I$ in a number field is mapped by a canonical embedding $\sigma$ to a lattice in the Euclidean space, called \textit{ideal lattice}. In this subsection, we will go through some geometric quantities of $I$ (i.e., its ideal lattice $\sigma(I)$) including its determinant and minimum distance. The results in this subsection are directly related to the gap (or approximation) factors of hard ideal lattice problems. 
% Before we start, the reader may find it helpful to read through the trace and norm section in Appendix \ref{appen:ant} to get familiar with these two concepts in terms of elements in a number field. 

%We first state the main result of this section, which explicitly states the determinant of an ideal lattice. We defer the proof till the end of this section. The reader can refer to Proposition 4.26 in J. S. Milne's book or Corollary 10.6.2 in Ben Green's book. 

To begin with, we first state the main result that is directly relevant to the RLWE's\index{ring LWE} hardness proof. Recall that the minimum distance $\lambda_1(L)$ of a lattice $L$ is the length of the shortest non-zero vector in $L$, where the length is measured by $L_p$-norm as defined in Equation \ref{eq:lpNorm}.

\begin{lemma}
Let $I$ be a fractional ideal in an $n$-dimensional number field $K$, then its minimum distance measured by $L_p$-norm satisfies 
\begin{align}
\label{eq:minDistBound}
    n^{1/p} \cdot N(I)^{1/n} \le \lambda_1(I) \le n^{1/p} \cdot N(I)^{1/n} \cdot \sqrt{\Delta_K^{1/n}}.
\end{align}
\end{lemma}

Here, $N(I)$ is the norm of the fractional ideal and $\Delta_K$ is the discriminant of the number field $K$. We will introduce these concepts next, which not only helps to understand the lemma, but give insights about the algebraic structures of $\ok$ and its ideals under the canonical embedding. 

Given a subgroup $H$ of $G$, the Lagrange's Theorem says that the order of $G$ satisfies $|G|=|G:H||H|$, where $|G:H|$ is the index of $H$ that measures the number of cosets of $H$ in $G$. If $H$ is a normal subgroup, then the index is equivalent to the order of the quotient group $G/H$. Since an ideal $I$ of $\ok$ is an additive normal subgroup and it has a geometric interpretation due to the canonical embedding, we relate its index to the norm as next.    

\begin{definition}
\label{def:idealNorm}
\reversemarginpar
\marginnote{\textit{Ideal norm}}\index{ideal norm}
Let $I$ be a non-zero ideal of $\ok$. The \textbf{norm} of $I$, denoted by $N(I)$, is the index of $I$ as a subgroup of $\ok$, i.e., $N(I) = |\ok / I|$.
\end{definition}

As for the norm of number field elements (Appendix \ref{appen:ant}), the norm of ideals is also multiplicative. That is, $N(IJ) = N(I)N(J)$. If $I=J/d$ is a fractional ideal in $K$ with the integral ideal $J$, then its norm is
\begin{align}
\label{eq:idealNorm}
    N(I) = N(dI) / |N(d)|
\end{align}

\begin{example}
When $\ok=\Z$, the integral ideal $J=5\Z$ and the fractional ideal $I=J/4=\frac{5}{4}\Z$, the norm $N(I)=N(J)/|N(4)|=5/4$.
\end{example}

For the fractional ideal $I$ and integral ideal $dI$ with $d \in \ok$, we have $dx \in dI$ for any non-zero $x \in I$. Hence, when viewed as subgroups, their indices satisfies $[\ok:(dx)] \ge [\ok:dI]$ and it follows $N(dx)\ge N(dI)$. By Equation \ref{eq:idealNorm} and the multiplicity of norm, we have $N(x) \ge N(I)$ for any non-zero $x \in I$. Combine this with Equation \ref{eq:lpNorm} of $L_p$-norm, we can prove the lower bound of $\lambda_1(I)$. The upper bound is proved by the discriminant of $K$ and Minkowski's First Theorem\index{Minkowski} (\Cref{app thm:min 1st}; see also Lemma 6.1 of \citet{peikert2007lattices} for the proof of the upper bound).

The discriminant of a number field loosely speaking measures the size of the ring of integers $\ok$. Without loss of generality, for the basis elements $e_1, \dots, e_n$ of $K$, define the $n$ by $n$ matrix 
\begin{equation*}
M = 
\begin{pmatrix}
\sigma_1(e_1) & \sigma_1(e_2) & \cdots & \sigma_1(e_n) \\
\sigma_2(e_1) & \sigma_2(e_2) & \cdots & \sigma_2(e_n) \\
\vdots & \vdots & \cdots & \vdots \\
\sigma_n(e_1) & \sigma_n(e_2) & \cdots & \sigma_n(e_n) 
\end{pmatrix},
\end{equation*}
where $\sigma=(\sigma_1,\dots,\sigma_n)$ is the canonical embedding of $K$. By the same argument in the proof of Theorem \ref{thm:rngIntLat}, we know the determinant of $M$ is non-zero. We know this matrix is related to the basis matrix $N$ of the ideal lattice and their determinants satisfy Equation \ref{eq:detN}. This matrix looks just like the basis matrix for a lattice that was introduced in Section \ref{section:lattice theory}. Now we are ready to define the discriminant of $K$. 

\begin{definition}
Let $K$ be an $n$-dimensional number field with an integral basis $\{e_1, \dots, e_n\}$. 
The \textbf{discriminant} \reversemarginpar\marginnote{\textit{$\Delta_K$}}\index{$\Delta_K$}
of $K$ is 
\begin{equation*}
	\Delta_K = \text{disc}_{K/\Q}(e_1, \dots, e_n) = \det (M)^2.
\end{equation*}
\end{definition}

An important property of number field discriminant is that it is invariant under the choice of an integral basis. This can be seen from the following lemma and corollary. 

\begin{lemma}
Suppose  $x_1, \dots, x_n, y_1, \dots, y_n \in K$ are elements in the number field and they are related by a transformation matrix $A$, then 
\begin{equation*}
    \text{disc}_{K / \Q}(x_1, \dots, x_n) = det (A)^2 \text{disc}_{K / \Q}(y_1, \dots, y_n).
\end{equation*}
\end{lemma}

Since the change of integral basis matrix $A$ is an unimodular matrix, i.e., $\det A = \pm 1$, we conclude that discriminant is an invariant quantity. 
\begin{corollary}
\reversemarginpar
\marginnote{\textit{Invariant $\Delta(K)$}}
Suppose $\{e_1, \dots, e_n\}$ and $\{e'_1, \dots, e'_n\}$ are both integral bases of the number field $K$, then 
\begin{equation*}
    \text{disc}_{K / \Q}(e_1, \dots, e_n) =  \text{disc}_{K / \Q}(e'_1, \dots, e'_n).
\end{equation*}
\end{corollary}

We finish this subsection by making some observations about $\Delta_K$. First, the determinant of the basis matrix $M$ is equivalent to the fundamental domain of $\sigma(\ok)$. This entails that the absolute\footnote{Although it is defined as the square of a matrix determinant, discriminant can be negative as the matrix entries can be complex numbers.} discriminant of $K$ measures the geometric sparsity of $\ok$. Larger $|\Delta_K|$ implies larger $\det M$, so the more sparse the ideal lattice is. 

Second, equation \ref{eq:detN} says $|\det N| = \frac{1}{2^{s_2}} |\det M|$. Since $N$ is the basis matrix of the ideal lattice $\sigma(\ok)$, by definition of field discriminant, this equation implies 
\begin{align}
\label{eq:detOK}
    \det (\sigma(\ok)) = \frac{1}{2^{s_2}} \sqrt{|\Delta_K|}.
\end{align}

Finally, an integral lattice $I$ is an additive subgroup of $\ok$ so Lagrange's Theorem entails $|\ok|=|\ok:I||I|$. The canonical embedding $\sigma$ is an isomorphism between $\ok$ and $I$ to the corresponding ideal lattices. Moreover, $I$ being a subgroup is sparser than $\ok$ when mapped by $\sigma$, so has larger determinant. Hence, we have 
\reversemarginpar
\marginnote{\textit{Ideal lattice determinant}}
\begin{align}
\label{eq:detIdealLat}
    \det (\sigma(I)) &= [\sigma(\ok):\sigma(I)] \det (\sigma(\ok)) \nonumber \\ 
    &= N(I) \det (\sigma(\ok)) \nonumber \\
    &= \frac{1}{2^{s_2}} N(I) \sqrt{|\Delta_K|}  
\end{align}

Equation \ref{eq:detIdealLat} also holds for a fractional ideal $J=I/d$. Substitute the integral ideal $I=dJ$ into the equation will incur a factor $d$ on both sides, because $\det(\sigma(dJ))=d\det(\sigma(J))$ and $N(dJ)=N(d)N(J)=dN(J)$.

%--------------------------------------------------------------------

\subsection{Dual lattice in number field}
\label{subsec:dualLatInNumField}

%For the proof details and intuitions about dual lattice, the reader should refer to Conrad's lecture notes on ``Different ideal''.

In the previous subsection, we have built a connection between a number field $K$ and its image $H=\sigma(K)$ under the canonical embedding $\sigma$ and shown that $H \cong \R^n$. 
% Through this subsection, we speak of $K$ as if it is a geometric space isomorphic to $\R^n$. 
In this subsection, we discuss how dual lattices in $K$ are defined. The motivation is to understand the structure of dual lattices of an ideal lattice $\sigma(I)$. The notion of dual appears in crucial parts of the development of lattice-based cryptography, including the definition of smoothing parameters of a lattice (Definition \ref{def:smthPara}) and the general definition of RLWE distribution (Definition \ref{def:rlwe2}). 

\begin{definition}
\reversemarginpar
\marginnote{\textit{Lattice in $K$}}
A \textbf{lattice} in an $n$-dimensional number field $K$ is the $\Z$-span of a $\Q$-basis of $K$.  
\end{definition}

For lattices in $\R^n$, dot product is an obvious metric between two geometric vectors.
For lattices in a number field, we need a more general inner product that can be obtained through the trace operator.
\begin{definition}
% \label{app def:trcNorm2}
Given a canonical embedding of a number field $K$
\begin{align*}
    \sigma &: K \rightarrow \R^{s_1} \times \C^{2s_2} \\
    \sigma(\alpha) &\mapsto (\sigma_1(\alpha), \dots, \sigma_n(\alpha)),
\end{align*}
the \textbf{trace} of an element $\alpha \in K$ is defined as 
\reversemarginpar
\marginnote{\textit{Trace operator}}\index{trace operator}
\begin{align*}
    Tr_{K \setminus \Q}&: K \rightarrow \Q \\%\text{  s.t.  }  
    Tr_{K/\Q}(\alpha) &= \sum_{i =1}^n \sigma_i(\alpha). %, \\
%    N_{K \setminus \Q}&: K \rightarrow \Q \\%\text{ s.t. } 
%    N_{K / \Q}(\alpha) &= \prod_{i \in [n]} \sigma_i(\alpha). 
\end{align*}
\end{definition}
\noindent From that, we obtain the trace inner product\index{trace inner product} as follows:
\begin{align}
\label{equ:trace}
    Tr_{K/\Q}(xy) = \sum \sigma_i(xy) = \sum \sigma_i(x) \sigma_i(y) = \langle \sigma(x), \overline{\sigma(y)} \rangle.
\end{align}

\begin{definition}
\reversemarginpar
\marginnote{\textit{Dual lattice}}\index{dual lattice}
Let $L$ be a lattice in a number field $K$. Its \textbf{dual lattice} is 
\begin{equation*}
    L^{\vee} = \{x \in K \mid Tr_{K/Q}(xL) \subseteq \Z\}.
\end{equation*}
\end{definition}

% To check whether or not an element belongs to the dual, one can check its trace product with the lattice basis. This also gives a way of writing out the dual of a given lattice. 

\begin{example}
The lattice $L=\Z[i]$ in the number field $K=\Q(i)$ has a basis $B=\{1,i\}$. 
\iffalse
To find $\dual{L}$, take an element $a+bi \in K$ and solve its trace inner product with the basis  
\begin{align*}
    Tr_{K/\Q}(a+bi) &\in \Z \\
    Tr_{K/\Q}((a+bi)i) &\in \Z.
\end{align*}
Let $\alpha=a+bi$ and $\beta=-b+ai$. By Definition \ref{app def:trcNorm} of trace, we have $[m_{\alpha}] = \begin{pmatrix}
  a & -b\\ 
  b & a
\end{pmatrix}$ and 
$[m_{\beta}] = \begin{pmatrix}
  -b & -a\\ 
  a & -b
\end{pmatrix}$. For both traces to be integers, it entails $2a \in \Z$ and $2b \in \Z$, so 
\fi
The dual lattice $L^{\vee}=\frac{1}{2}\Z[i]$ with a basis $B^{\vee}=\{\frac{1}{2},\frac{i}{2}\}$.
\end{example}

\iffalse
% From the example, it can be seen that the basis and the dual basis satisfy $Tr(e_i e_j^{\vee}) = \delta_{ij}$. This gives rise to the following theorem that states the dual of a number field lattice is also a lattice. 

\begin{theorem}
\label{thm:dualBasis}
\reversemarginpar
\marginnote{\textit{$\dual{L}$ is lattice}}
For an $n$-dimensional number field $K$ and a lattice $L$ in $K$ with a $\Z$-basis $\{e_1, \dots, e_n\}$, the dual $L^{\vee}=\bigoplus \Z e_i^{\vee}$ is a lattice with a dual basis  $\{e_1^{\vee}, \dots, e_n^{\vee}\}$ satisfying $Tr_{K/\Q}(e_i e_j^{\vee}) = \delta_{ij}$.\marginpar{what is $\delta_{ij}$}
\end{theorem}
\fi

The dual of a number field lattice is also a lattice.
Here are some properties of the dual in $\R^n$ that also hold true for dual in number fields.

\begin{corollary}
For lattices in a number field $K$, the following hold: 
\begin{enumerate}
    \item $L^{\vee \vee}=L$,
    \item $L_1 \subseteq L_2 \iff \dual{L_2} \subseteq \dual{L_1}$,
    \item $\dual{(\alpha L)} \iff \frac{1}{\alpha}\dual{L}$, for an invertible element $\alpha \in K$.
\end{enumerate}
\end{corollary}

The following theorem relates the dual lattice to differentiation and provides an easier way of computing the dual basis and dual lattice from a given lattice. 

\begin{theorem}
\label{thm:dualLatDiff}
\reversemarginpar
\marginnote{\textit{Dual basis}}\index{dual basis}
Let $K=\Q(\alpha)$ be an $n$-dimensional number field with a power basis $\{1, \alpha, \dots, \alpha^{n-1}\}$ and $f(x) \in \Q[x]$ be the minimal polynomial of the element $\alpha$, which can be expressed as 
\begin{equation*}
    f(x) = (x-\alpha)(c_0 + c_1 x + \dots + c_{n-1} x^{n-1}).
\end{equation*}
Then the dual basis to the power basis relative to the trace product is $\left\{\frac{c_0}{f'(\alpha)}, \dots, \frac{c_{n-1}}{f'(\alpha)}\right\}$.
In particular, if $K=\Q(\alpha)$ and the primitive element $\alpha \in \ok$ is an algebraic integer, then the lattice $L=\Z[\alpha]=\Z + \Z\alpha + \dots + \Z \alpha^{n-1}$ and its dual are related by the first derivative of the minimal polynomial, that is, 
\begin{equation*}
    \dual{L} = \frac{1}{f'(\alpha)}L.
\end{equation*}
\end{theorem}

\begin{example}
\label{ex:dualL}
An important application of this theorem in RLWE\index{Ring LWE} is when $K=\Q[\zeta_m]$ is the m-th cyclotomic number field, where $m=2n=2^k>1$ is a power of 2. Let the lattice $L=\ok=\Z[\zeta_m]$. The minimal polynomial of $\zeta_m$ is $f(x)=x^n+1$, whose derivative is $f'(x)=nx^{n-1}$. By Theorem \ref{thm:dualLatDiff},
\begin{equation*}
    \dual{L}=\dual{(\Z[\zeta_m])} = \frac{1}{f'(\zeta_m)} \Z[\zeta_m] = \frac{1}{n\zeta_m^{n-1}} \Z[\zeta_m] = \frac{1}{n} \zeta_m^{n+1} \Z[\zeta_m]=\frac{1}{n}L.
\end{equation*}
The second last equality is because the roots of unity form a cyclic group so $\zeta_m^{-(n-1)} = \zeta_m^{n+1}$.
\end{example}
This example shows an essential property of cyclotomic number fields when choosing appropriate parameter settings. It says the ideal lattice $\sigma(\ok)$ and its dual are related by only a scaling factor, so there is no difference working in either domain when defining the RLWE problem. We will see more detail in the next section. 

We further study the ideal lattice $\ok$ in a general number field. By definition, the dual of $\ok$ is 
\begin{equation*}
    \dual{\ok} = \{x \in K \mid Tr_{K/\Q}(x \ok) \subseteq \Z\}.
\end{equation*}
Since each element in $\ok$ is an algebraic integer, in that has an integer trace.\footnote{This can be verified by taking the power basis $\{1, r, \dots, r^{n-1}\}$ of $K$ which is also a $\Z$-basis of $\ok$. Each $x \in \ok$ can be written as $x=c_0+c_1 r + \dots + c_{n-1}r^{n-1}$. By definition, only $Tr(c_0) \in \Z$ and the rest are 0.} So on the one hand, $\ok \subseteq \dual{\ok}$. On the other hand, not all elements with integer traces are in $\dual{\ok}$. The next theorem shows that these elements need to form a fractional ideal. 

\begin{theorem}
\reversemarginpar
\marginnote{\textit{$\dual{\ok}$ is frac ideal}}
The dual lattice $\dual{\ok}$ is the largest fractional ideal in $K$ whose elements have integer traces. 
\end{theorem}
\iffalse
\begin{proof}
Let $I$ be a fractional ideal in $K$. By definition of fractional ideal, we have $I\ok=I$. Then we have 
\begin{align*}
    Tr(I) \subseteq \Z \iff Tr(I\ok)\subseteq \Z \iff I \subseteq \dual{\ok}
\end{align*}
The largest such fractional ideal is precisely the dual. If there is a set whose elements have all integer trace but not a fractional ideal, then the first ``iff'' does not follow.  If there is a larger fractional ideal who contains an element with non-integer trace, then it does not in the dual.
\end{proof}
\fi
% The next theorem shows the relation between $\dual{\ok}$ and a random fractional ideal dual. 

\begin{theorem}
\label{thm:fracIdealDual}
%\reversemarginpar
%\marginnote{\textit{Frac ideal dual}}
For a fractional ideal $I$ in $K$, its dual lattice is a fractional ideal satisfying the equation $\dual{I} = I^{-1} \dual{\ok}$.
\end{theorem}

We have seen the inverse of a fractional ideal in Equation \ref{equ:fracIdInv}, it is tempting to see if the inverse of the dual $\dual{\ok}$ (which is also a fractional ideal) is any special. By definition of fractional ideal inverse (Equation \ref{equ:fracIdInv}), we have 
\begin{align*}
    \inv{(\ok)} &= \{x \in K \mid x \ok \subseteq \ok \} = \ok \\
    \inv{(\dual{\ok})} &= \{x \in K \mid x \dual{\ok} \subseteq \ok \}.
\end{align*}
Since $\ok \subseteq \dual{\ok}$, their inverses satisfy $(\dual{\ok})^{-1} \subseteq \ok$. Unlike the dual which is a fractional ideal and not necessarily within $\ok$, this inclusion makes $\inv{(\dual{\ok})}$ an integral ideal, which is also called the \textbf{different ideal}.
\reversemarginpar
\marginnote{\textit{Different ideal}}\index{different ideal}
% \footnote{To be clear. Some refer $\DD_K$ as the different ideal of $K$ and the notation suggests it too. But $K$ is a field which has exactly two ideals, the zero ideal and itself, so $\DD_K$ is not an ideal of $K$ but of $\ok$.} 
For example, let $K=\Q(i)$ and $\ok=\Z[i]$. The dual ideal is $\dual{\ok}=\dual{\Z[i]}=\frac{1}{2}\Z[i]$, so the different ideal is $\DD_K=\inv{(\frac{1}{2}\Z[i])}=2\Z[i]$.

In the special case when $\ok$ has a power basis, Theorem \ref{thm:dualLatDiff} can also be expressed in terms of different ideal because  
\begin{align*}
    \dual{\ok} &= \frac{1}{f'} \ok \\
    \implies f' \inv{\ok} &= (\dual{\ok})^{-1} \\
    \implies (f') &= \DD_K
\end{align*}
When $f=x^n+1$, the last equality implies $\DD_K=n\ok$.% that is also consistent with Example \ref{ex:dualL}'s conclusion $\dual{\ok}=\frac{1}{n}\ok$. 
See Theorem \ref{app thm:difIdeal1} % and Lemma \ref{app lm:difIdeal} 
in Appendix \ref{appen:ant} for formal statements of these results.

\begin{lemma}
\label{lm:difIdeal}
\reversemarginpar
\marginnote{\textit{$\DD_K = n \OO_K$}}
For $m=2n=2^k \ge 2$ a power of 2, let $K=\Q(\zeta_m)$ be an $m$th cyclotomic number field and $\OO_K=\Z[\zeta_m]$ be its ring of integers. The different ideal satisfies $\DD_K = n \OO_K$.
\end{lemma}
This lemma plays an important role in RLWE\index{ring LWE} in the special case where the number field is an $m$-th cyclotomic field. It implies that the ring of integers $n^{-1}\OO_K=\dual{\OO_K}$ and its dual are equivalent by a scaling factor. Hence, the secret polynomial $\vc{s}$ and the random polynomial $\vc{a}$ can both be sampled from the same domain $R_q$, unlike in the general context where the preference is to leave $\vc{s} \in \dual{R_q}$ in the dual.

\newpage
\section{Ring Learning with Errors}

\label{section:rlwe}

In \Cref{section:lwe}, we have sketched the key steps of LWE's hardness proof by reductions from two standard lattice problems (i.e., GAPSVP and SIVP) using a combination of quantum and classical reductions. The benefit of reducing an arbitrary instance to all instances of some (highly conjectured) worst-case lattice problems sparked many LWE-based cryptosystems, including some developments in quantum-resistant cryptosystems and homomorphic encryption schemes. In addition to being worst-case hard, smaller public key size and ciphertext expansion are also the main motivations for basing a scheme on LWE over the SIS problem. However, the quadratic key size (in the security parameter $n$) is still a serious constraint for practical LWE-based schemes. 

In this section, we will introduce a variation of LWE, called \textbf{ring learning with errors} or ring-LWE (\textbf{RLWE}), which entails multiple benefits over LWE in terms of key size and computational efficiency. The problem originated from LWE, but is defined in terms of ideal lattices that were discussed in the previous section. Recall that a fractional ideal of a number field $K$ is mapped to a metric space by an embedding, so that it makes sense to talk about the distance between two ideal elements as well as define distance-based lattice problems on fractional ideals. As we will see in this section, these special lattices have additional algebraic structures that allow the public key size to be further reduced to $\Tilde{O}(n)$ while retaining almost identical provable security.

\begin{figure}[ht]
    \centering
    \includegraphics[page=9]{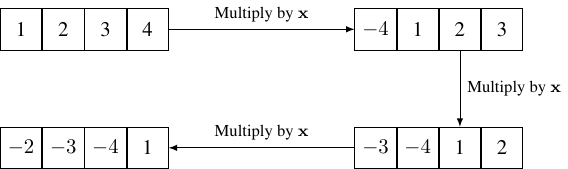}
    \caption{Let $R=\Z[x]/(x^4+1)$. Given the polynomial $\vc{a}=1+2x+3x^2+4x^3$, the nega-cyclic action is equivalent to multiplying $\vc{a}$ by $\vc{x}$, which yields $\vc{a} \star \vc{x} = x+2x^2+3x^3+4x^4=-4+x+2x^2+3x^3$. After $n=4$ rounds of anti-cyclic actions, we get back to $-\vc{a}$.} 
    \label{fig:antiCyc}
\end{figure}

% motivation of the following is unsure  
Here is an example of an additional algebraic structure in the RLWE setting. Ideal lattices are images of fractional ideals under the canonical (or coefficient) embedding. Furthermore, fractional ideals are closed under multiplications by the ring elements. So this structure is preserved by the embedding, which endows the corresponding lattice with an additional algebraic structure. A concrete example is an ideal of the ring $\Z[x]/(x^n-1)$ is closed under multiplication by the polynomial $x$ in the ring. Under the coefficient embedding, this multiplication by $x$ corresponds to rotating the coefficient vector components by one place to the right, so the corresponding ideal lattice is a \textit{cyclic lattice}. Another example which relates to the RLWE problem is  when the ring is $\Z[x]/(x^n+1)$. Multiplying ideal elements by $x$ corresponds to the cyclic lattice rotation and negate the first component as shown in \Cref{fig:antiCyc}.
The believe is that these special lattice problems are still hard because there is currently no known way to exploit the extra structure to reduce the run time for solving them compared to their more general counterparts, with the exception of the GAPSVP problem on ideal lattices, which is known to be easy.
This is the reason why RLWE hardness is based on the K-SVP and K-SIVP problems, but not their gap variants. 

\subsection{Some ideal lattice problems}

We first re-define some lattice problems in terms of an ideal lattice in a number field which is going to be our working domain for the following proofs. 
Recall that the canonical embedding enables us to talk about geometric norms of number field elements by mapping them to elements in the canonical space which is isomorphic to $\R^n$. Hence, we can define the $L_p$-norm of an element $x \in K$ as 
\begin{equation*}
    ||x||_p = ||\sigma(x)||_p = 
    \begin{cases}
     \left( \sum_{i \in [n]} |\sigma_i(x)|^p \right)^{1/p} & \text{ if $p < \infty$},  \\
    \max_{i \in [n]} |\sigma_i(x)| & \text{ if $p = \infty$}.
    \end{cases}
\end{equation*}
With geometric norm, it makes sense to compare the lengths of two elements in a number field. 

\begin{tcolorbox}
\noindent
\textbf{The $\gamma$-Shortest Vectors Problem in $K$ (K-SVP$_{\gamma}$)}\\
Let $K$ be an $n$-dimensional number field. Given a fractional ideal\index{fractional ideal} $I$ of $K$, find a non-zero element $x \in I$ such that $||x||_p \le \gamma(n) \cdot \lambda_1(I)$.
\end{tcolorbox}

\begin{tcolorbox}
\noindent
\textbf{The $\gamma$-Shortest Independent Vectors Problem in $K$ (K-SIVP$_{\gamma}$)}\\
Let $K$ be an $n$-dimensional number field. Given a fractional ideal $I$ of $K$, find $n$ linearly independent non-zero elements $x_i, \dots, x_n \in I$ such that $\max_{i \in [1,n]} ||x_i||_p \le \gamma(n) \cdot \lambda_n(I)$.
\end{tcolorbox}

\begin{tcolorbox}
\noindent
\textbf{The $\alpha$-Bounded Distance Decoding in $K$ (K-BDD$_{\alpha}$)}\\
Let $K$ be an $n$-dimensional number field. Given a fractional ideal $I$ of $K$ and an element $y =x+e \in K$, where $x \in I$ and $||e||_{\infty} \le \alpha \cdot \lambda_1(I)$, find the element $x \in I$.
\end{tcolorbox}

\begin{tcolorbox}
\noindent
\textbf{The $\gamma$-Discrete Gaussian Sampling in $K$ (K-DGS$_{\gamma}$)}\\
Let $K$ be an $n$-dimensional number field. Given a fractional ideal\index{fractional ideal} $I$ of $K$ and a number $s \ge \gamma=\gamma(I)$, produce samples from the discrete Gaussian distribution $D_{I, s}$ over the ideal lattice $I$ with the scale $s$.
\end{tcolorbox}

%The above ideal lattice problems can be defined in terms of either fractional ideals or integral ideals in the number field. Given a fractional ideal $I \subseteq K$ with the denominator $d \in \ok$, the corresponding integral ideal is $dI \subseteq \ok$. Since the norm $N(d)=\prod_{i=1}^n \sigma_i(d)$ where $\sigma_i$ is a real or complex embedding in the canonical embedding, we have $N(d) \in \langle d \rangle$ is in the principal ideal, i.e., $N(d)=k  d$ for an element $k \in K$. This implies that $N(d) I =k(dI) \subseteq  \ok$ is a scaled integral ideal. 

\subsection{RLWE in general number field}

In this subsection, we define RLWE distribution in a (general) number field. The definition is similar to the LWE distribution definition, but with different domains for random samples and noise elements. With this definition, it is sufficient to prove the hardness of the (search) RLWE problem by drawing deductions from some ideal lattice problems introduced in the preceding subsection. The more specialized RLWE definition in a cyclotomic number field will be introduced in a later subsection in order to reduce the search to decision RLWE, which is more convenient to support the security of an encryption scheme. That being said, it may be useful to jump to the start of \Cref{subsec:rlweCyc} to see a concrete example of the ring $R=\Z[x]/(x^n+1)$ in order to have a more intuitive understanding of this domain before moving forward.

When presenting the generalized definition, \citet{lyubashevsky2010ideal} used the notation $K_{\C} = K \otimes_{\Q} \C$ to represent the tensor product\index{tensor product} between the number field $K$ and $\C$. % \footnote{In fact, the authors used $K_{\R} = K \otimes_{\Q} \R$. But I think $K_{\C}$ is a better one to use.} 
This tensor product $K_{\C}$ is where the RLWE errors are sampled from according to a certain error distribution $\psi$. For an $n$-dimensional separable (\Cref{def:sepExt}) 
%see JS Milne's ANT book, page 24 about why separability is essential
number field $K=\Q(\alpha)$ and the minimal polynomial $f(x) \in \Q[x]$ of the primitive element $\alpha$, we have the following isomorphisms. The first isomorphism is by the definition of number field and the second is by the definition of tensor product (see Page 21 of \citet{milneANT})
\begin{align*}
    K \otimes_{\Q} \C \cong \left( \Q[x]/(f(x)) \right) \otimes_{\Q} \C \cong \C[x] / (f(x)).
    % see JS Milne, page 23 for details of the first 2 isomorphisms
    % 1st isomorphism is by definition of K,
    % 2nd isomorphism is by definiton of tensor product, 
\end{align*}
It is often convenient to think of $K_{\C}$ as the canonical space $H$. This is because %where an isomorphism exists between them. Below, we give an intuition of the isomorphism without proving it. 
the minimal polynomial $f(x)=f_1(x) \cdots f_n(x)$ splits into irreducible factors in the complex space $\C$, so we have an isomorphism between $K_{\C}$ and the canonical space $H$ by the Chinese Remainder Theorem, because the principle ideals are coprime 
\begin{align*}
    K_{\C} = K \otimes_{\Q} \C \cong \prod_{i=1}^n \C[x]/(f_i(x))=H.
\end{align*}

The RLWE errors are sampled from $K_{\C}$ and followed by modulo $\dual{R}$ to reduce them to within the dual lattice. For a number field $K$ and its ring of integers $R=\ok$, let $R_q=R/qR$ and $\dual{R}_q=\dual{R}/q\dual{R}$ and $\T = K_{\C} / \dual{R}$ (a high-dimensional torus). The following RLWE definition generalizes Definition \ref{def:rlwe1} to an arbitrary number field. 

We use $\vc{f} \star \vc{g}$ to denote polynomial multiplication in order to distinguish it from vector dot product. From \Cref{subsec:canonical embedding}, we know that polynomial addition and multiplication can be done efficiently under the canonical embedding. 
%The $\bmod \,R_q$ operation is as defined in Equation~(\ref{eq:modulo lattice}).

\begin{definition}
\label{def:rlwe2}
Given the following parameters
\begin{itemize}\itemsep1mm\parskip0mm
    \item $n$ - the security parameter that satisfies $n=2^k$ for an integer $k \ge 0$,
    \item $q$ - a large (public) prime modulus that is polynomial in $n$ and satisfies $q = 1 \bmod 2n$,
\end{itemize}
for a fixed $\vc{s} \in \dual{R}_q$ and an error distribution $\psi$ over $K_{\C}$, the \textbf{RLWE distribution} $A_{s,\psi}$ over $R_q \times \T$,
\reversemarginpar
\marginnote{\textit{RLWE distribution}}  
is obtained by repeating these steps
\begin{itemize}\itemsep1mm\parskip0mm
    \item sample an element $\vc{a} \leftarrow R_q$,
    \item sample a noise element $\vc{\epsilon} \leftarrow \psi$ over $K_{\C} \cong H$,
    \item compute the polynomial $\vc{b} = (\vc{s} \star \vc{a})/q + \vc{\epsilon} \bmod \dual{R}$,
    \item output $(\vc{a},\vc{b})$.
\end{itemize} 
\end{definition}
%note, in his first paper, the rlwe distribution was defined by reducing b to mod dual(R), in a following paper, it was reduce to mod q*dual(R). the two should be equivalent, but the former requires a simple argument the modulo dual(R) is well defined, not sure why, probably because of the dividing by q part. 

%Here, we make some observations of this definition. First, as $R \subseteq \dual{R}$, the product $\vc{s} \star \vc{a} \in \dual{R}_q$ and $(\vc{s} \star \vc{a})/q \in \dual{R}$. The error $\vc{\epsilon}$ is taken from $K_{\C} \cong H$ but can be thought of as further reduced to within the dual $\dual{R}$ when computing $\vc{b}$ by $\bmod \dual{R}$. So the definition is in direct analogy of the LWE distribution. 

As will be seen later, \Cref{def:rlwe1} in cyclotomic field is a special case of the above. Although in this general setting, $\vc{a}$ and $\vc{s}$ are taken from $R_q$ and its dual $\dual{R}_q$ respectively, when $K$ is a cyclotomic field with the cyclotomic polynomial $\Phi_m(x)$ where $m$ is a power of 2, it has been shown in \Cref{ex:dualL} that 
\reversemarginpar
\marginnote{\textit{$R=n\dual{R}$}}
\begin{equation}
\label{eq:scaleRDual}
    R=n\dual{R}.
\end{equation}
% at the end of Section \ref{subsec:dualLatInNumField}. 
Hence, it makes no difference that $\vc{s}$ and $\vc{a}$ are sampled from different domains in the cyclotomic field case. This relationship between $R$ and $\dual{R}$ is essential when reducing the search to decision RLWE. 

The error distribution $\psi$ above is not a 1-dimensional Gaussian distribution any more. Unlike in the LWE case where the 1-dimensional error $\epsilon$ is added to the dot product $\vc{a}\cdot \vc{s}$, in RLWE the $n$-dimensional error $\vc{\epsilon}$ is added to the resulting polynomial $\vc{a}\star \vc{s}$. Depending on how a polynomial is represented, the number of parameters in the high-dimensional error distribution varies. In the coefficient representation, the $n$-dimensional Gaussian error distribution is parameterized by the $n \times n$ covariance matrix. In contrast, in the canonical embedding representation, the same Gaussian distribution  $D_{\vc{r}}$ is the product of $n$ independent 1-dimensional Gaussian with either the same or different scales $\vc{r}=(r_1, \dots, r_n)$. (This is another justification for using canonical embedding in RLWE.) When $\vc{r}$ is a constant vector, $D_{\vc{r}}$ is called a \textbf{spherical Gaussian distribution}, otherwise it is called an \textbf{elliptical Gaussian distribution}. % \kl{They can be made equivalent.}

An important observation when using a high-dimensional error distribution is when reducing  ideal lattice problems to RLWE. As remarked after the LWE hardness proof, in order to employ the assumed LWE oracle to solve BDD, one may need to adjust the embedded random noise magnitude to fulfil the oracle's requirement. This can be done relatively easier by adding additional controlled noise to meet the appropriate noise magnitude for the LWE oracle. But in the RLWE case, there is no straightforward error adjustment to meet the target high-dimensional error distribution for the RLWE oracle, so the proof has to assume the RLWE oracle works for a wide range of error distributions that are defined next. 

\begin{definition}
For $\alpha > 0$, the set $\Psi_{\le \alpha}$
\reversemarginpar
\marginnote{$\Psi_{\le \alpha}$ \textit{family}}
consists of all \textbf{elliptical Gaussian distributions} $D_{\vc{r}}$ over $K_{\C}$ such that each $D_{r_i}$ has scale $r_i \le \alpha$.
\end{definition}

With this family of error distributions, we can define the search RLWE problem as follows.

\begin{definition}
\reversemarginpar
\marginnote{\textit{Search RLWE}}
Given the parameter $q$ and the family of error distributions $\Psi_{\le \alpha}$, the \textbf{search RLWE} problem, denoted by \textbf{RLWE$_{q,\Psi_{\le \alpha}}$}, is to compute the secret key $\vc{s}$  given samples $\{(\vc{a}, \vc{b})\}$ from the RLWE distribution $A_{\vc{s},\psi}$ for an arbitrary $\vc{s} \in \dual{R}_q$ and $\psi \in \Psi_{\le \alpha}$.
\end{definition}

The decision RLWE is an average case problem for a random secret key and a random error distribution. The distribution for the secret key $\vc{s}$ is uniform over the dual lattice $\dual{R}$. The distribution $\Upsilon_{\alpha}$ over the elliptical Gaussian error distributions $\Psi_{\le \alpha}$ is chosen to be a Gamma distribution with shape 2 and scale 1.\footnote{\citet{lyubashevsky2010ideal} emphasized that any efficiently samplable continuous distributions can be used, e.g., Gaussian distribution.} Since the reduction from search to decision RLWE can only be made possible in cyclotomic number fields, we define $\Upsilon_{\alpha}$ specifically in these cyclotomic fields. Recall that for $m=2n=2^k > 2$, the canonical embedding for a cyclotomic number field $K=\Q(\zeta_m)$ consists only $n$ complex embeddings which are in $n/2$ conjugate pairs $\sigma_i=\overline{\sigma_{i+n/2}}$ for $i \in [1,n/2]$, so the scale parameters that correspond to a conjugate pair can be set identical. This gives rise to the next definition of the distribution $\Upsilon_{\alpha}$. 

\begin{definition}
For $m=2n=2^k > 2$ an integer power of 2, let $K=\Q(\zeta_m)=\Q[x]/(x^n+1)$ be the $m$th cyclotomic field. For a real $\alpha>0$, let $\Upsilon_{\alpha}$ be the \textbf{distribution over the family} $\Psi_{\le \alpha}$ 
\reversemarginpar
\marginnote{\textit{Distribution over} $\Psi_{\le \alpha}$}
of elliptical Gaussian distributions. Then every element $\psi$ sampled from $\Upsilon_{\alpha}$ is an elliptical Gaussian distribution $D_{\vc{r}}$ over $K_{\C}$ whose scale parameters satisfy $r_i^2 = r_{i+n/2}^2=\alpha^2(1+\sqrt{n} x_i)$, where $x_1, \dots, x_{n/2}$ are chosen independently from the Gamma distribution $\Gamma(2,1)$.
\end{definition}

Using this definition, we define the average-case decision version of RLWE as follows. 

\begin{definition}
\reversemarginpar
\marginnote{\textit{Decision RLWE}}
Given the parameter $q$ and a distribution $\Upsilon_{\alpha}$ over the family $\Psi_{\le \alpha}$ of elliptical Gaussian distributions, the \textbf{average-case decision RLWE} problem, denoted by \textbf{RDLWE}$_{q, \Upsilon_{\alpha}}$, is defined as follows: for a random choice of $(\vc{s},\psi) \leftarrow U(\dual{R}) \times \Upsilon_{\alpha}$, distinguish with non-negligible probability between samples from the RLWE distribution $A_{\vc{s},\psi}$  and  uniform samples over $R_q \times \T$.
\end{definition}

The mean of $\Gamma(2,1)$ is 2, by the above definition of $\Upsilon_{\alpha}$ we have $||r_i|| \approx O(\alpha n^{1/4})$. Recall that in the proof of LWE hardness, we discussed the upper bound of the scale parameter $\alpha$ in the Gaussian error distribution $\Psi_{\alpha}$ in order for $\Psi_{\alpha}$ to be distinguishable from the uniform distribution once reduced by $\bmod \,\Z_p^n$. The same argument carries over to the RLWE problem too, that is, $\psi \bmod \dual{R}$ and the uniform distribution over $\T=K_{\C}/\dual{R}$ should be distinguishable, for otherwise the decision RLWE is unsolvable. The difference is in the $n$th successive minima $\lambda_n(R)$. When $K$ is a cyclotomic number field, it has a power basis $\{1, \zeta, \dots, \zeta^{n-1} \}$, which is also a basis of $R$. Under the canonical embedding, each element $\zeta^k$ in the power basis is mapped to an element $(\sigma_1(\zeta^k), \dots, \sigma_n(\zeta^k))$ in the canonical space, where each $\sigma_i$ maps $\zeta^k$ to a different element in the power basis with $||\sigma_i(\zeta^k)||=1$. Hence, the Euclidean norm of $\zeta^k$'s image under the canonical embedding is $\sqrt{n}$ and $\lambda_n(R)=\sqrt{n}$. This implies the $n$th successive minima $\lambda_n(\dual{R})=1/\sqrt{n}$ and hence the upper bound of $\alpha$ in RLWE is $\alpha \le O(\sqrt{\log n/n})$ by \Cref{lm:smthParUpperBd}, which is smaller than $O(\sqrt{\log n})$  in LWE. 

%The above argument of $\lambda_n(R)=\sqrt{n}$ not only demonstrates from a theoretical perspective that the ring version can embed more noise than the ordinary LWE, but supports the claim that the working domain $\dual{R}$ is the correct choice for RLWE. The successive minima of the dual of $\dual{R}$ is just $\lambda_n(R)$. 

We now state the main theorem of decision RLWE in the context of cyclotomic field 
$K=\Q(\zeta_m)=\Q[x]/(x^n+1)$, where its ring of integers is $R=\ok=\Z[x]/(x^n+1)$. 

\begin{theorem}
\label{thm:svpToRLWE}
\reversemarginpar
\marginnote{\textit{SVP, SIVP to RDLWE}}
Let $K$ be defined above, $\alpha < \sqrt{\log n/n}$ and $q=q(n) \ge 2$ be a prime such that $q = 1 \bmod m$ and $\alpha q \ge \omega(\log n)$. There is polynomial time quantum reduction from the ideal lattice $\Tilde{O}(\sqrt{n}/\alpha)$-SIVP (or SVP) problem to 
\begin{itemize}
    \item RDLWE$_{q, \Upsilon_{\alpha}}$ or 
    \item RDLWE$_{q, D_{\xi}}$ given only $l$ samples, where $\xi=\alpha (nl / \log(nl))^{1/4}$ is the scale parameter for the spherical Gaussian error distribution.
\end{itemize}
\end{theorem}

The first reduction is to the decision RLWE with a random elliptical Gaussian error distribution, whilst the second is to the decision RLWE with a fixed spherical Gaussian error distribution but given only a small number of samples. We will make clear the connection between these two problems in a following subsection. % \kl{mention the ellipitcal gaussian incurs a factor $n^{1/4}$.}

The threshold $\alpha$ for the Gaussian distribution's scales is upper bounded to guarantee the solvability of the decision RLWE. In the meantime, the scales must also be sufficiently large to guarantee the sampled Gaussian noise once reduced to a smaller domain is almost uniformly distributed. See Section 4 of \citet{lyubashevsky2010ideal} for an additional explanation for the choice of $\alpha$.

\subsection{Hardness of search RLWE}

%x^n+1 should be irreducible over the integers, it can be reducible over p, that doesn't matter. 

\begin{figure}[hbt!]
    \centering
    \includegraphics[page=17]{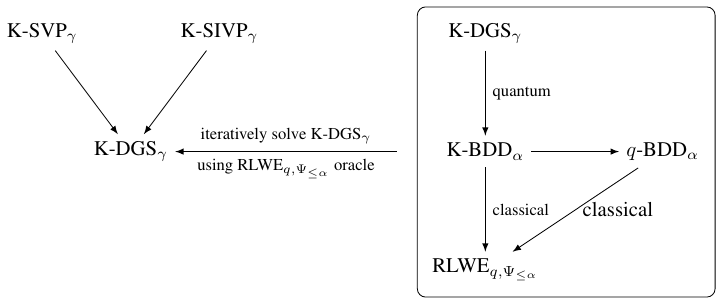}
    \caption{RLWE reductions}
    \label{fig:rlweReduction}
\end{figure}

Similar to the (search) LWE's hardness proof, the hardness of (search) RLWE relies on reductions from hard ideal lattice problems K-SVP$_\gamma$ and K-SIVP$_\gamma$, through the intermediate K-DGS problem. We omit the reductions from the two ideal lattice problems to K-DGS, but only focus on the classical part of the quantum reduction to RLWE. The following theorem states a quantum reduction, which can be separated into a quantum and a classical step. We emphasize again that the context of this reduction is for arbitrary number fields (not necessarily cyclotomic). 

In contrast to the small $o$ notation (i.e., $f(n)=o(g(n))$) that indicates an upper bound of a function's growth, the small omega notation (i.e., $f(n) = \omega(g(n))$) indicates a lower bound of the function's growth. More precisely, $f(n) = \omega(g(n))$ if for all $k>0$ there exists a threshold $n_0$ such that for all $n > n_0$ it satisfies $|f(n)|>k|g(n)|$. Throughout the proof, $\omega(\sqrt{\log n})$ is used to denote a function that grows asymptotically faster than  $\sqrt{\log n}$.

\begin{theorem}
\label{thm:rlweIter}
\reversemarginpar
\marginnote{\textit{K-DGS to RLWE}}
Let $\alpha=\alpha(n)>0$ and $q=q(n) \ge 2$ such that $\alpha q \ge 2 \omega(\sqrt{\log n})$. There is a PPT quantum reduction from K-DGS$_{\gamma}$ to RLWE$_{q,\Psi_{\le \alpha}}$, where 
\begin{equation}
\label{equ:gamma}
    \gamma = \max\{\eta_{\epsilon}(I) (\sqrt{2}/\alpha) \omega(\sqrt{\log n}), \sqrt{2n}/\lambda_1(\dual{I})\}.
\end{equation}
\end{theorem}

Given $\alpha < \sqrt{\log n / n}$ as stated in \Cref{thm:svpToRLWE} and the smoothing parameter $\eta_{\epsilon}(I) > 1/ \lambda_1(\dual{I})$ by Claim 2.13 of  \citet{regev2009lattices}, it always satisfies that $\gamma =\eta_{\epsilon}(I) (\sqrt{2}/\alpha) \omega(\sqrt{\log n})$ in the above theorem. 

Again, the motivation behind the theorem is to obtain discrete Gaussian samples over an ideal lattice $I$ (in $K$) with scale $s$ as close to the lower bound $\gamma$ as possible, so that certain standard ideal lattice problems can be solved with the help of these short discrete Gaussian samples. The feasibility of obtaining short samples can be proved using almost the same strategy as that in the BDD to LWE reduction. Recall that the BDD to LWE reduction gives rise to an iterative strategy to reduce the discrete Gaussian sample norms. In the RLWE setting, this means (as shown in \Cref{fig:rlweReduction}) to solve the K-BDD problem with an RLWE oracle and some discrete Gaussian samples with scale $r$, then feed the K-BDD output to a quantum algorithm to produce new discrete Gaussian samples with scale$r' < r/2$ half of the previous norms. We ignore the quantum step of the reduction (Lemma 4.4 \citep{lyubashevsky2010ideal}). The classical part is stated in the next lemma. 

%this is lemma 4.3 in lyubashevsky2010ideal
\begin{lemma}[Lemma 4.3 \citep{lyubashevsky2010ideal}]
\reversemarginpar
\marginnote{\textit{K-BDD to RLWE}}
Let $\alpha =\alpha(n)> 0$, $q =q(n) \ge 2$ be an integer with known factorization. Let $I$ be a fractional ideal of a number field $K$ and $r\ge \sqrt{2} q \eta_{\epsilon}(I)$ for some negligible $\epsilon=\epsilon(n)$. Given a discrete Gaussian oracle for $D_{I,r}$, there is a PPT reduction from K-BDD$_d$ in the dual lattice $\dual{I}$ where $d=\alpha q / (\sqrt{2}r)$ to RLWE$_{q,\Psi_{\le \alpha}}$.
\end{lemma}

To solve the K-BDD problem for an element in the ideal lattice $I$ of $K$, the same bit-by-bit strategy as in Lemma \ref{lm:lweCoeffModQ} can be applied. That is, find a solution in the scaled ideal lattice $qI$ and then iteratively build a solution in $I$ from the least to the most significant bit in the base $q$. Since Lemma \ref{lm:lweCoeffModQ} was proved for general lattices, it also holds for ideal lattices without re-proving. The K-BDD problem in a scaled ideal lattice $qI$ is called $q$-BDD. Hence, it remains to prove a solution for $q$-BDD with the help of an RLWE oracle and discrete Gaussian samples.

\begin{lemma}
\label{lm:qbdd2RLWE}
\reversemarginpar
\marginnote{\textit{$q$-BDD to RLWE}}
Assume there is an oracle for RLWE$_{q,\Psi_{\le \alpha}}$ and a discrete Gaussian oracle for generating samples from $D_{I,r}$ where $r \ge \sqrt{2} q \eta_{\epsilon}(I)$. Given a K-BDD$_{\dual{I},d}$ instance $\vc{y}=\vc{x}+\vc{e}$, where $\vc{x} \in \dual{I}$ and $||\vc{e}||_{\infty} \le d$, there is a polynomial time algorithm solves $q$-BDD$_{\dual{I},d}$, that is, finds $\vc{x} \bmod q\dual{I}$.
\end{lemma}

The proof of this lemma follows a similar strategy as that of Proposition \ref{prop:bddToLWE}. That is, construct RLWE samples for the oracle using the given K-BDD instance $\vc{y}$ and the discrete Gaussian samples over $I$. The proof, however, is more involved, because the solution of K-BDD is in $\dual{I}$ and discrete Gaussian noise elements are sampled from $I$, whilst the RLWE oracle works in  $R_q$ and its dual. Hence, it is necessary to be able to transform elements between these domains without losing their structures. To achieve this, %we isomorphisms stated in Lemma \ref{lm:clearIdeals2}. 
we re-state the following two important results that have been proved in \Cref{subsec:appInRLWE}, but in the context of a number field $K$ and its ring of integers $\ok$. 

% see lemma 5.2.2 in stein2012algebraic
\begin{lemma}
\label{lm:coprimeIdeals2}
If $I$ and $J$ are non-zero integral ideals of $R=\ok$, then there exists an element $t \in I$ such that $(t)\inv{I} \subseteq R$ is an integral ideal coprime to $J$. 
\end{lemma}

% state the lemma according to lemma 2.15 lyubashevsky2010ideal, which is a special case of prop 5.2.4 in stein2012algebraic.
\begin{lemma}
\label{lm:clearIdeals2}
Let $I$ and $J$ be ideals in $R=\ok$ and $M$ be a fractional ideal in the number field $K$. Then there is an isomorphism 
\begin{align*}
    M/JM \cong IM/IJM.
\end{align*}
\end{lemma}

To make the proof work, we only focus on special cases of \Cref{lm:clearIdeals2}. More precisely, let $J=(q)$ and $M=R$ be the ring of integers itself or $M=\dual{I}$ be the dual ideal. Given the prime factors of the integer $q$, say $q=ab$ where $a,b \in \Z$ are primes, the principal ideal can be written as $(q)=(a)(b)$ the product of prime ideals in $\Z$. Using a prime ideal factorization technique (will be briefly discussed in the next subsection), we can find the prime factors of $(a)$ and $(b)$ in $R$ hence $(q)$. It then follows from Lemma \ref{lm:coprimeIdeals2} that there is an element $t \in I$ to construct an ideal $(t)\inv{I}$ coprime to $J=(q)$ (see proofs of these lemmas in Section \ref{subsection:number field}, also see the proof of lemma 5.2.2 of \citet{stein2012algebraic} to see why we need to know the prime factors of the ideal $J$). Then the map 
\begin{align*}
    \theta_t: K &\rightarrow K \\
    u &\mapsto ut
\end{align*}
induces two important isomorphisms 
\begin{align}
\label{eq:theta1}
&R_q=R/(q)R \cong IR/I(q)R =I_q\\
\label{eq:theta2}
&\dual{I}_q=\dual{I}/(q)\dual{I} \cong I\dual{I}/I(q)\dual{I}=I \inv{I}\dual{R}/I(q)\inv{I}\dual{R} = \dual{R}/(q)\dual{R}=\dual{R}_q.
\end{align}

Both isomorphisms in Equation \ref{eq:theta1} and \ref{eq:theta2} are precisely what we need in order to prove \Cref{lm:qbdd2RLWE}. Below we state the process to build the reduction.
To construct $A_{\vc{s},\psi}$ samples from $\vc{y} \in K$, repeat the following steps: 
\begin{enumerate}
    \item Compute the element $t \in I$ such that $(t)\inv{I}$ and $(q)$ are coprime by Lemma \ref{lm:coprimeIdeals2}. Define the function $\theta_t(x) = xt$, which yields the two isomorphisms 
    \begin{align*}
        R_q & \cong I_q \\ 
        \dual{I}_q &\cong \dual{R}_q.
    \end{align*}
    
    \item Sample $\vc{z} \leftarrow D_{I, r}$ using the discrete Gaussian oracle, and compute 
    \begin{align*}
        \vc{a} =\inv{\theta}_t(\vc{z} \bmod qI) \in R_q.
    \end{align*}
    
    \item Sample $\vc{e}' \leftarrow D_{\alpha/\sqrt{2}}$ a continuous Gaussian noise, and compute 
    \begin{align*}
        \vc{b} = ((\vc{z} \bmod qI) \star \vc{y})/q + \vc{e}' \bmod \dual{R}.
    \end{align*}
    
    \item Output the pair $(\vc{a},\vc{b})$.
    
\end{enumerate}
Once the RLWE oracle is given the samples $\{(\vc{a},\vc{b})\}$, it produces the secret key $\vc{s} \in \dual{R}_q$ and output 
\begin{align*}
    \vc{x} \bmod q\dual{I} = \inv{\theta}_t(\vc{s}) \in \dual{I}_q.
\end{align*}

We now prove that $\{(\vc{a},\vc{b})\}$ are nearly genuine samples from the $A_{\vc{s},\psi}$ distribution,
% \kl{Note I used $\vc{z} \bmod qI$ instead of $\vc{z}$ to restrict to the correct domain $I_q$ for the isomorphism $\inv{\theta}_t: I_q \rightarrow R_q$, because I cannot follow the original proof by using only $\vc{z}$. This does not change the proof nor does it make the implementation of RLWE any harder.} 
hence the RLWE oracle produces a result for the $q$-BDD problem. The proof is structured as follows: first, show $\vc{a}$ distributes uniformly in $R_q$ and $\vc{b}$ follows  $\vc{b}=(\vc{a} \star \vc{s})/q + \vc{\epsilon} \bmod \dual{R}$; then show that the secret key in RLWE gives rise to the solution $\inv{\theta}_t(\vc{s}) = \vc{x} \bmod q\dual{I}$.  

\begin{proof}
Since $\vc{z}$ is sampled from the discrete Gaussian distribution $D_{I,r}$ with a large scale $r \ge \sqrt{2} q \eta_{\epsilon}(I)$, when reduced it by taking modulo $qI$, the reduced sample is almost uniformly distributed within $I_q$, and hence its image $\vc{a}$ under the isomorphism $\inv{\theta}_t$ is also uniformly distributed within $R_q$. %For a detailed argument, see the proof of Lemma 4.7 in \cite{lyubashevsky2010ideal}. 

% add lemma 2.3

For the second component, we can re-write it as 
\begin{align*}
    b &= ((\vc{z} \bmod qI) \star \vc{y})/q + \vc{e}' \bmod \dual{R} \\
    &= ((\vc{z} \bmod qI) \star (\vc{x} + \vc{e}) )/q + \vc{e}' \bmod \dual{R} \\
    &= ((\vc{z} \bmod qI) \star \vc{x})/q +  ((\vc{z} \bmod qI)/q) \star \vc{e} + \vc{e}' \bmod \dual{R}. \\
\end{align*}
The key is to show that the first term is identical to $(\vc{a} \star \vc{s})/q \bmod \dual{R}$ and the second and third terms combined is within negligible distance to the elliptical Gaussian $D_{\vc{r}}$ over $K_{\C}$.

Given $\vc{z} \bmod qI=\theta_t(\vc{a}) =\vc{a} \star \vc{t} \bmod qI$, we have 
\begin{align*}
    &\theta_t(\vc{a})-\vc{a} \star \vc{t} = 0 \bmod qI \\
    \implies &\theta_t(\vc{a})-\vc{a} \star \vc{t} \in qI\\
    \implies &(\theta_t(\vc{a})-\vc{a} \star \vc{t}) \star \vc{x} \in qI\dual{I}=qI\inv{I}\dual{R}=q\dual{R} \\
    \implies &\theta_t(\vc{a})\star \vc{x} = \vc{a} \star \vc{t} \star \vc{x} \bmod q\dual{R}.
\end{align*}
It follows from this and $\theta_t(\vc{x} \bmod q\dual{I}) =\vc{s}$ that 
\begin{align*}
    &(\vc{z} \bmod I_q) \star \vc{x} = \theta_t(\vc{a}) \star \vc{x} 
    = \vc{a} \star \vc{t} \star \vc{x} \bmod \dual{R}_q 
    = \vc{a} \star \vc{s} \bmod \dual{R}_q \\
    \implies &((\vc{z} \bmod I_q) \star \vc{x}) / q = (\vc{a} \star \vc{s})/q \bmod \dual{R}
\end{align*}
Therefore, we have proved that 
\begin{align*}
    b=(\vc{a} \star \vc{s})/q +  ((\vc{z} \bmod qI)/q) \star \vc{e} + \vc{e}' \bmod \dual{R}
\end{align*}
It remains to show the other parts combined is close to the discrete Gaussian $D_{\vc{r}}$ over $K_{\C}$. We skip this step, which is proved in Lemma 4.8 of  \citet{lyubashevsky2010ideal}. % and does not use much of the algebraic properties, so we skip the proof. 

We have shown that the samples $\{(\vc{a},\vc{b})\}$ follow the RLWE distribution and hence are legitimate inputs for the RLWE oracle. Since the oracle outputs the secret key $s \in \dual{R}_q$, by the induced isomorphism $\inv{\theta}_t: \dual{R}_q \rightarrow \dual{I}_q$, we have found $\inv{\theta}_t(\vc{s})=\vc{x} \bmod q\dual{I}$, the least significant digit of the K-BDD solution $\vc{s} \in \dual{I}$.
\end{proof}

To recap, we have shown in this subsection a polynomial time classical reduction from K-BDD to the search RLWE problem. In order for the reduction to work, we need to know the prime factorization of the integer $q = q(n) \ge 2$. The number field $K$ needs not be cyclotomic, so the result holds in general number fields.

\subsection{RLWE in cyclotomic field}
\label{subsec:rlweCyc}

In this subsection, we will re-state the RLWE problem in a special number field, i.e., the cyclotomic field, which is the most common setting for RLWE-based cryptosystems. It is the working domain for the search to decision reduction of the RLWE problem. 

Recall the $m$th cyclotomic polynomial $\Phi_m(x)$ is the polynomial whose roots are the primitive $m$th roots of unity. As we have seen in \Cref{rmk:speCycPoly}, when $m=2n=2^k \ge 2$ is a positive power of 2, the corresponding cyclotomic polynomial has the simple algebraic form $\Phi_m(x)=x^n+1$. Using this cyclotomic polynomial, we can define  
$R=\Z[x]/(\Phi_m(x))$ to be the ring of integer coefficient polynomials modulo (the principle ideal generated by) $\Phi_m(x)$. This is the primary domain where RLWE is defined in the special case. There are two way to interpret the ring $R$ stated below. 
\begin{enumerate}\itemsep1mm\parskip0mm
    \item $R = \Z[x]/(x^n+1)$ is a quotient ring where every polynomial in $R$ has integer coefficients and degree less than $n$.
    \item $R = \Z[x]/(\Phi_m(x))$ is isomorphic to $\Z[\zeta_m]$, the ring of integers $\ok$ for the $m$-th cyclotomic field $K=\Q(\zeta_m)$. This interpretation is supported by \Cref{thm:ring LWE isomoprhic 2}. The choices of $m$ and $n$ are motivated by \Cref{lm:difIdeal} that relates $\ok$ and its dual by a scaling factor, i.e., $\dual{\ok}=n^{-1} \ok$. This simplifies the RLWE definition by allowing the secret polynomial $\vc{s}$ to be sampled from the same domain as a public polynomial $\vc{a}$ as in \Cref{def:rlwe1}.)
    % Second, let $K=\Q(\zeta_m)$ be the $m$th cyclotomic field, then $R=\ok$ is the ring of integers obtained by adjoining $\zeta_m$ to $\Z$, i.e., $R=\Z[\zeta_m]$.
\end{enumerate} 
The first interpretation is the natural interpretation but the second interpretation is more useful when proving hardness result of RLWE. We have been through some important properties of $\ok$ such as its fractional ideals form a UFD and its geometric interpretation under the canonical embedding. 

To work in a finite domain, some elements in the following RLWE definition are taken from $R$ modulo a prime $q$, that is, $R_q=\Z_q[x]/(\Phi_m(x))$, where the polynomial coefficients are in $\Z_q$. This turns $R_q$ into a field of order $q^n$ because each coefficient has $q$ choices and there are $n$ coefficients, see \Cref{thm:quoRngIsField} for more details. %The field $R_q$ is the underlying domain for generating random RLWE elements. From here, we can define the RLWE distribution as follows. 

\begin{definition}
\label{def:rlwe1}
Given the following parameters
\begin{itemize}\itemsep1mm\parskip0mm
    \item $n$ - the security parameter that satisfies $n=2^k$ for an integer $k \ge 0$,
    \item $q$ - a large (public) prime modulus that is polynomial in $n$ and satisfies $q = 1 \bmod 2n$,
\end{itemize}
for a fixed $\vc{s} \in R_q$ and an error distribution $\chi$ over $R$ that is concentrated on ``small integer'' coefficients, the \textbf{RLWE distribution} 
\reversemarginpar
\marginnote{\textit{RLWE distribution}}
over $R_q \times R_q$, denoted by 
\begin{equation*}
    RLWE(n,q,\chi) := \{(\vc{a},\vc{b})\}
\end{equation*}
is obtained by repeating these steps 
\begin{itemize}\itemsep1mm\parskip0mm
    \item sample an element $\vc{a} \leftarrow R_q$,
    \item sample a noise element $\vc{\epsilon} \leftarrow \chi$ over $R$,
    \item compute the polynomial $\vc{b} = \vc{s} \star \vc{a} + \vc{\epsilon} \bmod R_q$,
    \item output $(\vc{a},\vc{b})$.
\end{itemize}
\end{definition}

In a LWE-based cryptosystems, as shown in Section \ref{subsec:lweSecurity}, the public key is $(\vc{A},\vc{b})$ where $\vc{A} \in \Z_q^{n \times m}$ is a matrix that needs $O(mn)$ storage. 
For an RLWE-based cryptosystem, the public key size can be reduced to $O(n)$, which is a significant saving in terms of storage. 
The reason is because each sample from an RLWE distribution is a pair of $n$-degree polynomials (Definition~\ref{def:rlwe2}) that can replace $n$ samples from the standard LWE distribution (Definition~\ref{def:lweDist}).

%There are two questions that we would like to investigate in this section. First, given the public key is not completely random in RLWE-based cryptosystems, do we still get the same level of protection as in LWE? Second, why does this not work in general lattices like in LWE? More specifically, if we also reuse each random vector $\vc{a} \in \Z_q^n$ in LWE, can it still provide the same level of protection as before? 

\subsection{Search to decision RLWE}

%As mentioned before, the RLWE reduction from search to decision relies on algebraic properties of cyclotomic fields. 
% Hence, the following notations are used throughout this subsection, denote $\zeta_m$ a primitive $m$th root of unity, $\Phi_m(x)$ the $m$th cyclotomic polynomial, $K=\Q(\zeta_m)$ the $m$th cyclotomic number field, $R=\ok=\Z[\zeta_m]$ the ring of integers of $K$. 
%The main theorem of this section is as follows. 

%this is a reduction from worst-case search to average-case decision

Recall that the reduction from search to decision LWE in \Cref{subsec:lweDist} used a simple argument by guessing each vector component of the secret key $\vc{s}$ using the decision LWE oracle. 
% There, the decision oracle was assumed to work for one component of the LWE samples. This essentially entails that the oracle also works for the other components for two reasons. First, an LWE sample is computed from integer vectors and the computations (i.e., addition and multiplication) are component-wise, so permuting the vector components does not affect these component-wise computations. Second, a random permutation of the LWE sample components remains in the same domain with the same LWE distribution, so the permuted sample is still a valid input for the decision oracle. 
We plan to use the same strategy to reduce the search to decision RLWE problem by calling the decision oracle to solve the RLWE problem component by component. 
Also recall that the connection between a number field and its geometrical embedding is via the canonical embedding (\Cref{subsec:canonical embedding}). The canonical embedding is chosen over the coefficient embedding for several reasons, including the equivalence between number field element multiplications and embedded canonical vectors' component-wise multiplications. 

A consequence of the component-wise operations is that a change in a single component of the secret polynomial $\vc{s}$ leads to a change in a single component of the polynomial $\vc{b}$ and vice versa. This is in contrast to the LWE case, where $b = \vc{s}\cdot \vc{a} + \epsilon$ is the vector dot product, so any change in $\vc{s}$ is not associated with a single component change in $b$ and vice versa. 
This raises the question of whether or not a RLWE oracle that is limited to discover a single component of the secret vector is able to discovery the entire $\vc{s}$.

%on suitable transformations of the given RLWE samples to guess each component of the secret polynomial, although the argument is not as straightforward as for LWE. % However, such a straightforward argument does not work for the two same reasons mentioned above. 

%A key issue in the RLWE setting %, as pointed out in \cite{lyubashevsky2010ideal}, is that the DLWE oracle may only have the ability to distinguish between a sample from an RLWE distribution and a sample from the uniform distribution relative to one component of the secret $s$, which affects only one component of $b$. (Recall that $b$ in RLWE is an $n$-dimensional vector, and changing one component of $s$ only affects $b$ in one component when additions and multiplications are component-wise. In contrast, $b$ in LWE is a single number and is affected by every change to a component of $s$ in LWE.)
% corresponds to an ideal factor $q_i$, for some $i$, of $(q)$. 
Hence, we need a way to leverage that oracle-distinguishable component to guess the value of all the other components of the secret $\vc{s}$, % at other comordinates corresponding to the other factors $q_j$, $j \neq i$, of $(q)$, 
by using the automorphisms\index{automorphism} of the underlying cyclotomic field to `shuffle' the components (\Cref{subsec:galois group cyclotomics}).
In addition, in shuffling the components and adding a guess for each component of the secret $s$, we need to make sure 
\begin{itemize}\itemsep1mm\parskip0mm
\item a new sample $(\vc{a}',\vc{b}')$ presented to the decision RLWE oracle obtained by transforming a given RLWE sample $(\vc{a},\vc{b})$ is close to a sample from an RLWE distribution when the guess is correct, and close to a sample from the uniform distribution when the guess is incorrect. % The latter is not obvious given that $b$ is an $n$-dimensional vector, and changing $s$ in one coordinate only affects $b$ in one coordinate. 
\item the noise vector in the transformed $b'$ value stays in the noise distribution family $\Psi_{\le \alpha}$. %, which is not obvious.
\end{itemize}
% Second, even with canonical representation, it is not obvious that a vector $e$ sampled from an elliptical Gaussian error distributions stays in its family $\Psi_{\le \alpha}$ after a permutation of $e$'s values so this needs extra care.
Below, we state the main theorem of this subsection. Its proof is divided into several parts in the rest of this subsection. For details of these proofs, see Section 5 of  \citet{lyubashevsky2010ideal}.

\begin{theorem}
\label{thm:rlweSearchToDecision}
Let $R$ be the ring of integers of a cyclotomic field $K$ and $q=q(n) = 1 \bmod m$ be a prime such that $\alpha q \ge \eta_{\epsilon}(\dual{R})$ for some negligible $\epsilon=\epsilon(n)$. There is a randomized polynomial time reduction from the search problem RLWE$_{q,\Psi_{\le \alpha}}$ to the average-case decision problem RDLWE$_{q,\upsilon_{\alpha}}$. 
\end{theorem}

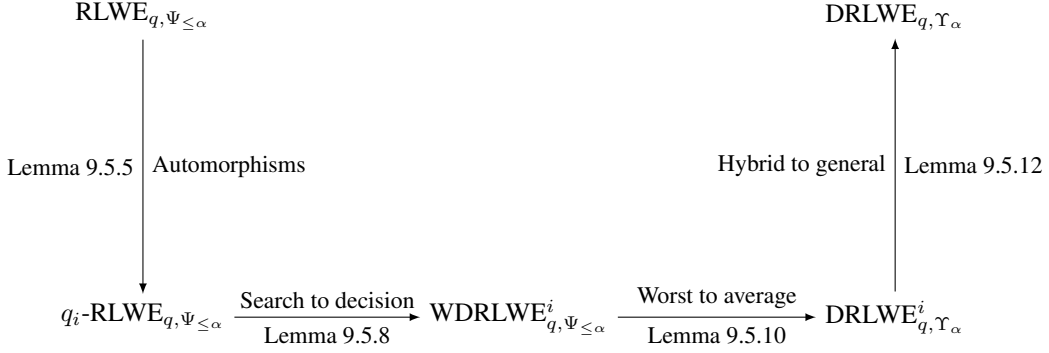
\begin{figure}
    \centering
    \begin{tikzpicture}[scale=1]
        \begin{scope}[>={Stealth[black]},every edge/.style={draw=black}]
        
        \node (rlwe) at (0,0) {RLWE$_{q,\Psi_{\le \alpha}}$};
        \node (qrlwe) at (0,-4) {$q_i$-RLWE$_{q,\Psi_{\le \alpha}}$};
        \node (wdrlwe) at (5,-4) {WDRLWE$_{q,\Psi_{\le \alpha}}^i$};
        \node (drlwei) at (10,-4) {DRLWE$_{q,\Upsilon_{\alpha}}^i$};
        \node (drlwe) at (10,0) {DRLWE$_{q,\Upsilon_{\alpha}}$};
        
        \path [>=latex, ->] (rlwe) edge node[right,midway,font=\footnotesize] {Automorphisms} node[left,midway,font=\footnotesize] {Lemma \ref{lm:rlweRed1}} (qrlwe);
        \path [>=latex, ->] (qrlwe) edge node[below,midway,font=\footnotesize] {Lemma \ref{lm:rlweRed2}} node[above,midway,font=\footnotesize] {Search to decision} (wdrlwe);
        \path [>=latex, ->] (wdrlwe) edge node[below,midway,font=\footnotesize] {Lemma \ref{lm:rlweRed3}} node[above,midway,font=\footnotesize] {Worst to average} (drlwei);
        \path [>=latex, ->] (drlwei) edge node[right,midway,font=\footnotesize] {Lemma \ref{lm:rlweRed4}} node[left,midway,font=\footnotesize] {Hybrid to general} (drlwe);
        
        \end{scope}
        
    \end{tikzpicture}
    \caption{A reduction map from the search to decision RLWE.}
    \label{fig:rlweToDrlwe}
\end{figure}

% For these reasons, 
The search to decision RLWE reduction is achieved by a combination of four separate reductions as shown in Figure \ref{fig:rlweToDrlwe}. The first reduction is from RLWE to component-wise RLWE in the canonical representation. 
% In other words, once we obtain the secret $\vc{s}$ in all CRT components, the entire $\vc{s}$ can be found using the CRT induced isomorphism. 
The second reduction is from a component-wise search oracle to a worst-case decision oracle. The third reduction is between a worst-case and average-case decision oracle. And the last reduction guarantees that given an overall decision oracle it also works for a particular component.  

% Next, we recall some technical aspects in prime ideal factorization then state the first reduction. 

Given a prime $q$ satisfying $q = 1 \bmod m$, 
the ideal $(q)$ in $R_q = \Z_q[x]/(\Phi_m(x))$ factors into $\varphi(m)$ distinct prime ideals: $(q) = \prod_{i\in\Z_m^*} \mathfrak{q}_i$. (See Example~\ref{ex:q ideal factorisation} for more details.)\index{ideal factorization}
Further, by Lemmas~\ref{lm:coprimeIdeals2}, \ref{lm:clearIdeals2} and (\ref{eq:ok prime ideal crt}), there is an efficiently computable isomorphism between $R_q^{\vee}$ and $\bigoplus_{i \in \Z_m^*} (R^{\vee}/\mathfrak{q}_iR^{\vee})$.
Given we are going to guess the secret key $\vc{s}$ one component at a time in the canonical representation, this gives rise to the restricted RLWE definition.\index{canonical embedding}

\begin{definition}
\reversemarginpar
\marginnote{\textit{$\mathfrak{q}_i$-RLWE}}
Given 
\begin{itemize}\itemsep1mm\parskip0mm
\item an oracle that generates samples from the RLWE distribution $A_{\vc{s},\psi}$, for an arbitrary $\vc{s} \in \dual{R}_q$ and $\psi \in \Psi_{\le \alpha}$, and
\item a prime ideal $\mfq_i$ in the factorisation of $(q)$, 
\end{itemize}
the $\mathfrak{q}_i$-RLWE$_{q,\Psi_{\le \alpha}}$ problem is to find $\vc{s} \bmod \mfq_i \dual{R}$.
\end{definition}

An important observation is that each prime ideal $\mfq_i$ is mapped by the automorphisms in the Galois group to a different prime ideal. Recall that the key result (\Cref{thm:galGrpCycField}) in \Cref{subsec:galois group cyclotomics} states that the Galois group of a cyclotomic field $K=\Q(\zeta_m)$ is isomorphic to the integer multiplicative group, i.e.,
\begin{align*}
    Gal(K/\Q) \cong (\Z/m\Z)^*.
\end{align*}
If we think each $i \in (\Z/m\Z)^*$ as a function of the roots of unity that is given by $i: \zeta_m \mapsto \zeta_m^i$, then each automorphism $\tau$ in the Galois group is uniquely mapped with a multiplicative integer $i$ if and only if $\tau(\zeta_m)=\zeta_m^i$. 
% Another way of interpreting these automorphisms is that they are permutations of. 

All of these come down to the observations that each automorphism $\tau \in Gal(K/Q)$ maps the ring of integers $R$ to itself and its dual $\dual{R}=\frac{1}{n}R$ to itself. More importantly, we have the next lemma. It enables us to transfer between different prime ideals $\mfq_i$ and $\mfq_j$. This is also known as the Galois automorphisms act \textbf{transitively} on the prime ideals $\mfq_j$. This helps with solving all components of the secret key $\vc{s}$ in the CRT-basis using a particular $\mfq_i$-RLWE oracle. In other words, once we have an oracle for a single CRT component, we can use this oracle to solve for all the other components too. 

\begin{lemma}
\reversemarginpar
\marginnote{\textit{$\tau_k(\mfq_i)=\mfq_{i/k}$}}
Let $\tau_k \in Gal(K/\Q)$ be an automorphism, then we have $\tau_k(\mfq_i)=\mfq_{i/k}$ for any $i, k \in \Z_m^*$.
\end{lemma}

For the proof of this lemma, see Lemma 2.16 of \citet{lyubashevsky2010ideal}. Since a cyclotomic field is also a Galois extension field, for a more general result see Theorem 9.2.2 of \citet{stein2012algebraic}, where $K$ is a Galois extension of $\Q$.

We have shown that both $R$ and $\dual{R}$ are closed under Galois automorphisms. To transfer a RLWE sample $(\vc{a},\vc{b})$ using an automorphism, we also need to make sure the family $\Psi_{\le \alpha}$ of elliptical Gaussian distributions is also closed under Galois automorphisms. This can be easily seen from the next lemma. 

\begin{lemma}
\reversemarginpar
\marginnote{\textit{$\Psi_{\le \alpha}$ is closed under $\tau$}}
For any $\alpha>0$, the family $\Psi_{\le \alpha}$ of elliptical Gaussian distributions is also closed under Galois automorphisms of $K$, that is, for any $\tau \in Gal(K/\Q)$ and any $\psi \in \Psi_{\le \alpha}$, we have $\tau(\psi) \in \Psi_{\le \alpha}$.
\end{lemma}

\begin{proof}
Given a $n$-dimensional $K=\Q(\zeta)$, it has a power basis $\{1, \zeta, \dots, \zeta^n\}$. We know each Galois automorphism of $K$ maps $\zeta$ to a different root of unity. Under the canonical embedding, this automorphism permutes the components of $\zeta$, so does it permutes the components of any element in $K$. Since each $D_{\vc{r}} \in \Psi_{\le \alpha}$ is a distribution over the space $K_{\C}$ that is isomorphic to the canonical space, $\tau(D_{\vc{r}})$ is still over the same space but with possibly an reordering of the scale vector $\vc{r}$. Hence, $\tau(D_{\vc{r}}) \in \Psi_{\le \alpha}$. 
\end{proof}

We are now ready to prove the following reduction. 

\begin{lemma}
\label{lm:rlweRed1}
\reversemarginpar
\marginnote{\textit{RLWE to $\mfq_i$-RLWE}}
For every $i \in \Z_m^*$, there is a deterministic polynomial time reduction from RLWE$_{q,\Psi_{\le \alpha}}$ to $\mfq_i$-RLWE$_{q,\Psi_{\le \alpha}}$.
\end{lemma}

\begin{proof}
Assume there is a  $\mfq_i$-RLWE$_{q,\Psi_{\le \alpha}}$ oracle that solves $s \bmod \mfq_i \dual{R}$ from $A_{\vc{s}, \psi}$ samples $\{(\vc{a},\vc{b})\} \subseteq R_q \times \T$ for arbitrary $\vc{s} \in \dual{R}_q$ and $\psi \in \Psi_{\le \alpha}$. We want to show that this oracle works for all CRT components, i.e., it solves $\vc{s} \bmod \mfq_j \dual{R}$ for all $j \in \Z_m^*$.

Let $k \in \Z_m^*$ such that $i=j/k$, then the automorphism $\tau_k \in Gal(K/\Q)$ maps a RLWE sample 
\begin{align*}
    (\vc{a},\vc{b}) \mapsto \tau_k((\vc{a},\vc{b})) &= (\tau_k(\vc{a}),\tau_k(\vc{b}))\\
    &= (\tau_k(\vc{a}),\tau_k((\vc{a} \star \vc{s})/q+\vc{\epsilon}))
\end{align*}
Since $R$, $\dual{R}$ and $\Psi_{\le \alpha}$ are closed under automorphisms, the transformed sample $\tau_k((\vc{a},\vc{b}))$ is also in the domain $R_q \times \T$, and most importantly distributed according to $A_{\tau_k(\vc{s}), \tau_k(\psi)}$.  
In addition, the prime ideal is mapped by $\tau_k(\mfq_j) = \mfq_{j/k}=\mfq_i$, we can then use the $\mfq_i$-RLWE$_{q,\Psi_{\le \alpha}}$ oracle to solve $\tau_k(\vc{s}) \bmod \mfq_i \dual{R}$ from the transformed RLWE samples, because it works for arbitrary secret key and error distribution. By taking the inverse of the automorphism $\tau_k$, we get an answer for the CRT component $\bmod \,\mfq_j \dual{R}$, that is, 
\begin{align*}
    \inv{\tau}_k \left(\tau_k(\vc{s}) \bmod \mfq_i \dual{R}\right) \mapsto \vc{s} \bmod \tau_k(\mfq_i)\tau_k(\dual{R})= \vc{s} \bmod \mfq_j \dual{R}.
\end{align*}
Since this works for every $j \in \Z_m^*$, we get all the CRT components. Since all the prime ideals $\mfq_i$ are also coprime and their product is the ideal $(q)$, by CRT we have an induced isomorphism 
\begin{align*}
    R/(q) &\cong \bigoplus_i (R/\mfq_i) \\
    \implies R/qR &\cong \bigoplus_i (R/\mfq_i R) \\
    \implies \dual{R}/q\dual{R} &\cong \bigoplus_i (\dual{R}/\mfq_i \dual{R}),
\end{align*}
where the last step is by the fact that $\dual{R}=(1/n) R$. Therefore, according to this isomorphism, we can compute the entire secret $\vc{s} \in \dual{R}_q$. % by using a CRT technique. 
\end{proof}

As we recover the secret key component by component in the CRT representation, we add an extra piece of information to an RLWE sample, not only at the component of interest, but all the components  before it. This gives rise to a new ``hybrid'' distribution as defined next and is used for the rest of the proof of Theorem \ref{thm:rlweSearchToDecision}.

\begin{definition}
\reversemarginpar
\marginnote{\textit{Hybrid distribution}}
For a given RLWE distribution $A_{\vc{s},\psi}$ and an integer $i \in \Z_m^*$ in the multiplicative group, the \textbf{hybrid RLWE distribution} $A_{\vc{s},\psi}^i$ over $\dual{R}_q \times \T$ is obtained by the following steps:
\begin{itemize}
    \item generate an RLWE sample $(\vc{a},\vc{b}) \leftarrow A_{\vc{s},\psi}$,
    \item generate $\vc{h} \leftarrow \dual{R}_q$ such that $\vc{h} \bmod \mfq_j \dual{R}$ is uniformly random and independent for $j \le i$ and $\vc{h} \bmod \mfq_i \dual{R}=0$ for $j > i$. That is, in its CRT representation $(h_1, \dots, h_i, \dots, h_n) \in \bigoplus_k \dual{R}/\mfq_k \dual{R}$, the components $h_1, \dots, h_i$ are uniformly random and independent and $h_{i+1}=\cdots =h_n=0$,
    \item output $(\vc{a},\vc{b}+\vc{h}/q)$.
\end{itemize}
\end{definition}

Note both indices $i$ and $j$ are integers coprime with $m$. Denote $i-$ the largest integer in $\Z_m^*$ that is smaller than $i$. By convention, denote $1-$ to be 0 and $A_{\vc{s},\psi}^{1-}=A_{\vc{s},\psi}^0=A_{\vc{s},\psi}$ the original RLWE distribution. 

\begin{definition}
\reversemarginpar
\marginnote{\textit{WDRLWE$_{q,\Psi_{\le \alpha}}^i$}}
For $i \in \Z_m^*$, the \textbf{worst-case decision RLWE relative to $\mfq_i$} problem, denoted WDRLWE$_{q,\Psi_{\le \alpha}}^i$, is to distinguish between the hybrid RLWE distributions $A_{\vc{s},\psi}^{i-}$ and $A_{\vc{s},\psi}^i$ for arbitrary $\vc{s} \in \dual{R}_q$ and $\psi \in \Psi_{\le \alpha}$.
\end{definition}

Now we state and prove the second reduction. It works in a similar fashion as the search to decision LWE reduction. That is, modify the original RLWE samples by adding an extra piece of information, which incorporates the guess of one particular CRT component $\vc{s} \bmod \mfq_i \dual{R}$.  

\begin{lemma}
\label{lm:rlweRed2}
For any $i \in \Z_m^*$, there is a PPT reduction from $\mfq_i$-RLWE$_{q,\Psi_{\le \alpha}}$ to WDRLWE$_{q,\Psi_{\le \alpha}}^i$.
\end{lemma}

\begin{proof}
Given an RLWE sample $(\vc{a},\vc{b}) \leftarrow A_{\vc{s},\psi}$, we can construct a hybrid RLWE sample $(\vc{a},\vc{b}+\vc{h}/q) \in A_{\vc{s},\psi}^{i-}$ by taking $\vc{h} \leftarrow \dual{R}_q$ such that $\vc{h} \bmod \mfq_j \dual{R}$ is uniformly random and independent for $j \le i-$ and $\vc{h} \bmod \mfq_i \dual{R}=0$ for $j \ge i$. This is further transformed by 
\begin{align*}
    (\vc{a},\vc{b}+\vc{h}/q) \mapsto (\vc{a}',\vc{b}') &= (\vc{a}+\vc{v}, \vc{b}+(\vc{v} \star \vc{g})/q)\\
    &=(\vc{a}+\vc{v},(\vc{a}' \star \vc{s} + \vc{h} + \vc{v} \star (\vc{g}-\vc{s}))/q + \vc{e}),
\end{align*}
where $\vc{v}\leftarrow R_q$ such that $\vc{v} \bmod \mfq_i$ is uniformly random and $\vc{v} \bmod \mfq_j=0$ for $j \neq i$. It is easy to see that the first part $\vc{a}+\vc{v} \in R_q$ is uniform. 

% the questions is: why do we need to make the RLWE sample hybrid first by randomized the <i components, then add a random guess to the ith component? why can't we just add a random guess to the ith component of the RLWE sample? 
The distribution of the second part $b'$ depends on whether or not $\vc{g}=\vc{s} \bmod \mfq_i \dual{R}$ is the correct guess of the CRT component. If it is, then $\vc{g}-\vc{s}$ is 0 at the $\mfq_i \dual{R}$ component, consequently $\vc{v} \star (\vc{g}-\vc{s})$ is 0 everywhere, so the distribution of the transformed sample stays as $A_{\vc{s},\psi}^{i-}$. If the guess is incorrect, then $\vc{v} \star (\vc{g}-\vc{s})$ is uniform at the $\mfq_i \dual{R}$ component and 0 everywhere else, so the transformed sample distributed as $A_{\vc{s},\psi}^i$. Given the WDRLWE$_{q,\Psi_{\le \alpha}}^i$ oracle can distinguish the two distributions, we can enumerate all possible values of $\vc{s} \bmod \mfq_i \dual{R}$ to make the correct guess. 
\end{proof}

We omit the worst-case to average-case decision RLWE relative to $\mfq_i$ reduction because the proof uses mostly probability tools, but only state the average-case definition and the reduction lemma. 

\begin{definition}
For $i \in \Z_m^*$ and a distribution $\Upsilon_{\alpha}$ over $\Psi_{\le \alpha}$, the \textbf{average-case decision RLWE relative to $\mfq_i$} problem, denoted DRLWE$_{q,\Upsilon}^i$, is to distinguish with a non-negligible probability the hybrid RLWE distributions $A_{\vc{s},\psi}^{i-}$ and $A_{\vc{s},\psi}^i$ over the random choice $(\vc{s}, \psi) \leftarrow U(\dual{R}_q) \times \Upsilon_{\alpha}$.
\end{definition}

\begin{lemma}
\label{lm:rlweRed3}

For any $\alpha > 0$ and every $i \in \Z_m^*$, there is a randomized polynomial time reduction from WDRLWE$_{q,\Psi_{\le \alpha}}^i$ to DRLWE$_{q,\Upsilon_{\alpha}}^i$.

\end{lemma}

Finally, the proof of Theorem \ref{thm:rlweSearchToDecision} comes down to the last step which shows that given a decision RLWE oracle, it solves the decision problem relative to $\mfq_i$. This relies on the fact that the hybrid distribution $A_{\vc{s},\psi}^{m-1}$ is within negligible distance to the uniform distribution over the same domain. 

\begin{lemma}
\label{lm:hybRLWECloseToUnif}
Let $\alpha \ge \eta_{\epsilon}(\dual{R})/q$ for some $\epsilon > 0$. For any $\vc{s} \in \dual{R}_q$ and error distribution $\psi \in \Psi_{\le \alpha}$ sampled according to the distribution $\Upsilon_{\alpha}$, the hybrid RLWE distribution $A_{\vc{s},\psi}^{m-1}$ is within statistical distance $\epsilon/2$ of the uniform distribution over $(R_q,\T)$.
\end{lemma}

With this lemma, we are able to prove the final step as given next. 

\begin{lemma}
\label{lm:rlweRed4}
There is a polynomial time reduction from DRLWE$_{q,\Upsilon_{\alpha}}^i$ to DRLWE$_{q,\Upsilon_{\alpha}}$ for some $i \in \Z_m^*$.
\end{lemma}

\begin{proof}
Given Lemma \ref{lm:hybRLWECloseToUnif}, it is not difficult to see this lemma follows. We know $A_{\vc{s},\psi}^{0}=A_{\vc{s},\psi}$ is the RLWE distribution and $A_{\vc{s},\psi}^{m-1}$ is nearly uniform, so the DRLWE$_{q,\Upsilon}$ oracle can distinguish the two. This is an easy task for the oracle. 

If we bring the two distributions closer, say for $i \in \Z_m^*$ and start with $i=1$, we ask the oracle to distinguish the two hybrid distributions $A_{\vc{s},\psi}^{i-}$ and $A_{\vc{s},\psi}^{i}$.
Intuitively, both distributions should be close to the RLWE distribution for small $i$ and to the uniform distribution for large $i$. So the oracle will not distinguish them. But there must be an index $i$ such that at that point $A_{\vc{s},\psi}^{i-}$ is closer to the RLWE distribution and $A_{\vc{s},\psi}^{i}$ is closer to the uniform distribution, so the oracle can easily distinguish them. This index $i \in \Z_m^*$ is what will be used for all the previous reduction steps that we have discussed.  
\end{proof}

% \begin{color}{red}
% $\alpha < \eta(\dual{R})$ but also $\alpha > \eta(\dual{R})/q$, the former is to ensure the decision RLWE is still a valid problem, the latter is to ensure $A^{m-1}$ is close to uniform. 

% need cyclotomic, because they split into factors of the same degree, so when rotate, the dimension of the noise doesn't change. 

% crt representation (i.e., evaluate polynomial in roots) makes things component wise, this is essential for proof, because we need to swap coponents in the proof. this also gives efficiency. 

% randomness implies average-case problem,
% \end{color}

\newpage

\subsection{An RLWE-based encryption scheme}
\label{subsec:lpr}
To end this section, we state a simple RLWE-based public-key encryption scheme presented by \citet{lyubashevsky2010ideal}. 

Let $R=\Z[x]/(x^n+1)$, where $n$ is taken to be a power of 2 to make the modulo polynomial cyclotomic, hence $R$ a cyclotomic field. This is the domain for the secret key and noise vectors that are sampled according to a specific distribution $\chi$. Restrict the public key and ciphertexts to be in the domain $R_q=\Z_q[x]/(x^n+1)$. The scheme is presented as follows with slight modifications to be consistent with the BFV scheme that will be presented in the next section.  

Decryption works if the parameters are properly set and polynomials sampled from $R$ have small coefficients (according to the distribution $\chi$). Because 
\begin{align}
\label{eq:lprScheme}
    \vc{u} + \vc{v} \cdot \vc{s} = \floor{q/2} \cdot \vc{m} + (\vc{e} \cdot \vc{r} + \vc{e}_1 + \vc{e}_2 \cdot \vc{s})  \bmod q.
\end{align}
If those polynomials are taken with large coefficients, after multiplications they will neither staying within modulo $q$, nor being rounded to 0. 

As for its security, the public key $(\vc{b}, \vc{a})$ is a RLWE sample with the secret vector $\vc{s}$, so it is pseudo-random which implies there no way to recover $\vc{s}$ because that requires a solution to the search RLWE problem. In terms of semantic security (\cref{def:semSec}), the pairs $(\vc{b},\vc{u}-\round{q/2} \cdot \vc{m} \bmod q)$ and $(\vc{a}, \vc{v})$ are also RLWE samples with the corresponding secret vector $\vc{r}$, so the ciphertext $\vc{c}$ is pseudo-random too, which implies semantic security. 

\begin{tcolorbox}
\noindent
\textbf{Private key}: Sample a private key $\vc{s} \leftarrow \chi$.\\

\textbf{Public key}: Sample random polynomials $\vc{a} \leftarrow R_q$ and $\vc{e} \leftarrow \chi$ and output the public key $(\vc{b} = -\sqbracket{ \vc{a} \cdot \vc{s} + \vc{e}}_q,\vc{a})$.\\

\textbf{Encryption:} Encrypt an $n$ bits message $\vc{m} \in \{0,1\}^n$ by computing 
\begin{align*}
    \vc{u} &= \vc{b} \cdot \vc{r} + \vc{e}_1 + \floor{q/2} \cdot \vc{m} \bmod q\\
    \vc{v} &= \vc{a} \cdot \vc{r} + \vc{e}_2 \bmod q,
\end{align*}
where $\vc{r},\vc{e}_1,\vc{e}_2 \leftarrow \chi$ are random samples. Then output the ciphertext $\vc{c}=(\vc{u},\vc{v})$.\\

\textbf{Decryption:} Decrypt the ciphertext $\vc{c}$ using the secret key by computing 
\begin{align*}
    m = \sqbracket{\round{\frac{2}{q} \sqbracket{\vc{u} + \vc{v} \cdot \vc{s}}_q}}_2.
\end{align*}
\end{tcolorbox}

\begin{figure}
\centering
\caption{A Sage implementation of the RWLE-based encryption scheme described above.\\ \textbf{Note:} This implementation is not suitable for use in real-world applications.}
\begin{tcolorbox}
\begin{verbatim}
#!/usr/bin/env sage

from sage.misc.prandom import randrange
import sage.stats.distributions.discrete_gaussian_integer as dgi

# Define parameters
def sample_noise(n, P):
    D = dgi.DiscreteGaussianDistributionIntegerSampler(sigma=1.0)
    return P([D() for i in range(n)])

q = 655360001
n = 2^10

P = QuotientRing(PolynomialRing(Integers(q), name="x"),
                 x^n + 1)
Q = PolynomialRing(Rationals(), name="y")
Z2 = Integers(2)

# Generate keys
secret_key = sample_noise(n, P)

e = sample_noise(n, P)
a = P.random_element()
b = -(a*secret_key) + e

public_key = (b,a)

# Encrypt Message
message = P([randrange(0,2) for i in range(n)])

r = sample_noise(n, P)
e1 = sample_noise(n, P)
e2 = sample_noise(n, P)

u = b*r + e1 + (q//2)*message
v = a*r + e2

ciphertext = (u,v)

# Decrypt Message
w1 = u + v*secret_key
w2 = (2/q) * Q(w1.list())

decrypted_message = P([Z2(w.round()) for w in w2.list()])

# Verification
print(decrypted_message == message)
\end{verbatim}
\end{tcolorbox}
\end{figure}
%%%%%%%%%%%%%%%%%%%%%%%%%%%%%%%%%%%%%%%%%%%%%%%%%%%%%%%%%%%%%%%%%%%%%%%%%%%%%%%%%%%%%%%%%%%%%%%%%%%
%%%%%%%%%%%%%%%%%%%%%%%%%%%%%%%%%%%%%%%%%%%%%%%%%%%%%%%%%%%%%%%%%%%%%%%%%%%%%%%%%%%%%%%%%%%%%%%%%%%

%%%%%%%%%%%%%%%%%%%%%%%%%%%%%%%%%%%%%%%%%%%%%%%%%%%%%%%%%%%%%%%%%%%%%%%%%%%%%%%%%%%%%%%%%%%%%%%%%%%

%\newpage
%\bibliography{references}
%\bibliographystyle{abbrvnat}

\newpage
\section{Homomorphic Encryption}

\label{sec:he}
% information introduction of HE, 

% \section{Homomorphic encryption}

Shortly after the RSA encryption scheme \citep{rivest1978method} was released, \citet{rivest1978data} raised the question of whether it is possible to perform arithmetic operations (e.g., addition and multiplication) on encrypted data without the secret key, and the results can be decrypted to the correct results if the same operations were performed on the unencrypted data. An encryption scheme possessing such a property is called a homomorphic encryption scheme. 

\subsection{Basic definitions}
% A homomorphic encryption scheme can be formally stated as follows in terms of public key encryption. 

% Having introduced the notations, 
We formally define here the sub-routines of a public key homomorphic encryption (HE) scheme. Similar to non-HE schemes, an HE scheme also has a key generation process, an encryption process, and a decryption process. The difference is that an HE scheme consists of an extra evaluation process that evaluates a function, which is often expressed as an arithmetic circuit on the ciphertexts, and produces an ``evaluated ciphertext''. % Denote by $f:\{0,1\}^l \rightarrow \{0,1\}$ a function that can be evaluated by HE schemes. 

\begin{definition}
A \textbf{homomorphic encryption scheme} \index{homomorphic encryption (HE)!scheme}
\reversemarginpar
\marginpar{HE scheme}
is a four tuple of PPT algorithms 
\begin{align*}
    \he=(\he.\keygen,\he.\enc,\he.\eval,\he.\dec)
\end{align*}
that takes the security parameter $\lambda$ as the input. Each of the PPT algorithms is defined as follows: 
\begin{itemize}
    \item \textbf{Setup}: Given the security parameter $\lambda$, generate a parameter set $\para=(n,q,N,\chi) \leftarrow \he.\setup(1^{\lambda})$ for the following steps.
    
    \item \textbf{Key generation}: Given the parameters generated above, the algorithm produces $(\pk,\sk,\evk) \leftarrow \he.\keygen(\para)$ a set of keys that consists of a public key, a secret key and an evaluation key.
    
    \item \textbf{Encryption}: The algorithm takes the public key and a plaintext $m$ (i.e., the secret message) to produce a ciphertext text $c \leftarrow \he.\enc(\pk,m, n,q,N)$. 
    
    \item \textbf{Evaluation}: Given the evaluation key, the evaluation function $f : \{0,1\}^l \rightarrow \{0,1\}$ and a set of ciphertexts, the algorithm produces an evaluated ciphertext $c_f \leftarrow \he.\eval(\evk,f,c_1,\dots,c_l)$. 
    
    \item \textbf{Decryption}: The algorithm decrypts the ciphertext using the secret key to find the corresponding plaintext $m_f \leftarrow \he.\dec(\sk,c_f)$. 

\end{itemize}
\end{definition}
This is a basic form of an HE scheme. A more complicated scheme may take extra input parameters for additional purposes such as reducing ciphertext noise magnitude and so on. 

% \begin{definition}
The plaintext $m_f$ corresponds to the function output of $f$ when applied to the plaintexts directly. If the decrypted ciphertext after evaluations does not match with $m_f$, the HE scheme is considered as unsuccessful. More formally, let $m_1$ and $m_2$ be two plaintexts, $pk$ and $sk$ be the public key and secret key for encryption and decryption, respectively. A homomorphic encryption scheme satisfies the property that for an operation $\diamond$ in the plaintext space, there is a corresponding operation $\bullet$ in the ciphertext space such that \index{homomorphic encryption (HE)! property} 
\begin{equation}
\label{eq:he}
    Dec(sk,Enc(pk,m_1) \bullet Enc(pk,m_2)) = m_1 \diamond m_2,
\end{equation}
% then the encryption scheme is said to be \textbf{homomorphic}. 
% \end{definition}

Most of the HE schemes have the same operations in both plaintext and ciphertext spaces. That is, additions of ciphertexts can be decrypted to additions of plaintexts. Similarly for multiplications. The name ``homomorphic'' is likely taken from the concept of \textit{homomorphism} in mathematics, which is a structure-preserving map between two algebraic structures. The analogy here is that the decryption function is a homomorphism from the ciphertext space to the plaintext space that preserves the same operations in the two spaces as stated in \Cref{eq:he}. 

\iffalse
The process is also shown in a commutative diagram in \cref{fig:he}. 

\begin{figure}[h]
    \centering
    %page3
    \includegraphics[page=18,scale=0.8]{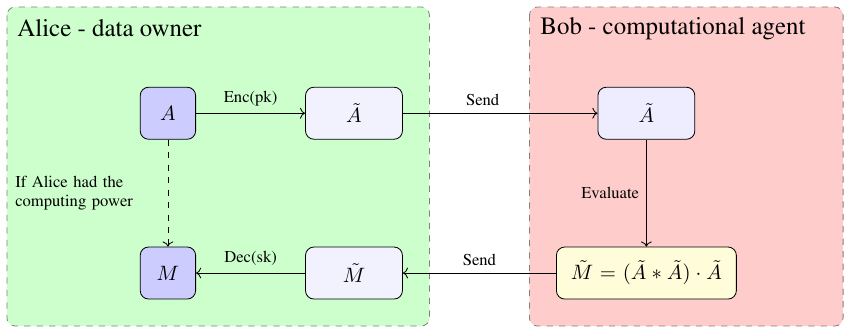}
    \caption{Diagram of a homomorphic encryption scheme. The data owner Alice encrypted the data $A$ using the public key and sent the ciphertext $\tilde{A}$ to the computational agent Bob, who then performed HE permitted operations on the ciphertext. We denote by $*$ and $\cdot$ the matrix multiplication and matrix element-wise multiplication, respectively. The resulting ciphertext $\tilde{M}$ was sent back to Alice for decryption using the secret key. Alice could obtain the same result $M$ from $A$ if she had the computing power to perform the same operations that Bob did on the ciphertext.
    \begin{color}{blue}
    KS: The expression "if Alice had the same computing power to perform the same operations" is a bit strange. The issue is usually about privacy, not computation. In fact, doing computations in the encrypted space always more expensive than doing computations in plantext space.
    \end{color}
    }
    \label{fig:he}
\end{figure}
\fi

It is important to note that the encryption function is not homomorphic, that is,  
\begin{equation*}
    \text{Enc}(\text{pk},m_1) \bullet \text{Enc}(\text{pk},m_2)) \neq \text{Enc}(\text{pk},m_1 \diamond m_2),
\end{equation*}
because encryptions in HE are non-deterministic 
% that injects random noise in ciphertext 
in order to satisfy semantic security (\Cref{def:semSec}). Recall that semantic security assures that given a ciphertext $c$ that encrypts one of the two messages $m_1$ and $m_2$, it is impossible for a PPT attacker to guess the source message from $c$ with a better chance than random guessing. 
% The noise needs to be well-controlled during encryption and evaluations in order to ensure the evaluated ciphertext can be decrypted to the correct result. 

\begin{example}
The RSA encryption system\index{RSA}, without message padding, is a homomorphic encryption system for multiplication. (Of course, without message padding\index{padding}, the RSA system is not semantically secure.)
\end{example}

\begin{example}
Here is a simple homomorphic encryption system given by \citet{brakerski2014efficient}.  
Let $\mathbf{s} \in \mathbb{Z}_q^n$ be the secret key. The private message $m \in \{0,1\}$ is encrypted by
\begin{align*}
    c = (\mathbf{a}, b=\mathbf{a} \cdot \mathbf{s} + 2 e + m) \in \mathbb{Z}_q^n \times \mathbb{Z}_q,
\end{align*}
where $e$ is a random noise with small magnitude. The decryption of this ciphertext with the secret key is done by 
\begin{align*}
    m = ((b - \mathbf{a} \cdot \mathbf{s}) \bmod q) \bmod 2,
\end{align*}
provided $e$ is small enough to ensure $b - \mathbf{a} \cdot \mathbf{s} = 2e+m$ is within $\mathbb{Z}_q$. 
% A poor setting is when $q=7$, $m=1$ and $e=4$, so $2e+m=9 = 2 \bmod 7$. Taking modulo 2 of this result gives $m=0$, so the decryption fails. 
%
Given two ciphertexts $c_1$ and $c_2$ that respectively encrypts the messages $m_1$ and $m_2$ as above, their sum can be easily computed by the bilinearity of dot product, so
\begin{align*}
    c_1+c_2&=(\mathbf{a}_1+\mathbf{a}_2,b_1+b_2)\\
    &=(\mathbf{a}_1+\mathbf{a}_2,(\mathbf{a}_1+\mathbf{a}_2)\cdot \mathbf{s} + 2(e_1+e_2) + (m_1+m_2)).
\end{align*}
Decryption proceeds as before and produces the sum of the two messages $m_1+m_2$, so the scheme is additive homomorphic. The scheme can also be shown to be multiplicative homomorphic. % as demonstrated in \citet{brakerski2014efficient}. %, the detail is skipped here.
\end{example}

In many homomorphic encryption systems, the ciphertext noise increases after each homomorphic evaluation operation, and if the overall noise is higher than a threshold called the \textit{noise ceiling}\index{noise ceiling} (e.g., the modulo $q$ in the above example), decryption can fail to output the correct result. Given a noise ceiling % (i.e., ciphertext domain modulo) 
and the noise bound (on which the noise distribution is supported), the number of homomorphic evaluations that can be performed on the ciphertexts is usually restricted. The breakthrough made by \citet{gentry2009fully} enables an unlimited number of homomorphic evaluations on ciphertexts using squashing and bootstrapping, which are described in the next subsection. 
%Many existing encryption schemes have been shown to be either additive or multiplicative homomorphic, with a few that can work with both operations. 
Below, we listed a few commonly mentioned HE categories, which are grouped by the class of arithmetic circuits they can evaluate.
\begin{itemize}
    \item Partially HE (PHE) \index{homomorphic encryption (HE)! partial} - Schemes that can evaluate circuits containing only one type of arithmetic gates, that is, either addition or multiplication, for unbounded circuit depth. 
    
    \item Leveled HE (LHE) \index{homomorphic encryption (HE)! leveled} - Schemes that can evaluate circuits containing both addition and multiplication gates, but only for a pre-determined multiplication depth $L$. % Moreover, bit length of the evaluation key depends on $L$, but not other keys.                      
    
    \item Somewhat HE (SHE) \index{homomorphic encryption (HE)! somewhat} - Schemes that can evaluate a subset of circuits containing both addition and multiplication gates, whose complexity grows with the circuit depth. SHE is more general than LHE. Examples include \citet{gentry2009fully,gentry2010computing}.        
    
    \item Leveled Fully HE \index{homomorphic encryption (HE)! leveled fully} - Almost identical to leveled HE, except these schemes can evaluate \textbf{all} circuits of depth $L$. Examples include \citet{brakerski2014efficient,brakerski2014leveled,brakerski2012fully}.
    
    \item Fully HE (FHE) \index{homomorphic encryption (HE)! fully} - Schemes that can evaluate all circuits containing both addition and multiplication gates for unbounded circuit depth. Examples include \citet{gentry2009fully} and \citet{brakerski2014efficient,brakerski2014leveled,brakerski2012fully} under the \textit{weak circular security}, which guarantees security when using only one pair of secret and public keys.
    
\end{itemize}

% \kl{Should include some definitions, including compactness, FHE, etc.}

\subsection{Gentry's original FHE using squashing and bootstrapping}
\label{subsec:gentry bootstrap}

As discussed above, noise growth needs to be well controlled during homomorphic evaluations in order to guarantee correct decryption. Under such a constraint, a scheme can only perform a certain number of arithmetic on ciphertexts, unless the ciphertext noise can be constantly reduced after evaluations. An obvious noise elimination method is ciphertext decryption that completely clears the embedded noise in the ciphertext. So the question is how to utilize a scheme's own decryption circuit to reduce noise growth and carry on more homomorphic evaluations.

Gentry's original construction to achieve FHE consists of three components. The first component is a SHE scheme that can handle both addition and multiplication for a non-trivial but limited number of steps. The second component is a squashing process to make the SHE scheme's decryption step easier in order to permit bootstrapping. The third component is the actual bootstrapping process that enables the evaluation of the scheme's own decryption circuit, plus an extra evaluation step. The key observation here is that during bootstrapping, a ciphertext will be doubly encrypted and  decrypted only from the inner layer. This is then followed by a single arithmetic step on the (singly encrypted) ciphertexts. The three components put together gives a scheme, whose ciphertext noise can be reduced before running the next arithmetic step, and consequently leads to FHE. A formal definition of bootstrappable is stated next. 

\begin{definition}
A scheme is \textit{$\mathcal{C}$-homomorphic} if it can evaluate any circuit in the class $\mathcal{C}$.
\end{definition}

\begin{definition}
\label{def:bootstrappable}
Let HE be a $\mathcal{C}$-homomorphic scheme and $f_{add}^{c_1,c_2}(s)$ and $f_{mult}^{c_1,c_2}(s)$ be two decryption functions augmented by an addition and an multiplication, respectively. Then HE is \textbf{bootstrappable} \index{bootstrappable} if $\{f_{add}^{c_1,c_2}(s), f_{mult}^{c_1,c_2}(s)\}_{c_1,c_2} \in \mathcal{C}$ the two augmented decryptions are in the class.
\end{definition}

The definition suggests that decryption needs to be simple enough so that not only it is in $\mathcal{C}$, but it needs to be followed by an arithmetic operation to allow further evaluation. To ensure this, Gentry added a ``hint'' to the ciphertext to make decryption simpler. This process is later known as \textit{squashing} \index{squashing}. Next, we restate the simple concrete HE scheme by \citet{dijk2010fully} that was also used by \citet{gentry2010computing} to illustrate the squashing and bootstrapping concept. 

Set the parameters $N=\lambda$, $P=\lambda^2$ and $Q=\lambda^5$ for the given security parameter $\lambda$. The (secret key) encryption scheme consists of the following steps:%The scheme can be turned into a public key scheme.  
\begin{itemize}
    \item \textbf{Key generation}: $p \leftarrow \keygen(\lambda)$, where $p$ is an odd integer of $P$-bit. 
    
    \item \textbf{Encryption}: To encrypt a message $m \in \{0,1\}$, choose an $N$-bit integer $m'$ such that $m'=m \bmod 2$. Then output the ciphertext $c=m'+pq \leftarrow \enc(p,m)$, where $q$ is a random $Q$-bit number.
    
    \item \textbf{Decryption}: To decrypt the ciphertext $c$, run the sub-routine $(c \bmod p) \bmod 2 \leftarrow \dec(p, c)$. It will output the correct message $m$, because $c \bmod p = m'$ which has the same parity as $m$ as chosen in the encryption step. 
\end{itemize}
The scheme is both additive and multiplicative homomorphic, given % where the additive ciphertext $c_1+c_2=(m_1'+m_2')+p \cdot (q_1+q_2)$ and the multiplicative ciphertext 
\begin{gather*}
  c_1+c_2=(m_1'+m_2')+p \cdot (q_1+q_2) \\
  c_1 \cdot c_2 = (m_1' \cdot m_2') + p \cdot (m_1' \cdot q_2 + m_2' \cdot q_1 + p \cdot q_1 \cdot q_2).
\end{gather*}  

%$m'=c \bmod p$ is the noise associated to this ciphertext. Its size is at most $N$-bit. Given two ciphertexts $c_1$ and $c_2$ with noises $k_1$ and $k_2$ bits respectively. Addition increases the noise to $\max\{k_1, k_2\}+1$ bits and multiplication increases the noise to $(k_1+k_2)$ bits. For decryption to work correctly, the noise in an evaluated ciphertext must be less than $p/2$, for otherwise the noise will interact with the term $pq'$ and hence outputs an incorrect result when applying modulo $p$. 

%Generalize this to a circuit with multiple inputs, which can also be represented by a multivariate function $f^{\dagger}(c_1, \dots, c_t)=f^{\dagger}(m_1', \dots, m_t')+pq'$. Hence, the above scheme can evaluate functions $f^{\dagger}$ that satisfies $|f^{\dagger}(m_1', \dots, m_t')| < p/2$, where each $m_i'$ is at most an $N$-bit integer. 

%As suggested in \citet{gentry2009fully} in a special example, this scheme can evaluate elementary symmetric polynomials with degree $d < P/(N \log t)$, for $t$ ciphertexts $c_1, \dots, c_t$.

However, it is not bootstrappable due to the complexity of the decryption step. More precisely, the decryption function $(c \bmod p) \bmod 2$ is equivalent to $\text{LSB}(c) \text{ XOR } \text{LSB}(\lfloor c/p \rceil)$, where LSB is the least significant bit. The most time-consuming step in the decryption function is the multiplication of two large numbers $c \cdot 1/ p$.
To simplify this multiplication, Gentry's idea is to replace $c \cdot 1/p$ by summing a small set of numbers, which is known as the \textit{sparse subset sum problem} (SSSP) \index{sparse subset sum problem (SSSP)}. This sum is the ``hint'' to decryption to reduce its running time and consequently permit bootstrapping. The modified scheme is as follows: 

\begin{itemize}
    \item \textbf{Key generation}: First, generate $(\pk,\sk) \leftarrow \keygen(\lambda)$, where $\sk=p$ is the odd integer. Then, generate a real vector $\vc{y} \in [0, 2)^{\beta}$ such that there exists a subset of indices $S \subseteq \{1, \dots, \beta\}$ of size $\alpha$ and $\sum_{i \in S} \vc{y}_i \approx 1/p \bmod 2$ can approximate the original secret key $\sk$. Finally, output the keys $(\pk^*, \sk^*)$, where $\pk^*=(\pk, \vc{y})$ and $\sk^*=S$. Here when $\alpha$ and $\beta$ are set properly, given the set $\vc{y}$ and $1/p$, it is hard to find the subset of indices $S$ that is the new secret key $\sk^*$. So the ``hint'' is added to the public key. 
    
    \item \textbf{Encryption}: First, compute $c \leftarrow \enc(\pk, m)$. Then, compute $\vc{z}_i = c \cdot \vc{y}_i$. Finally, output $c^*=(c, \vc{z})$. 
    
    \item \textbf{Decryption}: Run $\text{LSB}(c) \text{ XOR } \text{LSB}(\lfloor c/p \rceil)$. Here, we approximate $c/p$ by $\lfloor \sum_{i \in S} z_i \rceil$. From the key generation step, we know that $\sum_{i \in S} z_i = \sum_{i \in S} c \cdot \vc{y}_i = c \cdot \sum_{i \in S} \vc{y}_i \approx c \cdot 1/p \bmod 2$. The summation is over a small subset and is relatively easier to compute than the multiplication of two long numbers. 
\end{itemize}

This revised scheme is also both additive and multiplicative homomorphic, which can be achieved by extracting the ciphertext $c$ from $c^*$ then apply the addition and multiplication operations as in the original scheme. The cost of squashing decryption is the scheme's security, which is now also based on the hardness assumption of \textit{SSSP}, in addition to the scheme's original security assumption. In other words, the attacker is also given the encryption of the secret key by the corresponding public key. This situation is properly dealt with by the additional security assumption stated next and is necessarily assumed when pursuing for FHE.  
\begin{definition}
\label{def:weakCircularSec}
A public key encryption scheme is \textbf{weak circular secure} \index{secure! weak circular} if it is CPA\index{CPA} secure even in the presence of the encryption of the secret key bits. 
\end{definition}

It is worth keeping in mind that this concrete scheme is only a simplified illustration of Gentry's original SHE construction based on ideal lattices \citep{gentry2009fully}. Besides his breakthrough to achieve FHE using squashing and bootstrapping, Gentry's work also inspired a great number of subsequent developments in FHE, especially those that tried to improve efficiency without using squashing and bootstrapping. In the next few subsections, we will see a sequence of such works.

\subsection{\texorpdfstring{$\bvv$}{BV*} : SHE by relinearization}

We will cover the body of works in \citet{brakerski2014efficient,brakerski2014leveled,brakerski2012fully,fan2012somewhat} that were inspired by \citet{regev2009lattices}'s scheme. These second-generation homomorphic encryption schemes are more efficient than Gentry's original construction and also based on standard lattice problems via the learning with error problem.   

The first work in this line of research is \citet{brakerski2014efficient}. 
Without using bootstrapping, \citeauthor{brakerski2014efficient} were able to construct an SHE scheme $\bvv$\footnote{We name the scheme after the authors' surname initials.} that can perform a non-trivial number of homomorphic evaluations. With an additional dimension-modulus reduction step that we describe in Section~\ref{subsec:modulus reduction}, this scheme's efficiency can be further improved to allow it to achieve leveled FHE without using \citet{gentry2010computing}'s squashing idea, which needs an extra hardness assumption to guarantee a scheme's security. 

%\paragraph{Setup}
The scheme is similar to \citeauthor{regev2009lattices}'s scheme, which we describe in Section~\ref{subsec:regev scheme}, but with minor changes and an \textit{evaluation key} specifically for homomorphic multiplications. 
Given the security parameter $\lambda$,
%\footnote{The Security parameter $\lambda$ is just the underlying positive integer $k$ for determining $n$ in the LWE distribution.} 
$\bvv$ produces the parameters 
\begin{align*}
    \para=(n,q,N,\chi) \leftarrow \bvv.\setup(1^{\lambda})
\end{align*}
just as in $\reg$. One difference is that $q$ does not need to be a prime and is taken from a larger range $q \in [2^{n^{\epsilon}},2 \cdot 2^{n^{\epsilon}})$, which is subexponential in $n$ for a constant $\epsilon \in (0,1)$. Also, the LWE sample size $N \ge n \log q+2k$. Furthermore, the scheme has a pre-determined multiplication level for the arithmetic circuits that will be evaluated. This level parameter is approximately $L\approx \epsilon \log n$ for an arbitrary constant $\epsilon \in (0,1)$, and only related to the number of keys that needs to be generated. % as will be seen next. 

In the following, $\Z_q$ denotes the symmetric range $[-q/2, q/2) \cap \Z$, which is different from its standard use for representing the ring $\Z/\Z_q=[0,q)$. Also, $y=[x]_q$ denotes the reduction of $x$ to within $\Z_q$ such that $[x]_q = x \bmod q$. The modulo q reduction (i.e., $\bmod q$) is to be distinguished from $[x]_q$, where the former is reduction to $\Z/\Z_q$ and the latter is to $\Z_q$. For simplicity, (in particular in the BFV scheme) we use $r_q(x) = x \bmod q$ to denote the remainder. We use boldface to denote vectors and matrices. When working with matrices, all vectors are by default considered as column vectors. Vector multiplications are denoted by $\vc{a} \cdot \vc{b}$, whilst matrix (and sometimes scalar) multiplications are denoted without the ``dot'' in the middle. 

A distribution $\chi$ over the integers is $B$ bounded, denoted by $|\chi| \le B$, means $\chi$ is only supported on $[-B, B]$.

\subsubsection*{Key generation}

The important part of the key generation, which does not appear in Regev's scheme, is the generation of the evaluation key for relinearization, a term that will be explained in detail next. First, run Regev's secret key generation to produce a sequence of secret vectors 
\begin{align*}
    \vc{s}_0, \dots, \vc{s}_L \leftarrow \bvv.\seckeygen(n,q), \text{ where }
    \vc{s}_i = (1, \vc{t}_i) 
    % \vc{s}_i[0]=1 
    \text{ and } \vc{t}_i \leftarrow \Z_q^n, \forall i \in [0, L].
\end{align*}
Each of the $L$ secret keys will then be embedded in the evaluation key that is used for relinearizing quadratic terms that appear during homomorphic multiplications. 
In particular, the evaluation key is a set $\Psi=\{\psi_{l,i,j,\tau}\}$, $1 \le l \le L$, $0 \le i \le j \le n$, $0 \le \tau \le \lfloor \log q \rfloor$, where
\begin{align}
    \psi_{l,i,j,\tau} := \left( \vc{a}_{l,i,j,\tau}, b_{l,i,j,\tau}=\sqbracket{\vc{a}_{l,i,j,\tau} \cdot \vc{s}_l + 2 \cdot e_{l,i,j,\tau} + 2^{\tau} \cdot \vc{s}_{l-1}[i] \cdot \vc{s}_{l-1}[j]}_q \right) %\in \Z_q^n \times \Z_q.
\end{align}
is computed by sampling a random vector $\vc{a}_{l,i,j,\tau} \leftarrow \Z_q^n$ and a noise $e_{l,i,j,\tau} \leftarrow \chi$.
% To compute the evaluation key, sample a random vector $\vc{a}_{l,i,j,\tau} \leftarrow \Z_q^n$ and a noise $e_{l,i,j,\tau} \leftarrow \chi$. For all $1 \le l \le L$, $0 \le i \le j \le n$ and $0 \le \tau \le \lfloor \log q \rfloor$, compute the following two-tuple
%\begin{align}
%    \psi_{l,i,j,\tau} := \left( \vc{a}_{l,i,j,\tau}, b_{l,i,j,\tau}=\vc{a}_{l,i,j,\tau} \cdot \vc{s}_l + 2 \cdot e_{l,i,j,\tau} + 2^{\tau} \cdot \vc{s}_{l-1}[i] \cdot \vc{s}_{l-1}[j] \bmod q \right). %\in \Z_q^n \times \Z_q.
% \end{align}
One can interpret the first element $\vc{a}_{l,i,j,\tau}$ of this tuple as the ``public key'' and the second element $b_{l,i,j,\tau}$ as an noisy ``encryption'' under the secret key $\vc{s}_l$ of the message $2^{\tau} \cdot \vc{s}_{l-1}[i] \cdot \vc{s}_{l-1}[j]$. This ``encrypted'' message will be used to approximate a multiplicative ciphertext once it has gone through a multiplicative gate. (This will become clearer in Section~\ref{subsubsec:bvHomMul}.) Although the evaluation key is public, it is not needed to assume weak circular security to guarantee the scheme's security, because the secret key is not being encrypted by its corresponding public key. This also explains why the evaluation key is a series of key pairs rather than one pair. 

% Set the evaluation key $\Psi=\{\psi_{l,i,j,\tau}\}$ to be the set of two-tuples for all values of the parameters $l, i, j, \tau$. 
The parameter $\tau$ corresponds to each bit position of a random $\Z_q$ sample when represented in binary format. For example, if $h_{i,j} \in \Z_q$ then its binary form is $h_{i,j}=\sum_{\tau=0}^{\lfloor \log q \rfloor} 2^{\tau} h_{i,j,\tau}$, where $h_{i,j,\tau} \in \{0,1\}$ and $\lfloor \log q \rfloor$ is the maximum bit length minus 1. This particular set up is to reduce the relinearization error during homomorphic multiplications. It will also be discussed in more detail later. 

The rest of the key generation process is similar to the corresponding process in Regev's. The secret key of $\bvv$ for decryption is $\vc{s}_L$, the last secret vector in the sequence,
% \begin{align*}
%    \vc{s}_L \leftarrow \bvv.\seckeygen(n,q).
% \end{align*}\cdot
indicating the ciphertexts have gone through the complete evaluation circuit of max depth $L$.
% \footnote{In \citet{brakerski2014efficient}, 
%For simplicity, a ciphertext will only be decrypted once it has gone through the entire arithmetic circuit. 
%The functions being evaluated are assumed to have exactly $L$ multiplications. This corresponds to arithmetic circuits having exactly $L$ multiplicative depth.
 
Taking the first secret vector $\vc{t}_0$ generated above, the public key generation process adds an even integer noise vector to the ciphertext as in Regev's starred public key generation process to get the following public key in the matrix format
\begin{align*}
    \vc{P} =[\vc{b} \mid -\vc{A}] \leftarrow \bvv.\pubkeygen(n,q,N,\chi,\vc{t}_0), % =\reg.\pubkeygen^*(\vc{t}_0,\para).
\end{align*}
where $\vc{A} \leftarrow \Z_q^{N \times n}$, 
and $\vc{b} = \sqbracket{\vc{A} \vc{t}_0 + 2\vc{e}}_q$ for a random noise vector $\vc{e} \leftarrow \chi^N$, and $\vc{P} \in \Z_q^{N \times (n+1)}$ is the result of appending the column vector $\vc{b}$ to the front of the matrix $-\vc{A}$.

To summarise, the output of the key generation step is
\begin{align*}
%\label{eq:sh.keygen}
    (\pk, \sk, \evk) &\leftarrow \bvv.\keygen(1^{\lambda}), \text{ where }\\
    \pk = \vc{P} &\leftarrow \bvv.\pubkeygen(n,q,N,\chi,\vc{t}_0), \\
    \sk=\vc{s}_L &\leftarrow \bvv.\seckeygen(n,q), \\
    \evk=\Psi &\leftarrow \bvv.\evalkeygen(n,q,\chi).
\end{align*}

\subsubsection*{Encryption} 
The encryption function is similar to $\reg.\enc(\pk,m)$ but has a level tag to keep track of the number of evaluated multiplicative gates, starting from 0 till the maximum value $L$. 
To encrypt a message $m \in \{0,1\}$ using the public key, the algorithm concatenates $m$ with 0s to get a length $n+1$ vector $\vc{m}=(m, 0, \dots, 0)$. It then generates $\vc{r} \leftarrow \{0,1\}^N$ and outputs the ciphertext
\begin{align*}
%\label{eq:sh.enc}
    \vc{c}^l=\bracket{\vc{c}=\sqbracket{\vc{P}^T \vc{r}+ \vc{m}}_q, l} \leftarrow \bvv.\enc(\vc{P}, m,n,q,N,l) % = \reg.\enc(\vc{P},m,n,q,N)
\end{align*}
as a two-tuple, where the first element  $\vc{c}$ is a length $n+1$ vector. % \footnote{\citet{brakerski2014efficient} used vector (instead of matrix) notations to represent the keys and ciphertext. Later, they switched to matrix notation. For example, their ciphertext is $\vc{c}=(\vc{v}, w)$, where to match our notations, $\vc{v}=\vc{A}^T \cdot \vc{r}$ and $w=\vc{b}^T \cdot \vc{r}+m$. The two notations are equivalent in the way that our above ciphertext notation $\vc{c}=(w,\vc{v})$.} 

\subsubsection*{Decryption} The decryption is also identical to $\reg.\dec(\sk,\vc{c})$, but the rounding operation is omitted because of the setting $t=q$ so the noise can be eliminated by taking modulo 2. To decrypt the ciphertext $\vc{c}^L=(\vc{P}^T \vc{r}+ \vc{m},L)$, which has gone through the complete circuit, the  algorithm computes %\footnote{In \citet{brakerski2014efficient}, decryption only applies to ciphertexts that have been produced by evaluating the complete circuit. These ciphertexts always have the level tag $l=L$.} 
\begin{align*}%\begin{empheq}[box=\mymath]{align}
%\label{eq:sh.dec}
    \sqbracket{\sqbracket{\vc{c} \cdot \vc{s}_L}_q}_2 \leftarrow \bvv.\dec(\vc{s}_L,\vc{c},q). %=\reg.\dec(\vc{s}_L,\vc{c},q).
\end{align*}%\end{empheq}
Substitute terms into the dot product, we get
\begin{align*}
    \sqbracket{\vc{c} \cdot \vc{s}_L}_q &=\sqbracket{(\vc{b}^T \vc{r} + m) - \vc{t}_L^T  \vc{A}^T  \vc{r}}_q \\
    &=\sqbracket{((\vc{A} \vc{t}_L)^T \vc{r}+ 2 \vc{e}^T \vc{r} + m) -  \vc{t}_L^T \vc{A}^T \vc{r}}_q \\
    &=\sqbracket{m+2 \vc{e}^T \vc{r} +  \vc{t}_L^T \vc{A}^T \vc{r} - \vc{t}_L^T \vc{A}^T \vc{r}}_q \\ 
    &= \sqbracket{m +2 \vc{e}^T \vc{r}}_q
\end{align*}
As long as the noise is well controlled such that the whole term $m+2 \vc{e}^T \vc{r}$ is within the symmetric range $\Z_q$, the decryption process will output the correct message $m$, after taking modulo 2 to get rid of the noise.
Note the fresh ciphertext is encrypted under $\vc{s}_0$, but after it has gone through $L$ multiplications, it becomes a ciphertext encrypted under $\vc{s}_L$, which explains why we have $\vc{t}_L$ in the second equality in the above derivation.

\subsubsection*{Homomorphic evaluation} 
\label{subsubsec:bvHomMul}
The function $f:\{0,1\}^t \rightarrow \{0,1\}$ to be evaluated is represented as a binary arithmetic circuit. As multiplications incur most of the noise and a ciphertext contains a tag to track the multiplicative depth, it is convenient to construct the circuit with arbitrary fan-in for addition ``+'' and fan-in 2 for multiplication ``$\times$''. Furthermore, its layers are organized in a way that they contain only one type of arithmetic operations. That is, no layer contains both addition and multiplication operations. Finally, the circuit is assumed to have exactly $L$ multiplicative depth.\footnote{This circuit construction equalizes the number of multiplications and the multiplicative depth $L$. But in practice, what matters the most to the noise growth is the degree of the function being evaluated, not the number of multiplications. For example, both functions $f(a, b, c)=a \cdot b + b \cdot c$ and $g(a,b,c)=a\cdot b \cdot c$ contain two multiplications, but $g$ is a degree three polynomial, hence incurs more noise after being evaluated.}

For notational convenience, denote $f_{\vc{c}}(\vc{x}) := \sqbracket{\vc{c} \cdot \vc{x}}_q$ so that the evaluation of the function at $\vc{x}=\vc{s}$ is equivalent to decryption of the ciphertext under the secret key. The evaluation algorithm $\bvv.\eval(\evk, f, \vc{c}_1, \dots, \vc{c}_t)$ is defined separately for addition and multiplication as done next. 
The key thing to note is that the ciphertext after going through each circuit gate should satisfy the invariant property
\begin{align}
\label{eq:shEvalInvariant}
    f_{\vc{c}}(\vc{x}) := \sqbracket{\vc{c} \cdot \vc{x}}_q = \sqbracket{m + 2e}_q
\end{align}
for some noise term $e$ that is not too large to make the whole term exceeds the range $\Z_q$. If it is beyond the range, there will be no guarantee that the exact noise can be eliminated by taking modulo 2. If the invariant property is guaranteed through all circuit gates, the final evaluated output can then be decrypted to the correct message. Therefore, checking the evaluations are homomorphic becomes checking the invariant property is guaranteed throughout the arithmetic circuit. %which could also be obtained by evaluating the plaintexts at the same gate. 

\iffalse 
\begin{align*}
    f_{\vc{v},w}(\vc{x}) = w - \vc{v} \cdot \vc{x} \bmod q 
    = w-\sum_{i=1}^n \vc{v}[i] \cdot \vc{x}[i] \bmod q.
\end{align*}
This linear function embeds the ciphertext $\vc{c}=(\vc{v},w)$ as the coefficients of $\vc{x}$. Evaluate the function at $\vc{s}_l$ and then take modulo 2 is essentially the same as decrypt the ciphertext using the secret key $\sk=\vc{s}_l$. %Given two ciphertexts $c_1^l=((\vc{v}_1,w_1),l)$ and $c_2^l=((\vc{v}_2,w_2),l)$ and the ciphertext $c_3=((\vc{v}_3,w_3),l_3)=c_1 \bullet c_2$ after evaluating $c_1^l$ and $c_2^l$ at a circuit gate, if the scheme SH is homomorphic where ``$\bullet$'' is either ``$+$'' or ``$\times$'', it must satisfy that
%\begin{align}
%    f_{\vc{v}',w'}(\vc{s}_{l'}) \bmod 2= \left[f_{\vc{v}_1,w_1}(\vc{s}_l) \bullet f_{\vc{v}_2,w_2}(\vc{s}_l) \right] \bmod 2.
%\end{align}
To use this function to reconfirm the additive homomorphism, we have
For the purpose of analysing homomorphic evaluations, it is convenient to use a less compact form of ciphertext $\vc{c}=(\vc{v},w)$, where 
\begin{align}
%\label{eq:sh.enc1}
    \vc{v}&=\vc{A}^T \cdot \vc{r} \bmod q, \\
%\label{eq:sh.enc2}
    w&=\vc{b}^T \cdot \vc{r} + m \bmod q.
\end{align}
Putting them together, we also get that $[w \mid \vc{v}]=\sqbracket{\vc{P}^T \cdot \vc{r} + \vc{m}}_q$.
\fi

\paragraph{Homomorphic addition} The addition of arbitrarily many ciphertexts $\vc{c}_1, \dots, \vc{c}_t$ is performed by adding the ciphertexts component wise and leaving the level tag unchanged. That is, %Denote the addition gate by $\oplus$, the additive ciphertext is defined as
\begin{align}%\begin{empheq}[box=\mymath]{align}
\label{eq:bvv.add}
    &\vc{c}_{add}^l = (\vc{c}_{add},l) \leftarrow \bvv.\add(\vc{c}_1^l, \dots, \vc{c}_t^l,q), \text{ where } \nonumber \\
    &\vc{c}_{add}[i]=\sqbracket{\vc{c}_1[i]+\cdots+\vc{c}_t[i]}_q, \text{ for all } i \in [0,n].
\end{align}%\end{empheq}
To check that $\vc{c}_{add}^l$ satisfies the invariant \Cref{eq:shEvalInvariant}, we show that the decryption of the additive ciphertext equals the sum of the messages. That is,
\begin{align*}
    f_{\vc{c}_{add}}(\vc{s}_l) &= \sqbracket{\vc{c}_{add} \cdot \vc{s}_l}_q \\
    &=\sqbracket{(\vc{c}_1 +\cdots+ \vc{c}_t) \cdot \vc{s}_l}_q \\
    &=\sqbracket{\sqbracket{\vc{c}_1 \cdot \vc{s}_l}_q + \cdots + \sqbracket{\vc{c}_t \cdot \vc{s}_l}_q}_q\\
    &=\sqbracket{f_{\vc{c}_1}(\vc{s}_l) +\cdots+ f_{\vc{c}_t}(\vc{s}_l)}_q\\
    &=\sqbracket{(m_1 + \cdots + m_t) + \underbrace{2(e_1 + \cdots + e_t)}_{\text{noise}}}_q.
\end{align*}
So long as the aggregated noise is well controlled such that the entire term is still within $\Z_q$, the decryption step will output the correct summed message after a further reduction by modulo 2. 

\paragraph{Homomorphic multiplication} 
The homomorphic multiplication algorithm involves the important relinearization step which reduces a quadratic to a linear function by approximation. To prove multiplication is also homomorphic, we need to define $\vc{c}_{mult}$ and prove that $f_{\vc{c}_{mult}}(\vc{x})=\sqbracket{f_{\vc{c}_1}(\vc{x}) \cdot f_{\vc{c}_2}(\vc{x})}_q$ just as in the homomorphic addition case. The trouble is that when multiplying two functions of $\vc{x}[i]$, it becomes a quadratic function of $\vc{x}[i]$. More precisely, writing $f_\vc{c}(\vc{x}) = \sqbracket{\sum_{i=0}^n h_i \cdot \vc{x}[i]}_q$ as a function of $\vc{x}[i]$, where the coefficient set $(h_0, \dots, h_n)$ is the ciphertext $\vc{c}$, we have % When multiplying two functions, it becomes 
\begin{align}
\label{eq:homoMultQuad}
    \sqbracket{f_{\vc{c}_1}(\vc{x}) \cdot f_{\vc{c}_2}(\vc{x})}_q 
    = \sqbracket{\left(\sum_{i=0}^n h_i \cdot \vc{x}[i]\right) \left(\sum_{i=0}^n h_j \cdot \vc{x}[j]\right)}_q  
    = \sqbracket{\sum_{i,j=0}^n h_{i,j} \cdot \vc{x}[i] \cdot \vc{x}[j]}_q.
\end{align}
The number of coefficients, which is essentially the ciphertext size, has gone up to approximately $n^2/2$, as compared to $n+1$ coefficients in the previous linear function. 

\paragraph{Relinearization} One solution is to approximate the quadratic function by a linear function, known as \textit{relinearization}.\index{relinearization}
\reversemarginpar
\marginpar{relinearization}
It implies the quadratic terms will be replaced by their linear approximates, with proper protections such as ``encrypting'' $\vc{s}_l[i] \cdot \vc{s}_l[j]$ under a new secret key to make it a fresh linear ciphertext. More precisely, let the new secret key be $\dot{\vc{s}}=(1,\dot{\vc{t}})$ and the corresponding public key be $\dot{\vc{P}}$, then call the previous encryption subroutine to get the ``ciphertext''
\begin{align*}
    \dot{\vc{c}}_{i,j} &\leftarrow \bvv.\enc(\dot{\vc{P}},\vc{s}_l[i] \cdot \vc{s}_l [j],n,q,N,l), \text{ where }\\
    \dot{\vc{c}}_{i,j}&=
    %\sqbracket{\dot{\vc{P}}^T \vc{r} + (\vc{s}_l[i] \cdot \vc{s}_l [j],0,\dots,0)}_q\\
    \sqbracket{\dot{\vc{t}}^T (\vc{A}^T  \vc{r}) + \vc{s}_l[i] \cdot \vc{s}_l [j]+ 2 \vc{e}^T \vc{r} \mid -\vc{A}^T  \vc{r}}_q.
\end{align*}
The ciphertext can also be decrypted by taking dot product with the new secret vector, so we get
\begin{align*}
    f_{\dot{\vc{c}}_{i,j}}(\dot{\vc{s}})
    =\sqbracket{\dot{\vc{c}}_{i,j} \cdot \dot{\vc{s}}}_q 
    =\sqbracket{\xcancel{\dot{\vc{t}}^T  (\vc{A}^T  \vc{r})} + \vc{s}_l[i] \cdot \vc{s}_l[j]+ 2 \vc{e}^T \vc{r} -\xcancel{(\vc{A}^T \vc{r}) \cdot \dot{\vc{t}}}}_q.
\end{align*}
If the noise $2 \vc{e}^T \vc{r}$ has small magnitude, the quadratic term $\sqbracket{\vc{s}_l[i] \cdot \vc{s}_l[j]}_q \approx \sqbracket{\dot{\vc{c}}_{i,j} \cdot \dot{\vc{s}}}_q$ can be well approximated by the dot product. So the evaluation of \Cref{eq:homoMultQuad} at $\vc{x}=\vc{s}_l$ becomes a linear function of the new secret vector $\dot{\vc{s}}$ as shown below
\begin{align*}
    \sqbracket{f_{\vc{c}_1}(\vc{s}_l) \cdot f_{\vc{c}_2}(\vc{s}_l)}_q 
    = \sqbracket{\sum_{i,j=0}^n h_{i,j} \cdot \vc{s}_l[i] \cdot \vc{s}_l[j]}_q
    \approx \sqbracket{\sum_{i,j=0}^n h_{i,j} \cdot (\dot{\vc{c}}_{i,j} \cdot \dot{\vc{s}})}_q
    =\sqbracket{\sum_{k=0}^n \dot{h}_{k} \cdot \dot{\vc{s}}[k]}_q,
\end{align*}
with only $(n+1)$ coefficients, a considerable reduction from its original quadratic form. 

To further guarantee an accurate approximation of the quadratic function, it is necessary to keep each coefficient $h_{i,j}$ as small as possible, so that if $\sqbracket{\vc{s}_l[i] \cdot \vc{s}_l[j]}_q \approx \sqbracket{\dot{\vc{c}}_{i,j} \cdot \dot{\vc{s}}}_q$ is with small error, then the error stays small when multiplying each side by the coefficient $\sqbracket{h_{i,j} \cdot \vc{s}_l[i] \cdot \vc{s}_l[j]}_q \approx \sqbracket{h_{i,j} \cdot \dot{\vc{c}}_{i,j} \cdot \dot{\vc{s}}}_q$. To achieve this, turn the coefficient $h_{i,j}$ to its binary form  
\begin{align*}
    h_{i,j} 
    = \sum_{\tau=0}^{\lfloor \log q \rfloor} 2^{\tau} \cdot h_{i,j,\tau} \bmod q
    = \sqbracket{\sum_{\tau=0}^{\lfloor \log q \rfloor} 2^{\tau} \cdot h_{i,j,\tau}}_q,
\end{align*}
where each $h_{i,j,\tau} \in \{0,1\}$ and $\lfloor \log q \rfloor$ is the max bit length minus 1 for samples in $\Z_q$. The second equality is satisfied by definition of $[\cdot]_q$, in which $[x]_q=x\bmod q$. Substitute this into the ciphertext multiplication, the LHS of the above approximation becomes 
\begin{align}
\label{eq:homoMultQuad3}
    \sqbracket{f_{\vc{c}_1}(\vc{s}_l) \cdot f_{\vc{c}_2}(\vc{s}_l)}_q 
    = \sqbracket{\sum_{\substack{0 \le i,j \le n \\ 0 \le \tau \le \lfloor \log q \rfloor}} h_{i,j,\tau} \cdot (2^{\tau} \cdot \vc{s}_l[i] \cdot \vc{s}_l[j])}_q
\end{align}
and the new quadratic term to be approximated becomes 
\begin{align*}
    \sqbracket{2^{\tau} \cdot \vc{s}_l[i] \cdot \vc{s}_l[j]}_q \approx \sqbracket{\dot{\vc{c}}_{i,j} \cdot \dot{\vc{s}}}_q.
\end{align*}
By design, each element in the evaluation key\index{evaluation key} is in the following format  
\begin{align*}
        \psi_{l+1,i,j,\tau} := \left( \vc{a}_{l+1,i,j,\tau}, b_{l+1,i,j,\tau}=\sqbracket{\vc{a}_{l+1,i,j,\tau} \cdot \vc{s}_{l+1} + 2 \cdot e_{l+1,i,j,\tau} + 2^{\tau} \cdot \vc{s}_{l}[i] \cdot \vc{s}_{l}[j]}_q \right).
\end{align*}
By arranging terms, it implies  
\begin{align*}
    \sqbracket{2^{\tau} \cdot \vc{s}_l[i] \cdot \vc{s}_l[j]}_q \approx \sqbracket{b_{l+1,i,j,\tau} - \vc{a}_{l+1,i,j,\tau} \cdot \vc{s}_{l+1}}_q.
\end{align*}
By now, it should be clear why the evaluation key was set up in that particular form. With this approximation, when evaluating \Cref{eq:homoMultQuad3} at $\vc{x}=\vc{s}_l$, it follows that 
\begin{align}
\label{eq:homoMultLinear}
    \sqbracket{f_{\vc{c}_1}(\vc{s}_l) \cdot f_{\vc{c}_2}(\vc{s}_l)}_q 
    &= \sqbracket{\sum_{\substack{0 \le i,j \le n \\ 0 \le \tau \le \lfloor \log q \rfloor }} h_{i,j,\tau} \cdot (2^{\tau} \cdot \vc{s}_l[i] \cdot \vc{s}_l[j])}_q \nonumber \\
    &\approx \sqbracket{\sum_{\substack{0 \le i,j \le n \\ 0 \le \tau \le \lfloor \log q \rfloor }} h_{i,j,\tau} \cdot (b_{l+1,i,j,\tau} - \vc{a}_{l+1,i,j,\tau} \cdot \vc{s}_{l+1})}_q.
\end{align}
We are now ready to define the multiplicative ciphertext for the inputs $\vc{c}_1^l$ and $\vc{c}_2^l$ as follows 
\begin{align}\label{eq:cmult}
    \vc{c}_{mult}^{l+1} &= (\vc{c}_{mult}, l+1) \leftarrow \bvv.\mult(\evk=\Psi,\vc{c}_1^l, \vc{c}_2^l, q), \text{ where } \nonumber \\
    \vc{c}_{mult}&= \bracket{\sqbracket{\sum_{\substack{0 \le i,j \le n \\ 0 \le \tau \le \lfloor \log q \rfloor }} h_{i,j,\tau} \cdot b_{l+1,i,j,\tau}}_q,
    \sqbracket{\sum_{\substack{0 \le i,j \le n \\ 0 \le \tau \le \lfloor \log q \rfloor }} h_{i,j,\tau} \cdot \vc{a}_{l+1,i,j,\tau}}_q } \in \Z_q^{n+1}, 
    %w_{mult} &= \sum_{0 \le i,j \le n}^{0 \le \tau \le \lfloor \log q \rfloor} h_{i,j,\tau} \cdot b_{l+1,i,j,\tau} \bmod q.
\end{align}

To verify that $\vc{c}_{mult}$ satisfies the invariant property in \Cref{eq:shEvalInvariant}, we work through the following derivation  
\begin{align}
\label{eq:homoMultPf}
    &\sqbracket{\vc{c}_{mult} \cdot \vc{s}_{l+1}}_q \nonumber \\
    &= \sqbracket{\sum_{\substack{0 \le i,j \le n \\ 0 \le \tau \le \lfloor \log q \rfloor }} h_{i,j,\tau} \cdot b_{l+1,i,j,\tau} - \sum_{\substack{0 \le i,j \le n \\ 0 \le \tau \le \lfloor \log q \rfloor }} h_{i,j,\tau} \cdot \vc{a}_{l+1,i,j,\tau} \cdot \vc{s}_{l+1}}_q \nonumber \\
    &= \sqbracket{\sum_{\substack{0 \le i,j \le n \\ 0 \le \tau \le \lfloor \log q \rfloor }} h_{i,j,\tau} \cdot \left(b_{l+1,i,j,\tau} - \vc{a}_{l+1,i,j,\tau} \cdot \vc{s}_{l+1}\right)}_q \nonumber \\
    &= \sqbracket{\sum_{\substack{0 \le i,j \le n \\ 0 \le \tau \le \lfloor \log q \rfloor }} h_{i,j,\tau} \cdot \left(2 e_{l+1,i,j,\tau} + 2^{\tau} \cdot \vc{s}_l[i] \cdot \vc{s}_l[j] \right)}_q \nonumber \\
    &= \sqbracket{f_{\vc{c}_1}(\vc{s}_l) \times f_{\vc{c}_2}(\vc{s}_l) + \sum_{\substack{0 \le i,j \le n \\ 0 \le \tau \le \lfloor \log q \rfloor }} h_{i,j,\tau} \cdot 2 e_{l+1,i,j,\tau}}_q \nonumber \\
    &= \sqbracket{(m_1+2e_1) \times (m_2 +2e_2) + \sum_{\substack{0 \le i,j \le n \\ 0 \le \tau \le \lfloor \log q \rfloor }} h_{i,j,\tau} \cdot 2 e_{l+1,i,j,\tau}}_q \nonumber \\
    &= \sqbracket{m_1 \times m_2 + \underbrace{2 \left( m_1 \cdot e_2 + m_2 \cdot e_1 + 2 e_1 \cdot e_2 + \sum_{\substack{0 \le i,j \le n \\ 0 \le \tau \le \lfloor \log q \rfloor }} h_{i,j,\tau} \cdot e_{l+1,i,j,\tau} \right)}_{\text{noise}}}_q . 
\end{align}
Therefore, to guarantee the decryption can correctly produce $m_1 \times m_2$, it is necessary to keep the noise small enough so that the whole term in \Cref{eq:homoMultPf} is within $\Z_q$.

\subsection{BV : Leveled FHE by dimension-modulus reduction}\label{subsec:modulus reduction}

The $\bvv$ scheme presented above (with relinearization) produces a constant ciphertext $\vc{c}$ in the domain $\Z_q^{(n+1)}$, with the maximum bit length $(n+1)\log q$, which is considered quite large for large values of $n$ and $q$. To reduce it, consequently reduce the decryption complexity to make the scheme more bootstrappable (without the need for squashing), \citet{brakerski2014efficient} performed a dimension-modulus reduction at the completion of homomorphic evaluations. This reduction step was later used in \citet{brakerski2014leveled} and \citet{brakerski2012fully} to achieve fully leveled HE without using bootstrapping. Below, we discuss dimension-modulus reduction and how it helps to reduce ciphertext bit length.

\subsubsection{Modulus reduction to reduce ciphertext size}

The reduction step consists of two parts, the modulus reduction and the dimension reduction. The reduction in modulus is achieved by scaling down ciphertexts by the factor $p/q$ where $p<q$. The next definition defines the scale of an integer vector, which in our context is a ciphertext. 

\begin{definition}
Let $\vc{x}$ be an integer vector. For integers $m<p<q$, an integer vector $\vc{x}' \leftarrow \text{Scale}(\vc{x},q,p,r)$ is the \textbf{scale}\index{scaled ciphertext} of $\vc{x}$ 
\reversemarginpar
\marginnote{\textit{Scale}}
if it is the vector closest to $(p/q) \cdot \vc{x}$ that satisfies $\vc{x}'=\vc{x} \bmod r$.
\end{definition}

\begin{example}
Let $p=5,q=11,r=2$, then the scale of the vector $\vc{c}=(5,6)$ is $\vc{c}'=(3,2)$, because it is the closest integer vector to $(5/11) \cdot (5,6)$ and $\vc{c}'=\vc{c} \bmod 2$. 
\end{example}

The correctness of modulus reduction is captured in the following lemma, which is a special case of the first part of Lemma 5 of \citet{brakerski2014leveled}. The parameter $r=2$ implies $q=p=1 \bmod 2$ are odd integers. % The dimension $d=1$ implies $R=\Z$, so the expansion factor $\gamma_R \le \sqrt{d} = 1$.
% This lemma states that the two ciphertexts, although belonging to two different ciphertext spaces, can be decrypted to the same message by using the same secret key, provided the secret key satisfies a certain condition. % The lemma suggests that modulus reduction is a safe step to apply before decryption.
Below, we use $||\vc{x}||$ to denote the $l_1$-norm of the vector $\vc{x}$. 

\begin{lemma}
\label{lm:modSwitch}
\reversemarginpar
\marginnote{\textit{Modulus reduction}}
Let $q$ and $p$ be two odd moduli such that $p<q$. Let $\vc{c}$ be an integer vector and $\vc{c}' \leftarrow \text{Scale}(\vc{c},q,p,2)$ be the scale of $\vc{c}$. Then for any vector $\vc{s}$ with $||\sqbracket{\vc{c} \cdot \vc{s}}_q|| < q/2 - (q/p) \cdot ||\vc{s}||$, it satisfies 
\begin{align*}
    \sqbracket{\sqbracket{\vc{c}' \cdot \vc{s}}_p}_2  = \sqbracket{\sqbracket{\vc{c} \cdot \vc{s}}_q}_2. %\text{ and }\\
    %|| \sqbracket{\vc{c}' \cdot \vc{s}}_p || &< \frac{p}{q} \cdot  || \sqbracket{\vc{c} \cdot \vc{s}}_q || + l_1(\vc{s}),
\end{align*}
%where $l_1(\vc{s})$ is the $l_1$-norm of the vector $\vc{s}$.
\end{lemma}   
 
\begin{proof}
By definition of modulo operation, there exists a unique integer $k$ such that $\sqbracket{\vc{c} \cdot \vc{s}}_q = \vc{c} \cdot \vc{s} - kq \in [-q/2, q/2)$. Using the integer $k$, we can define a noise term
\begin{align*}
    e_p = \vc{c}' \cdot \vc{s} - kp \in \Z.
\end{align*}
By taking modulo $p$, the noise satisfies $e_p = \sqbracket{\vc{c}' \cdot \vc{s}}_p \bmod p$. If we can show $e_p = \sqbracket{\vc{c}' \cdot \vc{s}}_p$ without taking modulo $p$, it then follows that 
\begin{align*}
    \sqbracket{\vc{c}' \cdot \vc{s}}_p = e_p = \vc{c}' \cdot \vc{s} - kp = \vc{c} \cdot \vc{s} - kq = \sqbracket{\vc{c} \cdot \vc{s}}_q \bmod 2.
\end{align*}
To show $e_p = \sqbracket{\vc{c}' \cdot \vc{s}}_p$, it is sufficient to prove its norm satisfies $||e_p|| < p/2$. Re-write the noise as 
\begin{align*}
    e_p 
    = \vc{c}' \cdot \vc{s} + \frac{p}{q}\cdot (-kq) 
    = \vc{c}' \cdot \vc{s} + \frac{p}{q} \cdot (\sqbracket{\vc{c} \cdot \vc{s}}_q - \vc{c} \cdot \vc{s}) 
    =\frac{p}{q} \cdot \sqbracket{\vc{c} \cdot \vc{s}}_q + (\vc{c}'-\frac{p}{q}\vc{c})\cdot \vc{s}.
\end{align*}
We can show its norm satisfies 
\begin{align*}
    ||e_p|| &= || \frac{p}{q} \cdot \sqbracket{\vc{c} \cdot \vc{s}}_q + (\vc{c}'-\frac{p}{q} \cdot \vc{c})\cdot \vc{s}||\\
    &\le \frac{p}{q} \cdot ||  \sqbracket{\vc{c} \cdot \vc{s}}_q|| + ||(\vc{c}'-\frac{p}{q} \cdot \vc{c})\cdot \vc{s}||\\
    &\le \frac{p}{q} \cdot ||  \sqbracket{\vc{c} \cdot \vc{s}}_q|| + \sum_{i=1}^{n} ||(\vc{c}'[i]-\frac{p}{q} \cdot \vc{c}[i])|| \cdot ||\vc{s}[i]||\\
    &\le \frac{p}{q} \cdot ||  \sqbracket{\vc{c} \cdot \vc{s}}_q|| + 1 \cdot \sum_{i=1}^{n} ||\vc{s}[i]||\\
    &\le \frac{p}{q} ||\sqbracket{\vc{c} \cdot \vc{s}}_q|| + ||\vc{s}|| \\
    &< p/2.
\end{align*}
The last inequality follows from the assumption of the vector $\vc{s}$ as stated in the Lemma's premises. The third last inequality follows because $\vc{c}'$ is close to $(p/q)\cdot \vc{c}$ and they are congruent modulo 2. In this case, each element differs by at most 1. 
\end{proof}

\subsubsection{The BV scheme}
\label{subsubsec:bv}
The improved version of $\bvv$, named BTS in \citet{brakerski2014efficient}, employs $\bvv$ as its building block and reduces the ciphertext dimension and modulus by a reduction step. We rename BTS to $\bv$\index{HE scheme! BV} in this tutorial to make it more recognizable when comparing with subsequent works. 
The main benefit of adding the reduction step once a ciphertext has gone through the complete circuit is that $\bv$ becomes bootstrappable without the \textit{squashing} step % , e.g., the sparse subset sum problem 
used by \citet{gentry2009fully}. 
In addition to the parameters in $\para=(n,q,N,\chi)$ in $\bvv$, this improved scheme takes on three additional parameters $(k,p,\hat{\chi})$ to cope with the dimension-modulus reduction step. The parameters $k$ and $p$ are a smaller dimension and modulus, respectively. The new noise distribution $\hat{\chi}$ is over the smaller domain $\Z_p$ to produce smaller integer noise. The sub-routines of $\bv$ are listed as follows. 

\paragraph{Key generation}

The key generation first runs the sub-routine 
\begin{align*}
    (\vc{P}, \vc{s}_L, \Psi) \leftarrow \bvv.\keygen(\para).
\end{align*}
Its public key is set to $\vc{P}$. 
% so 
% \begin{align*}
%    \bv.\pubkeygen(\vc{t}_0,\para)=\reg.\pubkeygen^*(\vc{t}_0,\para).
% \end{align*}
The secret key is generated by
\begin{align*}
    \hat{\vc{s}} \leftarrow \bv.\seckeygen(k,p) % =\reg.\seckeygen(k,p).
\end{align*}
from a smaller domain $\Z_p^k$ with a lower dimension. This new secret key is to decrypt a ciphertext of reduced dimension and modulus. The $\bvv$ evaluation key $\Psi$ becomes part of the new evaluation key $(\Psi, \hat{\Psi})$ for $\bv$, because it is needed for homomorphic multiplication which runs $\bvv.\mult()$ as a sub-routine. The extra piece $\hat{\Psi}=\{\hat{\psi}_{i,\tau}\}_{i,\tau}$ ``encrypts'' the secret vector $2^{\tau} \cdot \vc{s}_L$ in a similar fashion as $\Psi$ ``encrypts'' $2^{\tau} \cdot \vc{s}_{l-1}[i] \cdot \vc{s}_{l-1}[i]$, % for approximating a quadratic by a linear function, 
except here the ``encryption'' of $2^{\tau} \cdot \vc{s}_L$ is for approximating a ciphertext by another ciphertext with smaller dimension and modulus. More precisely, % this ``encryption'' has the following form
\begin{align*}
    \hat{\psi}_{i,\tau} &= (\hat{\vc{a}}_{i,\tau},\hat{b}_{i,\tau}), \text{ where} \\
    \hat{\vc{a}}_{i,\tau} &\leftarrow \Z_p^k \\
    \hat{e}_{i,\tau} &\leftarrow \hat{\chi} \\
    \hat{b}_{i,\tau} &= \hat{\vc{a}}_{i,\tau} \cdot \hat{\vc{s}}+\hat{e}_{i,\tau}+\round{\frac{p}{q} \cdot (2^{\tau} \cdot \vc{s}_L[i])} \bmod p.
\end{align*}
The important observation is that both $\hat{\vc{a}}$ and $\hat{\vc{s}}$ are of dimension $k$ and modulus $p$, which are different from their counterparts in $\Z_q^n$ produced by $\bvv.\keygen(\para)$. These setups will lead to smaller ciphertexts as we will see later. Note the noise in $\hat{b}_{i,\tau}$ is not multiplied by 2. This does not pose an issue when eliminating the noise by modulo 2, because the whole noise term will be multiplied by 2 at a later stage. To summarise, the output of $\bv$'s key generation step is  
\begin{align*}
    (\pk=\vc{P}, \sk=\hat{\vc{s}}, \evk=(\Psi,\hat{\Psi})) \leftarrow \bv.\keygen(\para,k,p,\hat{\chi}).
\end{align*}

\paragraph{Encryption and decryption}
The encryption and decryption steps are identical to that of $\bvv$, but with a different decryption parameter.  
\begin{align*}
    \vc{c}^l=\left(\vc{c}=\sqbracket{\vc{P}^T \vc{r}+ \vc{m}}_q, l \right) &\leftarrow \bv.\enc(\vc{P},m,n,q,N)\\ %=\bvv.\enc(\vc{P},m,n,q,N) \\
    m=\sqbracket{\sqbracket{\vc{\hat{c}} \cdot (1, \vc{\hat{s}})}_p}_2
    &\leftarrow \bv.\dec(\hat{\vc{s}}, \hat{\vc{c}},p) % =\bvv.\dec(\hat{\vc{s}}, \hat{\vc{c}},p).
\end{align*}

\paragraph{Homomorphic evaluation}

The evaluation algorithm runs the following sub-routines
\begin{align*}
    \vc{c}^l &\leftarrow \bvv.\add(\evk=\Psi,\vc{c}_1^l,\dots,\vc{c}_t^l,q)\\
    \vc{c}^{l+1} &\leftarrow \bvv.\mult(\evk=\Psi,\vc{c}_1^l,\vc{c}_2^l,q).
\end{align*} 
Once the complete circuit has been evaluated, it is followed by a dimension-modulus reduction before decryption starts. 

\paragraph{Dimension-modulus reduction}
\label{paragraph:dimModReduction}
By \Cref{lm:modSwitch}, modulus reduction is a valid step that guarantees correct decryption. In \citet{brakerski2014efficient}, the modulus reduction is made possible by multiplying the decryption equivalent function $f_\vc{c}(\vc{x})=\vc{c} \cdot \vc{x}$ by the factor $p/q$ to scale its coefficients down to within the new domain to get a new decryption equivalent function\index{modulus reduction}
\reversemarginpar
\marginpar{modulus reduction}
\begin{align*}
    \phi(\vc{x}) = \sqbracket{\frac{p}{q} \cdot \bracket{\frac{q+1}{2} \cdot \bracket{\vc{c} \cdot \vc{x}}}}_p
    = \sqbracket{\sum_{i=0}^n h_i \cdot \bracket{\frac{p}{q} \cdot \vc{x}[i]}}_p.
\end{align*}
%The key factor is $p/q$ that scales the coefficients of $\vc{x}[i]$ to within the smaller domain modulo $p$. 
The fractional term $(q+1)/2$ is the inverse of 2 in modulo $q$. It is useful for getting rid of the coefficient in front of the encrypted message $m$. 

The reduction of ciphertext dimension is achieved by approximating the longer vector $\vc{x}$ by a shorter one. It follows a similar approximation strategy for the quadratic terms in $\bvv$. The first thing is to turn $h_i$ to its binary form to keep a smaller approximation error. The function then becomes 
\begin{align*}
    \phi(\vc{x}) = \sqbracket{\sum_{\substack{0\le i \le n \\ 0 \le \tau \le \lfloor\log q \rfloor}} h_{i,\tau} \cdot \bracket{\frac{p}{q} \cdot 2^{\tau} \cdot \vc{x}[i]}}_p.
\end{align*}
The term inside the bracket now looks like a part of $\hat{b}_{i,\tau}$ in the evaluation key $\hat{\psi}_{i,\tau}$, so the function can be approximated using the second half of the evaluation key as  
\begin{align*}
    \phi(\vc{x}) \approx \sqbracket{\sum_{\substack{0\le i \le n \\ 0 \le \tau \le \lfloor\log q \rfloor}} h_{i,\tau} \cdot \bracket{\hat{b}_{i,\tau} - \hat{\vc{a}}_{i,\tau} \cdot \hat{\vc{s}}}}_p.
\end{align*}
This gives rise to a revised ciphertext\index{dimension reduction} 
\reversemarginpar
\marginpar{dimension reduction}
\begin{align*}
    \hat{\vc{c}} = \bracket{\sqbracket{\sum_{\substack{0\le i \le n \\ 0 \le \tau \le \lfloor\log q \rfloor}} 2 \cdot h_{i,\tau} \cdot \hat{b}_{i,\tau}}_q, \sqbracket{\sum_{\substack{0\le i \le n \\ 0 \le \tau \le \lfloor\log q \rfloor}} 2 \cdot h_{i,\tau} \cdot \hat{\vc{a}}_{i,\tau}}_q}  \in  \Z_p^{k+1}
\end{align*}
in the domain $\Z_p^{k+1}$ with a smaller set $\Z_p$ and a lower dimension $k+1$. The new ciphertext bit length is therefore reduced to $(k+1)\log p$ from $(n+1) \log q$. In general, the use of ciphertexts of smaller dimension and modulus introduces an approximation error that is in addition to those incurred during homomorphic evaluations. This additional error, however, does not become an issue for decryption, so long as the ciphertext space is large enough to incorporate both types of errors. 

As it was proved in the analysis of $\bvv$, this dimension-modulus reduction also satisfies the invariant property stated by \Cref{eq:shEvalInvariant}. The detailed proof can be found at the end of Section 4.2 of \citet{brakerski2014efficient}. 
% We also skip the homomorphic properties and the security of $\bvv$ and $\bv$, which can be found in  
The homomorphic properties can be proved by showing that the evaluated ciphertexts after running $\bvv$'s evaluation process and $\bv$'s dimension-modulus reduction process are still within $\Z_q$, provided the parameters are set at the appropriate values. 
Details can be found in Section 4.3 and Section 4.4 of \citet{brakerski2014efficient}.

\subsubsection{$\bv$ is bootstrappable}

To see $\bv$ is bootstrappable\index{bootstrapping} and hence can be made fully HE within a pre-determined level (i.e., leveled FHE), we introduce the function class $\text{Arith}[L,T]$ that consists of arithmetic circuits\index{arithmetic circuit} over the message space $\{0,1\}$ with only addition and multiplication gates such that each circuit has $2L+1$ layers, where the odd layers contain only the add gates with fan-in $T$ and the even layers contain only the multiply gates with fan-in 2. The following theorem states that $\bv$ and $\bvv$ are capable of evaluating certain size arithmetic circuits. 

\begin{theorem}[Theorem 4.3 \citep{brakerski2014efficient}]
Let $n=n(\lambda) \ge 5$ be a polynomial of the security parameter, $q \ge 2^{n^{\epsilon}} \ge 3$ be an odd modulus for $\epsilon \in (0,1)$, $\chi$ be an $n$-bounded distribution and $N=(n+1) \log q + 2 \lambda$. Furthermore, let $k=\lambda$, $p=16nk\log(2q)$ be odd and $\hat{\chi}$ be a $k$-bounded distribution. Then $\bvv$ and $\bv$ are both $\text{Arith}[L=\Omega(\epsilon \log n), T=\sqrt{q}]$-homomorphic.   
\end{theorem}

As it was further proved by Lemma 4.6 of \citet{brakerski2014efficient} that $\bv$'s decryption is a circuit with 2 fan-in and $O(\log k + \log \log p)$ depth, the decryption circuit is in $\text{Arith}[O(\log k), 1]$, even with an augmented addition or multiplication gate. Hence, as long as the parameter $n$ is made sufficiently large, the decryption circuit is included in the class $\text{Arith}[L=\Omega(\epsilon \log n), T=\sqrt{q}]$, which implies the encryption scheme $\bv$ is bootstrappable and can be made leveled FHE. The generation of the relinearization key requires the circuit maximum level to be pre-specified. It constraints the scheme from getting to \textit{(non-leveld) FHE}. This situation can be avoided by assuming weak circular security, which then simplifies the size of the relinearization key to just one pair of keys and hence gets rid of the prerequisite for $L$ being pre-determined.

\iffalse 
Some remarks: 
\begin{itemize}
    \item in BTS, the noise in the relinearization key does not multiply by 2 because we need to do modulus reduction, 2 is multiplied later once the reduction is done. 
    
    \item addition of two ciphertexts with noise B incurs noise to 2B, multiplication incurs to $B^2$.
    
    \item Fresh ciphertext has noise $B$, and after $L$ multiplications, the noise increases to $B^{2^L}$. This requires a very large modulus $q \approx B^{2^L}$ in order to guarantee correct decryption. 
    
    \item the deimension and modulus reductions in BTS ensures that BTS has shorter ciphertext length and lower decryption complexity which is essential in proving it is bootstrappable and hence can be made fully HE. 
\end{itemize}
\fi

\subsection{Additional tools for computational efficiency}

% \kl{What benefit does it have?}

\subsubsection{Noise management by modulus switching}
\label{subsubsec:modulusSwitching}
A subsequent work inspired by $\bv$ proposed a more efficient encryption scheme, namely BGV \citep{brakerski2014leveled} (again by the authors' surname initials), which can achieve leveled FHE without going through the computationally expensive bootstrapping step. This scheme applies a modulus switching (similar to modulus reduction) step after each homomorphic addition and multiplication in order to reduce the accumulated noise magnitude. The advantage of this noise reduction is not only on its absolute magnitude, but the deceleration of the gap reduction between the noise level and noise ceiling, so that a scheme combined with the modulus switching can handle more ciphertext multiplications before decryption fails. 

Take the following case as an example. Let the ciphertext space modulus be $q=x^{16}$ for some $x$, which is also the noise ceiling.\index{noise ceiling} If two ciphertexts have a noise magnitude $x$, their multiplication produces a ciphertext with noise magnitude roughly $x^2$. After 4 multiplications, the ciphertext has noise magnitude $x^{16}$, which has reached the ceiling. So the scheme can handle circuits with multiplicative depth at most 4. If at the end of each ciphertext multiplication, the modulus is switched to a smaller modulus $p=q/x$, although the noise ceiling is also reduced in the mean time, the scheme can now handle 16 multiplications. More precisely, after the first multiplication, the ciphertext has noise magnitude $x^2$, which is then scaled down to $x$ by modulus switching to the ciphertext space $\Z_p$ with $p=q/x$. In the mean time, the noise ceiling is scaled down to $p=q/x=x^{15}$. Repeat modulus switching 16 times, the noise level remains at $x$ and noise ceiling meets the noise level at $x$, so the scheme reaches its maximum number of multiplications. Therefore, without relying on bootstrapping, the combined scheme with modulus switchings can handle a decent number of multiplications.

The noise reduction property of modulus switching\index{modulus switching} is captured in Lemma~\ref{lm:modSwitch} presented earlier for modulus reduction.\index{modulus reduction}

\subsubsection{Vector decomposition} Vector decomposition consists of two functions. The first function, $\bitdecomp()$, decomposes a vector of length $n$ to a vector of length $nl$, where $l$ is the maximum bit length in the domain $\Z_q$. The benefit of decomposing an integer vector is to minimize the error when switching the ciphertext from one secret key to another. The second function, PowersOfTwo(), is defined in relation to the first one, so that the dot product of these two functions preserves the dot product of the original two vectors. % The two functions are formally defined as follows.

%\begin{itemize}
\paragraph{BitDecomp$_q(\vc{x})$} Let $\vc{x} = (x_1, \dots, x_n) \in \Z^n$ and $l = \lceil \log q \rceil$. % be the bit length of $q$. 
Each $x_i \bmod q$ can be written in binary representation (from least significant bit to most significant bit) as follows  
\begin{align*}
    x_1 &= (x_{1,0}, \dots, x_{1,l-1} )\\
    \vdots \\
    x_n &= (x_{n,0}, \dots, x_{n,l-1} ).
\end{align*}
Let $\vc{w}_i=(x_{1,i},\dots,x_{n,i})$ be the set of i-th binary bits. The bit decomposition function is defined as
\begin{align*}
    \bitdecomp_q(\vc{x}) \rightarrow (\vc{w}_0, \dots, \vc{w}_{l-1}). %\in \{0,1\}^{n l} 
\end{align*}
The $\vc{w}_i$'s so-constructed thus satisfy $\vc{x} = \sum_{i=0}^{l-1} 2^i \cdot \vc{w}_i \bmod q$.
%    \text{ s.t. } \vc{x} = \sum_{i=0}^{l-1} 2^i \cdot \vc{w}_i \bmod q.
    
For example, consider the case when $\vc{x}=(1,3) \in \Z^2$, $q=4$, and $l=\lceil \log 4 \rceil =2$. The decomposed vectors are $\vc{w}_0=(1,1)$ and $\vc{w}_1=(0,1)$, and they satisfy \begin{align*}
    \sum_{i=0}^1 2^i \cdot \vc{w}_i = 1 \cdot (1,1) + 2 \cdot (0,1) = (1,3)=\vc{x} \bmod 4.
\end{align*}
So $\bitdecomp_q(\vc{x}) = (1,1,0,1) \in \{0,1\}^4$.

\paragraph{PowersOfTwo$_q(\vc{y})$} Let $\vc{y} \in \Z^n$, the powers of two function produces a vector by multiplying $\vc{y}$ with $2^i$ in modulo $q$ for each $i \in [0, l-1]$. That is, 
\begin{align*}
    \text{PowersOfTwo}_q(\vc{y}) \rightarrow [(\vc{y}, \vc{y} \cdot 2, \dots, \vc{y} \cdot 2^{l-1})]_q \in \Z_q^{n l}.
\end{align*}
If $\vc{y}=(3,2)$, then $\powersoftwo_4(\vc{y})=(3,2,2,0)$.
%\end{itemize}

It is not hard to see the next equality. That is, the dot product of the two vectors is congruent to the dot product of the two functions modulo $q$, which then leads to the dot product of the two functions in the range $\Z_q$.  
\iffalse 
\begin{align*}
    &\bitdecomp_q(\vc{x}) \cdot \text{PowersOfTwo}_q(\vc{y}) 
    %&=  (\vc{w}_0, \dots, \vc{w}_l) \cdot [(\vc{y}, \vc{y} \cdot 2, \dots, \vc{y} \cdot 2^l)]_q 
    %=\sum_{i=0}^{l-1} \vc{w}_i \cdot (\vc{y} \cdot 2^i) \\
    =&  \sum_{i=0}^{l-1} (\vc{w}_i \cdot 2^i) \cdot \vc{y}_i 
    =  \sum_{i=0}^{l-1} \vc{x}_i \cdot \vc{y}_i \bmod q 
    =&\vc{x} \cdot \vc{y} \bmod q.
\end{align*}
\fi 

\begin{align*}
    \vc{x} \cdot \vc{y} 
    %&= \left(\sum_{i=0}^{l-1} 2^i \cdot \vc{w}_i \right) \cdot \vc{y}  
    %&=  (\vc{w}_0, \dots, \vc{w}_l) \cdot \vc{y} \cdot (2^0, \dots, 2^{l-1}) \bmod q\\
    = \bitdecomp_q(\vc{x}) \cdot \text{PowersOfTwo}_q(\vc{y}) \bmod q = \sqbracket{\bitdecomp_q(\vc{x}) \cdot \text{PowersOfTwo}_q(\vc{y})}_q
\end{align*}
% This observation was also shown as Lemma 2 in \citep{brakerski2014leveled}.

\subsubsection{Key switching} The key switching process is to transform a ciphertext encrypted under a secret key $\vc{s}_1=(1,\vc{t}_1) \in \Z^{n_1+1}_q$ to a different ciphertext encrypted under a secret key $\vc{s}_2=(1,\vc{t}_2) \in \Z_q^{n_2+1}$, while preserving the secret message. Note that $n_1 \neq n_2$ in general. There are two functions in this process. The first function hides $\vc{s}_1$ under $\vc{s}_2$. The second function uses the auxiliary information from the first function to transform a ciphertext to under the secret key $\vc{s}_2$. %Let $\vc{t}_1 \in \Z^{n_1}$ and $\vc{t}_2 \in \Z^{n_2}$ be the ``source'' and ``target'' secret keys. % , where $n_1$ and $n_2$ are their dimensions respectively. 
In the following description, $N_1 = (n_1+1) \cdot \lceil \log q \rceil$.
% As above, the output of PowersOfTwo$_q(\vc{s}_1)$ is a vector of length $N_1 = (n_1+1) \cdot \lceil \log q \rceil$, where $n_1=|\vc{t}_1|$. % denoted by $\hat{n}_1$ for simplicity. 
% We are now ready to define the two functions of key switching. 

\paragraph{SwitchKeyGen$_{q,\chi}(\vc{s}_1,\vc{s}_2)$} This function % is constructed in a similar way as
encrypts the value of PowersOfTwo$_q(\vc{s}_1)$ under the secret key $\vc{s}_2 = (1,\vc{t}_2)$. The steps are as follows. Sample a matrix $\vc{A}_{\vc{s}_1:\vc{s}_2} \leftarrow \Z_q^{N_1 \times n_2}$ and a noise vector $\vc{e}_{\vc{s}_1:\vc{s}_2} \leftarrow \chi^{N_1}$. Then compute 
\begin{align*}
    \vc{b}_{\vc{s}_1:\vc{s}_2} = [\vc{A}_{\vc{s}_1:\vc{s}_2} \vc{t}_2 + \vc{e}_{\vc{s}_1:\vc{s}_2} + \text{PowersOfTwo}_q(\vc{s}_1)]_q \in \Z_q^{N_1}
\end{align*}
and publish the concatenated matrix 
\begin{align*}
    \vc{P}_{\vc{s}_1:\vc{s}_2} = [\vc{b}_{\vc{s}_1:\vc{s}_2} \mid -\vc{A}_{\vc{s}_1:\vc{s}_2}] \in \Z_q^{N_1 \times (n_2+1)}.
\end{align*}
Despite the fact that the encrypted message is PowersOfTwo$_q(\vc{s}_1)$, the output matrix $\vc{P}_{\vc{s}_1:\vc{s}_2}$ looks exactly like a public key in the Regev's scheme. This auxiliary information is precisely what enables the ciphertext transformation between different secret keys. 

\paragraph{SwitchKey$_q(\vc{P}_{\vc{s}_1:\vc{s}_2},\vc{c}_{s_1})$} To transform a ciphertext $\vc{c}_{s_1} \in \Z_q^{n_1+1}$ to a new one encrypted under the secret key $\vc{s}_2$, compute\index{key switching}
\begin{align*}
    \vc{c}_{s_2} = [\bitdecomp_q(\vc{c}_{s_1})^T \vc{P}_{\vc{s}_1:\vc{s}_2}]_q \in \Z_q^{n_2+1}.
\end{align*}
To verify that this transformation preserves the secret message (as proved by Lemma 3 of \citet{brakerski2014leveled}), we see that for $\vc{s}_i = (1, \vc{t}_i)$
\begin{align*}
    [\vc{c}_{s_2} \cdot \vc{s}_2]_q &=  [[\bitdecomp_q(\vc{c}_{s_1})^T \vc{P}_{\vc{s}_1:\vc{s}_2}]_q \cdot \vc{s}_2]_q\\
    &= [\bitdecomp_q(\vc{c}_{s_1})^T (\vc{P}_{\vc{s}_1:\vc{s}_2} \vc{s}_2)]_q \\
    &= [\bitdecomp_q(\vc{c}_{s_1})^T (\vc{b}_{\vc{s}_1:\vc{s}_2} - \vc{A}_{\vc{s}_1:\vc{s}_2} \cdot \vc{t}_2)]_q \\
    &= [\bitdecomp_q(\vc{c}_{s_1})^T (\vc{e}_{\vc{s}_1:\vc{s}_2} + \text{PowersOfTwo}_q(\vc{s}_1))  ]_q\\
    &= [\bitdecomp_q(\vc{c}_{s_1}) \cdot \vc{e}_{\vc{s}_1:\vc{s}_2} + \bitdecomp_q(\vc{c}_{s_1}) \cdot \text{PowersOfTwo}_q(\vc{s}_1)]_q \\
    &= [\vc{c}_{s_1} \cdot \vc{s}_1 + \underbrace{\bitdecomp_q(\vc{c}_{s_1}) \cdot \vc{e}_{\vc{s}_1:\vc{s}_2}}_{\text{error}}]_q.
\end{align*}
\iffalse 
It shows that 
the dot product of the new ciphertext with the new secret key is almost equal to the dot product of the original ciphertext with the source key
\begin{align}
\label{eq:cipherTrans}
    \vc{c}_{s_2} \cdot \vc{s}_2 = \vc{c}_{s_1} \cdot \vc{s}_1 + \underbrace{\vc{e}_{\vc{t}_1:\vc{t}_2} \cdot \bitdecomp_q(\vc{c}_{s_1})}_{\text{error}} \bmod q,
\end{align}
\fi 
The error is of small magnitude because BitDecomp$_q(\vc{c}_{s_1})$ is a binary vector. This also reveals the motivation of defining the vector decomposition procedure. 

The security of the key switching procedure needs both functions to be secure. The second function SwitchKey$_q(\vc{P}_{\vc{s}_1:\vc{s}_2},\vc{c}_{s_1})$ is obviously semantically secure, because its output is a transformation of the original ciphertext, which is encrypted by a semantically secure procedure. If it is not semantically secure, it becomes a PPT algorithm to solve the LWE problem. The first function's output is the auxiliary information $\vc{P}_{\vc{s}_1:\vc{s}_2}$, so its security means this output must be computationally indistinguishable from a uniform matrix sampled from the same domain $\Z_q^{N_1 \times (n_2+1)}$. This again relies on the result that DLWE is hard to solve. See Lemma 3.6 of \citet{brakerski2012fully} or Lemma 4 of \citep{brakerski2014leveled} for a more formal statement of SwitchKeyGen$_{q,\chi}(\vc{s}_1,\vc{s}_2)$'s security.

\subsection{BGV : Leveled FHE by modulus and key switching}

As mentioned above, the $\bgv$ scheme can be made leveled FHE without using the computationally expensive bootstrapping step. This is achieved by iteratively refreshing an evaluated (especially multiplicative) ciphertext by modulus switching. 
The $\bgv$ scheme also uses Regev's encryption scheme as its building block. %, but with the parameter $t=q$ and even integer noise in $\reg$'s encryption process. 
The security assumption, however, is based the hardness of either LWE or RLWE. The two problems are summarized as \textbf{General LWE (GLWE)}, with a binary indicator $b=0$ indicates LWE and $b=1$ indicates RLWE. For this reason, the encryption scheme needs a slightly different parameter set $\para=(n,d,q,N,\chi)$ to incorporate the RLWE problem, where $d$ corresponds to the quotient polynomial degree in RLWE. 

% Recall the Ring-LWE (RLWE) problem is a generalization of LWE to the ring of polynomials $R=\Z[x]/f(x)$, where the polynomials are of integer-coefficients (an alternative interpretation of $R$ is a ring of algebraic integers), where $f(x)=x^d+1$ is the quotient polynomial whose degree $d=d(\lambda)$ is a power of 2. For a polynomial $r \in R=\Z[x]/(x^d+1)$, let $||r||$ denote the norm of its coefficient vector. Define $\gamma_R=\max\{\frac{||a \cdot b||}{||a|| \cdot ||b||} \mid a, b \in R\}$ as the expansion factor of the polynomial ring. By Cauchy-Schwarz inequality, the expansion factor satisfies $\gamma_R \le \sqrt{d}$. Apart from all these, RLWE and LWE share the rest of the parameters. 

Below we present each step of the $\bgv$ scheme, after a brief note on tensor products.

For $n$-dimensional vectors $\vc{x}$ and $\vc{y}$, 
\reversemarginpar
\marginnote{\textit{Tensor product}}
their tensor product $\vc{x} \otimes \vc{y}$\index{tensor product} is a $n \times n$ matrix or an $n^2$-dimensional vector, where each element has the form $\vc{x}[i] \cdot \vc{y}[j]$. For example, for the vectors $\vc{x}=(x_1,x_2)$ and $\vc{y}=(y_1,y_2)$, their tensor product is the 2 by 2 matrix 
\begin{equation*}
\vc{x} \otimes \vc{y} =
\begin{pmatrix}
x_1 y_1 & x_1 y_2 \\
x_2 y_1 & x_2 y_2
\end{pmatrix}.
\end{equation*}
The notion of tensor product will appear in ciphertext multiplications, which result in functions of the tensor product elements $\vc{x}[i] \cdot \vc{y}[j]$. 
\iffalse
Here is the reason why. %=(\vc{b}^T \cdot \vc{r}+m, -\vc{A}^T \cdot \vc{r})
Jumping ahead, let $\vc{c}$ be a ciphertext and $\vc{s}=(1, \vc{t})$ be a secret key, where $\vc{t} \leftarrow \Z_q^n$ and $\vc{s}[0]=1$. Both $\vc{c}$ and $\vc{s}$ are length $(n+1)$ vectors. To decrypt the ciphertext using the secret key, it is equivalent to compute the inner product  (or dot product) $\innerprod{\vc{c},\vc{s}}=\sum_{i=0}^n h_i \cdot \vc{s}[i]$, which is expressed as a function of $\vc{s}[i]$ whose coefficient set is the ciphertext $\vc{c}$. When decrypting the product of two ciphertexts $\vc{c}_1$ and $\vc{c}_2$, we compute the product of the two inner products $\innerprod{\vc{c}_1, \vc{s}} \cdot \innerprod{\vc{c}_2, \vc{s}}=\sum_{i,j=0}^n h_{i,j} \cdot \vc{s}[i] \cdot \vc{s}[j]$, which becomes a quadratic function of $\vc{s}[i] \cdot \vc{s}[j]$. Hence, ciphertext multiplications can be expressed as functions of tensor product elements $\vc{x}[i] \cdot \vc{y}[i]$.
\fi
A property of the tensor product that will be useful later is $\langle \vc{x} \otimes \vc{y}, \vc{v} \otimes \vc{w} \rangle = \langle \vc{x}, \vc{v} \rangle \cdot \langle \vc{y}, \vc{w} \rangle$. This relation is particularly useful when decrypting a ciphertext tensor using a secret key tensor $\langle \vc{c}_1 \otimes \vc{c}_2, \vc{s}_1 \otimes \vc{s}_2 \rangle = \langle \vc{c}_1, \vc{s}_1 \rangle \cdot \langle \vc{c}_2, \vc{s}_2 \rangle$, where the decryption can be done separately. % \footnote{When the vector space is $\R^n$, the inner product is equivalent to the dot product between the vectors. When the vector space is of matrices, the inner product is the \textbf{Frobenius inner product} defined by $\langle \vc{A}, \vc{B} \rangle_F = \text{Tr}(\overline{\vc{A}^T} \vc{B}) = \sum_{i,j}(\overline{A_{i,j}} B_{i,j})$. Hence, the above relation between inner product and tensor product of matrices can be verified.} 

\paragraph{Setup} Given the security parameter $\lambda$, arithmetic circuit's multiplicative depth $L$ and the GLWE indicator $b \in \{0,1\}$, the encryption scheme starts by choosing appropriate parameter values to ensure the specific GLWE problem is $2^{\lambda}$-secure. Furthermore, it specifies an extra parameter $\mu=\mu(\lambda,L,b)=\theta(\log \lambda + \log L)$ that decides the size of the modulus $q$. More precisely, at each level $j \in \{L, L-1, \dots, 0\}$, % starting from the input level $j=L$, 
the $\setup$ step generates a sequence of parameter sets
\begin{align*}
    \para_j \leftarrow \bgv.\setup(1^{\lambda},1^{(j+1)\cdot \mu},b), %= \reg.\setup(1^{\lambda},1^{(j+1)\cdot \mu}, b)
\end{align*}
including a sequence of moduli $q_L, \dots, q_0$, whose sizes decrease from $(L+1) \cdot \mu$ bits to $\mu$ bits. These moduli will be used in modulus switching to manage ciphertext noise. 

\paragraph{Key generation} For $j=L$ to 0, generate % using Regev's secret key generation process 
a sequence of secret vectors as the secret key for $\bgv$ as follows: %. That is, 
\begin{align*}
    \sk = \{\vc{s}_L, \dots, \vc{s}_0\} \leftarrow \bgv.\seckeygen(\{ n_j,q_j \}_j),
    % , \text{ where for all } j \in [L,0] \\
    % \vc{s}_j &\leftarrow \reg.\seckeygen(n_j,q_j).\footnote{Note that in each secret vector $\vc{s}_i=(1,\vc{t}_i)$, the random vector $\vc{t}_i \leftarrow \chi^n$ is sampled from the domain $R_q^n$ according to the noise distribution. This is different from setting in Regev's building block scheme, but it does not incur any security risk.}
\end{align*}
where $\vc{s}_j = (1,\vc{t}_j), \vc{t}_i \leftarrow \Z^{n_j}_{q_j}$ for LWE and $\vc{t}_i \leftarrow \chi^{n_j}$ from the domain $R^{n_j}_{q_j}$ for RLWE.
% Note that in each secret vector $\vc{s}_i=(1,\vc{t}_i)$, the random vector $\vc{t}_i \leftarrow \chi^n$ is sampled from the domain $R_q^n$ according to the noise distribution. This is different from setting in Regev's building block scheme, but it does not incur any security risk.

These secret vectors will be used in key switching, where a ciphertext is transformed to another ciphertext under a different secret key. 
To allow key switching, compute the tensor product of each $\vc{s}_j$ with itself to get 
\begin{align*}
    \vc{s}_j' = \vc{s}_j \otimes \vc{s}_j % \in R_{q_j}^{n_j'}, \text{ where } n_j'=\binom{n_j+1}{2}.
\end{align*}
% The dimension $n_j'$ is counted by enumerating the selection of two elements from the vector $\vc{s}_j$. 
For all $j\in [L-1,0]$, ``encrypt'' the tensor product $\vc{s}_{j+1}'$ under the next secret vector $\vc{s}_j$ by running the key switching sub-routine to produce the auxiliary information
\begin{align*}
    \tau_{\vc{s}_{j+1}'\rightarrow \vc{s}_j} \leftarrow \switchkeygen(\vc{s}_{j+1}', \vc{s}_j).
\end{align*}
Finally, we use Regev's % starred 
public key generation step to produce a sequence of random matrices as part of the public key for $\bgv$. % \kl{I don't understand why generate a sequence of random matrices $A_j$. It seems only $A_L$ is used to produce fresh ciphertexts. I think I understand now, the sequence of public and secret keys are need without the weak circular security assumption. Also, the below notation may be incorrect, $A$ or $P$?}
\begin{align*}
    \vc{P}_j = [\vc{b}_j \mid -\vc{A}_j] \leftarrow \bgv.\pubkeygen(\vc{s}_j=(1,\vc{t}_j), N, \chi,\para_j), \text{ for all } j \in [L,0],
\end{align*}
where $A_j \leftarrow \Z_{q_j}^{N\times n_j}$, and $\vc{b}_j = [\vc{A}_j \vc{t}_j + 2\vc{e}]_{q_j}$ for a random noise vector $\vc{e} \leftarrow \chi^N$.

In summary, the public key of the BGV scheme is  
\begin{align*}
    \pk = \{\vc{P}_L,\dots, \vc{P}_0, \tau_{\vc{s}'_{L} \rightarrow \vc{s}_{L-1}}, \dots, \tau_{\vc{s}_{1}' \rightarrow \vc{s}_0}\} \leftarrow \bgv.\pubkeygen(\sk, \para).
\end{align*}

\paragraph{Encryption} The encryption of a message $m\in \{0,1\}$ is identical to Regev's encryption, that is, generate a random vector $\vc{r} \leftarrow \{0,1\}^{N}$ then compute the ciphertext 
\begin{align*}
    \vc{c} =\sqbracket{\vc{P}_L^T \vc{r} + \vc{m}}_{q_L} \leftarrow \bgv.\enc(\vc{P}_L,m,n_L,q_L,N) % =\reg.\enc(\vc{A}_L,m,n_L,q_L,N)
\end{align*}

\paragraph{Decryption} The decryption of a ciphertext that is encrypted under the secret key $\vc{s}_j$ is also identical to Regev's decryption 
\begin{align*}
    \sqbracket{\sqbracket{\vc{c} \cdot \vc{s}_j}_{q_j}}_2 \leftarrow \bgv.\dec(\vc{s}_j, \vc{c}, q_j) % =\reg.\dec(\vc{s}_j, \vc{c}, q_j).
\end{align*}

\paragraph{Homomorphic evaluation} Given two ciphertext $\vc{c}_1$ and $\vc{c}_2$ that are encrypted under the same secret key $\vc{s}_j$, the addition and multiplication of the two ciphertexts respectively produce the evaluated ciphertext 
\begin{align*}
    \vc{c}_{add} &= \vc{c}_1 + \vc{c}_2 \\ % \text{ and } \\
    \vc{c}_{mult} &=\vc{c}_1 \cdot \vc{c}_2,
\end{align*}
where addition is performed component wise as in \Cref{eq:bvv.add} and multiplication is the expansion of the ciphertext multiplication as in \Cref{eq:homoMultQuad}. Both evaluated ciphertexts are the coefficient vectors of the linear equations over the tensor product $\vc{x} \otimes \vc{x}$, so they can be decrypted by the secret key $\vc{s}_j' = \vc{s}_j \otimes \vc{s}_j$.

\paragraph{Refresh} The key component of $\bgv$ is the refresh step that is done after each homomorphic evaluation. It contains two sub-routines. 
\begin{enumerate}
    \item Switch key: The first sub-routine transforms a ciphertext to another ciphertext, both encrypt the same message but under different secret keys. Denote $\vc{c}_{q_j}^{\vc{s}_{j}'} \in \{\vc{c}_{add}, \vc{c}_{mult}\}$ a ciphertext that is encrypted under the secret key $\vc{s}'_j$, then    
    \begin{align*}
        \vc{c}_{q_j}^{\vc{s}_{j-1}} \leftarrow \switchkey_{q_j}(\tau_{\vc{s}'_j \rightarrow \vc{s}_{j-1}}, \vc{c}_{q_j}^{\vc{s}_{j}'}).
    \end{align*}
    
    \item Switch modulus: The second sub-routine reduces the ciphertext modulus in order to increase the gap between the noise ceiling and the ciphertext noise, while reducing both values at the same time. It runs the scale function to produce 
    \begin{align*}
        \vc{c}_{q_{j-1}}^{\vc{s}_{j-1}} \leftarrow \text{Scale}(\vc{c}_{q_j}^{\vc{s}_{j-1}}, q_j, q_{j-1}, 2).
    \end{align*}
\end{enumerate}

The $\bgv$ scheme is simpler than $\bv$, in the sense that it does not relinearize quadratic ciphertext. 
In addition, the scheme is leveled FHE with no \textit{bootstrapping} and its hardness is based on GLWE. % \footnote{The base on RLWE is not as straightforward as LWE, because the quotient polynomial degree $d$ is dependent on the level. Hence, special technique is needed to reduce the polynomial degree. See Section 4 in \citet{brakerski2014leveled}.} % I don't understand why $d$ is dependent on $L$ in RLWE.} 
The correctness of $\bgv$ is proved separately for each step by Lemma 6, 7, 8, 9 and 10 of \citet{brakerski2014leveled} respectively. Most of the hard work for these correctness proofs have been done in the correctness proofs of the building block encryption scheme, the modulus switching and key switching routines. The intuition is identical to correctness of previous schemes, that is, so long as the noise is well controlled and does not wrap around the modulus $q_j$ (i.e., noise ceiling), decryption will produce the correct message. In Section 5.4, \citet{brakerski2014leveled} guaranteed the parameters of $\bgv$ can be set to achieve such a goal.  

%\kl{start here next time...30 march 2022. Next time, talk about achieve leveled FHE based on RLWE, section 4 \citet{brakerski2014leveled}.}

In addition to removing dependence on bootstrapping, BGV can also reduce per-gate computation by basing its security on the RLWE problem. The per-gate computation is measured by the time taken to compute on ciphertexts to the time taken to compute on plaintexts. 
For security parameter $\lambda$ and circuit multiplicative depth $L$, the per-gate computation $\tilde{\Omega}(\lambda^4)$ in BV is reduced to $\tilde{O}(\lambda \cdot L^3)$ in BGV, and could be further reduced to $\tilde{O}(\lambda^2)$ when using bootstrapping as an optimization technique. % \kl{Further batching technique can reduce the complexity down to $\tilde{O}(\lambda)$. Both security are based on RLWE.} 

\subsection{The B scheme: scale invariant}

As a further simplification and improvement of their previous works, \citet{brakerski2012fully} proposed an encryption scheme that works with a fixed modulus $q$, but scales down a ciphertext by a factor $q$ each time. We call this scheme $\brak$ after the sole author's surname initial. The name ``scale invariant'' suggests the scheme does not decrease the moduli as in $\bgv$. Given a ciphertext $\vc{c} \in \Z_q$, the fractional ciphertext $\hat{\vc{c}} = \vc{c}/q \in \Z_1$ is within the symmetric range $[-1/2,1/2)$. 
The benefits of working with fractional ciphertexts are threefolds. First, it simplifies the scheme by not having a series of moduli and switching them iteratively. Second, it makes the evaluation noise grows linearly in the noise distribution bound $B$ and consequently requires a smaller noisy ceiling $q$ to guarantee decryption. For this matter, fractional ciphertexts appear only in homomorphic multiplications. Finally, on the security contribution, this work enables a \textbf{classical reduction} from the GAPSVP$_{n^{O(\log n)}}$ problem with a quasi-polynomial approximation factor. This is an improvement over \citet{peikert2009public}, in which the classical reduction can only be built for the same modulus size $q \approx 2^{n/2}$ from GAPSVP$_{2^{\Omega(n)}}$ with an exponential factor, which makes this lattice problem easy and hence unusable by HE schemes that want to rely on a classical reduction from lattice problems. 

We now state the procedures of $\brak$, which uses the same building blocks as previous schemes.  
\paragraph{Setup} The parameters are the same as $\bv$. That is, it has a pre-determined level $L=L(n)$ for the arithmetic circuits that will be evaluated and a parameter set $\para=(n,q,N,\chi)$.

\paragraph{Key generation} In this scheme, the fresh ciphertexts go into circuit level 0 and the completely evaluated ciphertexts are produced at level $L$. Sample a sequence of secret vectors 
\[ \vc{s}_0, \dots, \vc{s}_L \leftarrow \brak.\seckeygen(n,q) \] 
where $\vc{s}_i=(1,\vc{t}_i)$ with a random vector $\vc{t}_i \leftarrow \Z_q^n$. 
Generate a public key as usual by 
\begin{align*}
    \vc{P}_0=[\vc{b} \mid -\vc{A}] \leftarrow \brak.\pubkeygen(\vc{t}_0, \para) 
    %= \reg.\pubkeygen(\vc{t}_0, \para).
\end{align*}
where $\vc{A} \leftarrow \Z_{q}^{N\times n}$, and $\vc{b} = [\vc{A} \vc{t}_0 + \vc{e}]_{q}$ for a random noise vector $\vc{e} \leftarrow \chi^N$.
% Note, a difference from $\bv$ and $\bgv$ is the use of noise vector $\vc{e}$ rather than its multiple of 2 (which appears in the starred version $\bgv.\keygen^*$). 
Furthermore, to allow key switching during homomorphic evaluation, first compute the tensor product of each secret vector $\vc{s}_{i-1}$ with itself for $i \in [1,L]$
\begin{align*}
    \tilde{\vc{s}}_{i-1} = \bitdecomp(\vc{s}_{i-1}) \otimes \bitdecomp(\vc{s}_{i-1}),
\end{align*}
then compute the auxiliary information  
\begin{align*}
    \vc{P}_{(i-1):i} \leftarrow \switchkeygen(\tilde{\vc{s}}_{i-1}, \vc{s}_i).
\end{align*}
The final output of the key generation process is 
\begin{align*}
    (\pk, \sk, \evk) \leftarrow \brak.\keygen(\para), \text{ where }\\
    \pk = \vc{P}_0, \sk=\vc{s}_L, \evk=\{\vc{P}_{(i-1):i}\}_{i \in [1,L]}.
\end{align*}

\paragraph{Encryption and decryption} The two processes are identical to Regev's encryption and decryption, respectively. That is, 
\begin{align*}
    \vc{c} =\sqbracket{\vc{P}_0^T \vc{r} + \floor{\frac{q}{2}} \cdot \vc{m}}_{q} &\leftarrow \brak.\enc(\vc{P}_0,m,n,q,N) \\ % =\reg.\enc(\vc{P}_0,m,n,q,N) \\
    m = \sqbracket{\round{\frac{2}{q} \cdot \sqbracket{\vc{c} \cdot \vc{s}_L}_{q}}}_2 &\leftarrow \brak.\dec(\vc{s}_L, \vc{c}, q) %=\reg.\dec(\vc{s}_L, \vc{c}, q).
\end{align*}

\paragraph{Homomorphic evaluation} Addition and multiplications are defined separately, but both follow a two-step process. The first step is to produce an intermediate ciphertext in the powers of two format: 
\begin{align*}
    \tilde{\vc{c}}_{add} &= \powersoftwo(\vc{c}_1+\vc{c}_2) \otimes \powersoftwo((1,0,\dots,0)), \\
    \tilde{\vc{c}}_{mult} &= \round{\frac{2}{q} \cdot \powersoftwo(\vc{c}_1) \otimes \powersoftwo(\vc{c}_2)}. \\
\end{align*}
The tensor product with a dummy vector in the additive ciphertext is to ensure correct decryption when taking dot product with the corresponding secret vector in the following key switch process. 
% For an additive ciphertext, the fractional ideal is taken control noise growth. This is the key technical component if this scheme. \kl{Missing insight on why fractional ciphertext gives better noise growth.}
At gate $i$, the input ciphertexts are decryptable by $\vc{s}_{i-1}$. So these intermediate tensored ciphertexts are decryptable by the tensor secret vector $\tilde{\vc{s}}_{i-1}$. The second step is to transform an intermediate ciphertext to another ciphertext (non in tensor product format) under a new secret vector $\vc{s}_i$. That is, for $\tilde{\vc{c}} \in \{\tilde{\vc{c}}_{add}, \tilde{\vc{c}}_{mult}\}$, this is achieved by computing
\begin{align*}
    \vc{c} = \switchkey(\vc{P}_{(i-1):i}, \tilde{\vc{c}}).
\end{align*}

The scheme is thus completed, and as claimed it is a simpler construction than previous HE schemes. The homomorphic properties and security can be proved similarly as for previous schemes, see Theorem 4.2 and Lemma 4.1 of \citet{brakerski2012fully}. Furthermore, the scheme is leveld FHE without bootstrapping and can be made non-leveld by assuming weak circular security (Corollary 4.5 \citep{brakerski2012fully}) as in the BV scheme.

\subsection{The BFV scheme}

We finish this section by introducing the BFV\index{HE scheme!BFV} scheme \citep{fan2012somewhat}, whose security is solely based on the RLWE\index{RLWE} problem. Despite its similarity to the aforementioned schemes, it makes HE schemes practical by explicitly stating the specific parameters need to achieve a certain security level.%\kl{Double check this!} 
Therefore, we will emphasize on analysing the noise bounds of ciphertexts output by different encryption scheme subroutines, rather than presenting the homomorphic operations most of which have been discussed in preceding subsections.  

BFV is built upon the RLWE-based encryption scheme, named LPR \citep{lyubashevsky2010ideal} that was stated at the end of the previous section. Its plaintext space is generalized to $R_t$ from $R_2$ as in the simplified scheme. This also implies the fractional factor is now $\Delta=\floor{q/t}$ rather than $\floor{q/2}$. 
Besides that, the underlying domain $R_q = \Z_q[x]/(\Phi_m(x))$ is generalized to an arbitrary $mth$ cyclotomic field for a suitable modulus $q$ and cyclotomic polynomial  $\Phi(m)$, although the preferred one is still  $\Phi(m)=x^n+1$ for $m$ being a power of 2 and $n=m/2$.

A technical term that often appears in the analysis of BFV's noise bounds is expansion factor. When multiplying two polynomials $\vc{a} = a_0 + a_1 \cdot x + \cdots a_d \cdot x^d$ and $\vc{b} = b_0 + b_1 \cdot x + \cdots b_d \cdot x^d$, the coefficient of $x^i$ can be larger than $a_i + b_i$ due to the fact that there may be more than one term in $\vc{a} \cdot \vc{b}$\footnote{We also use boldface to represent polynomials and $\cdot$ to represent polynomial multiplications.} with the degree $i$. For this reason, we define the 
\reversemarginpar
\marginpar{expansion factor}
\textbf{expansion factor}\index{expansion factor} of the polynomial ring as $\gamma_R=\max\{||\vc{a} \cdot \vc{b}||/(||\vc{a}|| \cdot ||\vc{b}||) \mid \vc{a}, \vc{b} \in R\}$, where $||\vc{a}||=\max_i |a_i|$ is the maximum coefficient of the polynomial. It is worth mentioning that expansion factor appears only when analysing noise bounds in the polynomial coefficient embedding context, not in the canonical embedding context in which multiplications are element-wise. 

%As shown in \Cref{subsec:lpr}, decryption works by computing 
%\begin{align*}
%    \sqbracket{\round{\frac{t}{q} \sqbracket{\vc{u} + \vc{v} \cdot \vc{s}}_q}}_t.
%\end{align*}
Let $\vc{c}_i=(\vc{u}_i,\vc{v}_i)$ be a ciphertext. 
%If $\vc{e},\vc{e}_1,\vc{e}_2,\vc{r},\vc{s}$ in  all have small coefficients, their product is still within modulo $q$. 
Decryption works by first computing 
\begin{align}
\label{eq:bfvDec1}
    \sqbracket{f_{\vc{c}_i}(\vc{s})}_q &=\sqbracket{\vc{u}_i + \vc{v}_i \cdot \vc{s}}_q \\
\label{eq:bfvDec2}
    &= \Delta \cdot \vc{m}_i + \vc{e}'_i,
\end{align}
where $\vc{e}_i'=\vc{e} \cdot \vc{r} +\vc{e}_1+\vc{e}_2 \cdot \vc{s}$, as shown in \Cref{eq:lprScheme}, followed by the multiplication of a fractional, rounding and modulo $t$, that is, 
\begin{align*}
    \dec(\vc{s},\vc{c}_i) = \sqbracket{\round{\frac{t \cdot \sqbracket{f_{\vc{c}_i}(\vc{s})}_q}{q}}}_t.
\end{align*}
The bound on the noise's coefficients is 
\begin{align*}
    ||\vc{e}_i'|| \le 2 \cdot \delta_R \cdot B^2 + B,
\end{align*}
where $\delta_R$ is the expansion factor of $R$ and $[-B,B]$ is the support of the noise distribution $\chi$ over $R$. 
\reversemarginpar
\marginpar{encryption noise}
In \citet{fan2012somewhat}, the bound is further reduced to $2\cdot \delta_R \cdot B + B$ by taking $\vc{r}$ and $\vc{s}$  from $\{0,1\}^n$, with only a minor security implication (Optimization/Assumption 1 \citet{fan2012somewhat}). 

Next we jump straight to the homomorphic operations. For simplicity, we analyse the operations for two ciphertexts $\vc{c}_1=(\vc{u}_1,\vc{v}_1)$ and $\vc{c}_2=(\vc{u}_2,\vc{v}_2)$.  

\paragraph{Homomorphic addition}
Homomorphic addition is defined as component-wise addition. That is, 
\begin{align*}
    \vc{c}_{add}= \bracket{\sqbracket{\vc{u}_1+\vc{u}_2}_q, \sqbracket{\vc{v}_1+\vc{v}_2}_q} \leftarrow \bfv.\add(\vc{c}_1, \vc{c}_2).
    %&\vc{c}_{add}. % (\vc{c}_1[0]+\vc{c}_2[0], \vc{c}_1[1]+\vc{c}_2[1]).
\end{align*}
It is easy to see that addition is correct because 
\begin{align*}
    \sqbracket{(\vc{u}_1 + \vc{u}_2) + (\vc{v}_1 + \vc{v}_2) \cdot \vc{s}}_q
    = \sqbracket{(\vc{u}_1 + \vc{v}_1 \cdot \vc{s})+(\vc{u}_2 + \vc{v}_2 \cdot \vc{s})}_q
    = \sqbracket{\Delta \cdot (\vc{m}_1 + \vc{m}_2) + \vc{e}'_1+\vc{e}'_2}_q.
\end{align*}
To transform $\Delta \cdot (\vc{m}_1 + \vc{m}_2)$ to be in the plaintext space $R_t$, We notice that $\vc{m}_1 + \vc{m}_2 = \sqbracket{\vc{m}_1 + \vc{m}_2}_t + t \cdot \vc{r}_t$ for a polynomial $\vc{r}_t$ whose coefficients satisfy $||\vc{r}_t|| \le 1$, because $||\vc{m}_1 + \vc{m}_2|| \le 2t$ and $||\sqbracket{\vc{m}_1 + \vc{m}_2}_t|| \le t$. Let $\epsilon=q/t-\Delta = r_t(q)/t < 1$, we get 
\begin{align*}
    \sqbracket{\Delta \cdot (\vc{m}_1 + \vc{m}_2) + \vc{e}'_1+\vc{e}'_2}_q
    &=\sqbracket{\Delta \cdot \sqbracket{\vc{m}_1+\vc{m}_2}_t + \Delta \cdot t \cdot \vc{r}_t + \vc{e}'_1 + \vc{e}'_2}_q\\
    &=\Delta \cdot \sqbracket{\vc{m}_1+\vc{m}_2}_t + \vc{e}'_1 + \vc{e}'_2 - (q-\Delta \cdot t) \cdot \vc{r}_t\\
    &= \Delta \cdot \sqbracket{\vc{m}_1 + \vc{m}_2}_t + \underbrace{\vc{e}'_1 + \vc{e}'_2 - \epsilon \cdot t \cdot \vc{r}_t}_{\text{noise}}.
\end{align*}
After multiplying with $t/q$, the coefficient $t/q \cdot \Delta$ rounds to 1 and $(t/q \cdot \text{noise})$ rounds to 0. So decryption is guaranteed correct after taking the final $[\cdot]_t$. 
\reversemarginpar
\marginpar{addition noise}
Notice homomorphic addition only incurs an extra additive noise by a factor of $t$ because $||\vc{r}_t||\le 1$ and $\epsilon<1$ by construction. The incurred noise is usually much smaller than the noise ceiling $q$.

\paragraph{Homomorphic multiplication}
Similar to the previous schemes, much of the effort in BFV's construction deals with relinearization after homomorphic multiplications. The noise growth after a ciphertext multiplication is bounded by $2 \cdot t \cdot \delta_R^2 \cdot ||\vc{s}||$, which is better than quadratic growth (Lemma 2 \citep{fan2012somewhat}). 

It takes several steps to see how the noise bound is obtained. We have known from previous sections that a direct ciphertext multiplication produces a quadratic function as follows 
\begin{align}
\label{eq:bfvQuard1}
    &f_{\vc{c}_1}(\vc{s}) \cdot f_{\vc{c}_2}(\vc{s})=\vc{h}_0 + \vc{h}_1 \cdot \vc{s} + \vc{h}_2 \cdot \vc{s}^2, \text{ where }\\
    &\vc{h}_0= \vc{u}_1 \cdot \vc{u}_2, \text{ }
    \vc{h}_1= \vc{u}_1 \cdot \vc{v}_2 + \vc{u}_2 \cdot \vc{v}_1, \text{ }
    \vc{h}_2= \vc{v}_1 \cdot \vc{v}_2. \nonumber
\end{align}
%The three ``coefficients'' become the multiplicative ciphertext.
By looking at \Cref{eq:bfvDec2}, it is not hard to see that when multiplying $f_{\vc{c}_1}(\vc{s}) \cdot f_{\vc{c}_2}(\vc{s})$, it will results a term $\Delta^2 \cdot \vc{m}_1 \cdot \vc{m}_2$ and several other terms with $q$ being part of their coefficients. To get the message product back to $\Delta \cdot \vc{m}_1 \cdot \vc{m}_2$ in order to allow decryption to work, one way is to multiply it by $1/\Delta$. But this can cause round problem in other terms that contain $q$ as part of their coefficients. Let $\epsilon = q/t - \Delta$ be the rounding error, so the term $q/\Delta = q/(q/t - \epsilon)$. The problem with this is that it does not always round up back to $t$. For example, with $q=17$ we have $q/\Delta=2.15$ when $t=2$ and $q/\Delta \approx 5.67$ when $t=5$. So the later creates a rounding error that becomes problematic in subsequent steps. Hence, an alternative solution is to multiplying all the terms by $t/q$ then applying rounding. This is straightforward for the terms with $q$ being part of the coefficients. For the message product term, it gives $(t/q \cdot \Delta) \cdot (\Delta \cdot \vc{m}_1 \cdot \vc{m}_2)$ and $t/q \cdot \Delta = t/q \cdot (q/t - \epsilon) = 1 - (t/q) \cdot \epsilon \in  (0.5,1.5)$ as $|\epsilon|\le 1/2$ and $t \le q$ with equality implies $\epsilon=0$. Hence, multiply \Cref{eq:bfvQuard1} with the fraction, we get 
\begin{align*}
    \frac{t}{q} \cdot f_{\vc{c}_1}(\vc{s}) \cdot f_{\vc{c}_2}(\vc{s}) =\frac{t}{q} \cdot (\vc{h}_0 + \vc{h}_1 \cdot \vc{s} + \vc{h}_2 \cdot \vc{s}^2).
\end{align*}
As shown above, the coefficients need to be rounded to get the ciphertext back on track for decryption, so the above equation can be re-written as 
\begin{align}
\label{eq:bfvQuard2}
    \frac{t}{q} \cdot f_{\vc{c}_1}(\vc{s}) \cdot f_{\vc{c}_2}(\vc{s}) = &\round{\frac{t}{q} \cdot \vc{h}_0} + \round{\frac{t}{q} \cdot \vc{h}_1} \cdot \vc{s} + \round{\frac{t}{q} \cdot \vc{h}_2} \cdot \vc{s}^2 
    + \bracket{\frac{t}{q} \cdot \vc{h}_0 - \round{\frac{t}{q} \cdot \vc{h}_0}} \nonumber\\ 
    &+ \bracket{\frac{t}{q} \cdot \vc{h}_1 - \round{\frac{t}{q} \cdot \vc{h}_1}} \cdot \vc{s} + \bracket{\round{\frac{t}{q} \cdot \vc{h}_2} - \round{\frac{t}{q} \cdot \vc{h}_2}} \cdot \vc{s}^2 \nonumber\\
    = &\round{\frac{t}{q} \cdot \vc{h}_0} + \round{\frac{t}{q} \cdot \vc{h}_1} \cdot \vc{s} + \round{\frac{t}{q} \cdot \vc{h}_2} \cdot \vc{s}^2 + \vc{r}_a.
\end{align}
The three updated ``coefficients'' make the appropriate multiplicative ciphertext
\begin{align*}
    \vc{c}_{mult}=\bracket{\mathfrak{h}_0, \mathfrak{h}_1, \mathfrak{h}_2}
    :=\bracket{\sqbracket{\round{\frac{t}{q}\cdot \vc{h}_0}}_q,\sqbracket{\round{\frac{t}{q}\cdot \vc{h}_1}}_q,\sqbracket{\round{\frac{t}{q}\cdot \vc{h}_2}}_q}
    \leftarrow \bfv.\mult(\vc{c}_1, \vc{c}_2).
\end{align*}
Since rounding error is at most $1/2$ between integer coefficients, the approximation error satisfies $||\vc{r}_a|| < 1/2 + 1/2 \cdot ||\vc{s}|| \cdot \delta_R+1/2 \cdot ||\vc{s}|| \cdot \delta_R^2$. The bound can be made further loose to be $||\vc{r}_a|| < 1/2\cdot (1+||\vc{s}|| \cdot \delta_R)^2$ in order to be used by the following homomorphic multiplication noise bound analysis. 

By moving the approximation error $\vc{r}_a$ to the LHS of \Cref{eq:bfvQuard2} and reducing both sides to $R_q$, we get 
\begin{align}
\label{eq:bfvQuard3}
    \sqbracket{\frac{t}{q} \cdot f_{\vc{c}_1}(\vc{s}) \cdot f_{\vc{c}_2}(\vc{s}) - \vc{r}_a}_q = \sqbracket{\round{\frac{t}{q} \cdot \vc{h}_0} + \round{\frac{t}{q} \cdot \vc{h}_1} \cdot \vc{s} + \round{\frac{t}{q} \cdot \vc{h}_2} \cdot \vc{s}^2}_q. 
\end{align}
To derive the multiplication noise bound, we explicitly write out all the terms in $f_{\vc{c}_1}(\vc{s}) \cdot f_{\vc{c}_2}(\vc{s})$ using \Cref{eq:bfvDec2}, so we get  
\begin{align*}
    f_{\vc{c}_1}(\vc{s}) \cdot f_{\vc{c}_2}(\vc{s}) = &(\Delta \cdot \vc{m}_1 + \vc{e}'_1 + q \cdot \vc{r}_{q,1}) \cdot (\Delta \cdot \vc{m}_2 + \vc{e}'_2 + q \cdot \vc{r}_{q,2})\\
    = &\Delta^2 \cdot \vc{m}_1 \cdot \vc{m}_2 + \Delta \cdot (\vc{m}_1 \cdot \vc{e}_2' + \vc{m}_2 \cdot \vc{e}_1') + \Delta \cdot q \cdot (\vc{m}_1 \cdot \vc{r}_{q,2}+\vc{m}_2 \cdot \vc{r}_{q,1}) \\
    &+ q \cdot (\vc{r}_{q,1} \cdot \vc{e}_2' + \vc{r}_{q,2} \cdot \vc{e}_1')  + q^2 \cdot \vc{r}_{q,1} \cdot \vc{r}_{q,2} + \vc{e}_1' \cdot \vc{e}_2'.
\end{align*}
Same as in homomorphic addition, we want to express the product of secret messages in the plaintext space $R_t$, so we can write $\vc{m}_1 \cdot \vc{m}_2 = \sqbracket{\vc{m}_1 \cdot \vc{m}_2}_t + t \cdot \vc{r}_t$, where $||\vc{r}_t|| < t \cdot \delta_R/4$.%\footnote{This is different from the bound in \citet{fan2012somewhat}.}
Multiplying the above equation by $t/q$ on both sides, we get 
\begin{align*}
    \frac{t}{q} \cdot f_{\vc{c}_1}(\vc{s}) \cdot f_{\vc{c}_2}(\vc{s})
    = &\frac{t\cdot \Delta^2}{q} \cdot (\sqbracket{\vc{m}_1 \cdot \vc{m}_2}_t + t \cdot \vc{r}_t) + \frac{t \cdot \Delta}{q} \cdot (\vc{m}_1 \cdot \vc{e}_2' + \vc{m}_2 \cdot \vc{e}_1') \\
    &+ t \cdot \Delta \cdot (\vc{m}_1 \cdot \vc{r}_{q,2}+\vc{m}_2 \cdot \vc{r}_{q,1}) 
    + t \cdot (\vc{r}_{q,1} \cdot \vc{e}_2' + \vc{r}_{q,2} \cdot \vc{e}_1') \\
    &+ t\cdot q \cdot \vc{r}_{q,1} \cdot \vc{r}_{q,2} + \frac{t}{q} \cdot \vc{e}_1' \cdot \vc{e}_2'.
\end{align*}
Since modulo $q$ will be applied onto this as shown in \Cref{eq:bfvQuard3} followed by rounding, it is convenient to split the above into terms with and without integer coefficients. To do so, we can substitute $t \cdot \Delta = q - r_t(q)$ into the above equation. After re-arranging the terms, we get 
\begin{align*}
    \frac{t}{q} \cdot f_{\vc{c}_1}(\vc{s}) \cdot f_{\vc{c}_2}(\vc{s})
    = &\Delta \cdot \sqbracket{\vc{m}_1 \cdot \vc{m}_2}_t 
    + (\vc{m}_1 \cdot \vc{e}_2' + \vc{m}_2 \cdot \vc{e}_1') 
    + (q-r_t(q)) \cdot (\vc{r}_t + \vc{m}_1 \cdot \vc{r}_{q,2}+\vc{m}_2 \cdot \vc{r}_{q,1}) \\
    &+ t \cdot (\vc{r}_{q,1} \cdot \vc{e}_2' + \vc{r}_{q,2} \cdot \vc{e}_1') 
    + q \cdot t \cdot \vc{r}_{q,1} \cdot \vc{r}_{q,2} + \vc{r}_{\Delta} \\
    &+\underbrace{\frac{t}{q} \cdot \sqbracket{\vc{e}'_1 \cdot \vc{e}_2'}_{\Delta} 
    -\frac{r_t(q)}{q} \cdot (\Delta \cdot \vc{m}_1 \cdot \vc{m}_2 + (\vc{m}_1 \cdot \vc{e}_2' + \vc{m}_2 \cdot \vc{e}_1') + \vc{r}_{\Delta})}_{\vc{r}_r}.
\end{align*}
All the terms except $\vc{r}_r$ have integer coefficients, so they will not be affected by rounding. Substitute this into \Cref{eq:bfvQuard3}, we get 
\begin{align*}
    &\sqbracket{\round{\frac{t}{q} \cdot \vc{h}_0} + \round{\frac{t}{q} \cdot \vc{h}_1} \cdot \vc{s} + \round{\frac{t}{q} \cdot \vc{h}_2} \cdot \vc{s}^2}_q\\
    = &\Delta \cdot \sqbracket{\vc{m}_1 \cdot \vc{m}_2}_t \\
    &+ (\vc{m}_1 \cdot \vc{e}_2' + \vc{m}_2 \cdot \vc{e}_1') 
    - r_t(q) \cdot (\vc{r}_t + \vc{m}_1 \cdot \vc{r}_{q,2}+\vc{m}_2 \cdot \vc{r}_{q,1}) \\
    &+ t \cdot (\vc{r}_{q,1} \cdot \vc{e}_2' + \vc{r}_{q,2} \cdot \vc{e}_1') 
    + (\vc{r}_r - \vc{r}_a)\\
    = &\Delta \cdot \sqbracket{\vc{m}_1 \cdot \vc{m}_2}_t + \vc{e}'_3. 
\end{align*}
\reversemarginpar
\marginpar{multiplication noise}
Using the bounds proved above, it can be shown that $||\vc{e}'_3|| < 2 \cdot \delta_R \cdot t \cdot E \cdot (\delta_R \cdot ||\vc{s}|| + 1) + 2 \cdot t^2 \cdot \delta_R^2 \cdot (||\vc{s}||+1)^2$, which is dominated by $2 \cdot t^2 \cdot \delta_R^2 \cdot ||\vc{s}||^2$.

\paragraph{Relinearization}
As discussed in BV's relinearization, the problem with the direct multiplicative ciphertext is its increased length from 2 to 3 ``coefficients''. To overcome this, \citet{fan2012somewhat} presented two methods to relinearize the ciphertext with only two new coefficients and a small noise.
\begin{align*}
    \sqbracket{\mathfrak{h}_0+\mathfrak{h}_1 \cdot \vc{s} + \mathfrak{h}_2 \cdot \vc{s}^2}_q = \sqbracket{\mathfrak{h}'_0+\mathfrak{h}'_1 \cdot \vc{s} + \vc{err}}_q.
\end{align*}
% with new coefficients and a small error term $\vc{err}$. 

The first method,
\reversemarginpar
\marginpar{\textit{relinearization version 1}}
which is similar to the relinearization process in the BV scheme, produces a relinearization key $\{ \rlk_{\tau} \}$
\begin{align*}
    \rlk_{\tau} = \left(\vc{b}_{\tau}=\sqbracket{-(\vc{a}_{\tau} \cdot \vc{s} + \vc{e}_{\tau}) + T^{\tau} \cdot \vc{s}^2}_q, \vc{a}_{\tau} \right)
\end{align*}
that looks almost like the evaluation key in BV, except that the coefficient 
\begin{align*}
    \mathfrak{h}_2=\sum_{\tau=0}^l T^{\tau} \cdot \mathfrak{h}_2^{(\tau)} \bmod q
\end{align*}
is written in $T$-nary representation, where $l =\floor{\log_T q}$ and $\vc{h}_2^{(\tau)} \in R_T$. The polynomials were sampled by $a_{\tau} \leftarrow R_q$ and $\vc{e}_{\tau} \leftarrow \chi$.
The purpose of expressing $\mathfrak{h}_2$ in $T$-nary representation is to reduce the amplification effect on ciphertext noise after multiplications. % multiplying by $\vc{h}_2$ the linear ciphertext plus the noise. 
The same idea was also discussed in \Cref{subsubsec:bvHomMul}, which used $T=2$ to minimize the relinearization noise for BV. 

The main difference from the aforementioned schemes is that $\vc{s}^2$ is encrypted by the corresponding public key $\vc{a}_{\tau}$ in the same $(\pk,\sk)$ pair, while in the BV scheme for example, each quadratic secret key is encrypted by the next public key. So for this relinearization step to be secure, the weak circular security assumption (\Cref{def:weakCircularSec}) is needed. This is also why the BFV uses only a single secret key and a single public key instead of a series of keys. 

Given the relinearization key $\{ \rlk_{\tau} = (\vc{b}_{\tau}, \vc{a}_{\tau}) \mid  \tau \in [0, l]\}$, the two new coefficients are set to 
\begin{align*}
    \mathfrak{h}_0'=\sqbracket{\mathfrak{h}_0+\sum_{\tau=0}^l \vc{b}_{\tau} \cdot \mathfrak{h}_2^{(\tau)}}_q \text{ and }
    \mathfrak{h}_1'=\sqbracket{\mathfrak{h}_1+\sum_{\tau=0}^l \vc{a}_{\tau} \cdot \mathfrak{h}_2^{(\tau)}}_q.
\end{align*}
To check that they are the correct choices, we get 
\begin{align*}
    \sqbracket{\mathfrak{h}_0'+\mathfrak{h}_1' \cdot \vc{s}}_q = \sqbracket{\mathfrak{h}_0 + \mathfrak{h}_1 \cdot \vc{s} + \mathfrak{h}_2 \cdot \vc{s}^2 - \underbrace{\sum_{\tau=0}^l \mathfrak{h}_2^{(\tau)} \cdot \vc{e}_{\tau}}_{\vc{err}_1}}_q,
\end{align*}
where the relinearization noise's coefficients bound is $||\vc{err}_1|| \le (l+1) \cdot T \cdot B \cdot \delta_R/2$. So the larger $T$ is, the larger the error will be. However, $T$ should also be set not too small in order to match the noise magnitude after one ciphertext multiplication. % \kl{I don't understand why.} 

The second method
\reversemarginpar
\marginpar{\textit{relinearization version 2}}
relies on the noise reduction effect by modulus reduction as shown in \Cref{subsubsec:bv}. The motivation is to still be able to approximate a quadratic ciphertext by a linear one, but without slicing the coefficient $\vc{h}_2$ into many pieces which potentially increases the relinearization space and time. The idea is to encrypt a scaled quadratic secret key $p \cdot \vc{s}^2$ in the larger domain $\Z_{p\cdot q}$ for an integer $p$, then scale it down to within $\Z_q$ by dividing it by $p$. More precisely, randomly sample $\vc{a} \leftarrow R_{p \cdot q}$ and $\vc{e} \leftarrow \chi'$ from a different noise distribution, then output the relinearization key 
\begin{align*}
    \rlk = \left(\vc{b}=\sqbracket{-(\vc{a} \cdot \vc{s} + \vc{e}) + p \cdot \vc{s}^2}_{p\cdot q}, \vc{a} \right).
\end{align*}
Given this relinearization key, the two new coefficients are constructed by
\begin{align*}
    \mathfrak{h}_0' = \mathfrak{h}_0 + \sqbracket{\round{\frac{\mathfrak{h}_2 \cdot \vc{b}}{p}}}_q \text{ and } 
    \mathfrak{h}_1' = \mathfrak{h}_1 + \sqbracket{\round{\frac{\mathfrak{h}_2 \cdot \vc{a}}{p}}}_q.
\end{align*}
Again, to make sure these new coefficients can lead to the correct decryption, we get 
\begin{align*}
\sqbracket{\mathfrak{h}_0'+\mathfrak{h}_1' \cdot \vc{s}}_q 
=&\sqbracket{\mathfrak{h}_0 + \mathfrak{h}_1 \cdot \vc{s} + 
\round{\frac{\mathfrak{h}_2 \cdot \vc{b}}{p}}  + \round{\frac{\mathfrak{h}_2 \cdot \vc{a}}{p}} \cdot \vc{s}}_q\\
=&\left[\mathfrak{h}_0 + \mathfrak{h}_1 \cdot \vc{s} + 
\frac{\mathfrak{h}_2 \cdot \vc{b}}{p} + \frac{\mathfrak{h}_2 \cdot \vc{a}}{p} \cdot \vc{s}\right.\\
&\left.+ \bracket{\round{\frac{\mathfrak{h}_2 \cdot \vc{b}}{p}} - \frac{\mathfrak{h}_2 \cdot \vc{b}}{p}} + \bracket{\round{\frac{\mathfrak{h}_2 \cdot \vc{a}}{p}} - \frac{\mathfrak{h}_2 \cdot \vc{a}}{p}} \cdot \vc{s}\right]_q\\
=&\left[\mathfrak{h}_0 + \mathfrak{h}_1 \cdot \vc{s} + \frac{\mathfrak{h}_2 \cdot (-(\vc{a}\cdot \vc{s} + \vc{e}) + p \vc{s}^2 + p\cdot q \cdot \vc{r}_{pq} + \vc{a} \cdot \vc{s})}{p}\right.\\
&\left.+ \bracket{\round{\frac{\mathfrak{h}_2 \cdot \vc{b}}{p}} - \frac{\mathfrak{h}_2 \cdot \vc{b}}{p}} + \bracket{\round{\frac{\mathfrak{h}_2 \cdot \vc{a}}{p}} - \frac{\mathfrak{h}_2 \cdot \vc{a}}{p}} \cdot \vc{s}\right]_q\\
=&\left[\mathfrak{h}_0 + \mathfrak{h}_1 \cdot \vc{s} + \mathfrak{h}_2 \cdot \vc{s}^2\right. \\
&\left.+\underbrace{\frac{-\mathfrak{h}_2 \cdot \vc{e}}{p} 
	+ \bracket{\round{\frac{\mathfrak{h}_2 \cdot \vc{b}}{p}} - \frac{\mathfrak{h}_2 \cdot \vc{b}}{p}} + \bracket{\round{\frac{\mathfrak{h}_2 \cdot \vc{a}}{p}} - \frac{\mathfrak{h}_2 \cdot \vc{a}}{p}} \cdot \vc{s}}_{\vc{err_2}}\right]_q.
\end{align*}
So the second relinearization generates noise of magnitude $||\vc{err}_2|| < \frac{q\cdot B \cdot \delta_R}{p} + \frac{1}{2} + \frac{1}{2} \cdot ||\vc{s}|| \cdot \delta_R$.

The combined noise magnitude of homomorphic addition and multiplication with each relinearization step were stated in Lemma 3 of \citet{fan2012somewhat}. Given the fact that relinearization noises can be managed by setting parameters $T$ (version 1) and $p$ (version 2) at appropriate values, Theorem 1 of \citet{fan2012somewhat} proved the maximum multiplicative depth of the evaluated circuit according to the other parameter values. 

Finally, the scheme can be made bootstrappable by simplifying the decryption algorithm. The simplification can be done before the scheme evaluating its own decryption by a modulus switching from the original modulus $q$ to a smaller modulus $q'=2^n$ by scaling the ciphertext $(\vc{u},\vc{v})$ to get  
\begin{align*}
    \vc{u}' = \round{2^n/q \cdot \vc{u}} \text{ and } \vc{v}' = \round{2^n/q \cdot \vc{v}}.
\end{align*}
This is because if $q'=2^n$ and set $t=2^{n-k}$, then $\Delta=\floor{q/t}=2^k$. So in the decryption step, $t/q \cdot \sqbracket{f_{\vc{c}}(\vc{s})}_q$ becomes $1/\Delta \cdot \sqbracket{f_{\vc{c}}(\vc{s})}_q$ and division by $\Delta$ is efficient (Section 5.2 \citep{fan2012somewhat}).

Below, we summarize the BFV scheme and provide an implementation in Sage. 

\begin{tcolorbox}
\noindent
\textbf{Private key}: Sample a private key $\vc{s} \leftarrow R_2$.\\

\textbf{Public key}: Sample random polynomials $\vc{a} \leftarrow R_q$ and $\vc{e} \leftarrow \chi$ and output the public key $(\vc{b} = -\sqbracket{ \vc{a} \cdot \vc{s} + \vc{e}}_q,\vc{a})$.\\

\textbf{Relinearization key:} For a positive integer $T$, let $l=\floor{\log_T q}$. Let $\vc{a}_{\tau} \leftarrow R_q$, $\vc{e}_{\tau} \leftarrow \chi$, $\vc{a} \leftarrow R_{p \cdot q}$ and $\vc{e} \leftarrow \chi'$ a different noise distribution. Generate two sets of relinearization keys
\begin{align*}
    \rlk_1 &=\left\{\rlk_{\tau} = (\vc{b}_{\tau}, \vc{a}_{\tau}) \mid  \tau \in [0, l]\right\}, \text{ where } \vc{b}_{\tau}=\sqbracket{-(\vc{a}_{\tau} \cdot \vc{s} + \vc{e}_{\tau}) + T^{\tau} \cdot \vc{s}^2}_q\\
    \rlk_2 &= \left(\vc{b}=\sqbracket{-(\vc{a} \cdot \vc{s} + \vc{e}) + p \cdot \vc{s}^2}_{p\cdot q}, \vc{a} \right).
\end{align*}

\textbf{Encryption:} Encrypt a message $\vc{m} \in R_t$ by computing 
\begin{align*}
    \vc{u} &= \sqbracket{\vc{b} \cdot \vc{r} + \vc{e}_1 + \floor{q/t} \cdot \vc{m}}_q\\
    \vc{v} &= \sqbracket{\vc{a} \cdot \vc{r} + \vc{e}_2}_q,
\end{align*}
where $\vc{r} \leftarrow R_2$ and $\vc{e}_1,\vc{e}_2 \leftarrow \chi$ are random samples. Then output the ciphertext $\vc{c}=(\vc{u},\vc{v})$.\\

\textbf{Decryption:} Decrypt the ciphertext $\vc{c}$ using the secret key by computing 
\begin{align*}
    m = \sqbracket{\round{\frac{t}{q} \sqbracket{\vc{u} + \vc{v} \cdot \vc{s}}_q}}_t.
\end{align*}

\textbf{Homomorphic operations:} Given ciphertexts $\vc{c}_i=(\vc{u}_i,\vc{v}_i)$ for $i\in [1,2]$, 
\begin{align*}
    \vc{c}_{add}&= \bracket{\sqbracket{\vc{u}_1+\vc{u}_2}_q, \sqbracket{\vc{v}_1+\vc{v}_2}_q}\\
    \vc{c}_{mult}&=\bracket{\mathfrak{h}_0, \mathfrak{h}_1,\mathfrak{h}_2}. 
\end{align*}

\textbf{Relinearization:} Re-write $\mathfrak{h}_2=\sum_{\tau=0}^l T^{\tau} \cdot \mathfrak{h}_2^{(\tau)} \bmod q$. Choose one method from the following two. Use the corresponding key $\rlk_1$ and $\rlk_2$ for the two methods respectively.  
\begin{align*}
    &\text{ Method 1: } \mathfrak{h}_0'=\sqbracket{\mathfrak{h}_0+\sum_{\tau=0}^l \vc{b}_{\tau} \cdot \mathfrak{h}_2^{(\tau)}}_q \text{ and }
    \mathfrak{h}_1'=\sqbracket{\mathfrak{h}_1+\sum_{\tau=0}^l \vc{a}_{\tau} \cdot \mathfrak{h}_2^{(\tau)}}_q.\\
    &\text{ Method 2: } \mathfrak{h}_0' = \mathfrak{h}_0 + \sqbracket{\round{\frac{\mathfrak{h}_2 \cdot \vc{b}}{p}}}_q \text{ and } 
    \mathfrak{h}_1' = \mathfrak{h}_1 + \sqbracket{\round{\frac{\mathfrak{h}_2 \cdot \vc{a}}{p}}}_q.
\end{align*}
Output the relinearized ciphertext $(\mathfrak{h}'_0, \mathfrak{h}'_1)$.

\end{tcolorbox}

% \begin{figure}
% \centering
% \caption{A Sage implementation of the BFV cryptosystem described above.\\ \textbf{Note:} This implementation is not suitable for use in real-world applications.}
% \begin{tcolorbox}
% \begin{verbatim}
% #!/usr/bin/env sage

% from sage.misc.prandom import randrange
% import sage.stats.distributions.discrete_gaussian_integer as dgi

% # Define parameters
% def sample_noise(n, P):
%     D = dgi.DiscreteGaussianDistributionIntegerSampler(sigma=1.0)
%     return P([D() for i in range(n)])

% q = 655360001
% n = 2^10
% t = 37

% P = QuotientRing(PolynomialRing(Integers(q), name="x"),
%                  x^n + 1)
% Q = PolynomialRing(Rationals(), name="y")
% Zt = Integers(t)

% # Generate keys
% secret_key = sample_noise(n, P)

% e = sample_noise(n, P)
% a = P.random_element()
% b = -(a*secret_key) + e

% public_key = (b,a)

% # Encrypt Message
% message = P([randrange(0,t) for i in range(n)])

% r = sample_noise(n, P)
% e1 = sample_noise(n, P)
% e2 = sample_noise(n, P)

% u = b*r + e1 + (q//t)*message
% v = a*r + e2

% ciphertext = (u,v)

% # Decrypt Message
% w1 = u + v*secret_key
% w2 = (t/q) * Q(w1.list())

% decrypted_message = P([Zt(w.round()) for w in w2.list()])

% # Verification
% print(decrypted_message == message)
% \end{verbatim}
% \end{tcolorbox}
% \end{figure}

\subsection{Closing thoughts on HE developments}

To end this section, we provide some closing thoughts on the developments of HE and refer the reader to some recent works in the field. 

First of all, the LWE and RLWE-based HE schemes presented in this section are natural extensions of the building block encryption schemes Regev (\Cref{subsec:regev scheme}) and LPR (\Cref{subsec:lpr}). They inherit and preserve the additive and multiplicative homomorphic properties of these building block encryption schemes.
The reason addition and multiplication are preserved is because, in all these encryption schemes, the ciphertext is constructed from the plaintext and the LWE / RLWE samples using simple linear algebra operations. Take the LPR encryption scheme as an example. Its ciphertext $(\vc{u},\vc{v})$ is created by computing
\begin{align*}
    \vc{u} &= \vc{b} \cdot \vc{r} + \vc{e}_1 + \floor{q/2} \cdot \vc{m} \bmod q\\
    \vc{v} &= \vc{a} \cdot \vc{r} + \vc{e}_2 \bmod q.
\end{align*}
The pair $\vc{u}$ without the message part and $\vc{v}$ are RLWE samples. 

Secondly, the schemes presented here followed just one narrow path of HE developments, which is also referred as the second generation of HE developments in \citet{halevi2017homomorphic}. However, their simplicity of not needing to perform bootstrapping to reach FHE within a pre-determined multiplication depth % let them be well recognized and often 
have led to some practical implementations, including % as practical HE schemes, in particular the BFV scheme. 
% Examples of some well known implementations include 
some standalone open-source libraries such as Microsoft \citep{sealcrypto2022}, IBM HElib \footnote{\url{https://github.com/homenc/HElib}}, PALISADE \footnote{\url{https://palisade-crypto.org/}}, NFLlib \footnote{\url{https://github.com/CryptoExperts/FV-NFLlib}}, and some open-source R and Python libraries such as \textit{HomomorphicEncryption} \citep{aslett2015review}, pyFHE \citep{erabelli2020pyfhe} and PySEAL \footnote{\url{https://github.com/Lab41/PySEAL}}. Although some of these libraries' documentations have recommended parameter choices to achieve efficient HE encryption for certain security levels, for standardized HE schemes, parameters definitions and selections, the reader is referred to the \textit{Homomorphic Encryption Standard} \citep{HomomorphicEncryptionSecurityStandard2018}.

Thirdly, although HE continues to attract tremendous attention among researchers and practitioners alike, % (including some groundbreaking works) since \citet{gentry2009fully}'s breakthrough of achieving fully HE using bootstrapping, 
its adoption in secure data computation is still not a mainstream affair.
% it is still not the mainstream for secure data computation. 
There are at least a few reasons for this. 
\begin{itemize}
    \item An important issue is the high space requirements for storing and processing the ciphertexts, which can be large even under the relatively efficient RLWE-based schemes. To encrypt even binary plaintexts, the ciphertext space $\Z_q$ or $R_q=\Z[x]/(\Phi(x))$ needs to be large enough to allow a decent number of homomorphic multiplications. The ciphertext space size is directly influenced by the modulus $q$, which then affects the bit length of ciphertexts. Under reasonable security parameters, the ciphertext size can be up to 100 times larger than the plaintext.
    
    \item A direct consequence of large ciphertexts is longer ciphertext computations, which is another limitation of HE's practicality. 
    
    \item An inherent limitation of HE is that conditional statements like $\mathit{if}\, x \,\mathit{then}\, y \,\mathit{else}\, z$ and $\mathit{while}\, x\, \mathit{do}\,y$ cannot be evaluated easily in encrypted space. In the $\mathit{if}\, x \,\mathit{then}\, y \,\mathit{else}\, z$ case, we cannot simplify the statement to either $y$ or $z$ in the encrypted space because while we can compute the encrypted value of $x$, we cannot know what it is. 
        Similarly, the $\mathit{while}\, x\, \mathit{do}\,y$ statement cannot be executed in encrypted space because we cannot know when to stop.
        % as part of algorithm conditions. 
        % Some direct consequences of this include unable to compute an integer raise to the power of a ciphertext and ciphertexts comparisons. 
        This limitation is inherited from the semantic security property of HE and cannot be solved within an HE scheme itself, although one can sometimes use secure multi-party computation techniques in combination with HE to evaluate these conditionals; see, for example, \citet{chialvaD18}.        % It has a big impact on how applicable HE is to existing computer science algorithms, which often contains equality test (e.g., $x=y?$), zero test (e.g., $x=0?$), comparison of two numbers, and so on.
    
    \item Many other common operations cannot be done efficiently or purely in HE. For example, statements like $x = y$ or $x < y$ usually can only be evaluated by turning $x$ and $y$ into suitable binary representations that are then processed using logical gates that can be evaluated in HE. Integer divisions, in particular, have proved difficult and known schemes like those in \citet{veugen14} require two-party protocols. 
\end{itemize}
Although there are now several significant niche applications of HE, all the above limitations make it challenging to run many existing algorithms on homomorphically encrypted data, sometimes turning linear-time algorithm to high polynomial algorithms.

An early paper by \citet{naehrig2011can} discussed some concrete application scenarios, where only somewhat HE schemes are sufficient to fulfil these applications, and experimentally argued the newly developed scheme (at the time) BV \citep{brakerski2012fully} was an efficient candidate. A decade later, there have been numerous contributions that advanced the development of HE schemes, including reduction of their computational cost and ciphertext size expansion, and permission of arithmetic operations over encrypted real and complex numbers. Besides these performance improvements, there has been an increasing trend, together with the explosion of other computer science areas (e.g., machine learning and artificial intelligence), of applying HE under the current state of affairs, especially when combining with optimized data processing techniques (e.g., single instruction, multiple data SIMD) or other cryptographic primitives, which remarkably improve HE's computational overhead. 

Some examples of more recent applications including HE's combination with secure multiparty computation to achieve efficient (less communication overhead) and secure arithmetic circuits computation \citep{damgaard2012multiparty}; with batching, hashing, modulus switching and other data processing optimization techniques for efficient private set intersection where one set's size is significantly smaller than the other \citep{chen2017fast}; predicting homomorphically encrypted data using neural networks with encoding and parallel computing techniques that are based Chinese Remainder Theorem \citep{gilad2016cryptonets}. More HE applications in training machine learning models (e.g., logistic regression, decision tree, naive Bayes, etc) or applying them on homomorphically encrypted data have surveyed in \citet{wood2020homomorphic}.

%\citet{cramer1997secure} is an example of HE (Elgamal encryption) for secure electronic voting. 
\subsection{A Sage Implementation of the BFV Cryptosystem}

We present an implementation of the BFV cryptosystem in Sage. Note that this implementation is intended for pedagogical purposes and is not suitable for use in real-world applications.

\subsubsection{Package Imports}
We begin by importing two generic packages.
\begin{tcolorbox}
\begin{verbatim}
import numpy as np
import sage.stats.distributions.discrete_gaussian_integer as dgi
\end{verbatim}
\end{tcolorbox}

\subsubsection{Define Parameters}
Recall that the BFV cryptosystem is defined in terms of several parameters. These determine the ring over which the cryptographic operations will be performed, how many messages will be operated on in each ``batch'', and the distribution from which noise will be drawn during encryption.
Here we define a suitable set of parameters that can be used to generate a secret key / public key pair, encrypt a message, and decrypt a ciphertext. Other parameters are required to perform the relinearization operations which are needed for homomorphic multiplication operations. See Section \ref{subsubsection:homomorphic_multiplication} for details.

\begin{tcolorbox}
\begin{verbatim}
# Define parameters for encryption/decryption
q = 6620830889
n = 1024
t = 83
delta = q//t

P = PolynomialRing(Integers(), name="x")
f = x^n + 1
R = QuotientRing(P, f)

sigma = 1.0
D = dgi.DiscreteGaussianDistributionIntegerSampler(sigma=sigma)

parameters = (q,n,t,R,D)
\end{verbatim}
\end{tcolorbox}

\newpage

\subsubsection{Utility Functions}
We will frequently use a symmetric representation of the rings $\mathbb{Z}/q\mathbb{Z}$. That is, we represent elements in this ring as integers $x$ where $-q/2 \le x < q/2$. The function \texttt{symmetrize} is used to compute these representations. We also need to perform the operation $\left\lfloor \frac{t}{q} [x]_q \right\rceil$ during decryption. The function \texttt{multiply\_round} performs this operation.

\begin{tcolorbox}
\begin{verbatim}
def symmetrize(a,b):
    '''
    Convert integer polynomial coefficients to the symmetric
    representation of elements in Z/bZ.
    '''
    A = np.array(vector(a))
    A = A % b
    mask = A >= b/2
    A[mask] -= b
    return R(list(A))

def multiply_round(x, r, parameters):
    '''
    Multiply integer coefficients by a rational number
    and then round to the nearest integer.
    '''
    q,n,t,R,D = parameters
    temp = r * vector(Rationals(), x)
    return R([k.round() for k in temp])
\end{verbatim}
\end{tcolorbox}

\subsubsection{Noise Samplers}
Here we define functions to draw random values from various distributions. To generate keys for BFV encryption/decryption operations we use \texttt{sample\_e} to draw a random element from the error distribution $D$, \texttt{sample\_2} to sample an $n$-long binary vector,  and \texttt{sample\_r} to draw a random element of the ring $\mathbb{Z}[x]\,\big/\,(x^n + 1)$.
\begin{tcolorbox}
\begin{verbatim}
def sample_e(n,D):
    P = PolynomialRing(Integers(), name="x")
    f = x^n + 1
    R = QuotientRing(P, f)
    return R([D() for _ in range(n)])

def sample_2(n):
    P = PolynomialRing(Integers(), name="x")
    f = x^n + 1
    R = QuotientRing(P, f)
    return R([randint(0,1) for _ in range(n)])

def sample_r(n):
    P = PolynomialRing(Integers(), name="x")
    f = x^n + 1
    R = QuotientRing(P, f)
    return R.random_element()
\end{verbatim}
\end{tcolorbox}

\newpage

\subsubsection{Basic Cryptographic Operations}
Here we define functions to generate key pairs, encrypt messages, and decrypt ciphertexts. In the text, we will use the symbol $\mathcal{E}$ to represent encryption and $\mathcal{D}$ to represent decryption.
\begin{tcolorbox}
\begin{verbatim}
# Functions for encryption/decryption.
def generate_keys(parameters):
    q,n,t,R,D = parameters
    secret_key = sample_2(n)
    a = symmetrize(sample_r(n), q)
    e = symmetrize(sample_e(n, D), q)
    b = symmetrize(-(a*secret_key + e), q)
    public_key = (b,a)
    return secret_key, public_key

def encrypt(message, public_key, parameters):
    q,n,t,R,D = parameters
    delta = q//t
    b,a = public_key
    r = sample_2(n)
    e1 = sample_e(n,D)
    e2 = sample_e(n,D)
    u = symmetrize(b*r + e1 + delta*message, q)
    v = symmetrize(a*r + e2, q)
    return (u,v)

def decrypt(ciphertext, secret_key, parameters):
    q,n,t,R,D = parameters
    u,v = ciphertext
    temp = symmetrize(u + v*secret_key, q)
    temp = multiply_round(temp, t/q, parameters)
    return symmetrize(temp, t)
\end{verbatim}
\end{tcolorbox}

\paragraph{Usage Example}
Here we demonstrate how to generate keys, encrypt a random message, and decrypt the resulting ciphertext. We verify that $\mathcal{D}(\mathcal{E}(m, k_p), k_s) = m$ for a given public key $k_p$, secret key $k_s$, and a random message $m$.
\begin{tcolorbox}
\begin{verbatim}
# Usage example and verification of correctness
secret_key, public_key = generate_keys(parameters)

message = R([randrange(0,t) for i in range(n)])
ciphertext = encrypt(message, public_key, parameters)
decrypted_message = decrypt(ciphertext, secret_key, parameters)

print(symmetrize(message, t) == decrypted_message)
\end{verbatim}
\end{tcolorbox}

\newpage

\subsubsection{Homomorphic Addition}
We define the function $f$ that combines two ciphertexts, $c_1 = \mathcal{E}(m_1, k_p)$ and $c_2 = \mathcal{E}(m_2, k_p)$, such that $\mathcal{D}(f(c_1, c_2), k_s) = m_1 + m_2$.
\begin{tcolorbox}
\begin{verbatim}
def add_ciphertexts(c1, c2):
    u1,v1 = c1
    u2,v2 = c2
    u_sum = symmetrize(u1 + u2, q)
    v_sum = symmetrize(v1 + v2, q)
    return (u_sum, v_sum)
\end{verbatim}
\end{tcolorbox}

\paragraph{Usage Example}
We verify that if $m_1$ and $m_2$ are random messages and $c_i = \mathcal{E}(m_i, k_p)$, then $\mathcal{D}(f(c_1, c_2), k_s) = m_1 + m_2$.
\begin{tcolorbox}
\begin{verbatim}
message_1 = R([randrange(0,t) for i in range(n)])
message_2 = R([randrange(0,t) for i in range(n)])
ciphertext_1 = encrypt(message_1, public_key, parameters)
ciphertext_2 = encrypt(message_2, public_key, parameters)

ciphertext_sum = add_ciphertexts(ciphertext_1, ciphertext_2)
decrypted_message_sum = decrypt(ciphertext_sum,
                                secret_key,
                                parameters)

message_sum = symmetrize(message_1 + message_2, t)
print(message_sum == decrypted_message_sum)
\end{verbatim}
\end{tcolorbox}

\subsubsection{Homomorphic Multiplication} \label{subsubsection:homomorphic_multiplication}
We define a function $g$ that combines two ciphertexts $c_1 = \mathcal{E}(m_1, k_p)$ and $c_2 = \mathcal{E}(m_2, k_p)$, such that $\mathcal{D}(g(c_1, c_2), k_s) = m_1 \cdot m_2$. 
\begin{tcolorbox}
\begin{verbatim}
def multiply_ciphertexts(c1,c2,parameters):
    '''
    Compute product of two ciphertexts in the ciphertext domain.

    This produces a three-coefficient ciphertext that cannot
    be decrypted using the standard decryption function.
    '''
    q,n,t,R,D = parameters
    u1,v1 = c1
    u2,v2 = c2
    temp = multiply_round(u1*u2, t/q, parameters)
    hh0 = symmetrize(temp, q)
    temp = multiply_round(u1*v2 + u2*v1, t/q, parameters)
    hh1 = symmetrize(temp, q)
    temp = multiply_round(v1*v2, t/q, parameters)
    hh2 = symmetrize(temp, q)
    return (hh0, hh1, hh2)
\end{verbatim}
\end{tcolorbox}

\newpage

\subsubsection{Relinearization}
We implement Method 2 described above to relinearize the product of two ciphertexts. The relinearization operation is defined in terms of both the parameters used for the basic BFV cryptographic operations described above and two additional parameters, a second (large) prime number $p$ and a second noise distribution $D2$.
\begin{tcolorbox}
\begin{verbatim}
p = 655360001
sigma2 = 2.0
D2 = dgi.DiscreteGaussianDistributionIntegerSampler(sigma=sigma2)
\end{verbatim}
\end{tcolorbox}

We define a function that generates the relinearization key $k_r$ and another function $\mathcal{L}$ that applies $k_r$ to convert a three-coefficient ciphertext product into a two-coefficient ciphertext that can be decrypted.
\begin{tcolorbox}
\begin{verbatim}
def generate_relinearlization_key(secret_key, parameters, p, D2):
    q,n,t,R,D = parameters
    a = symmetrize(sample_r(n), p*q)
    e = symmetrize(sample_e(n, D2), p*q)
    b = symmetrize(-(a*secret_key + e) + p*secret_key^2, p*q)
    relinearization_key = (b,a)
    return relinearization_key

def relinearize(ciphertext_product,
                relinearization_key,
                parameters):
    q,n,t,R,D = parameters
    hh0, hh1, hh2 = ciphertext_product
    b,a = relinearization_key
    u = hh0 + multiply_round(hh2*b, 1/p, parameters)
    v = hh1 + multiply_round(hh2*a, 1/p, parameters)
    return (symmetrize(u, q), symmetrize(u, q))
\end{verbatim}
\end{tcolorbox}

\paragraph{Usage Example}
We verify that $\mathcal{D}\big(\mathcal{L}\big(g(c_1, c_2), k_r\big), k_s\big) = m_1 \cdot m_2$ for random messages $m_1$ and $m_2$ and their corresponding ciphertexts $c_1 = \mathcal{E}(m_1, k_p)$ and $c_2 = \mathcal{E}(m_2, k_p)$.
\begin{tcolorbox}
\begin{verbatim}
relinearization_key = generate_relinearlization_key(secret_key,
                                                    parameters,
                                                    p,
                                                    D2)

ciphertext_product = multiply_ciphertexts(ciphertext_1,
                                          ciphertext_2,
                                          parameters)
relinearized_ciphertext = relinearize(ciphertext_product,
                                      relinearization_key,
                                      parameters)
decrypted_message = decrypt(relinearized_ciphertext,
                            secret_key,
                            parameters)

message_product = symmetrize(message_1 * message_2, t)
print(message_product == decrypted_message_product)
\end{verbatim}
\end{tcolorbox}

% possibly at the cost of making a linear time algorithm cubic for example, let alone the extended running time of ciphertext computations. 
%%%%%%%%%%%%%%%%%%%%%%%%%%%%%%%%%%%%%%%%%%%%%%%%%%%%%%%%%%%%%%%%%%%%%%%%%%%%%%%%%%%%%%%%%%%%%%%%%%%
%%%%%%%%%%%%%%%%%%%%%%%%%%%%%%%%%%%%%%%%%%%%%%%%%%%%%%%%%%%%%%%%%%%%%%%%%%%%%%%%%%%%%%%%%%%%%%%%%%%

%%%%%%%%%%%%%%%%%%%%%%%%%%%%%%%%%%%%%%%%%%%%%%%%%%%%%%%%%%%%%%%%%%%%%%%%%%%%%%%%%%%%%%%%%%%%%%%%%%%

%\newpage
%\bibliography{references}
%\bibliographystyle{abbrvnat}

\newpage
\appendix
\section{Abstract Algebra}

\label{appen:abstract algebrac}

This section
%\footnote{This section is part of the work \textit{A Tutorial Introduction to Lattice-based Cryptography and Homomorphic Encryption} by the authors Yang Li, Kee Siong Ng, Michael Purcell from the School of Computing, Australian National University @2022.} 
introduces the basics of abstract algebra, including groups, rings, modules, fields, and ideals.  %These are essential concepts that before discussing algebraic number theory and its connection with lattices. 
The material covered are standard in algebra textbooks like \citet{artin}. 
For students who want to learn how to think about abstract algebra, we recommend \citet{alcock21}.

\subsection{Group theory}
\label{subsection:group theory}
There are at least two motivations to study group theory for lattice-based cryptography. First, more advanced algebraic structures such as rings and fields are build upon the concepts of groups. Second, it provides a different view of lattices which are additive subgroups of $\R^n$. 

\begin{definition}
A \textbf{group} $G=(S,\cdot)$ \reversemarginpar
\marginnote{Group}
is a set of elements together with a binary operator ``$\cdot$'' such that 
\begin{itemize}
    \item closed: for all $a,b \in S$, we have $a \cdot b \in S$,
    \item unique identity element: there exists a unique identity element $e \in S$ with respect to the binary operator,
    \item associative: for all $x, y, z \in S$, we have $(x\cdot y) \cdot z = x \cdot (y \cdot z)$,
    \item unique inverse element: for all $x \in S$, there exists an element $y \in S$ such that $x \cdot y = e$. %every element in $S$ also has an inverse in $S$.
\end{itemize}
\end{definition}
A group is an abstract algebraic structure. Elements in $S$ can be integers, fractions, matrices, functions, etc. The group operator can be addition, multiplication, matrix multiplication, function composition, etc. The pair forms a group as long as the four groups axioms are satisfied. 

When dealing with binary operators, one often wonders whether or not the same result will be produced if switching the order of the two inputs. That is, does $x \cdot y = y \cdot x$ for all $x, y \in S$? For some groups this is true, but not in general. For example, the condition is true for the additive group of integers $(\Z, +)$), but not the multiplicative group of $n \times n$ integer matrices $(M, \times)$. Such a property is called abelian or commutative. 

\begin{definition}
A group $(G,\cdot)$ is \textbf{abelian} (or \textbf{commutative}) if $x \cdot y = y \cdot x$ for all $x,y \in G$. 
\end{definition}
In cryptography, we almost always work with abelian groups such as the integer group or the polynomial group. 

The number of elements in a group can be finite or infinite. For groups with finitely many elements, we can definite the group order and element order as follows.

\begin{definition}
The \textbf{order} of a group $G$ is the number of elements in $G$.\reversemarginpar
\marginnote{Order}\index{order}
\end{definition}

\begin{definition}
For an element $a$ in a group $(G,\cdot)$, if there exists a positive integer $k$ such that $\underbrace{a \cdots a}_{k}=e$ is the group identity, then the element $a$ has \textbf{order} $k$. If no such an integer $k$ exists, then $a$ has infinite order. 
\end{definition}

Orders of groups and group elements are useful when working with finite groups. Every non-zero element in $(\Z, +)$ has infinite order. Let $\Z/3\Z=\{0,1,2\}$ be the group of integers modulo 3. The order of the group $(\Z/3\Z, +)$ is 3. The orders of the elements 0, 1, 2 are 1, 3, 3, respectively.

Some important examples of groups are: 
\begin{itemize}
    \item \textbf{Symmetric group $S_n$}: the set of all permutations of the indices $[n]:=\{1,\dots,n\}$. The group has order $|S_n|=n!$.
    
    \item \textbf{Cyclic group}: a group that is generated by a single element. For example, $(\Z, +)$ is an infinite cyclic group that is generated by $1$. Another example is $(\Z/n\Z, +)$ which is a finite cyclic group of order $n$ that is generated by $1$. The element $g \in G$ that generates the entire group $G$ is called a \textbf{generator}. The common notation is $G=\langle g\rangle $ or $G=C_n$ if $G$ has a finite order $n$.
    
    \item \textbf{Dihedral group $D_n$}: a group of symmetries - reflection $f$ and rotation $r$ - of a regular $n$-gon. For example, $D_4=\{e, f, r, r^2, r^3, fr, fr^2, fr^3\}$. The group operation is function composition. 
    
    \item \textbf{Klein four group $K_4$ or $V_4$} - a group of 4 elements in which each non-identity element has order 2 and the composition of two non-identity elements produces the third one. The Klein four group is isomorphic to the product of two cyclic groups of order 2, i.e., $V_4 \cong C_2 \times C_2$.
    
    %\item \textbf{free groups}
    
    %\item \textbf{simple groups}

\end{itemize}

\begin{definition}
Let $(G, \cdot)$ be a group. A subset $H$ of $G$ is a \textbf{subgroup} of $(G,\cdot)$ if $H$ forms a group with $G$'s operator. 
\end{definition}

Sometimes we omit the group operator for simplicity. An important type of subgroups is normal subgroup. 

\begin{definition}
Let $G$ be a group. A subgroup $N$ of $G$ is \textbf{normal} if $N$ is invariant under group conjugation.\reversemarginpar
\marginnote{Normal subgroup}\index{normal subgroup}
That is, for all elements $g \in G$ and all elements $h \in N$, we have $g^{-1}hg \in N$. 
\end{definition}

The notation for normal subgroups is $H \triangleleft G$ (or $H \trianglelefteq G$). Normal
subgroups are important because they partition a group $G$ into \textbf{cosets}, i.e.,
quotient group or factor group, which is important toward learning quotient rings. In addition, quotient groups regroup elements into non-overlapping classes which may help to reveal underlying structures of the original group that are difficult to be seen without the action of grouping.  

To introduce quotient groups, we first introduce equivalence relations, based on which group elements are put together. 

\begin{definition}
A binary relation $\sim$ on a set $S$ is said to be an \textbf{equivalence relation} if it satisfies the following axioms for all $a,b,c \in S$:
\begin{itemize}
    \item reflexive: $a \sim a$,
    \item symmetric: $a \sim b$ if and only if $b \sim a$,
    \item transitive: if $a \sim b$ and $b \sim c$, then $a \sim c$.
\end{itemize}
\end{definition}
% Equality ``$=$'' is an equivalence relation, but greater than ``$\ge$'' is not because it is not symmetric. The following definition of coset helps defining quotient groups easily. 

\begin{definition}
Given a subgroup $H$ of $G$, we can define a \textbf{left coset}\reversemarginpar
\marginnote{Left coset}\index{left coset} of $H$ in $G$ as the set of elements obtained by applying a fixed element of $G$ (under the group operation) on the left of $H$. That is,  for each element $g \in G$, the left coset of $H$ is 
\begin{equation*}
    gH = \{gh \mid h \in H\}. 
\end{equation*}
\end{definition}

The \textbf{right coset}\index{right coset} is defined respectively. Let $G=(\Z,+)$ and $H=(2\Z,+)$. The left
\reversemarginpar
\marginnote{\textit{Right coset}}
cosets of $H$ in $G$ are $0+2\Z$ and $1+2\Z$, because any additional cosets constructed by the other elements of $G$ will be identical to these two. We denote the cosets by $\bar{0}$ and $\bar{1}$, respectively. 

Each coset is an equivalence class with the equivalence relation ``belong to the same coset''. This can be checked easily. For elements $a,b \in G$, they belong to the same coset (i.e., $aH=bH$) if and only if $b^{-1}a \in H$. Given a normal subgroup $H \triangleleft G$, it divides $G$ into several equal-sized equivalence classes.

\begin{definition}
\reversemarginpar
\marginnote{Quotient group}\index{quotient group}
The \textbf{quotient group} of $G$ by a normal subgroup $H \triangleleft G$, denoted by $G/H$, is the set of cosets of $H$ in $G$.
\end{definition}

An important observation is that the set of cosets forms a group with the group operation in $G$. The identity element in the quotient group is precisely the normal subgroup $H$. That is why $G/H$ is called a quotient GROUP. For example, the set $\{\bar{0}, \bar{1}\}$ and addition form a group, in which $\bar{0}$ is the identity. It can be checked that the normal subgroup assumption is necessary because it ensures the set of cosets forms a group. This is not always true if $H$ is just an ordinary subgroup of $G$. %\textcolor{red}{Added proof?}

Given a subgroup $H$ of $G$, all cosets of $H$ have the same size, so we have a quantity, namely
\reversemarginpar
\marginnote{\textit{Index}}\index{index}
the \textbf{index} of $H$ in $G$ and denoted by $|G:H|$, that is defined as the number of
coset of $H$ in $G$. If $H$ is a normal subgroup of $G$, then the index $|G:H|=|G/H|$ is equal to the order of the quotient group. 

We sometimes have a function $f$ acts on a group $(G,\cdot)$ by mapping elements of $G$ to another set $H$. In that case, we would like to know whether or not the same group structure is preserved in $H$ by the function $f$. This function is formally defined as a group homomorphism.

\begin{definition}
A \textbf{homomorphism} \reversemarginpar
\marginnote{Group homomorphism}\index{homomorphism}\index{group homomorphism}
from a group $(G,\cdot)$ to a group $(H,*)$ is a function $f:G \rightarrow H$ such that for all elements $a, b \in G$ it holds that 
\begin{equation*}
    f(a \cdot b) = f(a) * f(b).
\end{equation*}
\end{definition}

In other words, the relationship between the two elements in $G$ are mapped to the relationship between the two corresponding elements in $H$. There are different types of group homomorphisms, depending on the function type and the function's codomain. The two important groups homomorphisms are isomorphisms and automorphisms. 

\begin{definition}
A homomorphism is called an \textbf{isomorphism}\reversemarginpar
\marginnote{Isomorphism}\index{isomorphism} if it is bijective. 
\end{definition}
If there is an isomorphism between two groups $(G,\cdot)$ and $(H,*)$, then they are isomorphic and denoted by $(G,\cdot) \cong (H,*)$. Isomorphisms are important because they tell you when two groups are identical. In addition, knowing one group will tell you everything about the other. An example of a group isomorphism is $f: (\R, +) \rightarrow (\R^+, \times)$ given by the function $f(x) = e^x$. A special case of isomorphism is between a group and itself, which we will see when introducing Galois theory. 

\begin{definition}
A homomorphism is called an \textbf{automorphism} if it is an isomorphism such that the domain and codomain are the same. That is, an isomorphism $f: G \rightarrow G$.
\end{definition}

\subsection{Ring theory}
\label{subsection:ring theory}
Unlike groups, rings are algebraic structures associate with two binary operators, addition and multiplication such that ring axioms are satisfied. 

\begin{definition}
A \textbf{ring} $R=(S,+, \times)$\reversemarginpar
\marginnote{Ring}\index{ring}
 is a set with two operations, namely addition and multiplication, such that the following ring axioms are satisfied:
\begin{itemize}
    \item $(S,+)$ is an abelian group under addition, 
    \item $(S,\times)$ is closed under multiplication, associative and contains the unique multiplicative identity $1$,
    \item multiplication is distributive with respect to addition, i.e., $a \times (b + c) = a \times b + a \times c$ for all $a, b, c\in S$.
\end{itemize}
\end{definition}

A ring $R$ is \textbf{commutative} (called commutative ring) if multiplication is also commutative in $R$. For example, the set of integers forms a commutative ring with integer addition and multiplication. However, none of the integers except 1 has a multiplicative inverse in the integer set. The set of $n \times n$ (real or integer) matrices forms a non-commutative ring with matrix addition and multiplication. Not all matrices have inverses. 
An important ring in lattice-based cryptography is the \textbf{ring of polynomials} or \textbf{polynomial ring} $\Q[x]$ or $\Z[x]$ with polynomial addition and multiplication as the ring operations. Again, not all polynomials in the ring $\Q[x]$ and $\Z[x]$ have inverses in the same ring. 

The pair $(S, \times)$ in a ring $R$ almost forms a multiplicative group, but it lacks of multiplicative inverses in general. Without multiplicative inverses (of non-zero elements), division cannot be carried out in rings. For this purpose, we introduce division rings. 
\begin{definition}
A \textbf{unit} in a ring $R$ is any element that has a multiplicative inverse in $R$.
\end{definition}
For example, $1$ is the only unit in the ring of integers. But 1, 2 are both units in the ring $(\mathbb{Z}_3,+,\times)$. 

\begin{definition}
A \textbf{division ring}\reversemarginpar
\marginnote{Division ring}\index{division ring}
 is a ring $R$ in which every non-zero element is a unit. That is, every non-zero element has a multiplicative inverse in $R$. 
\end{definition}

In a division ring, the pair $(S, \times)$ forms a multiplicative group, but not necessary abelian. If it is abelian, the ring is a field, which will be introduced in the next subsection. Similar to a group and its subgroups, subrings can be defined with respect to a ring. 

\begin{definition}
Let $(R,+,\times)$ be a ring. A subset $S \subset R$ is a \textbf{subring} if $(S,+,\times)$ forms a ring with the ring's addition and multiplication.
\end{definition}

The concept of a vector space can be generalized to a \textit{module} which is defined similarly, but over a ring instead of a field. The main difference is that every element in a field has a multiplicative inverse, so a vector in a vector space can be scaled up or down by a scalar and its multiplicative inverse. However, not every element in a ring has a multiplicative inverse, so an element in a module cannot always be scaled up and down. 

\begin{definition}
Let $R$ be a ring and 1 being its multiplicative identity. A \textbf{left $R$-module} $M$ \reversemarginpar
\marginnote{\textit{Module}}\index{module}
consists of an abelian group $(M, +)$ and an operation $\cdot: R \times M \rightarrow M$ such that for all $r,s \in R$ and $x,y\in M$, the following are satisfied:
\begin{itemize}
    \item $r \cdot (x+y) = r\cdot x + r \cdot y$
    \item $(r + s) \cdot x = r \cdot x + s \cdot x$
    \item $(rs) \cdot x = r \cdot (s \cdot x)$
    \item $1 \cdot x = x$
\end{itemize}
\end{definition}

The concept of a \textbf{right $R$-module} is defined similarly. The distinction between a left and right module arises from the fact that the underlying ring $R$ is not necessary commutative. In general, unless mentioned otherwise, we always refer a module to a left module. A $Z$-module is a module over the integer ring $Z$. It is both a left and right module as $Z$ is commutative. In \Cref{section:rlwe}, we will talk about the ring of integers of a number field. Without stating the proper definition here, the ring of integers is a the ring of all algebraic integers in a number field, where an algebraic integer is a root of an integer coefficient polynomial. It is not hard to see that the ring of integers form an abelian group under addition, as the sum of two algebraic integers is still an algebraic integer. For specific purposes, we often say the ring of integers is also a $\Z$-module, as the above conditions are all satisfied. 

\begin{definition}
Suppose $M$ is a left $R$-module and $N$ is a subgroup of $M$. Then $N$ is an \textbf{$R$-submodule} (or just \textbf{submodule}) if for any $n \in N$ and any $r \in R$, we have $r \cdot n \in N$.
\end{definition}
The definition of submodule is similar to subspace of a vector space, where the subspace is closed under addition and scalar multiplication. A important type of module is called a free module. 

\begin{definition}
A \textbf{free module} \reversemarginpar
\marginnote{\textit{Free module}}\index{free module}
is a module that has a basis. 
\end{definition}
Here a basis is a set of linearly independent vectors that generates $M$. That is, every element of $M$ can be written as a linear combination of the set of linearly independent vectors, where the coefficients are taken from the underlying ring $R$. So a \textbf{free $\Z$-module} is a module with a basis such that every element in the module is an integer combination of the basis.

\reversemarginpar
\marginnote{Ideals}
Similar to a normal subgroup, an ideal can partition a ring into cosets which form a ring with less elements, known as the  \textit{quotient ring}. As noted, not all subgroups can partition a group into a quotient group. Similarly, an ideal must have some special properties in order to construct a quotient ring. 

First, a ring is an additive group with an extra operation, an ideal of the ring should be a normal subgroup under addition (in fact, being a subgroup is enough as a ring is an abelian group under addition which implies normality), so an ideal must be closed under addition. Second, for cosets to be closed under multiplication, ideals must be closed under multiplication by any ring elements. 
More specifically, an ideal $I$ partitions a ring $R$ into a set of equivalence classes, each denoted by $[a] := a + I = \{a + r \mid r \in I\}$. Since we want this set of equivalence classes to form a ring, it must satisfy
\begin{itemize}
    \item $[a]+[b] = (a + I)+(b+I)=(a+b)+(I+I) = (a+b)+ I = [a+b]$ 
    \item $[a] [b] = (a+I)(b+I) = ab + aI + bI + I I = ab + I = [ab]$.
\end{itemize}
So we can see that ideals have to satisfy at least three criteria. First, closed under addition by itself. Second, closed under multiplication by itself. Third, closed under addition by all elements in the ring. Noted that the third criterion includes the second, so at least two criteria need to be satisfied. 
% In fact, it is enough to just meet these two criteria in order for mathematicians to prove some great results, e.g., $\text{Ideals}(\OO_K)$ = UFD. 
The formal definition of an ideal is stated as below. 

\begin{definition}
For an arbitrary ring $(R, +, \times)$, the subset $I \subset R$ is a \textbf{left ideal} of the ring if it satisfies: 
\begin{itemize}
    \item $(I, +)$ is an additive subgroup of the group $(R, +)$, 
    \item $I$ is closed under left multiplication by all elements of $R$. That is, for every $r \in R$ and every $x \in I$, their product $rx \in I$. 
\end{itemize}
\end{definition}
An \textbf{right ideal} is defined respectively. If $I$ is both a left and right ideals, then it is a two-sided ideal of the ring. Again, since most rings considered in cryptography are commutative, we do not distinguish left and right ideals. Throughout, we use the term ideals for two-sided ideals unless mentioned otherwise. 
For example, the set of even integers form an ideal in the integer ring, because even integers are closed under addition and any integer multiplied by an even integer is still even.  

Note that although an ideal is closed under addition and multiplication, it is not a ring because it does not necessary have a multiplicative identity, which is required by our definition of rings.% \footnote{Some define rings without requiring the existence of the multiplicative identity. These rings are sometimes denoted as RNG.}

Ideals can be generated by a set of elements $a_1,\ldots,a_n \in R$, denoted by
\[ (a_1,\ldots,a_n) = \{ r_1 a_1 + \cdots + r_n a_n : r_i \in R \}, \]
with the special case of $(a) = aR = Ra = \{ ra : r\in R\}.$
A \textbf{zero ideal} is an ideal contains only the zero element, i.e., $\{0\}$ or $(0)$. A \textbf{unit ideal} is the ring itself. A \textbf{proper ideal} is a non-unit ideal.

Intuitively, one can think of an ideal of a ring $R$ as a subset of $R$ that absorbs $R$, so it is closed under addition, and multiplication by ring elements. Ideal is an important concept that will frequently appear in lattice-based cryptography. It helps to build a quotient ring or even a field if the ideal used is maximal. This is similar to the construction of quotient groups via normal subgroups.

\begin{definition}
\label{def:quoRng}
\reversemarginpar
\marginnote{Quotient ring}\index{quotient ring}
The \textbf{quotient ring} of a ring $R$ by an ideal $I$, denoted by $R/I$, is the set of cosets of $I$ in $R$.
\end{definition}
The quotient ring $R/I$ has the additive identity $\Bar{0}=0+I$ (similar to a normal subgroup being the identity of the quotient group) and the multiplicative identity $\Bar{1}=1+I$.

Some ideals have additional properties that can make the corresponding quotient rings special. 
Below we introduce three special ideals.

\begin{itemize}
    \item Prime ideal $\rightarrow$ integral domain
    \item Principal ideal $\rightarrow$ principal ideal domain
    \item Maximal ideal $\rightarrow$ (residual) field
\end{itemize}

A prime ideal can be thought as a generalization of a prime number. Recall that if $p$ is a prime number and $p | ab$ for integers $a$ and $b$, then either $p | a$ or $p | b$. 
\begin{definition}
An ideal $P$ of a ring $R$ is \textbf{prime}\reversemarginpar\marginnote{Prime ideal}\index{prime ideal}
 if it satisfies the following two properties:
\begin{itemize}
    \item $P \neq R$, 
    \item for any two elements $a,b \in R$, if their product $ab \in P$, then either $a \in P$ or $b \in P$.
\end{itemize}
\end{definition}
The set of even integers in the ring of integers is a prime ideal.
To see why prime ideals are important, we introduce the concept of integral domains that are defined upon commutative rings.  

\begin{definition}
An \textbf{integral domain}\reversemarginpar
\marginnote{Integral domain}\index{integral domain}
 is a non-zero commutative ring in which the product of two non-zero elements is non-zero. 
\end{definition}
Integral domains are generalizations of the rings of integers of algebraic number fields that will be discussed in a later section. Integral domains provide a natural setting to study division, because they allow the cancellation of a non-zero factor $a$ in an equation like $ab=ac$. 
%Integral domain has the same property but for more abstract elements. 
%Being a field implies its an integral domain. 

\begin{proposition}
If $I \subsetneq R$ is a prime ideal, then the quotient ring $R/I$ is an integral domain.
\end{proposition}
\begin{proof}
$I$ being a prime ideal implies that no two elements that are not in $I$ can be multiplied to an element in $I$. Since $I$ is the additive identity in the quotient ring $R/I$, it is the zero element in the quotient ring. This implies that no two non-zero elements (i.e., elements not in $\bar{0}$) can be multiplied to a zero element (i.e., an element in $\bar{0}$). 
\end{proof}
For example, $12\mathbb{Z}$ is not a prime ideal, so the quotient ring $\mathbb{Z}/12\mathbb{Z}$ is not an integral domain because $3 \cdot 4 = 12 = 0 \bmod 12$. But $\mathbb{Z}/5\mathbb{Z}$ is an integral domain. 
Another example is the ring of polynomials whose coefficients come from an integral domain. 

\begin{proposition}
If $R$ is an integral domain, then the ring of polynomials $R[x]$ is also an integral domain.
\end{proposition}
\begin{proof}
$R$ is integral domain, the product of the leading coefficients of two non-zero polynomials is also non-zero, so $R[x]$ is an integral domain. 
\end{proof}

%Since integral domain is a generalization of the integers, we can generalize the concept of \textbf{prime} to an integral domain. 
%\begin{definition}
%An element is \textbf{prime} in an integral domain if it is not the zero element nor a unit and if %divides the product of two elements in the integral domain, then it divides at least one of them. 
%\end{definition}

\begin{definition}
An ideal in a ring $R$ is \textbf{principal} \reversemarginpar
\marginnote{Principal ideal}
if it can be generated by a single element of $R$ through multiplication by every element of $R$.  
\end{definition}
For example, $2\Z$ is a principle ideal in the integer ring, because it can be generated by $2$ multiplying every element of $\Z$. 

\begin{definition}
A \textbf{principal ideal domain (PID)} is an integral domain in which every ideal is principal. 
\end{definition}

As will be explained in detail later, fields are commutative division rings that possess nice properties for building cryptosystems. Given a ring $R$, one can construct a field by taking the quotient ring with a maximal ideal of $R$. 
\begin{definition}
A \textbf{maximal ideal}\reversemarginpar
\marginnote{Maximal ideal}\index{maximal ideal}
 in a ring is an ideal that is maximal among all the proper ideals of the ring. 
\end{definition}
In other words, if $I$ is a maximal ideal in a ring $R$, then $I$ is contained in only two ideals of $R$, i.e., $I$ itself and the entire ring $R$. 
An important observation is that every maximal ideal is a prime ideal. This can be easily seen if we define the divisibility of ideals.

\begin{proposition}
\label{prop:quotRngIsField}
If $I$ is a maximal ideal of a commutative ring $R$, then the quotient ring $R/I$ is a field. 
\end{proposition}
\begin{proof}
(Sketch) $I$ being a prime ideal is not sufficient to construct a field. Because the quotient ring $R/I$ may have a proper ideal that is not the trivial ideal. That is, there may be an ideal $I'$ in $R/I$ that is not equal to $\{0\}$ or $R/I$. Hence, multiplication of an element in $I'$ by an element not in $I'$ will only get to elements in $I'$. This implies that not all non-zero elements in $R/I$ have multiplicative inverses. 
\end{proof}

The quotient ring $R/I$ constructed using the maximal ideal is called a \textbf{residual field}.

Another concept that will be mentioned later and could help to understand the structure of fields are the characteristic of a ring. 
If it helps, the characteristic of a ring can be thought as the cyclic period of a ring. For example, the ring $\Z/4\Z$ has a characteristic 4 which is the rings cyclic period. 

\begin{definition}
The \textbf{characteristic} of a ring $R$,\reversemarginpar\marginnote{Characteristic}\index{characteristic!of a ring} denoted by $char(R)$, is the smallest number of times that the ring's multiplicative identity 1 can be added to itself to get the additive identity 0. If the ring's multiplicative identity can never be summed to get 0, then the ring has a characteristic zero. 
\end{definition}
The characteristic of a ring $R$ may also be taken as the smallest positive integer $n$ such that $\underbrace{a + \dots + a}_n = 0$ for every element $a \in R$ (if the characteristic exists). For example, the characteristic of $\Z_3$ is 3 because $1+1+1=3 = 0 \bmod 3$ or $2+2+2=6 = 0 \bmod 3$. We will talk more about the characteristics of fields in the following subsection. 

% So far, we have not defined arithmetic operations (e.g., sum and product) on ideals, because they are not essential to understand the rest of this section. We will say a bit more about ideal operations when discussing ideals of number fields. 

The First Isomorphism Theorem for rings is the fundamental method for identifying quotient rings. 
In the below, ring homomorphism\index{ring homomorphism}\index{homomorphism!ring} is defined analogously to group homomorphism\index{group homomorphism}\index{homomorphism!group}, and the kernel\reversemarginpar
\marginnote{kernel}\index{kernel}
 of a map $\varphi : R \to S$ is the subset of $R$ that map to the zero element in $S$: $ker(\varphi) = \{ r \in R : \varphi(r) = 0 \}$.
\begin{theorem}\label{thm:first isomorphism theorem}
Let $R$ and $S$ be rings and let $\varphi: R \to S$ be a ring homomorphism. Then \reversemarginpar\marginnote{First Isomorphism Theorem}\index{First Isomorphism Theorem}
\begin{enumerate}\itemsep1mm\parskip0mm
    \item the kernel\index{kernel} of $\varphi$ is an ideal of $R$;
    \item the image of $\varphi$ is a subring of $S$; and
    \item $R/ker(\varphi)$ is isomorphic to the image of $\varphi$.  
\end{enumerate}
\end{theorem}

\subsection{Field theory}
\label{subsection:field theory}

A field is a commutative division ring. That is, a field is a ring if $(S^*,\times)$ is an abelian group under multiplication, where $S^* := S \setminus \{0\}$ is the set of non-zero elements. More formally, we have the next definition.

\reversemarginpar
\marginnote{Field}
\begin{definition}
A \textbf{field} $F=(S, +, \times)$ is a set with two binary operators, addition and multiplication, such that the following field axioms are satisfied:  
\begin{itemize}
     \item $(S,+)$ is an abelian group under addition, 
    \item $(S^*,\times)$ is an abelian group under multiplication,
    \item multiplication is distributive with respect to addition, that is, $a \times (b + c) = a \times b + a \times c$ for all $a, b, c\in S$.
\end{itemize}
\end{definition}
Examples of fields are the field of rational numbers, real numbers and complex numbers. The smallest field is $\F_2 =\Z/2\Z = \{0,1\}$, because a field must contain at least two distinct elements 0 and 1. 

A field is an integral domain, because non-zero elements have multiplicative inverses, which eliminates the possibility that their product is zero. 

Sometimes, it is easier to construct a field from a given commutative ring rather than build it from scratch. One can construct a field from a commutative ring in two ways, by building the field of fractions or by quotienting the commutative ring by a maximal ideal as discussed earlier in Proposition \ref{prop:quotRngIsField}. 

\reversemarginpar
\marginnote{Field of fractions}
\begin{definition}
Let $R$ be an integral domain. The \textbf{field of fractions} $Frac(R)$ is the set of equivalence classes on $R \times (R \setminus \{0\})$ defined by 
\begin{equation*}
    Frac(R) = \{(p,q) \in R \times (R \setminus \{0\}) \mid (p,q) \sim (r,s) \iff ps = qr\}.
\end{equation*}
\end{definition}

This definition generalizes the idea of creating fractions from integers. For example, if $R=\mathbb{Z}$ then $\frac{p}{q} \in [(p,q)] \subseteq Frac(\mathbb{Z}) = \mathbb{Q}$. More precisely, let $p=5,q=20$ then $5/20$ is an element in the equivalence class consists of  $\{1/4,5/20,25/100,\dots\}$, which is also called the set of all equivalent fractions. The reason for $R$ being an integral domain is because we can have the usual addition and multiplication in the field of fractions without running into the trouble of having a zero divisor. For example, $\frac{a}{b}+\frac{c}{d} = \frac{ad+bc}{bd}$, since $R$ is an integral domain it is guaranteed that $bd \neq 0$.

%Below is a result that is related to Proposition \ref{proposition:residual field}. That is, if a commutative ring is quotient by its maximal ideal, the resulting quotient ring is a field. More importantly, this field has no other ideals besides the zero ideal and the unit ideal.    
\begin{proposition}
A non-zero commutative ring $R$ is a field if and only if it has no ideals other than $(0)$ and $R$.
\end{proposition}
\begin{proof}
If $R$ is a field, then every non-zero element has a multiplicative inverse. If $I$ is a non-zero ideal of $R$ and $a \in I$, then $a^{-1} a = 1 \in I$. So $I=R$. If $R$ has no proper non-zero ideal, then the ideal $I=R$ is a principal ideal. That is, $I=(a)$ for $a \neq 0$. Hence, there must exist an element $b \in R$ such that $ab=1$. Hence, $R$ is a field. 
\end{proof}

This proposition implies an important property of a field: its only ideals are the zero ideal and the field itself. 

One type of fields that is essential in cryptography is called \textbf{finite fields}.\reversemarginpar
\marginnote{Finite field}\index{finite field} These are fields with finitely many elements. The number of elements in a finite field is the \textbf{order} of the field (just like the order of a group). For example, $\Z_2 = \{0,1\}$ is a finite field of order 2.

Field characteristics\index{characteristics!of a field} is an important concept that can be used to decide the separability of extension fields. We will see more about the connection between field characteristic and separability in a later section.  

\reversemarginpar
\marginnote{$Char(F)=0$ or prime}

\begin{lemma}
The characteristic of any field is either 0 or a prime number. 
\end{lemma}
\begin{proof}
Let $n$ be the characteristic of the field $F$. It is easy to see that $n \neq 1$, because a field is not a trivial ring, so $1 \neq 0$. Assume $n=pq$ is a composite number, where $1 <p,q<n$. This implies that $\underbrace{(1+\cdots + 1)}_p \underbrace{(1+\cdots + 1)}_q=\underbrace{1+\cdots + 1}_n=0$. Hence, we have $pq=0$ which contradicts with the fact that the field is also an integral domain.
\end{proof}

\begin{corollary}
This lemma implies that the characteristic of any finite field is a prime number. 
\end{corollary}

\begin{corollary}
The characteristic of a subfield is the same as the characteristic of the field. 
\end{corollary}

\begin{theorem}
In a field of characteristic $p$ where $p$ is prime,\index{prime characteristic} the only $p$-th roots of unity is 1.
\end{theorem}
In a field of prime characteristic $p$, we have $x^p-1 = (x-1)^p$ because after expanding $(x-1)^p$, all terms except $x^p$ and $-1^p$ have coefficients that are multiples of $p$, which vanish when taking modulo $p$. Hence, solving $x^p-1 = 0$ is equivalent to solving $(x-1)^p = 0$, where the only solution is $x=1$.

So far in this section, we have introduced the concepts of groups, rings, fields and other related concepts. These will serve as a foundation for studying the Galois theory and algebraic number theory. 

%%%%%%%%%%%%%%%%%%%%%%%%%%%%%%%%%%%%%%%%%%%%%%%%%%%%%%%%%%%%%%%%%%%%%%%%%%%%%%%%%%%%%%%%%%%%%%%%%%%
%%%%%%%%%%%%%%%%%%%%%%%%%%%%%%%%%%%%%%%%%%%%%%%%%%%%%%%%%%%%%%%%%%%%%%%%%%%%%%%%%%%%%%%%%%%%%%%%%%%

%%%%%%%%%%%%%%%%%%%%%%%%%%%%%%%%%%%%%%%%%%%%%%%%%%%%%%%%%%%%%%%%%%%%%%%%%%%%%%%%%%%%%%%%%%%%%%%%%%%

%\newpage
%\bibliography{references}
%\bibliographystyle{abbrvnat}

\newpage
\section{Galois Theory}

%\section{Galois Theory (medium)}
\label{appen:galois theory}

In the previous section, we have introduced some basics about group, ring and field theories. We start this section 
%\footnote{This section is part of the work \textit{A Tutorial Introduction to Lattice-based Cryptography and Homomorphic Encryption} by the authors Yang Li, Kee Siong Ng, Michael Purcell from the School of Computing, Australian National University @2022.}
by introducing field extension that is fundamental to understand number field. All things lead to the Galois group in the end, which is interesting in itself as well as gives insights of cyclotomic number field that is widely used across recent lattice-based cryptography and homomorphic encryption developments.

\subsection{Field extension}
\label{subsection:field extension}
The concept of field extensions is fundamental in solving polynomials, especially polynomials with rational coefficients, denoted by $\Q[x]$. The first attempt to solve these polynomials is to find  their roots in the field of rationals $\Q$. For some rational (coefficient) polynomials, however, their roots only exist beyond $\Q$. For example, the polynomial $x^2-2$ has two irrational roots $\pm \sqrt{2}$. For this reason, we need to construct a field that is larger than $\Q$ so that it includes all roots of the polynomial $x^2-2$, but not too large that includes many unnecessary values. To achieve this goal, we first define extension fields.      

\begin{definition}
If a field $F$ is contained in a field $E$, then $E$ is called an \textbf{extension field} of $F$. 
\end{definition}
If $E$ is an extension (field) of $F$, then $F$ is a \textbf{subfield} of $E$. This pair of fields is called a \textbf{field extension} and denoted by $E/F$.\reversemarginpar
\marginnote{\textit{Field extension}}\index{field extension}

For the above example $x^2-2$, we can \textbf{adjoin} to $\Q$ the roots of this polynomial  to get a  larger field that includes all the roots of $x^2-2$, denoted by $\Q(\pm \sqrt{2}) := \{ a \pm b\sqrt{2} \,:\, a,b \in \Q\}$. Note that since an extension field is also a field, it is sufficient to adjoin only $\sqrt{2}$. Being a field also implies the extension $Q(\sqrt{2})$ includes more elements such as $1+\sqrt{2}$, $5\sqrt{2}$ and so on. 

\reversemarginpar
\marginnote{\textit{$F$-vector space}}
Given a field extension $E/F$, the larger field $E$ forms a vector space over $F$, which is also known as an \textbf{$F$-vector space}. The larger field $E$ consists of the ``vectors'' in the vector space and the smaller field $F$ consists of the scalars for multiplying with the vectors. For example, $\Q(\sqrt{2})$ forms a $\Q$-vector space, because the extension $\Q(\sqrt{2})$ is closed under addition (satisfying commutativity, associativity, additive identity and inverse) and scalar multiplication with $\Q$ (satisfying compatibility, scalar identity in $\Q$, distributivity of scalar multiplication w.r.t. scalar addition in $\Q$ or addition in $\Q(\sqrt{2})$). 

\reversemarginpar
\marginnote{\textit{Field extension degree}}
Since an extension forms a vector space over the base field, it makes sense to talk about the degree of an extension. 
\begin{definition}
Give a field extension $E/F$, the \textbf{degree} of the extension field $E$, denoted by $[E:F]$, is the dimension of the vector space formed by $E$ over $F$.
\end{definition}
An extension $E$ is \textbf{finite} if its degree is finite. Otherwise, it is infinite. 
There are at least two ways of counting the dimension of an extension. One way is through the degree of the minimal polynomial of a primitive element that generates the extension. This will be discussed in more detail in subsequent subsections.

The other way of counting the dimension of the extension field is by counting the number of linearly independent vectors in its basis (same as for vector spaces in linear algebra). Hence, one could specify a basis of the extension over the base field in order to get the degree of the extension. For example, the degree $[\mathbb{Q}(\sqrt{2}):\mathbb{Q}]=2$,  $[\mathbb{Q}(\sqrt{2},\sqrt{3}):\mathbb{Q}]=4$, $[\C:\mathbb{R}]=2$ because the corresponding basis for each  extension field is $\{1, \sqrt{2}\}, \{1, \sqrt{2},\sqrt{3},\sqrt{6}\}, \{1,i\}$ respectively.

Similar to Lagrange's theorem in group theory, the degrees of extensions follow the ``Tower Law''. 
\begin{proposition} (The Tower Law)
\label{prop:tower law}
If $L/M$ and $M/K$ are field extensions (finite or infinite), then the degrees of the extensions satisfy 
\begin{equation*}
    [L:K] = [L:M][M:K].
\end{equation*}
\end{proposition}
Intuitively, $L$ forms a $M$-vector space and $M$ forms a $K$-vector space, so $L$ also forms a $K$-vector space. Each dimension in $L$ over $M$ is again a $[M:K]$-dimensional vector space.  

The following subsections introduce some special types of field extensions that eventually lead to Galois extensions and Galois groups. 

\subsubsection{Algebraic extension}
Historically, solving mathematical equations with rational coefficients was a natural but challenging task. This lead to the definition of algebraic numbers that are roots of non-zero rational polynomials. More formally, 

\reversemarginpar
\marginnote{\textit{Algebraic number}}
\begin{definition}
A complex number is \textbf{algebraic} (over the rationals $\mathbb{Q}$) if it is a root of a non-zero polynomial whose coefficients are rational numbers. That is, $r \in \C$ is an algebraic number if it satisfies $f(r)=0$ for some non-zero polynomial $f(x) \in \Q[x]$.
\end{definition}

All rational numbers are algebraic because they can be written in a linear equation $x-r$ for all $r \in \Q$. The irrational number $\sqrt{2}$ is algebraic because it is a root of $x^2-2$. The complex number $i$ is also algebraic because it is a root of $x^2+1$. Complex numbers that are not algebraic are called \textbf{transcendental}. In other words, transcendental numbers are not roots of any rational coefficient polynomials. For example, the number $\pi$ or $e$. 

Almost all real numbers are not algebraic. The set of real numbers is uncountable, but the set of algebraic numbers are countable. That is, there is a one-to-one correspondence between all the algebraic numbers and the natural numbers. 

When developing cryptosystems, we almost always work with integer (coefficient) polynomials $\Z[x]$. Within $\Z[x]$, monic polynomials are of special interest due to their computational efficiency. A polynomial is \textbf{monic} if the coefficient of its leading term (i.e., the term with the highest degree) is one. For example, when dividing polynomials, it is convenient to work with integer polynomials with leading coefficient one. In most cases, we work with polynomials defined over a field (e.g., $\Z_p[x]$ for prime $p$), so even if it is not monic, it can always made monic by dividing its coefficients with the leading term's coefficient. %\textcolor{red}{This is contradictory.  We say that we almost always work in $\mathbb{Z}[x]$ but then say that we almost always work over polynomials defined in a field.} KL: What I said was a bit confusing, by filed I mean Z_p, so still integer coefficients, but from a field, rather than a ring Z.

\reversemarginpar
\marginnote{\textit{Algebraic integer}}
\begin{definition}
A complex number is an \textbf{algebraic integer} if it is a root of a monic polynomial with integer coefficients. 
\end{definition}

Algebraic integers are generalization of ordinary integers which we call rational integers. 
Similar to numbers, field extensions can be algebraic or transcendental too. 

\reversemarginpar
\marginnote{\textit{Algebraic extension}}
\begin{definition}
A field extension $E/F$ is \textbf{algebraic} if every element in the extension field $E$ is algebraic. 
\end{definition}
Since all rational numbers are algebraic, a field extension $\Q(\alpha)$ is algebraic if all the additional elements are algebraic. 

All transcendental extensions are of infinite degree. For example, the transcendental extension $Q(\pi)$ has a basis $\{1, \pi, \pi^2, \pi^3, \dots\}$ of infinite linearly independent vectors. The above statement also implies that all finite extensions are algebraic. This is also proved in the following proposition.  

\begin{proposition}
\label{prop:finite implies algebra}
Every finite extension is algebraic.
\end{proposition}
\begin{proof}
Let $E$ be an extension over $F$ with a finite degree $[E:F]=n$. For an element $x \in E$, the elements $1, x, x^2, \dots, x^n \in E$ because $E$ is a field. These $n+1$ elements are also in the $n$-dimensional vector space over $F$, so must be linear dependent. Hence, there exists a set of non-zero coefficients $\{a_0,\dots,a_n\}$ such that $1+a_1 x+a_2 x^2+\cdots+a_n x^n=0$. This implies that $x$ is algebraic. 
\end{proof}

\reversemarginpar
\marginnote{\textit{Algebraic closed}}
\begin{definition}
A field $F$ is \textbf{algebraically closed} if for any polynomial $f(x) \in F[x]$, all of its roots are in the field $F$. 
\end{definition}
Obviously $\mathbb{Q}$ and $\mathbb{R}$ are not algebraically closed, but $\mathbb{C}$ is. This is the \textbf{Fundamental Theorem of Algebra}. It implies that all polynomials can be completely solved or factored into linear factors in the complex field $\mathbb{C}$. 

As mentioned earlier, given a field extension $\Q(r)/\Q$, another way of identifying the degree of the extension is by identifying the degree of the minimal polynomial of $r$ over $\Q$. To finish off this subsection, we define what minimal polynomial is.

\begin{definition}
A polynomial $f(x) \in F[x]$ is \textbf{reducible} over the field $F$ if it can be factored into polynomials with smaller degrees. Otherwise, it is \textbf{irreducible}.\reversemarginpar
\marginnote{\textit{Irreducible polynomial}}\index{irreducible polynomial}
\end{definition}

\begin{example}
Given the following polynomials over the field of rationals $\Q$: 
\begin{align*}
    f_1(x)&=x^2+4x+4=(x+2)(x+2),\\
    f_2(x)&=x^2-4=(x+2)(x-2),\\
    f_3(x)&=9x^2-3=(3x+\sqrt{3})(3x-\sqrt{3}),\\
    f_4(x)&=x^2+1=(x+i)(x-i),
\end{align*}
the polynomials $f_1(x)$ and $f_2(x)$ are reducible over $\Q$ whilst the other two are irreducible over $\Q$. The polynomials $f_3(x)$ and $f_4(x)$ are reducible over $\R$ and $\C$, respectively. The polynomial $f_4(x)$ is irreducible over $\R$.

\end{example}
%\kl{The irreducibility of a function can be check by Eisenstein’s Criterion. Details are needed?}

\begin{theorem}
\label{thm:quoRngIsField}
Let $p$ be a prime and $f(x) \in \F_p[x]$ be a monic irreducible polynomial of degree $n$. The quotient ring $\F_p[x]/f(x)$ is a field of order $p^n$. (Each polynomial in $\F_p[x]/f(x)$ has coefficients taken from the field $\F_p$ and the polynomial degree is at most $n-1$.)
\end{theorem}
% When interpreting the ring of polynomials $\F_p[x]/f(x)$, it means each polynomial in this quotient ring has coefficients taken from the field $\F_p$ and the polynomial degree is at most $n-1$. 
% We will see this notation  more often when introducing \textit{Algebraic Number Theory}, which has a different interpretation. 
% Theorem~\ref{thm:quoRngIsField} is needed when defining the underlying ring of the ring LWE problem\index{ring LWE}.
\begin{proof}
Each coset in the quotient ring $\F_p[x]/f(x)$ has the form $a_0 + a_1 x + \cdots + a_{n-1} x^{n-1}$, where $a_i \in \F_p$. So there are $p^n$ different cosets. The polynomial $f(x)$ is irreducible implies the quotient ring is also a field. 
\end{proof}

\begin{definition} 
\reversemarginpar
\marginnote{\textit{Minimal polynomial}}\index{minimal polynomial}
Let $E/F$ be a field extension. If $r$ is algebraic over $F$, its \textbf{minimal polynomial} over $F$ is the irreducible monic polynomial $f(x) \in F[x]$ of the least degree satisfying $f(r)=0$.
\end{definition}

It is necessary for $r$ to be algebraic, for otherwise it is not a root of any polynomial in $F[x]$.
%\textcolor{red}{This should follow the definition of a minimal polynomial rather than precede it.} KL: moved, thanks.

\reversemarginpar
\marginnote{\textit{Uniqueness}}
Note the minimal polynomial of an algebraic number over a base field is unique up to scalar multiplication. A simple argument is as the following. Let $J_r = \{f(x) \in F[x] \mid f(r)=0\}$ be the set of all polynomials in $F[x]$ where $r$ is a root, then $J_r$ is an ideal of the polynomial ring $F[x]$ (easy to verify). Let $p, q \in J_r$ be two monic polynomials of least degree $n > 0$, then $p-q \in J_r$ because $J_r$ is an ideal. Also $p-q$ has degree less than $n$ because $p,q$ are monic. This contradicts with $p,q$ being least degree polynomials in $J_r$, unless $p=q$.   

For different base fields, the minimal polynomial of a number could be different. Here is an example. Given the field extension $\R/\Q$, the minimal polynomial of $\sqrt{2}$ over $\Q$ is $x^2-2$ because this polynomial is monic, irreducible and has the least degree over the base field $\Q$ where $\sqrt{2}$ is a root. However, in the field extension $\R/\R$, the minimal polynomial for $\sqrt{2}$ is $x-\sqrt{2}$.

The degree of an extension $E=F(r)$ is the degree of the minimal polynomial of $r$ over $F$. This is formally proved by \Cref{thm:fieldExtEquiv} in the next subsection. In the above example, the degree $[\Q(\sqrt{2}):\Q]=2$, because the minimal polynomial of $\sqrt{2}$ over $\Q$ is $x^2-2$. 

%Minimal polynomials will be mentioned again when introducing simple extension fields. 

\subsubsection{Simple extension}

%\textcolor{red}{In cryptography, people are interested in algebraic extensions only?} There are different types of extensions, some of which are essential for building cryptosystems. They will be introduced later in order to understand Galois theory.

\reversemarginpar
\marginnote{\textit{Simple extension}}
\begin{definition}
An extension field $E$ over $F$ is \textbf{simple} if there exists an element $r \in E$ with $E = F(r)$. 
\end{definition}
The simple extension $F(r)$ is the smallest extension over $F$ that contains $F$ and $r$. The number $r$ can be either transcendental or algebraic, but we are only interested in algebraic simple extensions. 

In the previous section, we mentioned that if $r$ is an algebraic number over the base field $F$ then its unique minimal polynomial $p(x)$ always exists. 
In addition, since $p(x)$ is irreducible over $F$, the principal ideal $\langle p(x)\rangle $ is also maximal in $F[x]$. This gives us a way of building the extension field $F(r)$ from the polynomial ring $F[x]$ using the principal ideal by Proposition \ref{prop:quotRngIsField} as stated in the following theorem.  

\begin{theorem}
\label{thm:fieldExtEquiv}
Let $E/F$ be a field extension and $r \in E$ be an algebraic number over $F$ with minimal polynomial $p(x) \in F[x]$ of degree $n$, then 
\begin{enumerate}
    \item $F(r) \cong F[x] / \langle p(x)\rangle$.
    \item $\{1,r,r^2,\dots,r^{n-1}\}$ is a basis of the vector space $F(r)$ over $F$. 
    \item $[F(r):F] = deg(p)$.
\end{enumerate}
\end{theorem}
The first part of Theorem~\ref{thm:fieldExtEquiv} is a direct consequence of the First Isomorphism Theorem (Theorem~\ref{thm:first isomorphism theorem}).\index{First Isomorphism Theorem}
%The theorem says that the simple extension $F(r)$ can be built by taking the quotient of the polynomial ring $F[x]$ with the principle ideal $\langle p(x) \rangle$ generated by the minimal polynomial of an algebraic number $r \in E$. Since $p(x)$ is irreducible, the principle ideal is also a maximal ideal. By , the quotient ring $F[x]/\langle p(x) \rangle$ is a field. Proof is omitted.  \kl{introduce power basis here or later?}
%
% \reversemarginpar\marginnote{\textit{Uniqueness}}
An important observation as stated in the following corollary of the above theorem is that if two algebraic numbers have the same minimal polynomial, then the simple extensions generated by them are isomorphic. This tells us that simple algebraic extension of an algebraic number is unique. 

\begin{corollary}
Let $E/F$ be a field extension. If two algebraic numbers $\alpha, \beta \in E$ over $F$ have the same minimal polynomial in $F[x]$, then there is an isomorphism $\phi: F(\alpha) \rightarrow F(\beta)$ with $\phi |_F = I$.
\end{corollary}

\subsubsection{Splitting field}

One way of building the smallest field extension for solving a polynomial is to look at the splitting field of the polynomial. 

Solving a degree $n$ polynomial $f(x) \in F[x]$ for its roots can be done by rewriting it as the product of linear factors in an appropriate extension field $E$. That is,  
\begin{equation*}
    f(x) = c \prod_{i=1}^n (x-a_i),
\end{equation*}
where $c \in F$ is a constant and $x-a_i \in E[x]$ is a linear factor. This rewriting process is also known as \textbf{splitting} a polynomial. %Then another way of constructing the smallest field extension for solving a polynomial is by looking into its splitting field. %We can see from the previous example that it is not possible to split $x^2-2$ into linear factors in $\Q$. Hence, the following definition specifically defines the field, in which a polynomial can be split into its linear factors. 

\begin{definition}
Let $F$ be a field and $f(x) \in F[x]$ be a polynomial. The extension field $E$ is a \textbf{splitting field}\reversemarginpar
\marginnote{\textit{Splitting field}}\index{splitting field} of $f(x)$ over $F$ if
\begin{itemize}
    \item $f(x)$ splits over $E$ and 
    \item if $F \subseteq L \subsetneq E$, then $f(x)$ does not split over $L$.
\end{itemize}
\end{definition}
By definition, a splitting field of $f(x)$ is the smallest extension that contains all the roots of $f(x)$. Alternatively, we say that the extension $E$ is generated by the roots of $f(x)$. That is, if $r_1, \ldots,r_n$ are the roots of $f(x)$ and $E$ is the splitting field of $f(x)$ then $E = F(r_1, \ldots, r_n)$. For example, the extension $\Q(\sqrt{2})$ is the splitting field of $x^2-2 \in \Q[x]$, because the polynomial splits into $(x+\sqrt{2})(x-\sqrt{2})$ in it. But $\C$ is not a splitting field of $x^2-2$, because it is not the smallest. 

The following theorems state that the splitting field of a polynomial always exists and is unique up to isomorphism. 

\reversemarginpar
\marginnote{\textit{Existence}}
\begin{theorem}(Existence)
Let $F$ be a field and $f(x) \in F[x]$ be a polynomial of degree $n > 0$. Then there exists a splitting field $K$ of $f(x)$ over $F$ with degree $[K:F] \le n!$. 
\end{theorem}
The construction of a splitting field can be done by taking the quotient of $F[x]$ with the principle ideal $\langle f(x)\rangle$ where $f(x)$ is irreducible. If it is reducible, we can factor it into irreducible factors and take the same process repeatedly until $f(x)$ splits. %See Wikipedia for more details. 

\reversemarginpar
\marginnote{\textit{Uniqueness}}
\begin{theorem}(Uniqueness)
Let $\phi:F \rightarrow E$ be an isomorphism, $f(x) \in F[x]$ be a polynomial and $\phi(f(x)) \in E[x]$ be the corresponding polynomial in $E[x]$. If $K$ and $L$ are the splitting fields of $f(x)$ and $\phi(f(x))$ over $F$ and $E$ respectively, then $\phi$ extends to an isomorphism $K \cong L$.
\end{theorem}

\subsubsection{Normal extension}

Sometimes we prefer to work with an algebraic extension that includes all the roots of a polynomial, so that we do not need to adjoin more roots to the extension. For this purpose, we define the following.

\reversemarginpar
\marginnote{\textit{Normal extension}}
\begin{definition}
An algebraic extension $E$ over $F$ is \textbf{normal} if whenever an irreducible polynomial over $F$ has a root in $E$, then it splits in $E$. 
\end{definition}
From splitting field, we know that an extension is normal if whenever it contains one root of a polynomial, it contains all roots of the polynomial. 
The most important result about normal extension is its connection with splitting field.

\reversemarginpar
\marginnote{\textit{Normal iff splitting}}
\begin{theorem}
\label{theorem:normal iff splitting}
A finite algebraic extension $E$ over $F$ is normal if and only if it is the splitting field of some polynomial $f(x) \in F[x]$.
\end{theorem}
The theorem implies that if $E$ is the splitting field of one polynomial over $F$, then it is the splitting field of every other polynomial over $F$ with one root in $E$. %In other words, a polynomial over $F$ either has no root in $E$ or splits completely in $E$. 

\subsubsection{Separable extension}

In addition to normal extensions, it is also convenient when a polynomial has distinct roots, so we do not need to worry about duplicated roots. This is especially the case when working with Galois groups that consist of automorphisms between polynomial roots. Before introducing separable extensions, we define what it means for a polynomial to be separable and how separability can be tested. 

\reversemarginpar
\marginnote{\textit{Separable polynomial}}
\begin{definition}
A polynomial over a field $F$ is \textbf{separable} if the number of its distinct roots in a splitting field is equal to the degree of the polynomial. 
\end{definition}

\begin{example}
The polynomial $x^2-2$ has two distinct roots $\pm \sqrt{2}$, so it is separable. 
The polynomial $(x^2-1)^2$ is not separable, because both roots $\pm 1$ have multiplicity 2. 
\end{example}

\reversemarginpar
\marginnote{\textit{Test separability}}
One way of testing separability is to check whether or not a polynomial is coprime with its \textit{formal derivative}\footnote{Formal derivative is similar to derivative in calculus, but for elements of a polynomial ring.}. 
\begin{lemma}
A polynomial $f(x) \in F[x]$ is separable if and only if $\gcd(f,f')=1$.
\end{lemma}
\begin{proof}
Let $K$ be the splitting field of $f(x)$ and $r \in K$ is a root of $f(x)$. The re-write the polynomial as 
\begin{equation*}
    f(x)=(x-r)^m g(x)
\end{equation*}
with $m \ge 1$ and $g(r) \neq 0$. Take the formal derivative, we get
\begin{equation*}
    f'(x) = m(x-r)^{m-1} g(x) + (x-r)^m g'(x) = (x-r)^{m-1} [mg(x) + (x-r)g'(x)].
\end{equation*}
Evaluating the second factor $mg(x) + (x-r)g'(x)$ at $r$ gives $mg(r) + 0 = 0 \iff m = 0$ because $g(r) \neq 0$.  

If $f(x)$ is separable, by definition $m = 1$ and $f'(x) = g(x) + (x-r)g'(x)$. So $f'(r) \neq 0$ and none of the two factors of $f(x)$ divides $f'(x)$. This implies they are coprime. 

If $f(x)$ is not separable, then $m > 1$ and $f'(r)=0$. Hence, $x-r$ is a common factor of $f$ and $f'$, so they are not coprime. 
\end{proof}

\begin{example}
In the examples above, $f(x)=x^2-2$ is separable, because its formal derivative $f(x)'=(x^2-2)' = 2x$ and $\gcd(f,f')=1$. If $f(x)=(x^2-1)^2$, then its formal derivative $f'(x)=((x^2-1)^2)'=4x(x^2-1)$ and $\gcd(f,f')=x^2-1$, so the polynomial $(x^2+1)^2$ is not separable. 
\end{example}

\reversemarginpar
\marginnote{\textit{Separable extension}}
\begin{definition}
\label{def:sepExt}
An algebraic extension $E$ over $F$ is \textbf{separable} if for every element $\alpha \in E$, its minimum polynomial over $F$ is separable. 
\end{definition}

The Fundamental Theorem of Galois Theory states a correspondence between intermediate field extensions and subgroups of a Galois group. Hence, we would like to know the separability of the intermediate field extensions between a base field and a separable extension.    

\reversemarginpar
\marginnote{\textit{Intermediate extensions are separable}}
\begin{theorem}
Given field extensions $L/M/K$. If $L/K$ is separable, then the intermediate extensions $L/M$ and $M/K$ are also separable. 
\end{theorem}
%M/K is separable is trivial to prove, the proof of L/M should be easy,

\reversemarginpar
\marginnote{\textit{$char(F)=0 \implies$ separable}}
In the previous section, we stated that a field characteristic is either 0 or a prime. The following results connect the characteristic of a polynomial to its separability.

\begin{theorem}
\label{thm:minPolyIsSepInChar0Field}
Every irreducible polynomial over a field of characteristic zero is separable, and hence every algebraic extension is separable. 
\end{theorem}

\begin{proof}
Let $E/F$ be a field extension with $char(F)=0$, and $f(x) \in F[x]$ be the minimal polynomial of $\alpha \in E$ over $F$. Assuming $f(x)$ is not separable. That is, without loss of generality, there is a root $\beta$ with multiplicity 2. Then $f(\beta)=0$ and its formal derivative $f'(\beta)=0$, because $f(x)$ has a factor $(x-\beta)^2$, which becomes $2(x-\beta)$ in $f'(x)$. 

However, $f'(x)$ does not have zero coefficients, because it is over a field of zero characteristic. The fact that $f(x)$ is a minimal polynomial implies it is irreducible, and $f'(x)$ has a lower degree than $f(x)$ imply that $\gcd(f,f')=1$. Hence, there are $a,b \in F[x]$ such that $af(x)+bf'(x)=1$. Substituting $x=\beta$, we get a contradiction, so $f(x)$ cannot be non-separable. Hence, every irreducible polynomial over $F$ is separable. This implies every algebraic extension is separable and every finite extension is also separable because every finite extension is algebraic by Proposition \ref{prop:finite implies algebra}.       
\end{proof}

A similar but more general result is the following theorem. 

\begin{theorem}
\label{theorem:char prime implies separable}
Let $f \in F[x]$ be an irreducible polynomial of degree $n$. Then $f$ is separable if either of the following conditions is satisfied: 
\begin{itemize}
    \item the field $F$ has characteristic $0$ or 
    \item the field $F$ has characteristic $p$ where $p$ is prime and $p \nmid n$.
\end{itemize}
\end{theorem}
The same argument can be used here to prove the second condition. Since $f(x)$ is a degree $n$ polynomial, its formal derivative $f'(x)$ much contain a term $n a_n x^{n-1}$, in which the coefficient $n a_n \neq 0$ in the field $F$ as $char(F)=p$ is prime and $p \nmid n$. So $\gcd(f,f')=1$ and the same contradiction can be reached is $f(x)$ is assumed to be non-separable. 

The intuition behind both theorems is that if the characteristic of the field $F$ does not satisfy either condition, then the coefficients of $f'(x)$ may be all zero. So $f'(x)=0$ cannot lead to the same contradiction when assuming $f(x)$ non-separable. 

\subsection{Galois extension and Galois group}
In the preceding subsections, we have defined different types of field extensions, finite, algebraic, simple, normal and separable. This section will connect some of these extensions to an important field extension, called \textit{Galois extension} and will define the \textit{Galois groups} of Galois extensions. 

\reversemarginpar
\marginnote{\textit{Group action}}
To start with, we introduce group action on a set. One way to define a group action on a set is by the following definition. 

\begin{definition}
A group $(G,*)$ \textbf{acts} on a set $S$ if there is a map  
\begin{equation*}
    \mu:G \times S \rightarrow S
\end{equation*}
such that 
\begin{itemize}
    \item for all $s \in S$, we have $\mu(e,s)=s$,
    \item for all $x,y \in G$ and $s\in S$, we have $\mu(x * y, s) = \mu(x,\mu(y,s))$.
\end{itemize}
\end{definition}

For simplicity, we write $\mu(x,s)$ as $x(s)$. Another way of defining group action is by a group homomorphism.

\begin{definition}
A group $G$ \textbf{acts} on a set $S$ if there is a homomorphism 
\begin{equation*}
    \phi:G \rightarrow Sym(S)
\end{equation*}
from the group to the symmetric group (or the permutation group $Perm(S)$) of $S$. 
\end{definition}

In this case, we say $\phi$ is the group action of $G$ on $S$. Each element of $G$ is mapped to a certain permutation of the set $S$ by the action. For example, when the Dihedral group
\begin{equation*}
    D_4=\langle r,f\rangle=\{e,r,r^2,r^3,f,fr,fr^2,fr^3)
\end{equation*}
acts on itself, 
% \textcolor{red}{I think that this example is unnecessarily complicated. Part of the problem is that the dihedral group is playing two roles here. Is there some reason we can't use another eight-element set?},
each element in $D_4$ is mapped to a certain permutation of the set $S=D_4$. For example, the elements \textit{rotation} $r$ and \textit{reflection} $f$ correspond to the following permutations of $D_4$
% \textcolor{red}{What is the homomorphism here?  It looks like in the "rotate" map we have $\phi(g,s) = g \cdot s$ while in the "flip" line we have $\phi(g,s) = s \cdot g$. Because $D_4$ is nonabelian, these are different functions. Right?} \kl{Yes, I think this example could be a bit complicated. I realized I made a mistake by mixing the left and right action. I think one is transferable to another, so it's sufficient to consider one action in most cases. I'll see if I can make a simpler example. I think most people may consider the left action. The first homomorphism is $\phi_1: r \rightarrow (e \text{ } r)(r\text{ }r^2) (r^2\text{ }r^3) (r^3\text{ }e) (f\text{ }fr^3) (fr\text{ }f) (fr^2\text{ }fr) (fr^3\text{ }fr^2)$ and the second homomorphism is $\phi_2: f \rightarrow (e \text{ } f)(r\text{ }fr) (r^2\text{ }fr^2) (r^3\text{ }fr^3) (f\text{ }e) (fr\text{ }r) (fr^2\text{ }r^2) (fr^3\text{ }r^3)$. They are different homomorphisms.}
\begin{align*}
    r: \{e,r,r^2,r^3,f,fr,fr^2,fr^3) \mapsto \{r,r^2,r^3,e, rf=fr^3,rfr=f,rfr^2=fr, rfr^3=fr^2) \\
    f: \{e,r,r^2,r^3,f,fr,fr^2,fr^3) \mapsto \{f,fr,fr^2,fr^3,e,r,r^2,r^3).
\end{align*}
The action of $D_4$ only gives rise to certain permutes of $D_4$. In other words, there are 8 elements in $D_4$ and the symmetric group has size $|Perm(D_4)|=8!$, the homomorphism $\phi$ is injective, which we call faithful as stated next. 

\reversemarginpar
\marginnote{\textit{Faithful action}}
\begin{definition}
A group action $\phi$ of $G$ on a set $S$ is \textbf{faithful} if $\phi$ is injective. That is, for every two distinct elements $g, h \in G$, there exists an element $s \in S$ such that $g(s) \neq h(s)$. 
\end{definition}
If a group action is faithful, then we can think the group $G$ embeds into the permutation group of $S$, as in the above example of $D_4$, where each element of $G=D_4$ corresponds to a certain permutation of the set $S=D_4$.

Similarly, we can define a group $G$ acts on a ring $R$ (or a field $F$). The difference is that a ring has more algebraic structures than a set, so simple permutations of the ring elements do not necessarily preserve the ring structure. For this reason, we replace permutations by automorphisms, which are bijective ring homomorphisms between $R$ and itself. Let $\text{Aut}(R)$ be the \textbf{automorphism group} of $R$. 

\begin{definition}
An \textbf{action} of a group $G$ on a ring $R$ is a group homomorphism 
\begin{equation*}
    \phi: G \rightarrow \text{Aut}(R).
\end{equation*}
\end{definition}

\reversemarginpar
\marginnote{\textit{Fixed field}}
Some elements in the ring $R$ or field $F$ stay invariant under the action. They make up the fixed field. 

\begin{definition}
Given a field extension $E/F$ and a group action of $G$ on $E$, the \textbf{fixed field} of $E$ under the action of $G$ 
\begin{equation*}
    E^G = \{a \in E \mid g(a)=a, \forall g \in G\}.
\end{equation*}
is the set of elements in the extension field that are fixed point-wise by all automorphisms of $R$. 
\end{definition}

\reversemarginpar
\marginnote{\textit{Automorphism group}}
\begin{definition}
\label{def:automorphismGroup}
Let $E/F$ be a field extension. The \textbf{automorphism group} of the field extension 
\begin{align*}
    Aut(E/F) &= \{\alpha \in Aut(E) \mid \alpha(x) = x, \text{ } \forall x \in F\} \\
    &=\{\alpha \in Aut(E) \mid \alpha_F = Id_F\}
\end{align*}
is the set of automorphisms that fixes $F$ when acting on $E$.
\end{definition}
The automorphism group is a group with function composition as the group operator. It is a subgroup of the group of automorphisms of $E$, i.e., $Aut(E/F) \subseteq Aut(E)$. Now, we are ready to define the Galois group of a field extension.  

\begin{definition}
The \textbf{Galois group} of a field extension $E/F$, denoted by $Gal(E/F)$, is the automorphism group of the field extension.\reversemarginpar
\marginnote{\textit{Galois group}}\index{Galois group of a field extension}
That is, 
\begin{equation*}
    Gal(E/F):=Aut(E/F) = \{\alpha \in Aut(E) \mid \alpha_F = Id_F\}.
\end{equation*}
\end{definition}

By definition, the Galois group is a subset of the automorphism group or permutation group (or symmetric group) of the extension $E$. 

As explained in the previous section that an extension field can be viewed as a vector space over the base field, so when working with Galois groups, instead of thinking where all elements in the extension are mapped to, it is convenient to know where the basis vectors are mapped to by the automorphisms.

Let us work through some simple examples. 
\begin{example}
Let the field extension be $\Q(\sqrt{2})/\Q$. It is a 2-dimensional $\Q$-vector space with a basis $\{1,\sqrt{2}\}$. The Galois group must fix the base field, so it contains the identity map $I$. In addition, it should contain another automorphism $\sigma$ that maps $\sqrt{2}$ to another element $a$ in the extension whiling fixing $\Q$. Since $\sigma$ is an automorphism, it must satisfy $a^2=\sigma(\sqrt{2})^2=\sigma((\sqrt{2})^2)=\sigma(2)=2$. So whatever $\sigma(\sqrt{2})=a$ is, it must satisfy $a^2-2=0$ in the extension, which means $a=\pm \sqrt{2}$. Since the identity map is already included, it entails $\sigma(\sqrt{2})=-\sqrt{2}$. Hence, the Galois group $Gal(\Q(\sqrt{2})/\Q) = \{I, \sigma:\sqrt{2} \mapsto -\sqrt{2}\} \cong C_2$ which is isomorphic to the cyclic group of order 2. 
\end{example}

\begin{example}
\label{exam:galoisgroup2}
Let the field extension be $\Q(\sqrt{2},i)/\Q$. This is a 4-dimensional $\Q$-vector space with a basis $\{1,\sqrt{2},i,\sqrt{2}i\}$. The minimal polynomials over $\Q$ for $\sqrt{2}$ and $i$ are $x^2-2$ and $x^2+1$, respectively. The Galois group of the field extension contains all the automorphisms that fix $\Q$ while permuting roots in each minimal polynomial. That is, it contains a map $\tau$ that permutes $\{\sqrt{2},-\sqrt{2}\}$ and a map $\sigma$ that permutes $\{i,-i\}$.
We can identify these automorphisms as shown in Table \ref{tab:galoisgroup}. The Galois group is isomorphic to the Klein four group $V_4 = C_2 \times C_2$. 
\begin{table}[h!]
\centering
\begin{tabular}{|c|c|c|c|c|}
\hline
 & 1 & $\sqrt{2}$ & $i$ & $\sqrt{2}i$ \\ \hline
$I$ & 1 & $\sqrt{2}$ & $i$ & $\sqrt{2}i$  \\ \hline
$\sigma$ & 1 & $\sqrt{2}$ & $-i$ & $-\sqrt{2}i$ \\ \hline
$\tau$ & 1 & $-\sqrt{2}$ & $i$ & $-\sqrt{2}i$ \\ \hline
$\sigma \tau$ & 1 & $-\sqrt{2}$ & $-i$ & $\sqrt{2}i$ \\ \hline
\end{tabular}
\caption{The Galois group of the extension $\Q(\sqrt{2},i)$. It is isomorphic to the Klein four group $V_4=C_2\times C_2$.}
\label{tab:galoisgroup}
\end{table}
\end{example}

It is important to note that not all automorphisms (or permutations) that fix the base field are in the Galois group. From the above two examples, we can see that the Galois group only contains those automorphisms that permute roots of the same minimal polynomial while fixing the base field. In Example \ref{exam:galoisgroup2}, $\sqrt{2}$ and $-\sqrt{2}$ come from the minimal polynomial $x^2-2$ in $\Q$ and $i$ and $-i$ come from the minimal polynomial $x^2+1$ in $\Q$. Let us take a look at a counter example. 

\begin{example}
Let the field extension be $\Q(\sqrt{2},\sqrt{3})/\Q$. The permutation $\phi:\sqrt{2} \mapsto \sqrt{3}$ is not in the Galois group. Assuming it is, then $\phi(\sqrt{2}) = \sqrt{3}$ implies $\phi(\sqrt{2})^2 = 3$. By the definition of homomorphism, $\phi(\sqrt{2})^2 = \phi(\sqrt{2}^2)=\phi(2)=2$ because $\phi$ fixes $\Q$. This implies $2=3$. 
\end{example}

\begin{example}
A slightly more complicated example is with a field extension $\Q(\sqrt[4]{2},i)/\Q$. The roots $\sqrt[4]{2}$ and $i$ have the minimal polynomials $x^4-2$ and $x^2+1$ over $\Q$, respectively. The polynomial $x^4-2$ has four roots $\pm \sqrt[4]{2}$ and $\pm i\sqrt[4]{2}$. The polynomial $x^2+1$ has two roots $\pm i$. The Galois group should contain automorphisms that permutes roots for each polynomial. The process of finding the automorphisms is more or less trial and error.\footnote{Perhaps there are better ways of finding the Galois group, but they are not in the scope of this material.} Let 
\begin{align*}
    \sigma(\sqrt[4]{2}) &= i\sqrt[4]{2} \text{ and } \sigma(i)=i,\\
    \tau(i)&=-i \text{ and } \tau(\sqrt[4]{2})=\sqrt[4]{2}.
\end{align*}
Then we have 
\begin{align*}
    \sigma^2(\sqrt[4]{2}) &= -\sqrt[4]{2} \text{ and } \sigma^2(i)=i,\\
    \sigma^3(\sqrt[4]{2}) &= -i\sqrt[4]{2} \text{ and } \sigma^3(i)=i,\\
    \sigma^4(\sqrt[4]{2}) &= \sqrt[4]{2} \text{ and } \sigma^4(i)=i,\\
    \tau^2(i)&=i \text{ and } \tau^2(\sqrt[4]{2})=\sqrt[4]{2}.
\end{align*}
So the orders of $\sigma$ and $\tau$ in the Galois group are 4 and 2, respectively. Hence, the Galois group is $\{I, \sigma,\sigma^2,\sigma^3,\tau,\sigma\tau,\sigma^2\tau,\sigma^3\tau\}$.
\end{example}

Combining the definitions of fixed field and Galois group, we know that for a field extension $E/F$, the fixed field by the Galois group %$E^{Gal(E/F)} = \{a \in E \mid g(a) = a, \forall g \in Gal(E/F)\}$ 
should at least contain the base field $F$. Because all automorphisms in the Galois group at least fix $F$, though they may fix more than $F$. 
Hence, we can define what it means for a field extension to be Galois. 

\reversemarginpar
\marginnote{\textit{Galois extension}}
\begin{definition}
A field extension $E/F$ is an \textbf{Galois extension} if the fixed field by the Galois group $Gal(E/F)$ is exactly $F$. That is, $E^{Gal(E/F)} = F$.
\end{definition}

In other words, the Galois group has to fix exactly the base field, nothing more nothing less. An important theorem that characterizes Galois extension using previously defined extension types is the following. 

\reversemarginpar
\marginnote{\textit{Normal and separable $\implies$ Galois}}
\begin{theorem}
\label{theorem:separable and normal implies galois}
An algebraic field extension is a \textbf{Galois extension} if it is normal and separable. 
\end{theorem}
This theorem says that for an algebraic field extension to be a Galois extension, any polynomial that has a root in the extension must have all its roots in the extension and these roots must be all distinct. The requirement of being normal and separable is a sufficient condition for a field extension to be Galois. %That is, the Galois group $Gal(E/F)$ (i.e., collection of automorphisms between $E$) only fixes the base field $F$, nothing else.  

\begin{example}
The Galois group $Gal(\Q(\sqrt[3]{2})/\Q) = \{I\}$ contains only the identity map. If $\phi(\sqrt[3]{2})=a$ is another automorphism, then it must satisfy $a^3-2=0$. So $\phi$ must map $\sqrt[3]{2}$ to a root of the minimal polynomial $a^3-2=0$ in the extension. But the only root that is in the extension is $\sqrt[3]{2}$, because the other two roots are complex. So $\phi$ is the identity map. Given the Galois group contains only the identity map, the fixed field is $\Q(\sqrt[3]{2})$ not $\Q$, so the field extension is not Galois. By Theorem \ref{theorem:separable and normal implies galois}, the extension is not both normal and separable. In fact, this is true, because the extension does not contain the two complex roots of the minimal polynomial $x^3-2$.
\end{example}

The example suggests that a field extension can have a Galois group, but it is not necessarily a Galois extension. 

Since a Galois extension is normal and separable, we would expect the number of automorphisms in the Galois group to be related to the number of roots of a minimal polynomial. The next lemma connects the number of automorphisms in the Galois group to the degree of a Galois extension. 
\begin{lemma}
If a finite field extension $E/F$ is Galois, then the number of elements in the Galois group is the degree of the field extension. That is, $|Gal(E/F)|=[E:F]$. 
\end{lemma}
For example, the field extension $Q(\sqrt{2},i) / Q$ has degree 4 (as it is a 4 dimensional vector space over $Q$) and there are 4 automorphisms in the Galois group as stated in Table \ref{tab:galoisgroup}. 

The next theorem is the most important theorem in Galois Theory. It builds a connection between subgroups of a Galois group and field extensions of a base field. The theorem is important in the sense that it provides a way of understanding field extensions from group's perspective, which is relatively well studied. In the most basic form, it states that if $L/M/K$ is a finite Galois extension, then there is a one-to-one correspondence between an intermediate extension and a subgroup of the Galois group $Gal(L/K)$. The next theorem explicitly defines what it means for a one-to-one correspondence between the two different algebraic structures. 

\begin{figure}[ht]
	\centering
	\begin{subfigure}[t]{0.45\textwidth}
		\centering
		\includegraphics[page=10]{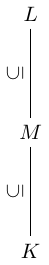}
		\caption{A finite Galois extension.}
		\label{subfig:finite galois ext}
	\end{subfigure}
	\begin{subfigure}[t]{0.45\textwidth}
		\centering
		\includegraphics[page=11]{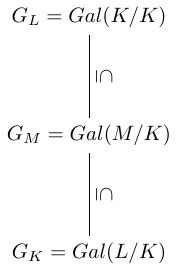}
		\caption{Subgroups of the Galois group $G_K=Gal(L/K)$.}
		\label{subfig:subgroups of galois group}
	\end{subfigure}
	\caption{A finite Galois extension and the corresponding Galois groups.}
	\label{fig:galois ext and groups example1}
\end{figure}

\reversemarginpar
\marginnote{\textit{Fundamental Theorem of Galois Theory}}
\begin{theorem}(Fundamental Theorem of Galois Theory)
Suppose $L/M/K$ is a finite Galois extension with the corresponding Galois group $G_K=Gal(L/K)$. 
\begin{enumerate}
    \item There is an inclusion reversing correspondence between an intermediate field $M$ of $L/K$ and a subgroup $G_M \subseteq G_L$ given as follows: 
    \begin{align*}
        %\{\text{intermediate fields $M$ between } K \text{ and } L\} & \leftrightarrow \{\text{subgroups } H \text{ of } G\} \\
        M &\rightarrow G_M = \{\phi \in Aut(L) \mid \phi_M = Id_M\} \\
        G_M &\rightarrow L^{G_M}=M.
    \end{align*}
    
    \item The degrees of the field extensions are given by 
    \begin{align*}
        [L:M] = |G_M| \text{ and }
        [M:K] = \frac{|G_K|}{|G_M|}.
    \end{align*}
    
    \item The intermediate field extension $M/K$ is Galois if and only if $G_M \triangleleft G_K$ is a normal subgroup.  In this case, the corresponding Galois group is given by 
    \begin{equation*}
        Gal(M/K) \cong G_K/G_M.
    \end{equation*}
\end{enumerate}
\end{theorem}

The first point of the theorem says that if $M$ is an intermediate extension between $L/K$, then $M$ corresponds to the set of automorphisms of $L$ that fixes $M$. If $M=K$, then $M$ corresponds to the set of automorphisms of $L$ that fixes $K$, which is the entire $Gal(L/K)$. If $M=L$, then $M$ corresponds to the set of automorphisms of $L$ that fixes $L$, which is identity map. 

The second point says the degree of the $M$-vector space $L$ equals the number of automorphisms of $L$ that fix $M$. If $M=K$ or $M=L$, then the degrees $[L:M]=[L:K]=|G_K|=Gal(L/K)$ or $[L:M]=[L:L]=|G_L|=1$, respectively.
Combining the two qualities, we get $[L:M][M:K]=|G_K|=[L:K]$ which is consistent with the Tower Law in Proposition \ref{prop:tower law}.

% \kl{I don't have a good intuition for the third point.}

\begin{figure}[!ht]
	\centering
	\begin{subfigure}[t]{0.48\textwidth}
		\centering
		\includegraphics[page=12]{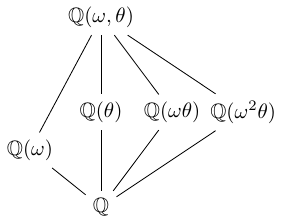}
		\caption{A finite Galois extension and the intermediate extensions.}
		\label{subfig:finite galois ext 2}
	\end{subfigure}\quad
	\begin{subfigure}[t]{0.48\textwidth}
		\centering
		\includegraphics[page=13]{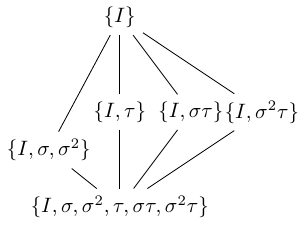}
		\caption{Subgroups of the Galois group $Gal(\Q(\omega,\theta)/\Q)$.}
		\label{subfig:subgroups of galois group 2}
	\end{subfigure}
	\caption{A finite Galois extension $\Q(\omega,\theta)/\Q$ and the corresponding Galois groups, where $\omega=\frac{-1}{2}+i\frac{\sqrt{3}}{2}$ and $\theta=\sqrt[3]{2}$. Each structure is a lattice and there is a one-to-one correspondence between them.}
	\label{fig:galois ext and groups example2}
\end{figure}

\begin{example}
%The polynomial $x^3-2$ is irreducible over $\Q$. To solve this polynomial, re-write $x^3=2$ as $(\theta\omega)^3=\theta^3 \cdot \omega^3 =2\cdot 1$, so $\theta=\sqrt[3]{2}$ and $\omega=\frac{-1}{2}\pm i\frac{\sqrt{3}}{2}$ is a cube root of unity but $\omega \neq 1$. Thus, 
Let the field extension be $\Q(\theta,\omega)/\Q$, where $\theta=\sqrt[3]{2}$ and $\omega=\frac{-1}{2}\pm i\frac{\sqrt{3}}{2}$. %is a splitting field of $x^3-2$. 
The extension is a 6-dimensional $\Q$-vector space with a basis $\{1,\theta,\theta^2,\omega,\theta\omega,\theta^2\omega\}$.
Define the automorphisms 
\begin{align*}
    \sigma(\theta)&=\omega\theta \text{ and } \sigma(\omega)=\omega,\\
    \tau(\theta)&=\theta \text{ and } \tau(\omega)=\omega^2.
\end{align*}
The two automorphisms in the Galois group have orders 3 and 2, respectively. It can be seen that they can make the entire Galois group $\{I,\sigma,\sigma^2,\tau,\sigma\tau,\sigma^2\tau\}$. The intermediate field extensions from $\Q$ to $\Q(\omega,\theta)$ are shown in Figure \ref{subfig:finite galois ext}. The extension $\Q(\omega)$ can be extended to $\Q(\omega,\theta)$ by adjoining $\theta$ and the other three extensions can be extended to $\Q(\omega,\theta)$ by adjoining $\omega$. The corresponding subgroups of the Galois group are shown in Figure \ref{subfig:subgroups of galois group}.

The two structures are lattices. % \textcolor{red}{Why is it important to mention that both structures are lattices here? } 
According to the Fundamental theorem of Galois Theory, they are in one-to-one correspondence. The automorphisms that fix $\Q(\omega)$ are $\{I,\sigma,\sigma^2\}$. The degree of the intermediate extension $\Q(\omega)$ is $[\Q(\omega,\theta):\Q(\omega)]=3$, because $\Q(\omega,\theta)$ has a basis $\{1,\theta,\theta^2\}$ over the field $\Q(\omega)$. Also, $[\Q(\omega,\theta):\Q]=[\Q(\omega,\theta):\Q(\omega)][\Q(\omega):\Q]=3\cdot 2=6$. The normal extensions are $\Q$, $\Q(\omega)$ and $\Q(\omega,\theta)$ because the corresponding subgroups $\{I,\sigma,\sigma^2,\tau,\sigma\tau,\sigma^2\tau\}$, $\{I,\sigma,\sigma^2\}$ and $\{I\}$ are normal subgroups of the Galois group $\{I,\sigma,\sigma^2,\tau,\sigma\tau,\sigma^2\tau\}$.
\end{example}

%%%%%%%%%%%%%%%%%%%%%%%%%%%%%%%%%%%%%%%%%%%%%%%%%%%%%%%%%%%%%%%%%%%%%%%%%%%%%%%%%%%%%%%%%%%%%%%%%%%
%%%%%%%%%%%%%%%%%%%%%%%%%%%%%%%%%%%%%%%%%%%%%%%%%%%%%%%%%%%%%%%%%%%%%%%%%%%%%%%%%%%%%%%%%%%%%%%%%%%

%%%%%%%%%%%%%%%%%%%%%%%%%%%%%%%%%%%%%%%%%%%%%%%%%%%%%%%%%%%%%%%%%%%%%%%%%%%%%%%%%%%%%%%%%%%%%%%%%%%

%\newpage
%\bibliography{references}
%\bibliographystyle{abbrvnat}

\newpage
\section{Algebraic Number Theory}

\label{appen:ant}

%\section{Algebraic Number Theory (medium+)}

This section introduces some of the basic results in \textit{Algebraic Number Theory} that will be used in lattice-based cryptography. In particular, we will focus on the ring of integers, their integral and fractional ideals. The aim is to build the important connection between ideals of a ring of integers and ideal lattices, which is the key in those homomorphic encryption schemes that are based on the ring learning with error (RLWE) problem. 

\subsection{Algebraic number field}
\label{app subsection:number field}

%\label{app subsection:algebraic number field}
%Some of the concepts in this section have been discussed in Section \ref{app app subsection:field extension}. 

Recall that an algebraic number (integer) is a complex number that is a root of a non-zero polynomial with rational (integer) coefficients. Below we define algebraic number fields, which are special cases of extension fields where the base field is the rationals $\Q$.

\begin{definition}
\reversemarginpar
\marginnote{\textit{Number field}}
An \textbf{algebraic number field} (or simply \textbf{number field}) is a finite extension of the field of rationals by algebraic numbers, i.e., $\mathbb{Q}(r_1, \dots, r_n)$, where $r_1, \dots, r_n$ are algebraic numbers.
\end{definition}

An 
\reversemarginpar
\marginnote{\textit{Cyclotomic field}}
nth root of unity $\zeta_n$ is an algebraic number, so the cyclotomic extension $\Q(\zeta_m)$ is also a number field that is called the 
\textbf{nth cyclotomic number field} (or \textbf{nth cyclotomic field}). 

A number field $K=\Q(r)$ forms a vector space over the base field $\Q$ with the basis $\{1,r,\dots,r^{n-1}\}$, which is called the 
\reversemarginpar
\marginnote{\textit{Power basis}}\index{power basis}
\textbf{power basis} of $K$ because it is formed by the powers of a number $r$. By the Primitive Element Theorem, it is always possible to get a power basis for a number field.

\begin{theorem}[\textbf{Primitive element theorem}]
\label{app thm:primEleThm}
\reversemarginpar
\marginnote{\textit{Primitive element}}
If $K$ is an extension field of $\mathbb{Q}$ and it has finite degree $[K:\mathbb{Q}] < \infty$, then $K$ has a
\textbf{primitive element} $r$ such that $r \notin \mathbb{Q}$ and $K=\mathbb{Q}(r)$.
\end{theorem}

\begin{example}
The number field $K=\Q(\sqrt{2})$ is a degree 2 $\Q$-vector space. It has a primitive element $\sqrt{2}$ and a basis $\{1, \sqrt{2}\}$.

The number field $K=\Q(\sqrt[3]{2})$ has degree 3. It has a primitive element $\sqrt[3]{2}$ and a basis $\{1,\sqrt[3]{2},\sqrt[3]{4}\}$. 

The number field $K=\Q(\sqrt{2},\sqrt{3})$ has degree 4. It has a primitive element $r=\sqrt{2}+\sqrt{3}$, so $K=\Q(\sqrt{2},\sqrt{3}) = \Q(\sqrt{2}+\sqrt{3})$. It has a power basis $\{1,r,r^2,r^3\}=\{1,\sqrt{2}+\sqrt{3},5+2\sqrt{6},11\sqrt{2}+9\sqrt{3}\}$. To see this is a basis, we know from field extension that $\{1,\sqrt{2},\sqrt{3},\sqrt{6}\}$ is a basis of $K$. This basis can be expressed in terms of the linear combinations of the power basis.
\end{example}

For a number field $K$, the set of all algebraic integers forms a ring under the usual addition and multiplication operations in $K$ (exercise).
This set generalizes the set of \textbf{rational integers} $\Z$. It is particularly important for the RLWE problem.

%\textcolor{red}{It perhaps has more deep motivations that I don't know of.}
\begin{definition}
\reversemarginpar
\marginnote{\textit{Ring of integers}}
The \textbf{ring of integers} of an algebraic number field $K$, denoted by $\OO_K$, is the set of all algebraic integers that lie in the field $K$. 
\end{definition}
For example, the set $\Z$ of rational integers is the ring of integers of the number field $\mathbb{Q}$, i.e., $\Z = \mathcal{O}_{\Q}$. 
Recall that an integral domain is a non-zero commutative ring in which the product of two non-zero elements is non-zero. $\Z$ is an integral domain, 
\reversemarginpar
\marginnote{\textit{$\OO_K$ is ID}}
so is its generalization $\OO_K$, because $\OO_K \subseteq K$ is in a number field which is an integral domain.
In general, determining the ring of integers of a number field is a difficult problem, unless the number field is quadratic that is a $\Q$-vector space of degree 2 as stated in the next theorem.

\begin{definition}
\reversemarginpar
\marginnote{\textit{Square free}}
A number is \textbf{squarefree} if its prime decomposition contains no repeated factors. 
\end{definition}
All prime numbers are squarefree. Some composite numbers are squarefree and some are not. For example, 4 is not squarefree, but 6 is. 

\begin{theorem}
\label{app thm:roiQuadField}
\reversemarginpar
\marginnote{\textit{$\OO_K$ in quadratic $K$}}
\label{app thm:OKQuadField}
%\label{app theorem:ring of integers quadratic field}
Let $K$ be a quadratic number field and $m$ be a unique squarefree integer such that $K=\Q(\sqrt{m})$. Then the set $\OO_K$ of algebraic integers in $K$ is given by 
\begin{equation*}
    \OO_K = 
    \begin{cases}
      \Z + \Z\sqrt{m}, & \text{if $m \not= 1 \bmod 4$} \\
      \Z + \Z\left( \frac{1+\sqrt{m}}{2}\right), & \text{if $m = 1 \bmod 4$}
    \end{cases}
\end{equation*}
\end{theorem}

For example, if $K=\Q(\sqrt{-7})$ then $\OO_K = \Z+\Z\left(\frac{1+\sqrt{-7}}{2} \right)$. 
If  $K=\Q(\sqrt{-5})$ then $\OO_K = \Z + \Z \sqrt{-5}$.

Since the set of rational integers $\Z \subseteq \OO_K$ is always contained in the ring of integers of a number field $K$ (of degree $n$), this makes $\OO_K$ a $Z$-module. Recall that a module is a generalization of a vector space where scalar multiplications are defined in a ring rather than a field. 
\reversemarginpar
\marginnote{\textit{$\OO_K$ is free $Z$-module}}
In fact, $\OO_K$ is a free $Z$-module, which means it has a basis $B=\{b_1, \dots, b_n\} \subseteq \OO_K$ such that every element in $\OO_K$ can be written as an integer linear combination of the basis. 
The basis is called a \textbf{$\Z$-basis} of $\OO_K$. It is also a \textbf{$\Q$-basis} of $K$, because every element $r \in K$ can be written as a linear combination $r=\sum_{i=1}^n a_i b_i$, where $a_i \in \Q$.  

More importantly, the basis $B$ is called an \textbf{integral basis} 
\reversemarginpar
\marginnote{\textit{Integral basis}}
of the number field $K$ (and of the ring of integers $\OO_K$ as used by Ben Green).
Note that although the ring of integers $\OO_K$ always has a basis, it does NOT always have a power basis. A special case is when $K$ is a cyclotomic number field. In this case, the power basis of $K$ is also an integral basis of $K$ (or $\OO_K$). 

The essential connection between $\OO_K$ and lattices is by relating the number field $K$ to the $n$-dimensional Euclidean space $\R^n$. This is done via an embedding of $K$ to a space $H$ that is isomorphic to $\R^n$. Suppose $K$ is a number field with degree $[K:\Q]=n$, then we have $n$ field embeddings (i.e., field or injective ring homomorphisms) $\sigma_i: K \rightarrow \C$ such that the base field $\Q$ is fixed by the embeddings. For a primitive element $r$ in $K$ but not in $\Q$, i.e., $K=\Q(r)$, each embedding $\sigma_i: K \rightarrow \C$ is given by the map from $r$ to a root of $r$'s minimal polynomial $f(x) \in \Q[x]$. The following proposition states that there are $n$ distinct such embeddings from $K$ to $\C$. 

\iffalse
By definition of the embeddings, we have 
\begin{enumerate}
    \item for a number field $K=\Q(r)$, each embedding $\sigma_i$ is uniquely determined by its image $\sigma_i(r)$, 
    \item if $f(x) \in \Q[x]$, then each $\sigma_i$ maps a root of $f(x)$ to another root of it i.e., if $f(r)=0$ then $f(\sigma_i(r))=0$, because $f(\sigma_i(r)) = \sigma_i(f(r)) = \sigma_i(0) = 0$ as ring homomorphism fixes 0.  
\end{enumerate}

Due to the above, the following proposition can be proved. 
\fi

\begin{proposition}
Let $K$ be an algebraic number field of degree $n$. Then there are precisely $n$ distinct field embeddings from $K$ to $\C$. 
\end{proposition}

The embeddings $\{\sigma_i\}_{i \in [n]}$ map the primitive element $r$ to different roots of $r$'s minimal polynomial $f(x)$, which is a collection of real and complex numbers. Hence, we can distinguish these embeddings as real and complex embeddings. 
\reversemarginpar
\marginnote{\textit{Real and complex embeddings}}
If $\sigma_i(K) \subseteq \R$ (or $\sigma_i(r) \in \R$) then it is a \textbf{real embedding}, otherwise it is a \textbf{complex embedding}. By Complex Conjugate Root Theorem\footnote{The complex roots of real coefficient polynomials are in conjugate pairs.},  the images of the complex embeddings are in conjugate pairs, so we only need to keep half of the complex embeddings and split each of them into the real and complex parts. Let $s_1$ be the number of real embeddings and $s_2$ be the number of conjugate pairs of complex embeddings, then the total number of embeddings is $n=s_1 + 2s_2$. In addition, let $\{\sigma_i\}_{i \in [s_1]}$ be the real embeddings, $\{\sigma_j\}_{j \in [s_1+1, n]}$ be the complex embeddings and $\sigma_{s_1 + j} = \overline{\sigma_{s_1 + s_2 + j}}$ be the conjugate pairs for $j \in [s_2]$, then we have the following definition of a canonical embedding of a algebraic number field.

\begin{definition}
\label{app def:canEmbd}
\reversemarginpar
\marginnote{\textit{Canonical embedding}}
A \textbf{canonical embedding} (or \textbf{Minkowski embedding}) $\sigma$ of an algebraic number field $K$ of degree $n$ to the $n$-dimensional complex plane $\C^n$ is defined as 
\begin{align*}
    \sigma: K &\rightarrow \R^{s_1} \times \C^{2s_2} \subseteq \C^n \\
    %\sigma(r) &\mapsto (\sigma_1(r), \dots, \sigma_{s_1}(r), \sigma_{s_1+1}(r),\dots, \sigma_{s_1+s_2}(r)),
    \sigma(r) &\mapsto (\sigma_1(r), \dots, \sigma_{s_1}(r), \sigma_{s_1+1}(r),\dots, \sigma_n(r)).
\end{align*}
\end{definition}

As mentioned above, the complex embeddings are in conjugate pairs so it is not necessary to keep both complex embeddings ini a conjugate pair. This gives rise to a different (and more practical)
\reversemarginpar
\marginnote{\textit{$\tau$ embedding}}
embedding  
\begin{align*}
    \tau:K &\rightarrow V\\
    \tau(r) &\mapsto (\sigma_1(r), \dots, \sigma_{s_1}(r), \sigma_{s_1+1}(r),\dots, \sigma_{s_1+s_2}(r)),
\end{align*}
where for all $i \in [s_1+s_2,n]$, each $\sigma_i$ separates the real and imaginary parts as  $\sigma_i(r)=\left(Re(\sigma_r(r)),Im(\sigma_i(r))\right)$, so the image of this embedding can be explicitly write out as  
\begin{align}
\label{app equation:minkowski embedding}
    \tau(r) = (&\sigma_1(r), \dots, \sigma_{s_1}(r), \nonumber \\  
    &Re(\sigma_{s_1+1}(r)),Im(\sigma_{s_1+1}(r)),\dots, Re(\sigma_{s_1+s_2}(r)),Im(\sigma_{s_1+s_2}(r))).
\end{align}

The canonical embedding maps a number field to an $n$-dimensional space, 
\reversemarginpar
\marginnote{\textit{Canonical space}}
named \textbf{canonical space} (or \textbf{Minkowski space}) and can be expressed as  
\begin{equation*}
H = \left\{(x_1, \dots, x_n) \in \R^{s_1} \times \C^{2s_2} \mid x_{s_1 + j} = \overline{x_{s_1 + s_2 + j}}, \forall j \in [s_2]\right\} \subseteq \C^n.   
\end{equation*}
%The \textbf{Minkowski space} $H_{\R}$ in \citep{mukherjee2016cyclotomic} is isomorphic to this canonical space $H$ with a minor difference. In $H_{\R}$, only one element from each complex conjugate pair is kept and it is written in two parts as $(Re, Im)$.
The canonical space $H$ can be verified to be isomorphic to $\R^n$ using the following steps. We can establish a one to one correspondence between the standard basis of $\C^n$ and an orthonormal basis of $H$. In detail, let $\{e_i\}_{i \in [n]}$ be the standard basis of $\C^n$ where in each $e_i$ the ith component is 1 and the rest are zero. Then we can build a basis $\{b_i\}_{i \in [n]}$ for $H$ such that 
\begin{itemize}
    \item for $j \in [s_1]$, let $h_j = e_j$  and 
    \item for $j \in [s_1+1, s_1 + s_2]$, let $h_j = \frac{1}{\sqrt{2}} (e_j + e_{j + s_2})$ and $h_{j+s_2} = \frac{i}{\sqrt{2}}(e_j - e_{j + s_2})$.
\end{itemize}
Similarly, we can prove the space $V$, to which $K$ is mapped to by the embedding $\tau$ is also isomorphic to $\R^n$. %This the isomorphism with $\R^n$ can be checked by a map $F(x_i) \mapsto x_i$ for $i \in [1, s_i]$ and $F(x_i) \mapsto Re(x_i)$ for $i \in [s_1+1, s_1+s_2]$ and $F(x_i) = Im(x_i)$ for $i \in [s_1+s_2+1,n]$. As shown in the example on Page 39 of \cite{mukherjee2016cyclotomic}.

In the next example, we will look at the canonical embedding of a cyclotomic number field and construct a basis of the canonical space by using the above rules. 
\begin{example}
Let  $K=\Q(\zeta_8)$ be a cyclotomic number field, where $\zeta_8 = \frac{\sqrt{2}}{2}+ i \frac{\sqrt{2}}{2}$ is an 8th primitive root of unity. The minimal polynomial of $\zeta_8$ is the 8th cyclotomic polynomial $\Phi_8(x) = x^4+1$ with degree $\varphi(8)=4$, whose roots are the 8th primitive roots
\begin{align*}
\zeta_8 &= \frac{\sqrt{2}}{2}+ i \frac{\sqrt{2}}{2}, \\
\zeta_8^3 &= -\frac{\sqrt{2}}{2}+ i \frac{\sqrt{2}}{2},\\
\zeta_8^5 &= -\frac{\sqrt{2}}{2}- i \frac{\sqrt{2}}{2}, \\
\zeta_8^7 &= \frac{\sqrt{2}}{2}- i \frac{\sqrt{2}}{2}.
\end{align*}
The degree of the cyclotomic field is $n=4$, so all 4 embeddings $\sigma_i: K \rightarrow \C^4$ are complex, that is, $s_1=0$ and $s_2=2$. The four complex embeddings are 
\begin{align*}
    \sigma_1\left(\frac{\sqrt{2}}{2}+ i \frac{\sqrt{2}}{2}\right) &= \frac{\sqrt{2}}{2}+ i \frac{\sqrt{2}}{2}, \\
    \sigma_2\left(\frac{\sqrt{2}}{2}+ i \frac{\sqrt{2}}{2}\right) &= -\frac{\sqrt{2}}{2}+ i \frac{\sqrt{2}}{2}, \\
    \sigma_3\left(\frac{\sqrt{2}}{2}+ i \frac{\sqrt{2}}{2}\right) &= \frac{\sqrt{2}}{2}- i \frac{\sqrt{2}}{2}, \\
    \sigma_4\left(\frac{\sqrt{2}}{2}+ i \frac{\sqrt{2}}{2}\right) &= -\frac{\sqrt{2}}{2}- i \frac{\sqrt{2}}{2}, 
\end{align*}
where $\sigma_1,\sigma_3$ and $\sigma_2,\sigma_4$ are in conjugate pairs. So the embedding by Equation \ref{app equation:minkowski embedding} is 
\begin{equation*}
    \tau\left(\frac{\sqrt{2}}{2}+ i \frac{\sqrt{2}}{2}\right)=\left(\frac{\sqrt{2}}{2}, \frac{\sqrt{2}}{2}, -\frac{\sqrt{2}}{2}, \frac{\sqrt{2}}{2} \right).
\end{equation*}

Let $x_1 = \zeta_8$, $x_2 = \zeta_8^3$, $x_3 = \zeta_8^7$, $x_4 = \zeta_8^5$, so $x_1 = \overline{x_3}$ and $x_2 = \overline{x_4}$ are in conjugate pairs. By definition of canonical space, we have $(x_1, x_2, x_3, x_4) \in H$ is an element of the space. According to the above basis construction, we get the basis $\{h_1, h_2,h_3,h_4\}$ for $H$ from the standard basis of $\R^4$, where
\begin{align*}
h_1 &= \frac{\sqrt{2}}{2} (e_1 + e_3), \\
h_2 &= \frac{\sqrt{2}}{2} (e_2 + e_4),\\
h_3 &= i\frac{\sqrt{2}}{2} (e_1 - e_3),\\
h_4 &= i\frac{\sqrt{2}}{2} (e_2 - e_4).    
\end{align*}
Hence, the element $(x_1, \dots, x_4) = h_1-h_2+h_3+h_4$ and its conjugate $\overline{(x_1, \dots, x_4)} = h_1-h_2-h_3-h_4$. The complex conjugation operator maps $H$ to itself by flipping the signs of the coefficients of $\{h_{s_1+s_2+1},\dots, h_n\}$ as shown in the example. 
\end{example}

Now we know a number field $K$ is mapped to a canonical space that is isomorphic to $\R^n$, we can defined the notion of geometric norm on the number field $K$ just as we did in $\R^n$. For any element $x \in K$, the \textbf{$L_p$-norm} of $x$ is defined as 
\reversemarginpar
\marginnote{\textit{$L_p$-norm}}
\begin{equation*}
    ||x||_p = ||\sigma(x)||_p = 
    \begin{cases}
     \left( \sum_{i \in [n]} |\sigma_i(x)|^p \right)^{1/p} & \text{ if $p < \infty$},  \\
    \max_{i \in [n]} |\sigma_i(x)| & \text{ if $p = \infty$}.
    \end{cases}
\end{equation*}

\begin{example}
We use this example to illustrate the $L_p$-norm of a root of unity in a cyclotomic number field. 

Let $\sigma: K(\zeta_n) \rightarrow H$ be the canonical embedding for the nth cyclotomic field. The minimal polynomial of $\zeta_n$ is the nth cyclotomic polynomial $\Phi_n(x)$ which has only complex roots for $n \ge 3$, because the two real roots are not primitive. The complex embeddings are given by $\sigma_i(\zeta_n) = \zeta_n^i$, where $i \in (\Z/n\Z)^*$, so $n = 2s_2 = |(\Z/n\Z)^*|$.

For any nth root of unity $\zeta_n^j \in K$, an embedding $\sigma_i(\zeta_n^j)$ is still a root of unity and hence has magnitude 1. So the $L_P$-norm of an nth root of unity $||\zeta_n^j||_p = n^{1/p}$ for $p < \infty$ and $||\zeta_m^j||_{\infty} = 1$.
\end{example}

We have specified the canonical embedding of a number field to a space that is isomorphic to $\R^n$. What we are really interested in is how the ring of integers is mapped by the embedding. The following theorem states that the canonical embedding maps $\OO_K$ to a full-rank lattice. Towards the end of this section, we will discuss the minimum distance (or the shortest vector) of this lattice and how the determinant of this lattice $\sigma(\OO_K)$ is related to a quantity of the number field, called the discriminant.

\begin{theorem}
\label{app thm:rngIntLat}
\reversemarginpar
\marginnote{\textit{$\tau(\OO_K)$ is lattice}}
Let $K$ be an $n$-dimensional number field and $\tau: K \rightarrow V \cong \R^n$ be the embedding of $K$ as defined in Equation \ref{app equation:minkowski embedding}, then $\tau$ maps the ring of integers $\OO_K$ to a full-rank lattice in $\R^n$. 
\end{theorem}

\begin{proof}
By definition, a lattice is a free $\Z$-module. Let $\{e_1,\dots,e_n\}$ be an integral basis of $\OO_K$, then every element $x \in \OO_K$ can be written as $x=\sum_{i=1}^n z_i e_i$, where $z_i \in Z$. The image of $x$ under the embedding is $\tau(x)=\sum_{i=1}^n z_i \tau(e_i)$, so $\tau(\OO_K)$ is $\Z$-module generated by $\{\tau(e_1),\dots,\tau(e_n)\}$. It remains to show the set is a basis of $\tau(\OO_K)$, which then leads to the conclusion that it is a free $\Z$-module, hence a lattice. To do so, define the following matrix and prove it has a non-zero determinant
\begin{equation*}
N^T = \left(
\begin{smallmatrix}
\sigma_1(e_1) & \cdots & \sigma_{r_1}(e_1) & Re(\sigma_{r_1+1}(e_1)) & Im(\sigma_{r_1+1}(e_1)) & \cdots & Re(\sigma_{r_1+r_2}(e_1)) & Im(\sigma_{r_1+r_2}(e_1)) \\
\vdots & & \vdots & \vdots & \vdots & & \vdots & \vdots \\
\sigma_1(e_n) & \cdots & \sigma_{r_1}(e_n) & Re(\sigma_{r_1+1}(e_n)) & Im(\sigma_{r_1+1}(e_n)) & \cdots & Re(\sigma_{r_1+r_2}(e_n)) & Im(\sigma_{r_1+r_2}(e_n)) \\
\end{smallmatrix}
\right).
\end{equation*}
It can be prove that $\det N $ is related to $\det M$, where $M$ is a matrix defined by using the canonical embedding $\sigma$ of $K$. In addition, $\det M \neq 0$, so $\det N \neq 0$. The details are skipped. See the proof of Lemma 10.6.1 on page 65 of Ben Green's book or the proof of Proposition 4.26 on page 80 of Milne's book. 
\end{proof}

\subsection{Ideals of ring of integers}
	
The ring of integers $\OO_K$ in a number field carries a lot of similarities to $\Z$, but it lacks an important property of being a unique factorization domain.

\begin{definition}
An integral domain $D$ is a 
\reversemarginpar
\marginnote{\textit{UFD}}
\textbf{unique factorization domain (UFD)} if every non-zero non-unit element $x \in D$ can be written as a product 
\begin{equation*}
    x= p_1 \cdots p_n
\end{equation*}
of $0<n<\infty$ irreducible elements $p_i \in D$ uniquely up to reordering of the irreducible elements.
\end{definition}

For example, $\Z$ is a UFD because every integer can be uniquely factored into prime factors. But the extension $\Z(\sqrt{5})$ is not a UFD, because $6=2 * 3 = (1+\sqrt{-5})(1-\sqrt{-5})$. UFD is essential for cryptography because if we assume factoring a large integer into prime factors is hard, we want to be sure that we are aware of all the factorizations. So it would be assuring if the factorization is unique. In addition, unique factorization implies unique divisibility.

For this reason, we do not work with individual elements in $\OO_K$ but study an enlarged world, the ideals of $\OO_K$, denoted as $\text{Ideals}(\OO_K)$, and prove that they can be uniquely factored into prime ideals. The general context of proving such a property and some other properties of ideals of $\OO_K$ is in a Dedekind domain. 
\reversemarginpar
\marginnote{\textit{Dedekind domain}}
A \textbf{Dedekind domain} is an integral domain in which every non-zero proper ideal factors into a product of prime ideals. The ring of integers $\OO_K$ is just a special case of a Dedekind domain as we will see at the end of this subsection once we have stated that the integral ideals of $\OO_K$ form a UFD. In addition, we introduce fractional ideals of $\OO_K$ and prove that they form a multiplicative group under ideal multiplication. %They are particularly important for RLWE because their images under the canonical embedding are ideal lattices. 
	
%The key thing to keep in mind while studying these concepts is that the canonical map sends a fractional ideal of $\OO_K$ (including $\OO_K$ and its integral ideals) to a full-rank lattice in the space $H$ (Proposition 3.5.1 \cite{mukherjee2016cyclotomic}). 
	
The RLWE problem is constructed based on ideal lattices, which are the images of the canonical embedding of integral (or fractional) ideals of $\OO_K$ (Proposition 3.5.1 \citep{mukherjee2016cyclotomic}, Proposition 4.26 of J. S. Milne's book \textit{Algebraic Number Theory}). 
Since integral and fractional ideals are related by an algebraic integer $d \in \OO_K$ (which is considered as the denominator), RLWE can be defined in either setting.

\subsubsection{Integral ideals}

We start this section by introducing the notion of ideal in $\OO_K$. The intuition is similar to an ideal in an ordinary ring. Recall that an ideal of a ring is an additive subgroup of the ring that is closed under multiplication by ring elements. Similarly, we can define an ideal of $\OO_K$. 
	
\begin{definition}
Given a number field $K$ and its ring of integers $\OO_K$, an 
\reversemarginpar
\marginnote{\textit{Integral ideal}}
\textbf{integral ideal} (or simply \textbf{ideal}) $I$ of $\OO_K$ is a non-empty (i.e., $I \neq \emptyset$) and non-trivial (i.e., $I \neq \{0\}$) additive subgroup of $\OO_K$ that is closed under multiplication by elements of $\OO_K$, i.e., for any $r \in \OO_K$ and any $x \in I$, we have $rx \in I$. 
\end{definition}

Since $\OO_K$ is commutative, we do not distinguish between left and right ideal. The above definition is consistent with ideals in ordinary rings, except that the zero ideal $\{0\}$ is excluded in order to define ideal division later. Since $\OO_K$ has a $\Z$-basis, its integral ideals have $\Z$-basis too. In other words, every non-zero integral ideal of $\OO_K$ is a free $\Z$-module. 
	
We can define a 
\reversemarginpar
\marginnote{\textit{Principle ideal}}
\textbf{principal ideal} in a similar way as an ideal that is generated by a single element via multiplications with all elements in $\OO_K$. That is, the principle ideal generated by an element $x \in \OO_K$ is 
\begin{equation*}
    (x) := \{\alpha x \mid \alpha \in \OO_K\}.
\end{equation*}
Given elements $x_1, \dots, x_r \in \OO_K$, the ideal \textbf{generated by} the $x_i$'s is 
\begin{equation*}
    (x_1,\dots, x_r) := \left\{\sum_{i \in [r]} \alpha_i x_i \mid \alpha_i \in \OO_K\right\}
\end{equation*} 
the set of linear combinations of the $x_i$'s, where the coefficients are taken from $\OO_K$.

We can also define some basics operations on ideals. If $I$ and $J$ are both integral ideals of $\OO_K$, their 
\reversemarginpar
\marginnote{\textit{Ideal sum}}
\textbf{sum} is defined as 
\begin{align*}
    I + J := \{x + y \mid x \in I \text{ and } y \in J\},
\end{align*}
which is still an ideal in $\OO_K$.\footnote{It can be proved that $I+J$ and $(I \cup J)$ are equivalent.} The sum ideal does not respect the additive structure on $\OO_K$. For example, if $I = J = (1)$, then $I+J= (1) \neq (1+1) = (2)$. The sum of two ideals is not so important, what more important for the following works is the product of two ideal. 
	
We would thought that the product set $S=\{xy \mid x \in I \text{ and } y \in J\}$ is also an ideal just like the sum but it is not, because it may not be closed under addition. For this reason, 
\reversemarginpar
\marginnote{\textit{Ideal product}}
the \textbf{product} of two ideals $I$ and $J$ is defined as 
\begin{align*}
    IJ := \left\{\sum_{i \in [r]} a_i b_i \mid a_i \in I \text{ and } b_i \in J\right\}. 
\end{align*}
It consists of all finite sums of the products of two ideal elements.\footnote{Again, it can be proved that $IJ$ and $(IJ)$ are equivalent.} 
By grouping all finite sums of products, the set is closed under addition. Closed under multiplication by elements in $\OO_K$ can be easily checked. Since $\OO_K$ is commutative, ideal multiplication is commutative too. 
	
\begin{example}
Given the ring of integers $\OO_K = \Z$ and two of its ideals $I = 2\Z = \{2, 4,6,8,\dots,\}$ and $J = 3\Z=\{3,6,9,12,\dots,\}$, their ideal product is $IJ=\{2\cdot 3, 2\cdot 6,2\cdot 3 + 2\cdot 6,\dots\}$.  
\end{example}

We have defined ideal multiplication, it is natural to also define ideal division, provided ideals of $\OO_K$ does not include the zero ideal according to the definition.  

\begin{definition}
Let $I$ and $J$ be two ideals of $\OO_K$. We say 
\reversemarginpar
\marginnote{\textit{Ideal division}}
$J$ \textbf{divides} $I$, denoted $J \mid I$, if there is an ideal $M \subseteq \OO_K$ such that $I = JM$.
\end{definition}

The following theorem gives a more intuitive way of thinking about ideal division by relating division with containment. 

\begin{theorem}
\label{app thm:divCont}
\reversemarginpar
\marginnote{\textit{Divisibility $\iff$ containment}}
Let $I$ and $J$ be two ideals of $\OO_K$. Then $J \mid I$ if and only if $I \subseteq J$. 
\end{theorem}
Divisibility implies containment, because if $J\mid I$ then $I=JK\subseteq J$, so $I\subseteq J$. The converse may not be true in general, but is certainly true in these ideals are in the ring of integers. Next, we define prime ideals in $\OO_K$ which is the same as how prime ideals are defined in rings. 

\begin{definition}
\reversemarginpar
\marginnote{\textit{Prime ideal}}
An ideal $I$ of $\OO_K$ is \textbf{prime} if  
\begin{enumerate}
	\item $I \neq \OO_K$ and 
	\item if $xy \in I$, then either $x \in I$ or $y \in I$. 
\end{enumerate}
\end{definition}

The next lemma gives an equivalent definition of prime ideals in terms of other ideals in $\OO_K$. 
\begin{lemma}
An ideal $I$ of $\OO_K$ is prime if and only if for ideals $J$ and $K$ of $\OO_K$, whenever $JK \subseteq I$, either $J \subseteq I$ or $K \subseteq I$. 
\end{lemma}
By the equivalence relation between division and containment, a prime ideal $I$ can be more intuitively defined as a proper ideal such that whenever $I \mid JK$, either $I \mid J$ or $I \mid K$. This is consistent with how prime numbers are defined in $\Z$.  

An important observation is that in $\OO_K$, prime ideals are also maximal. So we do not introduce maximal ideals separately. Recall that a maximal ideal in a ring is an ideal that is contained in exactly two ideals, i.e, itself and the entire ring. 

\begin{lemma}
\label{app lm:primeIsMax}
\reversemarginpar
\marginnote{\textit{Prime is maximal}}
In $\OO_K$, all prime ideals are maximal. 
\end{lemma}
The proof relies on the results that a commutative ring quotienting by a prime ideal gives an integral domain, quotienting by a maximal ideal gives a field. 
\begin{proof}
If $I$ is a prime ideal of $\OO_K$, then $\OO_K / I$ is an integral domain. In addition, the integral domain is finite. This implies that for every $x$ in the integral domain, it satisfies that $x^n = 1$ for some $n$, so $x \cdot (x^{n-1}) = 1$. Hence, every non-zero element in the integral domain has an inverse, which means the quotient ring $\OO_K / I$ is a field. Therefore, $I$ is maximal. 
\end{proof}

An important property of the ideals of $\OO_K$ is that they can be uniquely factorized into irreducible factors, in this case prime ideals. This is one of the main theorems in the course of Algebraic Number Theory. Note that it is not always true that $\OO_K$ is a unique factorization domain. As we have seen, an counter example is when $K=\Q(\sqrt{-5})$ and $\OO_K=\Z(\sqrt{-5})$, in which $6 = 2 * 3=(1+\sqrt{-5})*(1-\sqrt{-5})$.\footnote{It is also necessary to check that 2, 3, $1+\sqrt{-5}$ and $1-\sqrt{-5}$ are irreducible and are not associates of each other. For more details, see the example on Page 30 of Ben Green's notes on algebraic number theory.}

\begin{theorem}
\label{app thm:idealsOKUFD}
\reversemarginpar
\marginnote{$\text{Ideals}(\OO_K)$ is UFD}
For an algebraic number field $K$, every non-zero proper ideal $I$ of $\OO_K$ admits a unique factorization
\begin{equation*}
    I = P_1 \cdots P_k,
\end{equation*}
into prime ideals $P_i$ of $\OO_K$. 
\end{theorem}

\subsubsection{Fractional ideal}

% fractional ideal 
Another important concept in number fields is fractional ideal. It generalizes integral ideals in a number field, but is not an ideal in the number field or its ring of integers. The essential properties that are useful in proving RLWE are fractional ideals can be uniquely factorized into prime ideals and they form a multiplicative group. We first give a general definition of fractional ideals in an integral domain. We will then refine this definition in a number field. Let $R$ be an integral domain, recall a field of fractions of $R$ is 
\begin{equation*}
    Frac(R)=\{(p,q) \in R \times (R\setminus \{0\}) \mid (p,q) \sim (r,s) \iff ps=qr\}.
\end{equation*}
It is clear that $Frac(R)$ is an $R$-module and it contains $R$. Given an $R$-module $M$, recall a submodule $N$ of $M$ is a subgroup of $M$ that is closed under scalar multiplication by elements in $R$, that is, $ar \in N$ for any $a \in N$ and any $r \in R$. Now, we can define fractional ideal of an integral domain.

\begin{definition}
\label{app def:fracIdeal}
Let $R$ be an integral domain and $Q=Frac(R)$ be the field of fractions. A
\reversemarginpar
\marginnote{\textit{Frac ideal}}
\textbf{fractional ideal} $I$ of $R$ is an $R$-submodule of $Q$ such that there exists a non-zero element $d \in R$ satisfying $dI \subseteq R$. 
\end{definition}

$I$ is an $R$-submodule of $Q$ implies that $I$ is an (additive) subgroup of $Q$ and it is closed under multiplication by all elements in $R$. The existence of $d \in R$ can be thought as cancelling the denominator of $I$, which is also why $d$ needs to be non-zero. Combining with being an submodule, we have $rI \subseteq R$ is an integral ideal. As we will explain later that a fractional ideal is neither an ideal of $\OO_K$ nor $K$, so some prefer to call them ``fractional ideals in $K$'' while others refer to them as ``fractional ideals of $\OO_K$''. For simplicity, we sometimes refer to them just as fractional ideals without mentioning $\OO_K$ or $K$.

We further refine the definition for our purpose. In the context of a number field, $\OO_K$ is an integral domain and $K=Frac(\OO_K)$ is its field of fractions. By the above definition, a fractional ideal $I$ is an $\OO_K$-submodule of $K$ such that there exists a non-zero element $d \in \OO_K$ satisfying $dI \subseteq \OO_K$. Alternatively, we can just say that $dI$ is an integral ideal, which implies it is closed under addition and multiplication by the ring elements, hence equivalent as being a submodule. 
\begin{definition}
\label{app def:fracIdeal2}
Let $K$ be a number field and $\OO_K$ be its ring of integers. A \textbf{fractional ideal} $I$ of $\OO_K$ is a set such that $dI \subseteq \OO_K$ is an integral ideal for a non-zero $d \in \OO_K$.  
\end{definition}

Alternatively, given an integral ideal $J \subseteq \OO_K$ and an element $x \in K^{\times}$ (or an invertible element $x \in K$), the corresponding fractional ideal $I$ can be expressed as 
\begin{equation*}
I = x^{-1} J := \{x^{-1} a \mid a \in J\} \subseteq K.
\end{equation*}
From this expression, it is clearer that the non-zero element $d$ in the above definitions is for cancelling the denominator $x$ of in this expression. Note $x$ is in $K$ but not $\OO_K$ because it needs to be invertible. Since a non-zero integral ideal is a free $\Z$-module and a fractional ideal is related to an integral ideal by an invertible element, it follows that a fractional ideal is a 
\reversemarginpar
\marginnote{\textit{Free $\Z$-module}}
free $\Z$-module too. So it has a $\Z$-basis.   

Note that \textbf{a fractional ideal is not an ideal of $R$} (unless it is contained in $R$), because it is not necessarily a subset of the integral domain $R$. For example, as we will see in the following example, $\frac{5}{4}\Z \not\subseteq \OO_K$ is a fractional ideal of $\OO_K$. \textbf{Nor it is an ideal of the field of fractions $Frac(R)$}, because $Frac(R)$ is a field which has only zero and itself as ideals. 

\begin{example}
Let $K= \Q$ and $\OO_K= \Z$. Clearly, $\Q$ is a $\Z$-module. $I = \frac{5}{4}\Z$ is a $\Z$-submodule of $\Q$, because $I$ is an additive subgroup of $\Q$ and for all $x \in \Z$, we have $xI = I$. There exists an integer $4\in \Z$ such that $4 \cdot \frac{5}{4}\Z = 5\Z \subseteq\Z$ is an ideal. So  $I = \frac{5}{4}\Z$ is a fractional ideal of $\Z$. Alternatively, it can be expressed as $4^{-1}5\Z \subseteq \Q$, where $5\Z$ is an ideal of $\Z$.

A counter example is when $I=\Z[\frac{1}{2}]$. This is an $\OO_K$-submodule of $K=\Q$, but does not exists a denominator $d \in \OO_K$ such that $dI \subseteq \OO_K$ is an ideal. 
\end{example}

The product of two fractional ideals can be defined the same as the product of two 
\reversemarginpar
\marginnote{\textit{Product}}
integral ideals. That is, if $I$ and $J$ are both fractional ideals, then their product consists of all the finite sums $\sum_{i \in [n]} a_i b_i$, where $a_i \in I$ and $b_i \in J$. It is easy to check that the product of two fractional ideals is still a fractional ideal.

To reach the conclusion that fractional ideals form a multiplicative group, it remains to show that every fractional ideal has an inverse. This is done via the following two lemmas. The first lemma proves that every prime ideal of $\OO_K$ has an inverse. The second lemma proves that every non-zero integral ideal of $\OO_K$ has an inverse. 

\begin{lemma}
\reversemarginpar
\marginnote{\textit{Prime ideal inverse}}
If $P$ is a prime ideal in $\OO_K$, then $P$ has an inverse $P^{-1} = \{a \in K \mid a P \subseteq \OO_K\}$ that is a fractional ideal.
\end{lemma}

\begin{proof}
Since $\OO_K$ is a ring, it is closed under multiplication. This implies $\OO_K \subseteq P^{-1}$, so $P^{-1}$ is not an integral ideal of $\OO_K$. We want to show $P^{-1}$ is a fractional ideal of $\OO_K$. It is not difficult to see that $P^{-1}$ is a $\OO_K$-submodule of $K$. In addition, there is a $b \in \OO_K$ such that $bP^{-1}$ is an integral ideal of $\OO_K$, so by definition $P^{-1}$ is a fractional ideal of $\OO_K$. 

It remains to prove that $P^{-1}$ indeed is an inverse of $P$. We will not state the proof here. For details, see \textit{Proof of Theorem 3.1.8} on Page 45 in William Stein's \textit{Algebraic Number Theory}.
\end{proof}

\begin{example}
In the number field $K = \Q$, let $P = (2) = \{2, 4, 6, \dots\}$ be a prime ideal in $\OO_K = \Z$. Then its inverse $P^{-1} = \{\Z, \frac{\Z}{2}, \frac{\Z}{4}, \frac{\Z}{6}, \dots\}$ is a fractional ideal of $\Z$. 
\end{example}

Since a fractional ideal and the corresponding integral ideal can be obtained from each other, we can express a fractional ideal as $I=yJ$ for an integral ideal $J$ and an invertible element $y=x^{-1}$. To prove $I$ has an inverse $(yJ)^{-1}$, it is sufficient to show that the integral ideal $J$ has an inverse, because the principal ideal $(y)$ has an inverse $(1/y)$. 

\begin{lemma}
\reversemarginpar
\marginnote{Integral ideal inverse}
Every non-zero integral ideal of $\OO_K$ has an inverse. 
\end{lemma}
\begin{proof}
Prove by contradiction. Assume not every non-zero integral ideal of $\OO_K$ has an inverse. Let $I$ be the maximal non-zero integral ideal of $\OO_K$ that has no inverse. $P$ is still a prime ideal of $\OO_K$, then $I \subseteq P$. Multiplying both sides by $P^{-1}$, we get $I \subseteq P^{-1} I \subseteq P^{-1} P = \OO_K$. The key here is to show that $I \neq P^{-1} I$. Since $I$ is an integral ideal of $\OO_K$, the equality holds if $P^{-1} \subseteq \OO_K$ because an ideal is closed by multiplication with ring elements. But we already know from the above lemma that the inverse of a prime ideal is a fractional ideal of $\OO_K$ that is not in the ring, so $\OO_K \subseteq P^{-1}$. Hence, the equality cannot hold, that is we must have $I \subsetneq P^{-1} I \subseteq P^{-1} P = \OO_K$. Since $I$ is the maximal integral ideal in $\OO_K$ that does not have an inverse, the ideal $P^{-1} I$ must have an inverse $J$ such that $(P^{-1} I)J=\OO_K$, so $(P^{-1} J)I=\OO_K$ and $P^{-1}J$ is an inverse of $I$. 
\end{proof}

The two lemmas together prove that a fractional ideal has an inverse. 
See \textit{Proof of Theorem 3.1.8} on Page 46 in William Stein's \textit{Algebraic Number Theory} for more detail. To be more precise, the inverse 
\reversemarginpar
\marginnote{\textit{Frac ideal inverse}}
of a fractional ideal $I$ has the form 
\begin{equation}
\label{app equ:fracIdInv}
    I^{-1} = \{x \in K \mid xI \subseteq \OO_K\}.
\end{equation}
Given fractional ideals $I$ and $J$, if $IJ=(x)$ is a \textbf{principal fractional ideal}\footnote{Since both $I$ and $J$ are fractional ideals, their product is also a fractional ideal, which is not necessary an integral ideal, so it is named principal fractional ideal to differentiate it from a principal ideal.}, then its inverse is $I^{-1}=\frac{1}{x}J$.
It can be proved that this inverse is also a fractional ideal and it is unique for the given fractional ideal $I$. See Conrad's lecture notes on ``Ideal Factorization'' (Definition 2.5, Theorem 2.7 and Theorem 4.1). 

\begin{theorem}
\label{app thm:fracIdealGroup}
\reversemarginpar
\marginnote{\textit{Multiplicative group}}
The set of fractional ideals of the ring of integers $\OO_K$ of a number field $K$ is an abelian group under multiplication with the identity element $\OO_K$. 
\end{theorem}

The same theorem is also stated in \citet{alaca2004introductory}'s Theorem 8.3.4. Since fractional ideals include integral ideals, these two theorems are identical. 
\begin{theorem}
Let $K$ be an algebraic number field and $\OO_K$ be the ring of integers of $K$ . Then the set of all non-zero integral and fractional ideals of $\OO_K$ forms an abelian group with respect to multiplication.
\end{theorem}

Finally, we come to another important result of this section, which states that a fractional ideal can be uniquely factored into the product of prime ideals.

\begin{theorem}
\reversemarginpar
\marginnote{\textit{Unique factorization}}
If $I$ is a fractional ideal of $\OO_K$ then there exits prime ideals $P_1, \dots, P_n$ and $Q_1, \dots, Q_m$, unique up to order, such that 
\begin{equation*}
    I = (P_1 \cdots P_n)(Q_1 \cdots Q_m)^{-1}.
\end{equation*}
\end{theorem}
The theorem follows from the fact that a fractional ideal $I=J/a$, where $J$ is an integral ideal and $a \in \OO_K$. Since both $J$ and $(a)$ are ideals of $\OO_K$, Theorem \ref{app thm:idealsOKUFD} implies they have unique prime ideal factorization, so the theorem holds.

\subsubsection{Chinese remainder theorem}
\label{app subsubsec:crt}
Given that integral ideals form a UFD, the \textbf{Chinese Remainder Theorem (CRT)} carries over from rational integers to integral ideals of $\OO_K$. In this subsection, we state CRT in the general context of Dedekind domain, in which the ring of integers $\OO_K$ is a special case. This is to get the reader to be familiar with CRT in general, which will be used in latticed-based cryptography and homomorphic encryption. 

The classical form of CRT states that for integers $n_1, \dots, n_k$ that are pairwise 
\reversemarginpar
\marginnote{\textit{Classical CRT}}
coprime and integers $a_1, \dots, a_k$ such that $0 \le a_i < n_i$, the system of congruences 
\begin{align*}
    x &= a_1 \bmod n_1 \\
    x &= a_2 \bmod n_2 \\
    \vdots \\
    x &= a_k \bmod n_k
\end{align*}
has a unique solution $x$ up to congruent modulo $N = \prod_{i=1}^n n_i$, that is, if $y$ is another solution then $x = y \bmod N$.

Similarly, CRT can solve the problem of polynomial interpolation.\footnote{The example is taken from \url{https://math.berkeley.edu/~kmill/math55sp17/crt.pdf}} Given values $x_i, \dots, x_n, y_1, \dots, y_n \in \R$, there is a unique polynomial $p(x)$ satisfies 
\begin{align*}
    p(x_1) &=y_1 \\
    p(x_2) &= y_2 \\
    \vdots \\
    p(x_n) &= y_n.
\end{align*}
The problem can be solved in terms of CRT as finding a unique polynomial $p(x)$ that satisfies 
\begin{align*}
    p(x) & = y_1 \bmod x-x_1\\
    p(x) &= y_2  \bmod x-x_2\\
    \vdots \\
    p(x) &= y_n  \bmod x-x_n.
\end{align*}
We know from previous sections that $p(x) -y_i= 0 \bmod x-x_i$, which can also be expressed in quotient as $(p(x)-y_i) / (x-x_i)$, is the  extension field $\Q(x_i)$ over $\Q$ that contains the roots of $p(x) -y_i$ and $x_i$. % \kl{Why need this paragraph?}

A more abstract version of CRT states that if the $n_i$'s are pairwise coprime, the map 
\reversemarginpar
\marginnote{\textit{CRT in rings}}
\begin{equation*}
    x \bmod N \mapsto (x \bmod n_1, \dots, x\bmod n_k)
\end{equation*}
defines an isomorphism 
\begin{equation*}
    \Z / N\Z \cong \Z / n_1 \Z \times \cdots \times \Z / n_k \Z
\end{equation*}
between the ring of integers modulo $N$ and the direct product of the $k$ rings of integers modulo $n_i$. 

To generalize CRT to the ring of integers $\OO_K$, we define coprime ideals in $\OO_K$. Since ideals in $\OO_K$ can be uniquely factorized, it makes sense to talk about coprimality.

\begin{definition}
Let $I$ and $J$ be two integral ideals in $\OO_K$. Then $I$ and $J$ are \textbf{coprime} if they do
\reversemarginpar
\marginnote{\textit{Coprime ideals}}
not have any prime factors in common. That is, there is no prime ideal dividing both of them. 
\end{definition}

This definition relies on the notion of common factors of two ideals. 

\begin{definition}
\reversemarginpar
\marginnote{\textit{GCD of ideals}}
Let $I$ and $J$ be integral ideals of $\OO_K$, their \textbf{greatest common divisor (GCD)} $\gcd(I, J) = I+J$. 
\end{definition}

By definition of ideal GCD, we can re-define ideal coprimality as the next. 

\begin{definition}
Two ideals $I$ and $J$ in $\OO_K$ are \textbf{coprime} if $I+J=\OO_K$.
\end{definition}
In other words, two integral ideals are coprime if their sum is the entire ring of integers.
For example, the integral ideals $(2)$ and $(3)$ in $\Z$ are coprime because $(2)+(3)=(1)=\Z$. But the integral ideals $(2)$ and $(4)$ are not coprime because $(2)+(4)=(2) \neq \Z$. 

Now we have defined coprime ideals in $\OO_K$, we can state the Chinese Remainder Theorem in Dedekind domains.

\begin{theorem}
\reversemarginpar
\marginnote{\textit{CRT in $\OO_K$}}
Let $D$ be a Dedekind domain.
\begin{enumerate}
    \item Let $P_1,\dots, P_k$ be distinct prime ideals in $D$ and $b_1,\dots, b_k$ be positive integers. Let $\alpha_1,\dots,\alpha_k$ be elements of $D$. Then there exists an $\alpha \in  D$ such that  for all $i \in [1,k]$, it satisfies $\alpha = \alpha_i \bmod P_i^{b_i}$.
    
    \item Let $I_1,\dots, I_k$ be pairwise coprime ideals of $D$ and $\alpha_1, \dots, \alpha_k$ be elements of $D$. Then there exists an $\alpha \in D$ such that for all $i \in [1, k]$, it satisfies $\alpha = \alpha_i \bmod I_i$.
\end{enumerate} 

\end{theorem}

Another way of stating the second point above that is similar to the CRT in rings is the next theorem.

\begin{theorem}
   Let $I_1, \dots, I_k$ be pairwise corprime ideals in a Dedekind domain $D$ and $I = \prod_{i=1}^k I_i$. Then the map 
   \begin{equation*}
       D \rightarrow (D / I_1, \dots, D/I_k)
   \end{equation*}
   induces an isomorphism 
   \begin{equation*}
       D / I \cong D / I_1 \times \cdots \times D / I_k.
   \end{equation*}
\end{theorem}

To prove CRT in $\OO_K$, first prove the map is surjective. Then prove that the kernel of the map is $I_1 \cap \cdots \cap I_k$, which can be shown to be identical to $\prod_{i=1}^k I_i$ under the assumption that they are pairwise coprime. Then it follows from the First Isomorphism Theorem. 

The connection of this subsection to the RLWE result are the following two lemmas. The first lemma shows that given two ideals $I, J \subseteq R$ of a Dedekind domain $R$ (i.e., a ring of integers $\OO_K$ of a number field $K$), it possible to construct another ideal that is coprime with either one of them. 

\begin{lemma}
\label{app lm:coprimeIdeals}
If $I$ and $J$ are non-zero integral ideals of a Dedekind domain $R$, then there exists an element $a \in I$ such that $(a)I^{-1} \subseteq R$ is an integral ideal coprime to $J$. 
\end{lemma}

\begin{proof}
Since $a \in I$, the principal ideal $(a) \subseteq I$. By Theorem \ref{app thm:divCont}, we have $I \mid (a)$, that is, there is an ideal $M \subseteq R$ such that $IM=(a)$, so $M=(a)I^{-1} \subseteq R$ is an ideal of $R$. We skip the proof of coprimality. %because I don't fully understand. 
See Lemma 5.5.2 of \citet{stein2012algebraic}.
\end{proof}

The element $a \in I$ can be efficiently computable using CRT in $\OO_K$. Hence, given two ideals in $R$, we can efficiently construct another one that is coprime with either one of them. This corresponds to Lemma 2.14 of \citet{lyubashevsky2010ideal}. The next lemma is essential in the reduction from K-BDD problem to RLWE. 

% state the lemma according to lemma 2.15 lyubashevsky2010ideal, which is a special case of prop 5.2.4 in stein2012algebraic.
\begin{lemma}
Let $I$ and $J$ be ideals in a Dedekind domain $R$ and $M$ be a fractional ideal in the number field $K$. Then there is an isomorphism 
\begin{align*}
    M/JM \cong IM/IJM.
\end{align*}
\end{lemma}

\begin{proof}
Given ideals $I,J\subseteq R$, by Lemma \ref{app lm:coprimeIdeals} we have $tI^{-1} \subseteq R$ is coprime to $J$ for an element $t \in I$. Then we can define a map 
\begin{align*}
    \theta_t: K &\rightarrow K \\
    u &\mapsto tu.
\end{align*}
This map induces a homomorphism 
\begin{align*}
    \theta_t: M \rightarrow IM/IJM.
\end{align*}
First, show $ker(\theta_t)=JM$. Since $\theta_t(JM)=tJM \subseteq IJM$, then $\theta_t(JM)=0$. Next, show any other element $u \in M$ that maps to 0 is in $JM$. To see this, if $\theta_t(u)=tu=0$, then $tu \in IJM$. To use Lemma \ref{app lm:coprimeIdeals}, we re-write it as $(tI^{-1}) (uM^{-1})\subseteq J$. Since $tI^{-1}$ and $M$ are coprime, we have $uM^{-1}\subseteq J$, which implies $u\subseteq JM$. Therefore, $ker(\theta_t)=JM$ and
\begin{align*}
    \theta_t: M/JM \rightarrow IM/IJM
\end{align*}
is injective. 

Second, show the map is surjective. That is, for any $v \in IM$, its reduction $v \mod IJM$ has a preimage in $M/JM$. Since $tI^{-1}$ and $J$ are coprime, by CRT we can compute an element $c \in tI^{-1}$ such that $c = 1 \bmod J$. Let $a = cv \in tM$, then $a-v=cv-v=v(c-1) \in IJM$. Let $w=a/t \in M$, then $\theta_t(w)=t (a/t)=a = v \bmod IJM$. Hence, any arbitrary element $v \in IM$ satisfies the preimage of $v \bmod IJM$ is $w \bmod IM$. 
\end{proof}

In the hardness proof of RLWE as will be shown in the next section, we let $M=R$ or $M=\dual{I}=I^{-1}\dual{R}$ and $J=(q)$ for a prime integer $q$, then the isomorphism becomes 
\begin{align*}
    R/(q)R &\cong I/(q)I \text{ or } \\
    \dual{I} / (q)\dual{I} &\cong \dual{R} / (q)\dual{R}.
\end{align*}

\subsection{Trace and Norm}

As we have built a connection between a number field and a Euclidean space, we can relate more features of a Euclidean space to that of a number field. In this subsection, we will introduce two quantities, trace and norm, of elements in a number field. These quantities are useful to calculate the discriminant and determinant of elements in a number field. 
Recall that for a linear transformation $\phi:V \rightarrow V$ from a vector space $V$ to itself, we can write $\phi$ in its matrix representation $[\phi]$ by applying $\phi$ to a basis of $V$. That is, for each $e_j \in \{e_i\}_{i \in [n]}$ in a basis of $V$, we have $\phi(e_j) = \sum_{i \in [n]} a_{ij} e_i$ is the linear combination of the basis, so $[\phi] = (a_{ij})$ is the coefficient matrix. With this matrix representation of the linear map, we can 
%\reversemarginpar
%\marginnote{\textit{Trace and determinant of linear map}}
define its trace and determinant like in the context of linear algebra. 

\begin{example}
Let $\phi: \C \rightarrow \C$ be the complex conjugation. Take the basis $\{1, i\}$ for the complex space $\C$. Apply the complex conjugation to this basis, we get 
\begin{align*}
    \phi(1) &= 1 + 0 \cdot i, \\
    \phi(i) &= 0\cdot 1 + (-1) \cdot i.
\end{align*}
So the matrix representation of the complex conjugation is 
$[\phi] = \begin{pmatrix}
  1 & 0\\ 
  0 & -1
\end{pmatrix}$. 
Each column $j$ consists of the coefficients of $\phi(e_j)$. 
\end{example}

Since a number field $K$ is a $\Q$-vector space, we can speak of linear transformations on $K$ too. For any element $\alpha \in K$, we can define a map 
$m_{\alpha}(x) = \alpha x$  as a multiplication by $\alpha$ for all $x \in K$. It is easy to see that $m_{\alpha}$ is also a linear map from $K$ to itself, so there is a matrix representation of this linear map $m_{\alpha}$. 

\begin{example}
Let $K = \Q(\sqrt{2})$ be a number field with a basis $\{1, \sqrt{2}\}$. For $a, b \in \Q$, we have an element $\alpha = a + b\sqrt{2} \in K$ and its associated linear map $m_{\alpha}$. Apply this map to the basis of $K$, we get 
\begin{align*}
    m_{\alpha}(1) &= a \cdot 1 + b \cdot \sqrt{2},  \\
    m_{\alpha}(\sqrt{2}) &= 2b\cdot 1 + a \cdot  \sqrt{2}.
\end{align*}
So the matrix representation of the linear map is 
$[m_{\alpha}] = \begin{pmatrix}
  a & 2b\\ 
  b & a
\end{pmatrix}$. 
\end{example}

Now, we can define the trace and norm on a number field which will appear in the RLWE problem.
\begin{definition}
\label{app def:trcNorm}
The \textbf{trace} and \textbf{norm} of an element $\alpha$ in a number field $K$ are defined as 
\reversemarginpar
\marginnote{\textit{Trace and norm in $K$}}
\begin{align*}
    Tr_{K \setminus \Q}&: K \rightarrow \Q \\%\text{  s.t.  }  
    Tr_{K \setminus \Q}(\alpha) &= Tr([m_{\alpha}]) \in \Q, \\
    N_{K \setminus \Q}&:K \rightarrow \Q \\%\text{ s.t. } 
    N_{K \setminus \Q}(\alpha) &= \det([m_{\alpha}]) \in \Q.
\end{align*}
\end{definition}

\begin{example}
In the above example, the trace and norm of $m_{\alpha}$ are the trace and determinant of its matrix representation, i.e., $2a$ and $a^2 - 2b^2$, respectively. 
\end{example}

It is also possible to define trace and norm using the canonical embedding that was introduced in the previous section. This is due the the following theorem which states a connection between these two quantities and automorphisms in the Galois group of a general field extension. 

\begin{theorem}
If $E/F$ is a finite Galois extension, then the trace and norm of an element $\alpha \in E$ are 
\begin{align*}
    Tr_{E/F}(\alpha) &= \sum_{\sigma \in Gal(E/F)} \sigma(\alpha) \\
    N_{E/F}(\alpha) &= \prod_{\sigma \in Gal(E/F)} \sigma(\alpha). 
\end{align*}
\end{theorem}
The intuition is that when the extension field $E$ is Galois, each automorphism $\sigma(\alpha)$ in the Galois group is an eigenvalue of the linear transformation $m_{\alpha}$. Recall from linear algebra that the trace and determinant of a square matrix are the sum and product of its eigenvalues respectively. The connection with the canonical embedding is due to the following two observations:  
\begin{enumerate}
    \item the number field $K = \Q(r)$ is a Galois extension over $\Q$,
    \item each automorphism $\sigma_i \in Gal(E/F)$ in the Galois group is correspond to an element in the image of the canonical embedding $\sigma: K \rightarrow H$ in Definition \ref{app def:canEmbd}. 
\end{enumerate}
\noindent This gives rise to the following definitions of trace and norm of an element in a number field in terms of the canonical embedding, which appear in some books too. 

\begin{definition}
\label{app def:trcNorm2}
Given a canonical embedding of a number field $K$
\begin{align*}
    \sigma: K &\rightarrow \R^{s_1} \times \C^{2s_2} \\
    \sigma(\alpha) &\mapsto (\sigma_1(\alpha), \dots, \sigma_n(\alpha)),
\end{align*}
the \textbf{trace} and \textbf{norm} of an element $\alpha \in K$ are defined as 
\reversemarginpar
\marginnote{\textit{Trace and norm by canonical embedding}}
\begin{align*}
    Tr_{K \setminus \Q}&: K \rightarrow \Q \\%\text{  s.t.  }  
    Tr_{K/\Q}(\alpha) &= \sum_{i \in [n]} \sigma_i(\alpha), \\
    N_{K \setminus \Q}&: K \rightarrow \Q \\%\text{ s.t. } 
    N_{K / \Q}(\alpha) &= \prod_{i \in [n]} \sigma_i(\alpha). 
\end{align*}
\end{definition}

\begin{example}
In the same example where $K = \Q(\sqrt{2})$ and $\alpha = a + b\sqrt{2}$, the minimal polynomial of $\alpha$ over $\Q$ is $f(x) = (\frac{x-a}{b})^2 - 2$, which has two roots $ a \pm b \sqrt{2}$. So the canonical embedding $\sigma$ of $K$ maps $\alpha$ to each of these two roots. Hence, the trace of $\alpha$ is $Tr(\alpha)=(a+b\sqrt{2})+(a-b\sqrt{2}) = 2a$ and the norm is $N(\alpha)=(a+b\sqrt{2})(a-b\sqrt{2}) = a^2 - 2b^2$,  which are consistent with the results in the above example. 
\end{example}

Both definitions imply that trace is additive and norm is multiplicative, that is, $Tr(x+y) = Tr(x)+Tr(y)$ and $N(xy)=N(x)N(y)$. In addition, Definition \ref{app def:trcNorm2} entails that
\begin{align}
\label{app equ:trace}
    Tr(xy) = \sum \sigma_i(xy) = \sum \sigma_i(x) \sigma_i(y) = \langle \sigma(x), \overline{\sigma(y)} \rangle.
\end{align}
The second equality is due to the fact that each $\sigma_i$ is a homomorphism. The last equality is by definition of the inner product between complex vectors. %For complex vectors $x$ and $y$, their inner product $\langle x, y\rangle = \sum_i x_i \overline{y_i}$. 

%%%%%%%%%%%%%%%%%%%%%%%%%%%%%%%%%%%%%%%%%%%%%%%%%%%%%%%%%%%%%%%%%%%%%%%%%%%%%%%%%%%%%%%%%%%%%%%%%%%
%%%%%%%%%%%%%%%%%%%%%%%%%%%%%%%%%%%%%%%%%%%%%%%%%%%%%%%%%%%%%%%%%%%%%%%%%%%%%%%%%%%%%%%%%%%%%%%%%%%

\subsection{Ideal lattices}

% \kl{minimum distance of the ideal lattice, related to discriminant of K, upper and lower bounds of minimum distance of ideal lattice, see lemma 2.9 \citep{lyubashevsky2010ideal}.}
\iffalse
Recall that a canonical embedding $\sigma$ of an algebraic number field $K$ of degree $n$ to the canonical space $H$ is
\begin{align*}
    \sigma: K &\rightarrow H \cong \R^{s_1} \times \C^{2s_2} \subseteq \C^n \\
    \sigma(r) &\mapsto \left(\sigma_1(r), \dots, \sigma_{s_1}(r), \sigma_{s_1+1}(r),\dots, \sigma_n(r)\right),
\end{align*}
where $\sigma_{s_1 + j}(r) = \overline{\sigma_{s_1 + s_2 + j}(r)}$ are conjugate pairs for all $j \in [s_2]$ that correspond to the complex embeddings into $H$.

\fi

%We first state the main result of this section, which explicitly states the determinant of an ideal lattice. We defer the proof till the end of this section. The reader can refer to Proposition 4.26 in J. S. Milne's book or Corollary 10.6.2 in Ben Green's book. 

To start off this section, we state below some results in order to give some insights about the motivation of studying how ring of integers and its ideals are embedded in $\R^n$.

\begin{proposition}
\label{app prop:small norm}
\reversemarginpar
\marginnote{\textit{Small norm element}}
Let $K$ be a number field and $I$ be an integral ideal of $\OO_K$. Then there is some element $x \in I$ such that $|N_{K/\Q}(x)| \le M_K N(I)$.
\end{proposition}
Here, $M_K$ is the \textbf{Minkowski constant} defined as $M_K=\left(\frac{4}{\pi}\right)^{r_2} \frac{n!}{n^n}\sqrt{|\Delta_K|}$, where $n$ is the degree of $K$ and also the number of embeddings of $K$ with $n=r_1+2r_2$ for $r_1$ real embeddings and $r_2$ pairs of complex embeddings. $\Delta_K$ is the discriminant of the number field $K$, which will be introduced later.  

\begin{theorem}
\label{app thm:min 1st}
\reversemarginpar
\marginnote{\textit{Minkowski 1st Theorem}}\index{Minkowski}
Let $L$ be an $n$-dimensional lattice and $B\subseteq \R^n$ be a centrally symmetric, compact, convex body. Suppose $Vol(B) \ge 2^n \det(L)$, then $B$ contains a non-zero lattice vector of $L$.
\end{theorem}

To prove Proposition \ref{app prop:small norm}, it uses results from lattice theory and Theorem \ref{app thm:min 1st}. Given the canonical embedding $\sigma$ maps $K$ to a space isomorphic to $\R^n$, the first step is to prove $\OO_K$ is associated with a lattice in $\R^n$ and so are the ideals of $\OO_K$. Then it left to prove that the lattice associated with an ideal intersects with a bounded convex body in $\R^n$ by Theorem \ref{app thm:min 1st}, provided certain parameter conditions are satisfied. The first step is our focus in this section, so we do not discuss the second step.

Recall a canonical embedding $\sigma:K\rightarrow H \cong \R^n$ gives rise to another embedding $\tau:K \rightarrow V \cong \R^n$ as defined in Equation \ref{app equation:minkowski embedding}, which maps the ring of integers $\OO_K$ to a full-rank lattice as stated in Theorem \ref{app thm:rngIntLat}. This implies that the embedding $\tau$ maps a fractional (integral) ideal of $\OO_K$ to a full-rank lattice too.\footnote{See Corollary 10.6.2 of Ben Green's book \textit{Algebraic Number Theory} or Lemma 7.1.8 of \citet{stein2012algebraic}.} We give a name of such a lattice. 

\begin{definition}
The embedding $\tau:K\rightarrow V$ maps a fractional ideal of the ring of integers $\OO_K$ to a full-rank lattice, called the \textbf{ideal lattice}. 
\reversemarginpar
\marginnote{Ideal lattice}
\end{definition}

For the interest of building lattice-based cryptosystems, we study ideal lattices and their determinants. But for a general case, we state the next theorem. 

\begin{theorem}
\label{app thm:rngIntDet}
Let $\tau:K \rightarrow V$ be the embedding of the $n$-dimensional number field $K$ as defined in Equation \ref{app equation:minkowski embedding}. Then $\tau(\OO_K)$ is a full-rank ideal lattice in $\R^n$ and its determinant satisfies 
\reversemarginpar
\marginnote{$\det(\tau(\OO_K))$}
\begin{equation*}
    \det(\tau(\OO_K)) = \frac{1}{2^{r_2}}\sqrt{|\Delta_K|}.
\end{equation*}
\end{theorem}

Since we have proved in Theorem \ref{app thm:rngIntLat} that $\tau(\OO_K)$ is a full-rank lattice in $\R^n$, it remains to prove its determinant. There are two new quantities in the theorem that have not been introduced, the discriminant $\Delta_K$ of the number field $K$ and the norm $N(I)$ of an ideal I $\subseteq \OO_K$. So we delay the proof till the end of this subsection.

Recall from Section \ref{section:lattice theory} that an $n$-dimensional lattice $L$ is similar to a vector space $\R^n$ but with only discrete vectors. It is isomorphic to the group $(\Z^n,+)$. It shares many properties with $\R^n$ such as having a basis $\{v_1, \dots, v_n\}$. The determinant of a lattice is the size of its fundamental domain that is surrounded by its basis. This gives rise to the following equality 
\begin{equation*}
    \det(L) = Vol(F)=|\det(B)|,
\end{equation*}
where $F$ is the fundamental domain and $B$ is a basis matrix of $L$. An useful observation is that the determinant is an invariant quantity under the choice of a basis, because any two bases of $L$ are related by a unimodular matrix. 
	
Let $K$ be an algebraic number field of degree $n$ and $\sigma_i: K \rightarrow \C$ be a field homomorphism for all $i \in [n]$. For the elements $x_1, \dots, x_n \in K$, define the $n$ by $n$ matrix $M$ to be the linear map where $M_{ij} = \sigma_i(x_j)$, that is,
\begin{equation*}
M = 
\begin{pmatrix}
\sigma_1(x_1) & \sigma_1(x_2) & \cdots & \sigma_1(x_n) \\
\sigma_2(x_1) & \sigma_2(x_2) & \cdots & \sigma_2(x_n) \\
\vdots & \vdots & \cdots & \vdots \\
\sigma_n(x_1) & \sigma_n(x_2) & \cdots & \sigma_n(x_n) 
\end{pmatrix}.
\end{equation*}
It can be proved that the matrix is always non-singular if the elements $\{x_1, \dots, x_n\}$ form a basis of $K$ over $\Q$ (Lemma 1.7.1 Ben Green's \textit{Algebraic Number Theory}). Without loss of generality, assume $M=M(e_1,\dots,e_n)$ for a basis $\{e_1,\dots,e_n\}$ of a $n$-dimensional number field $K$.

\begin{definition}
Let $K$ be an $n$-dimensional number field with a basis $\{e_1, \dots, e_n\}$ and
\reversemarginpar
\marginnote{\textit{Element discriminant}}
$M$ be the matrix defined above. The \textbf{discriminant of the elements} is  
\begin{equation*}
    \text{disc}_{K / \Q}(e_1, \dots, e_n) = \det (M) ^2.
\end{equation*}
\end{definition}

Alternatively, the discriminant of elements in $K$ can be defined by their traces, because 
\begin{align*}
    \text{disc}_{K / \Q}(e_1, \dots, e_n) = \det (M)^2 = \det(M^T M)
\end{align*}
and the matrix entry $(M^T M)_{ij}=\sum_k \sigma_k(e_i) \sigma_k(e_j) = \sum_k \sigma_k(e_i e_j) = Tr_{K / \Q}(e_i e_j)$ as $\sigma_i$ is a homomorphism. Therefore, the discriminant of number field elements is equal to the determinant of the trace matrix as stated next in the equivalent definition. 

\begin{definition}
Let $K$ be an $n$-dimensional number field with a basis $\{e_1, \dots, e_n\} \in K$. The \textbf{discriminant of the elements} is  
\begin{equation*}
    \text{disc}_{K / \Q}(e_1, \dots, e_n) = \det \left( (Tr_{K / \Q}(e_i e_j))_{ij} \right).
\end{equation*}
	
\end{definition}

From the previous section, we know that the trace of an element is a rational number, so the discriminant is also a rational number. Note although it is defined as the square of a matrix determinant, discriminant can be negative as complex numbers are involved. From the discriminant of basis elements and the integral basis of a number field $K$, we can define the discriminant of $K$. 

\iffalse
Next, we define the discriminant of an ideal. 
	
\begin{definition}
Let $K$ be a number field of degree $n$ and $I$ be a non-zero ideal of $\OO_K$ \reversemarginpar
\marginnote{Ideal discriminant}
with $B=\{b_1, \dots, b_n\}$ being a basis of $I$. The \textbf{discriminant} of the ideal $I$ is 
\begin{equation*}
    D(I) = \text{disc}(b_1, \dots, b_n).
\end{equation*}
\end{definition}
\fi 

\begin{definition}
Let $K$ be an $n$-dimensional number field and $\{e_1, \dots, e_n\}$ be an
\reversemarginpar
\marginnote{\textit{$\Delta(K)$}}
integral basis of $K$. The \textbf{discriminant of the number field} $K$ is 
\begin{equation*}
	\Delta_K = \text{disc}_{K/\Q}(e_1, \dots, e_n)=\det \left(( Tr_{K / \Q}(e_i e_j))_{ij} \right) = \det (M)^2.
\end{equation*}
\end{definition}

The discriminant loosely speaking measures the size of the ring of integers $\OO_K$ in the number field $K$ and it is invariant under the choice of an integral basis, which is the same as the determinant of a lattice. This can be seen from the following Lemma and corollary. 

\begin{lemma}
Suppose  $x_1, \dots, x_n, y_1, \dots, y_n \in K$ are elements in the number field and they are related by a transformation matrix $A$, then 
\begin{equation*}
    \text{disc}_{K / \Q}(x_1, \dots, x_n) = det (A)^2 \text{disc}_{K / \Q}(y_1, \dots, y_n).
\end{equation*}
\end{lemma}

\begin{corollary}
\reversemarginpar
\marginnote{\textit{Invariant $\Delta(K)$}}
Suppose $\{e_1, \dots, e_n\}$ and $\{e'_1, \dots, e'_n\}$ are both integral bases of the number field $K$, then 
\begin{equation*}
    \text{disc}_{K / \Q}(e_1, \dots, e_n) =  \text{disc}_{K / \Q}(e'_1, \dots, e'_n).
\end{equation*}
\end{corollary}

From Theorem \ref{app thm:rngIntDet}, it can be seen that the (absolute) discriminant of a number field measures the geometric sparsity of its ring of integers, because the larger the discriminant, the larger the size of the fundamental region, hence the more sparse the ideal lattice. 

Another quantity appears in the theorem is the norm of an ideal. Recall that the index $|G:H|$ of a subgroup $H$ in $G$ is the number of cosets of $H$ in $G$. We define the norm of an ideal and its relation to the norm of an element in the following lemma (see Lemma 4.4.3 in Ben Green's book). 

\begin{definition}
\label{app def:idealNorm}
\reversemarginpar
\marginnote{\textit{Ideal norm}}
Let $I$ be a non-zero ideal of $\OO_K$. The \textbf{norm} of $I$, denoted by $N(I)$ (or sometimes $(\OO_K:I)$), is the index of $I$ as a subgroup of $\OO_K$, i.e., $N(I) = |\OO_K / I|$.
\end{definition}

\begin{lemma}
Suppose $I = (\alpha)$ is a principal ideal of $\OO_K$ for some non-zero $\alpha \in \OO_K$. Then $N(I) = |N_{K / \Q}(\alpha)|$. 
\end{lemma}

As for the norm of number field elements, the norm of ideals is also multiplicative. That is, $N(IJ) = N(I)N(J)$. In addition, if $I$ is a fractional ideal of $\OO_K$, then its norm satisfies $N(I) = N(dI) / |N(d)|$, where $d \in \OO_K$ is the element that makes $dI \in \OO_K$ an integral ideal.

\begin{proof}[Sketch proof of Theorem \ref{app thm:rngIntDet}]
To prove the determinant of the lattice $\tau(\OO_K)$, we know from the proof of Theorem \ref{app thm:rngIntLat} that $\{\tau(e_1),\dots,\tau(e_n)\}$ is a basis of the lattice and the basis matrix is 
\begin{equation*}
N^T = \left(
\begin{smallmatrix}
\sigma_1(e_1) & \cdots & \sigma_{r_1}(e_1) & Re(\sigma_{r_1+1}(e_1)) & Im(\sigma_{r_1+1}(e_1)) & \cdots & Re(\sigma_{r_1+r_2}(e_1)) & Im(\sigma_{r_1+r_2}(e_1)) \\
\vdots & & \vdots & \vdots & \vdots & & \vdots & \vdots \\
\sigma_1(e_n) & \cdots & \sigma_{r_1}(e_n) & Re(\sigma_{r_1+1}(e_n)) & Im(\sigma_{r_1+1}(e_n)) & \cdots & Re(\sigma_{r_1+r_2}(e_n)) & Im(\sigma_{r_1+r_2}(e_n)) \\
\end{smallmatrix}
\right),
\end{equation*}
so $\det(\tau(\OO_K))=|\det(N)|$. In addition, the canonical embedding $\sigma$ associates with the matrix 
\begin{equation*}
M^T = \left(
\begin{smallmatrix}
\sigma_1(e_1) & \cdots & \sigma_{r_1}(e_1) & \sigma_{r_1+1}(e_1) & \overline{\sigma_{r_1+1}(e_1)} & \cdots & \sigma_{r_1+r_2}(e_1) & \overline{\sigma_{r_1+r_2}(e_1)} \\
\vdots & & \vdots & \vdots & \vdots & & \vdots & \vdots \\
\sigma_1(e_n) & \cdots & \sigma_{r_1}(e_n) & \sigma_{r_1+1}(e_n) & \overline{\sigma_{r_1+1}(e_n)} & \cdots & \sigma_{r_1+r_2}(e_n) & \overline{\sigma_{r_1+r_2}(e_n)} \\
\end{smallmatrix}
\right),
\end{equation*}
whose determinant satisfies $\Delta_K=\det(M)^2$. It can be seen that the columns in $N^T$ correspond to the real (or complex) parts of the complex embeddings can be obtained from $M^T$ by adding (or subtracting) the complex conjugate columns. For example, expressing the matrices in column vector format, we get 
\begin{align*}
    N^T &= (\dots, Re(\sigma_{r_1+1}(e_1)), Im(\sigma_{r_1+1}(e_1)), \dots) \\
    &= (\dots, \frac{1}{2}(\sigma_{r_1+1}(e_1)+\overline{\sigma_{r_1+1}(e_1)}), \dots)\\
    &=-\frac{1}{2i}(\dots, \sigma_{r_1+1}(e_1),\overline{\sigma_{r_1+1}(e_1)}, \dots).
\end{align*}
Apply the same operations for all $r_2$ pairs of columns, we get $\det (N) = -\frac{1}{(2i)^{r_2}} \det M$. Hence, 
\begin{equation*}
    \det(\tau(\OO_K)) = |\det (N)| = \frac{1}{2^{r_2}} |\det M|=\frac{1}{2^{r_2}}\sqrt{|\Delta_K|}. 
\end{equation*}
\end{proof}

From Theorem \ref{app thm:rngIntDet}, it follows the determinant of an ideal lattice is also related to the discriminant of the number field. 

\begin{corollary}
\label{app cor:idealLatDet}
Let $I$ be an ideal of $\OO_K$. Then the ideal lattice $\tau(I)$ has determinant  
\reversemarginpar
\marginnote{$\det(\tau(I))$}
\begin{equation*}
    \det(\tau(I)) = \frac{1}{2^{r_2}}N(I) \sqrt{|\Delta_K|}.
\end{equation*}
\end{corollary}
We have stated that $\tau(I)$ is a lattice in $\R^n$ called ideal lattice. The same strategy can also be used to state the relationship between the associated matrix determinants $\det(N)$ and $\det(M)$. The only difference is that $I$ is a sublattice of $\OO_K$, so its determinant is larger than $\det(\OO_K)$. The scale is exactly the index of $I$ in $\OO_K$ as a subgroup, which is the norm of $I$ by Definition \ref{app def:idealNorm} of ideal norm.

\subsection{Dual lattice in number fields}
\label{app subsec:dualLatInNumField}
%\kl{Will come back to this if needed! still unsure about the motivation of dual ideal in K, dual ideal in general is used such as in introducing discrete gaussian, smoothing parameter, etc.}

For more detail of the proofs and intuitions in this subsection, the readers should refer to Conrad's lecture notes on ``Different ideal''.

\begin{definition}
\reversemarginpar
\marginnote{\textit{Lattice in $K$}}
A \textbf{lattice} in an $n$-dimensional number field $K$ is the $\Z$-span of a $\Q$-basis of $K$.  
\end{definition}
By the Primitive Element Theorem (Theorem \ref{app thm:primEleThm}), $K$ always has a power basis which is a $\Q$-basis. So the integer linear combination of the $\Q$-basis forms a lattice in $K$. For example, the ring of integers $\OO_K$ is a lattice in the number field $K$. Similar to lattices in general, number field lattices have dual too and share much of the same properties as the general dual lattices as we will see next. Unlike general lattices in $\R^n$ which equips with the dot product, the operator that equips with number field lattices is the trace as defined previously. More precisely, the dual lattice in a number field consists with elements that have integer \textit{trace product} with the given lattice by Equation \ref{app equ:trace}. 

\begin{definition}
\reversemarginpar
\marginnote{\textit{Dual lattice}}
Let $L$ be a lattice in a number field $K$. Its \textbf{dual lattice} is 
\begin{equation*}
    L^{\vee} = \{x \in K \mid Tr_{K/Q}(xL) \subseteq \Z\}.
\end{equation*}
\end{definition}

To check whether or not an element belongs to the dual, one can check its trace product with the lattice basis. This also gives a way of writing out the dual of a given lattice. 

\begin{example}
Let $K=\Q(i)$ and the lattice $L=\Z[i]$. Let $B=\{1,i\}$ be a basis of $L$. To find the dual of $L$, take an element $a+bi \in K$ and consider its trace product with the basis vector in $B$ and check if the trace products are integers. More precisely, we need to check the conditions under which
\begin{align*}
    Tr_{K/\Q}(a+bi) &\in \Z \\
    Tr_{K/\Q}((a+bi)i) &\in \Z.
\end{align*}
Let $\alpha=a+bi$ and $\beta=-b+ai$. By Definition \ref{app def:trcNorm} of trace, we have $[m_{\alpha}] = \begin{pmatrix}
  a & -b\\ 
  b & a
\end{pmatrix}$ and 
$[m_{\beta}] = \begin{pmatrix}
  -b & -a\\ 
  a & -b
\end{pmatrix}$. For both traces to be integers, we must have $2a \in \Z$ and $-2b \in \Z$, so the dual lattice $L^{\vee}=\frac{1}{2}\Z[i]$ and the basis of the dual is $B^{\vee}=\{\frac{1}{2},\frac{i}{2}\}$.
\end{example}

From the example, it can be seen that the basis and the dual basis satisfy $Tr(e_i e_j^{\vee}) = \delta_{ij}$. This gives rise to the following theorem that states the dual of a number field lattice is also a lattice. 

\begin{theorem}
\label{app thm:dualBasis}
\reversemarginpar
\marginnote{\textit{$\dual{L}$ is lattice}}
For an $n$-dimensional number field $K$ and a lattice $L \subseteq K$ with a $\Z$-basis $\{e_1, \dots, e_n\}$, the dual $L^{\vee}=\bigoplus \Z e_i^{\vee}$ is a lattice with a dual basis  $\{e_1^{\vee}, \dots, e_n^{\vee}\}$ satisfying $Tr_{K/\Q}(e_i e_j^{\vee}) = \delta_{ij}$.\marginpar{what is $\delta_{ij}?$}
\end{theorem}

Dual lattices in number fields share similar properties with dual lattices in general. We state a few of them in the following corollary. 

\begin{corollary}
For lattices in a number field, the following hold: 
\begin{enumerate}
    \item $L^{\vee \vee}=L$,
    \item $L_1 \subseteq L_2 \iff \dual{L_2} \subseteq \dual{L_1}$,
    \item $\dual{(\alpha L)} \iff \frac{1}{\alpha}\dual{L}$, for an element $\alpha \in K^{\times}$.
\end{enumerate}
\end{corollary}

The following theorem relates the dual lattice to differentiation and provides an easier way of computing the dual basis and dual lattice from a given lattice. 

\begin{theorem}
\label{app thm:dualLatDiff}
\reversemarginpar
\marginnote{\textit{Dual basis}}\index{dual basis}
Let $K=\Q(\alpha)$ be an $n$-dimensional number field with a power basis\index{power basis} $\{1, \alpha, \dots, \alpha^{n-1}\}$ and $f(x) \in \Q[x]$ be the minimal polynomial of the element $\alpha$, which can be expressed as 
\begin{equation*}
    f(x) = (x-\alpha)(c_0 + c_1 x + \dots + c_{n-1} x^{n-1}).
\end{equation*}
Then the dual basis to the power basis relative to the trace product is $\left\{\frac{c_0}{f'(\alpha)}, \dots, \frac{c_{n-1}}{f'(\alpha)}\right\}$.

In particular, if $K=\Q(\alpha)$ and the primitive element $\alpha \in \OO_K$ is an algebraic integer, then the lattice $L=\Z[\alpha]=\Z + \dots + \Z \alpha^{n-1}$ and its dual are related by the first derivative of the minimal polynomial, that is, 
\begin{equation*}
    \dual{L} = \frac{1}{f'(\alpha)}L.
\end{equation*}
\end{theorem}

\begin{example}
Let us work through an example to illustrate both theorems. Let the number field $K=\Q(\sqrt{d})$ and its lattice $L=\Z[\sqrt{d}]$.   

This is a 2-dimensional number field with the primitive element $\alpha = \sqrt{d}$ and the power basis $\{1, \sqrt{d}\}$. The minimal polynomial of $\alpha$ in $\Q[x]$ is $f(x) = x^2-d$ with the derivative $f'(x) = 2x$ so $f'(\alpha)=2\sqrt{d}$. Moreover, the minimal polynomial can be written as $f(x) = (x-\sqrt{d})(x+\sqrt{d})$. By Theorem \ref{app thm:dualLatDiff}, the dual basis is $\{\frac{1}{2}, \frac{1}{2\sqrt{d}}\}$. In addition, if $d \in \Z$ then $\alpha \in \OO_K$, so the dual lattice $\dual{L} = \frac{1}{2\sqrt{d}}L$. This is consistent with the dual basis obtained, because according to the dual basis, the dual lattice $\dual{L} = \Z \frac{1}{2}+\Z \frac{1}{2\sqrt{d}}= \frac{1}{2\sqrt{d}}(\Z+\Z \sqrt{d})=\frac{1}{2\sqrt{d}} L$.

To confirm the dual basis of $\{1, \sqrt{d}\}$ is $\{\frac{1}{2}, \frac{1}{2\sqrt{d}}\}$, we apply Theorem \ref{app thm:dualBasis} to check their trace products. We have 
\begin{align*}
    Tr(1 \cdot \frac{1}{2}) &= Tr(\sqrt{d} \cdot \frac{1}{2\sqrt{d}}) = Tr(\frac{1}{2}) = 1 \\
    Tr(1 \cdot \frac{1}{2\sqrt{d}}) &= Tr(\sqrt{d} \cdot \frac{1}{2}) = 0.
\end{align*}
\end{example}

\begin{example}
An important application of this theorem in our context is when the number field $K=\Q[\zeta_m]$ is the mth cyclotomic number field, where $m=2n=2^k>1$. The ring of integers is then $L=\OO_K=\Z[\zeta_m]$. The minimal polynomial of $\zeta_m$ is $f(x)=x^n+1$ with the derivative $f'(x)=nx^{n-1}$. According to the theorem, we have 
\begin{equation*}
    \dual{(\Z[\zeta_m])} = \frac{1}{f'(\zeta_m)} \Z[\zeta_m] = \frac{1}{n\zeta_m^{n-1}} \Z[\zeta_m] = \frac{1}{n} \zeta_m^{n+1} \Z[\zeta_m]= \left(\frac{1}{n}\right).
\end{equation*}
The second last equality is because the roots of unit form a cyclic group and hence $\zeta^{-(n-1)}=\zeta^{n+1} \in \OO_K$. 
\end{example}

As a special lattice in $K$, the ring of integers $\OO_K$ was further studied and the following theorems offer some useful observations of its dual. By definition, the dual of $\OO_K$ is 
\begin{equation*}
    \dual{\OO_K} = \{x \in K \mid Tr_{K/\Q}(x \OO_K) \subseteq \Z\}.
\end{equation*}
On the one hand, $\dual{\OO_K}$ is at least as large as $\OO_K$. Each element in $\OO_K$ is an algebraic integer that has an integer trace\footnote{This can be verified by taking the power basis $\{1, r, \dots, r^{n-1}\}$ of $K$ which is also a $\Z$-basis of $\OO_K$. An element $x \in \OO_K$ can be written as $x=c_0+c_1 r + \dots + c_{n-1}r^{n-1}$. By definition, only $Tr(c_0) \in \Z$ and the rest are 0.}, so $\OO_K \subseteq \dual{\OO_K}$ which happens when $x=1$. On the other hand, $\dual{\OO_K}$ is no larger than the set of elements in $K$ that have integer trace as shown in the next theorem. 

\begin{theorem}
\reversemarginpar
\marginnote{\textit{$\dual{\OO_K}$ is frac ideal}}
The dual lattice $\dual{\OO_K}$ is the largest fractional ideal in $K$ whose elements have integer traces. 
\end{theorem}

\begin{proof}
Let $I$ be a fractional ideal in $K$. As it is closed under multiplication by elements in $\OO_K$, we have $I\OO_K=I$. Hence, $Tr(I\OO_K)\subseteq \Z$ if and only if $Tr(I) \subseteq \Z$, which is equivalent to $I \subseteq \dual{\OO_K}$. From these relations, we know that the fractional ideal is in the dual lattice if its elements have integer traces, so the largest fractional ideal whose elements have integer traces is also in the dual. If an additional element is added into the largest fractional ideal that satisfies the condition, then it is not necessarily true that $I\OO_K=I$, so the above relations may not follow.  
\end{proof}

The next theorem reveals the role that $\dual{\OO_K}$ plays in the dual of an arbitrary fractional ideal, which is also a lattice in $K$. 

\begin{theorem}
\label{app thm:fracIdealDual}
\reversemarginpar
\marginnote{\textit{Frac ideal dual}}
For a fractional ideal $I$ in $K$, its dual lattice is a fractional ideal and satisfying $\dual{I} = I^{-1} \dual{\OO_K}$.
\end{theorem}

We have seen the inverse of a fractional ideal in Equation \ref{app equ:fracIdInv}, it is tempting to see if the inverse of the dual $\dual{\OO_K}$ (which is also a fractional ideal) is any special. By definition of fractional ideal inverse (Equation \ref{app equ:fracIdInv}), we have 
\begin{align*}
    \inv{(\OO_K)} &= \{x \in K \mid x \OO_K \subseteq \OO_K \} = \OO_K \\
    \inv{(\dual{\OO_K})} &= \{x \in K \mid x \dual{\OO_K} \subseteq \OO_K \}.
\end{align*}
Since $\OO_K \subseteq \dual{\OO_K}$, their inverses satisfy $(\dual{\OO_K})^{-1} \subseteq \OO_K$. Unlike the dual which is a fractional ideal and not necessarily within $\OO_K$, this inclusion makes $\inv{(\dual{\OO_K})}$ an integral ideal. Here, we give it a different name, \textbf{different ideal}
\reversemarginpar
\marginnote{\textit{Different ideal}}
and denote it by $\DD_K := \inv{(\dual{\OO_K})}$.\footnote{To be clear. Some refer $\DD_K$ as the different ideal of $K$ and the notation suggests it too. But $K$ is a field which has exactly two ideals, the zero ideal and itself, so $\DD_K$ is not an ideal of $K$ but of $\OO_K$.} For example, let $K=\Q(i)$ and $\OO_K=\Z[i]$. The dual ideal is $\dual{\OO_K}=\dual{\Z[i]}=\frac{1}{2}\Z[i]$, so the different ideal $\DD_K=\inv{(\frac{1}{2}\Z[i])}=2\Z[i]$.

The next theorem relates the different ideal with the differentiation of the minimal polynomial. It can be proved easily by applying Theorem \ref{app thm:dualLatDiff}.
\begin{theorem}\label{app thm:difIdeal1}
Let $\OO_K=\Z[\alpha]$ be the ring of integers of a number field $K$ and $f(x) \in \Z[x]$ be the minimal polynomial of $\alpha$, then the different ideal $\DD_K=(f'(\alpha))$.
\end{theorem}

As mentioned before, $\OO_K$ does not always have a power basis, so not all $\OO_K$ can be written as $\Z[\alpha]$. 
Let us look at a special case in the above example where $\OO_K=\Z[i]$, the minimal polynomial of $\alpha=i$ is $f(x)=x^2+1$ and its derivative is $f'(\alpha)=2i$. Hence, the different ideal $\DD_K=(2i)$ is a principal ideal of $\OO_K$, so $\DD_K=2i\cdot \Z[i]=2\Z[i]$. The example can be generalized to some special cyclotomic fields, in which there is an explicit relations between the different ideal and the ring of integers. It can be easily proved using the above theorem. 

\begin{lemma}
\label{app lm:difIdeal}
\reversemarginpar
\marginnote{\textit{$\DD_K = n \OO_K$}}
For $m=2n=2^k \ge 2$ a power of 2, let $K=\Q(\zeta_m)$ be an $m$th cyclotomic number field and $\OO_K=\Z[\zeta_m]$ be its ring of integers. The different ideal satisfies $\DD_K = n \OO_K$.
\end{lemma}
This lemma plays an important role in RLWE in the special case where the number field is an $m$ cyclotomic field. It implies that the ring of integers $n^{-1}\OO_K=\dual{\OO_K}$ and its dual are equivalent by a scaling factor. Hence, the secret polynomial $\vc{s}$ and the random polynomial $\vc{a}$ can both be sampled from the same domain $R_q$, unlike in the general context where the preference is to leave $\vc{s} \in \dual{R_q}$ in the dual. 

To finish off this subsection, we state the relation between the norm of the different ideal and the discriminant of the number field. See Theorem 4.6 in Conrad's lecture notes on ``different ideal''.

\begin{theorem}
For a number field $K$, its discriminant $\Delta_K$ and different ideal $\DD_K$ satisfies $N(\DD_K)=|\Delta_K|$.
\end{theorem}
\iffalse
\begin{proof}
To sketch the proof. Given an integral basis $\{e_1, \dots, e_n\}$ of $\OO_K$ and its dual basis $\{\dual{e_1}, \dots, \dual{e_n}\}$ for $\dual{\OO_K}$. Since $\DD_K$ is also an integral ideal of $\OO_K$ and the norm of an ideal is its index in the ring, so $N(\DD_K)=[\dual{\OO_K}:\OO_K]$ (derivation skipped). The index is 
\end{proof}
\fi 

%\newpage
%\bibliography{references}
%\bibliographystyle{abbrvnat}

\newpage
\section{Mind Maps}

\label{sec:mind maps}

\subsection{A mindmap for RLWE}

% \begin{tikzpicture}[mindmap, grow cyclic, every node/.style=concept, concept color=orange!40, 
%	level 1/.append style={level distance=5cm,sibling angle=90},
%	level 2/.append style={level distance=3cm,sibling angle=45},]
	
\begin{tikzpicture}[grow cyclic, text width=3cm, align=flush center,
	level 1/.style={level distance=3cm,sibling angle=90},
	level 2/.style={level distance=3.5cm,sibling angle=45}]	
\node{$\Z[x]/(\Phi_m(x))$ \\ $\cong \Z(\zeta_m) = O_{\Q(\zeta_m)}$}
child { node {Cyclotomics \& their Galois Groups}
	child { node {$m$-th Cyclotomic Polynomials for $m = 2^k = 2n$ [\ref{mindmap:rlwe defn},\ref{mindmp:decision to search rlwe}] }} 
	child { node {Automorphisms \&\\ Permutations of Polynomial Coeffs [\ref{mindmp:decision to search rlwe}]}}
}
child { node {Canonical Embedding}
	child { node {Ideal Lattices from Fractional Ideals [\ref{mindmap:rlwe defn}]}}
	child { node {Efficient Polynomial Multiplication [\ref{mindmap:rlwe computations}]}}
	child { node {Ideal Norms \\and Geometric Quantities [\ref{mindmap:rlwe defn},\ref{mindmap:search rlwe}]}}
%	child { node {Power Basis}}
}
child { node {Chinese Remainder Theorem}
	child { node {Isomorphisms \\between $R_q$ and $I_q$ [\ref{mindmap:search rlwe},\ref{mindmp:decision to search rlwe}]}}
	child { node {CRT Representations \\ of Polynomials [\ref{mindmap:rlwe computations},\ref{mindmp:decision to search rlwe}]}}	
}
child { node {Ideals and \\Fractional Ideals}
	child { node {Number Fields and Rings of Integers}}
	child { node {Unique Factorization into Prime Ideals [\ref{mindmp:decision to search rlwe}]}}
	child { node {Dual Lattices and Different Ideals [\ref{mindmap:rlwe defn}]}}
};
\end{tikzpicture}

\begin{enumerate}
\item Definition of RLWE and related ideal lattice problems \label{mindmap:rlwe defn}
\item Efficient computations in RLWE-based cryptosystems \label{mindmap:rlwe computations}
\item Hardness of Search RLWE \label{mindmap:search rlwe}
\item Decision to Search RLWE reduction \label{mindmp:decision to search rlwe}
\end{enumerate}

\newpage
\section{Notation}
\label{sec:notation}
We list here the key symbols and notations used in the tutorial.

\begin{table}[!htbp]
    \centering
    \begin{tabular}{|p{.31\textwidth}|p{.65\textwidth}|}
    \hline
    Symbol & Meaning \\
    \hline \hline
        $\Z$ & Integers \\
        \hline
        $\Q$ & Rational numbers \\
        \hline
        $\F_q$ for prime number $q$ & $\Z/q \Z= \{0, 1, 2, \ldots, q-1\}$   \\
        \hline
        $\Z[x]$ & Polynomials where the coefficients are integers \\
        \hline
        $F[x]$ & Polynomials where the coefficients take on values in $F$ \\
        \hline
        $\F_q[x]$ & Polynomials where the coefficients take on values in $\F_q$ \\
        \hline
        $\Z[\alpha]$ & the ring obtained by adjoining\index{adjoin} $\alpha$ to $\Z$ \\%, i.e. $\{ a + b\cdot\alpha : a,b \in \Z \}$ \\
        \hline
        $\Q(\alpha)$ & the smallest extension field of $\Q$ that contains $\alpha$ \\ %; when $\alpha$ is algebraic, this takes the simple form $\Q(\alpha) = \{ a + b\cdot\zeta : a,b \in \Q \}$ \\ 
        \hline
        $F[a]$ for a field $F$ & the set $\{ f(a) : f(x) \in F[x] \}$ \\
        \hline
        $F(a)$ for a field $F$ & the smallest extension field of $F$ that contains $a$ \\
        \hline
        $(a)$ for $a$ in ring $R$ & the ideal $\{ ar : r \in R\}$ \\ 
        \hline
        $(a_1,\ldots,a_n)$ for $a_i$ in ring $R$ & the ideal $\{ r_1 a_1 + \cdots + r_n a_n : r_i \in R \}$ \\
        \hline
        $R/I$ for a ring $R$ and an ideal $I$ & the quotient ring\index{quotient ring} of $R$ by $I$, which is the set of cosets of $I$ in $R$ \\
        \hline
        $\Z_n^*$ & multiplicative group modulo $n$; i.e. the set of all (multiplicatively) invertible elements in $\Z_m$; or equivalently $\{ k : k \in \{0,1,\ldots,n-1\}, \gcd(n,k) = 1 \}$ \\
        \hline
        $(\Z/n\Z)^*$ & same as $\Z_n^*$ \\
        \hline
        $E/F$ for fields $E$ and $F$ & a field extension\index{field extension}, where $F$ (the subfield) is contained in $E$ (the extension field) \\
        \hline
        $\zeta_n$ & the $n$-th root of unity \\
        \hline
        $\Phi_n(x)$ & the $n$-th cyclotomic polynomial \\
        \hline
        $\varphi(n)$ & Euler's totient function \\
        \hline
        $\lfloor x \rceil$ & rouding to the integer nearest to $x$ \\
        \hline 
        $[n]$ & $ \{ 1,2,\ldots, n \} $ \\
        \hline
        $a=b \bmod q$ & $a$ and $b$ are congruent modulo $q$ \\
        \hline
        $\Z_q$ & sometimes refer to the range $[-q/2,q/2) \cap \Z$\\
        \hline
        $[x]_q$ & the reduction of $x$ to the integer in $[-q/2,q/2)$ s.t. $[x]_q = x \bmod q$ \\
        \hline
    \end{tabular}
    \caption{List of key symbols}
    \label{tab:notation}
\end{table}

\newpage
\bibliography{references}

\begin{thebibliography}{55}
\providecommand{\natexlab}[1]{#1}
\providecommand{\url}[1]{\texttt{#1}}
\expandafter\ifx\csname urlstyle\endcsname\relax
  \providecommand{\doi}[1]{doi: #1}\else
  \providecommand{\doi}{doi: \begingroup \urlstyle{rm}\Url}\fi

\bibitem[Ajtai(1996)]{ajtai1996generating}
M.~Ajtai.
\newblock Generating hard instances of lattice problems.
\newblock In \emph{Proceedings of the 28th Annual ACM Symposium on Theory of
  Computing}, pages 99--108, 1996.

\bibitem[Ajtai et~al.(2001)Ajtai, Kumar, and Sivakumar]{ajtaiKS01}
M.~Ajtai, R.~Kumar, and D.~Sivakumar.
\newblock A sieve algorithm for the shortest lattice vector problem.
\newblock In J.~S. Vitter, P.~G. Spirakis, and M.~Yannakakis, editors,
  \emph{Proceedings on 33rd Annual {ACM} Symposium on Theory of Computing, July
  6-8, 2001, Heraklion, Crete, Greece}, pages 601--610. {ACM}, 2001.

\bibitem[Alaca and Williams(2004)]{alaca2004introductory}
{\c{S}}.~Alaca and K.~S. Williams.
\newblock \emph{Introductory algebraic number theory}.
\newblock Cambridge University Press Cambridge, 2004.

\bibitem[Albrecht et~al.(2018)Albrecht, Chase, Chen, Ding, Goldwasser,
  Gorbunov, Halevi, Hoffstein, Laine, Lauter, Lokam, Micciancio, Moody,
  Morrison, Sahai, and
  Vaikuntanathan]{HomomorphicEncryptionSecurityStandard2018}
M.~Albrecht, M.~Chase, H.~Chen, J.~Ding, S.~Goldwasser, S.~Gorbunov, S.~Halevi,
  J.~Hoffstein, K.~Laine, K.~Lauter, S.~Lokam, D.~Micciancio, D.~Moody,
  T.~Morrison, A.~Sahai, and V.~Vaikuntanathan.
\newblock Homomorphic encryption security standard.
\newblock Technical report, HomomorphicEncryption.org, Toronto, Canada,
  November 2018.

\bibitem[Alcock(2021)]{alcock21}
L.~Alcock.
\newblock \emph{How to think about Abstract Algebra}.
\newblock Oxford University Press, 2021.

\bibitem[Arora and Barak(2009)]{arora2009computational}
S.~Arora and B.~Barak.
\newblock \emph{Computational complexity: a modern approach}.
\newblock Cambridge University Press, 2009.

\bibitem[Artin(1991)]{artin}
M.~Artin.
\newblock \emph{Algebra}.
\newblock Prentice Hall, 1991.

\bibitem[Aslett et~al.(2015)Aslett, Esperan{\c{c}}a, and
  Holmes]{aslett2015review}
L.~J. Aslett, P.~M. Esperan{\c{c}}a, and C.~C. Holmes.
\newblock A review of homomorphic encryption and software tools for encrypted
  statistical machine learning.
\newblock \emph{arXiv preprint arXiv:1508.06574}, 2015.

\bibitem[Babai(1986)]{babai1986lovasz}
L.~Babai.
\newblock On lov{\'a}sz’lattice reduction and the nearest lattice point
  problem.
\newblock \emph{Combinatorica}, 6\penalty0 (1):\penalty0 1--13, 1986.

\bibitem[Bernstein et~al.(2009)Bernstein, Buchmann, and Dahmen]{bernstein09}
D.~J. Bernstein, J.~Buchmann, and E.~Dahmen.
\newblock \emph{Post-Quantum Cryptography}.
\newblock Springer, 2009.

\bibitem[Brakerski(2012)]{brakerski2012fully}
Z.~Brakerski.
\newblock Fully homomorphic encryption without modulus switching from classical
  {GapSVP}.
\newblock In \emph{Annual Cryptology Conference}, pages 868--886. Springer,
  2012.

\bibitem[Brakerski and Vaikuntanathan(2014)]{brakerski2014efficient}
Z.~Brakerski and V.~Vaikuntanathan.
\newblock Efficient fully homomorphic encryption from (standard) {LWE}.
\newblock \emph{SIAM Journal on Computing}, 43\penalty0 (2):\penalty0 831--871,
  2014.

\bibitem[Brakerski et~al.(2014)Brakerski, Gentry, and
  Vaikuntanathan]{brakerski2014leveled}
Z.~Brakerski, C.~Gentry, and V.~Vaikuntanathan.
\newblock ({L}eveled) fully homomorphic encryption without bootstrapping.
\newblock \emph{ACM Transactions on Computation Theory (TOCT)}, 6\penalty0
  (3):\penalty0 1--36, 2014.

\bibitem[Cai and Nerurkar(1997)]{cai1997improved}
J.-Y. Cai and A.~P. Nerurkar.
\newblock An improved worst-case to average-case connection for lattice
  problems.
\newblock In \emph{Proceedings of the 38th Annual Symposium on Foundations of
  Computer Science}, pages 468--477. IEEE, 1997.

\bibitem[Chen et~al.(2017)Chen, Laine, and Rindal]{chen2017fast}
H.~Chen, K.~Laine, and P.~Rindal.
\newblock Fast private set intersection from homomorphic encryption.
\newblock In \emph{Proceedings of the 2017 ACM SIGSAC Conference on Computer
  and Communications Security}, pages 1243--1255, 2017.

\bibitem[Chi et~al.(2015)Chi, Choi, Kim, and Kim]{chi15}
D.~P. Chi, J.~W. Choi, J.~S. Kim, and T.~Kim.
\newblock Lattice based cryptography for beginners.
\newblock \emph{{IACR} Cryptol. ePrint Arch.}, page 938, 2015.

\bibitem[Chialva and Dooms(2018)]{chialvaD18}
D.~Chialva and A.~Dooms.
\newblock Conditionals in homomorphic encryption and machine learning
  applications.
\newblock \emph{{IACR} Cryptol. ePrint Arch.}, page 1032, 2018.

\bibitem[Conrad(2009)]{conradcyclotomic}
K.~Conrad.
\newblock Cyclotomic extensions.
\newblock 2009.

\bibitem[Cormen et~al.(2001)Cormen, Leiserson, Rivest, and
  Stein]{cormen01introduction}
T.~H. Cormen, C.~E. Leiserson, R.~L. Rivest, and C.~Stein.
\newblock \emph{Introduction to Algorithms}.
\newblock The MIT Press, 2nd edition, 2001.

\bibitem[Damg{\aa}rd et~al.(2012)Damg{\aa}rd, Pastro, Smart, and
  Zakarias]{damgaard2012multiparty}
I.~Damg{\aa}rd, V.~Pastro, N.~Smart, and S.~Zakarias.
\newblock Multiparty computation from somewhat homomorphic encryption.
\newblock In \emph{Annual Cryptology Conference}, pages 643--662. Springer,
  2012.

\bibitem[Dijk et~al.(2010)Dijk, Gentry, Halevi, and
  Vaikuntanathan]{dijk2010fully}
M.~v. Dijk, C.~Gentry, S.~Halevi, and V.~Vaikuntanathan.
\newblock Fully homomorphic encryption over the integers.
\newblock In \emph{Annual international conference on the theory and
  applications of cryptographic techniques}, pages 24--43. Springer, 2010.

\bibitem[Erabelli(2020)]{erabelli2020pyfhe}
S.~Erabelli.
\newblock \emph{pyFHE-a Python library for fully homomorphic encryption}.
\newblock PhD thesis, Massachusetts Institute of Technology, 2020.

\bibitem[Fan and Vercauteren(2012)]{fan2012somewhat}
J.~Fan and F.~Vercauteren.
\newblock Somewhat practical fully homomorphic encryption.
\newblock \emph{IACR Cryptol. ePrint Arch.}, 2012:\penalty0 144, 2012.

\bibitem[Gentry(2009)]{gentry2009fully}
C.~Gentry.
\newblock Fully homomorphic encryption using ideal lattices.
\newblock In \emph{Proceedings of the 41st Annual ACM Symposium on Theory of
  Computing}, pages 169--178, 2009.

\bibitem[Gentry(2010)]{gentry2010computing}
C.~Gentry.
\newblock Computing arbitrary functions of encrypted data.
\newblock \emph{Communications of the ACM}, 53\penalty0 (3):\penalty0 97--105,
  2010.

\bibitem[Gilad-Bachrach et~al.(2016)Gilad-Bachrach, Dowlin, Laine, Lauter,
  Naehrig, and Wernsing]{gilad2016cryptonets}
R.~Gilad-Bachrach, N.~Dowlin, K.~Laine, K.~Lauter, M.~Naehrig, and J.~Wernsing.
\newblock Cryptonets: Applying neural networks to encrypted data with high
  throughput and accuracy.
\newblock In \emph{International conference on machine learning}, pages
  201--210. PMLR, 2016.

\bibitem[Halevi(2017)]{halevi2017homomorphic}
S.~Halevi.
\newblock Homomorphic encryption.
\newblock In Y.~Lindell, editor, \emph{Tutorials on the Foundations of
  Cryptography}. Springer, 2017.

\bibitem[Hoffstein et~al.(2008)Hoffstein, Pipher, and
  Silverman]{hoffstein2008introduction}
J.~Hoffstein, J.~Pipher, and J.~H. Silverman.
\newblock \emph{An introduction to mathematical cryptography}, volume~1.
\newblock Springer, 2008.

\bibitem[Katz and Lindell(2014)]{katz2014introduction}
J.~Katz and Y.~Lindell.
\newblock \emph{Introduction to modern cryptography}.
\newblock CRC press, 2014.

\bibitem[Khot(2005)]{khot05}
S.~Khot.
\newblock Hardness of approximating the shortest vector problem in lattices.
\newblock \emph{J. {ACM}}, 52\penalty0 (5):\penalty0 789--808, 2005.

\bibitem[Khot(2010)]{khot10}
S.~Khot.
\newblock Inapproximability results for computational problems on lattices.
\newblock In P.~Q. Nguyen and B.~Vall{\'{e}}e, editors, \emph{The {LLL}
  Algorithm - Survey and Applications}, Information Security and Cryptography,
  pages 453--473. Springer, 2010.

\bibitem[Korkine and Zolotareff(1873)]{korkine1873}
A.~Korkine and G.~Zolotareff.
\newblock Sur les formes quadratiques.
\newblock \emph{Mathematische Annalen}, 6:\penalty0 366--389, 1873.

\bibitem[Lyubashevsky et~al.(2010)Lyubashevsky, Peikert, and
  Regev]{lyubashevsky2010ideal}
V.~Lyubashevsky, C.~Peikert, and O.~Regev.
\newblock On ideal lattices and learning with errors over rings.
\newblock In \emph{Annual International Conference on the Theory and
  Applications of Cryptographic Techniques}, pages 1--23. Springer, 2010.

\bibitem[Micciancio and Goldwasser(2002)]{micciancio-goldwasser02}
D.~Micciancio and S.~Goldwasser.
\newblock \emph{Complexity of lattice problems - a cryptograhic perspective},
  volume 671 of \emph{The Kluwer international series in engineering and
  computer science}.
\newblock Springer, 2002.

\bibitem[Micciancio and Regev(2007)]{micciancio07worst}
D.~Micciancio and O.~Regev.
\newblock Worst-case to average-case reductions based on gaussian measures.
\newblock \emph{{SIAM} J. Comput.}, 37\penalty0 (1):\penalty0 267--302, 2007.

\bibitem[Micciancio and Regev(2009)]{micciancio2009lattice}
D.~Micciancio and O.~Regev.
\newblock Lattice-based cryptography.
\newblock In \emph{Post-quantum cryptography}, pages 147--191. Springer, 2009.

\bibitem[Milne(2020)]{milneANT}
J.~S. Milne.
\newblock Algebraic number theory (v3.08), 2020.
\newblock Available at www.jmilne.org/math/.

\bibitem[Mukherjee(2016)]{mukherjee2016cyclotomic}
T.~Mukherjee.
\newblock Cyclotomic polynomials in ring-lwe homomorphic encryption schemes.
\newblock Master's thesis, Rochester Institute of Technology, 2016.

\bibitem[Naehrig et~al.(2011)Naehrig, Lauter, and
  Vaikuntanathan]{naehrig2011can}
M.~Naehrig, K.~Lauter, and V.~Vaikuntanathan.
\newblock Can homomorphic encryption be practical?
\newblock In \emph{Proceedings of the 3rd ACM workshop on Cloud computing
  security workshop}, pages 113--124, 2011.

\bibitem[Nguyen and Vallée(2010)]{nguyen2010}
P.~Nguyen and B.~Vallée.
\newblock \emph{The LLL algorithm}.
\newblock Springer, Berlin, Heidelberg, 2010.

\bibitem[Peikert(2009)]{peikert2009public}
C.~Peikert.
\newblock Public-key cryptosystems from the worst-case shortest vector problem.
\newblock In \emph{Proceedings of the 41st annual ACM symposium on Theory of
  computing}, pages 333--342, 2009.

\bibitem[Peikert(2016)]{peikert16decade}
C.~Peikert.
\newblock A decade of lattice cryptography.
\newblock \emph{Found. Trends Theor. Comput. Sci.}, 10\penalty0 (4):\penalty0
  283--424, 2016.

\bibitem[Peikert and Rosen(2007)]{peikert2007lattices}
C.~Peikert and A.~Rosen.
\newblock Lattices that admit logarithmic worst-case to average-case connection
  factors.
\newblock In \emph{Proceedings of the 39th Annual ACM Symposium on Theory of
  Computing}, pages 478--487, 2007.

\bibitem[Pietrzak(2012)]{pietrzak12}
K.~Pietrzak.
\newblock Cryptography from learning parity with noise.
\newblock In M.~Bielikov{\'{a}}, G.~Friedrich, G.~Gottlob, S.~Katzenbeisser,
  and G.~Tur{\'{a}}n, editors, \emph{{SOFSEM} 2012: Theory and Practice of
  Computer Science - 38th Conference on Current Trends in Theory and Practice
  of Computer Science}, volume 7147 of \emph{Lecture Notes in Computer
  Science}, pages 99--114. Springer, 2012.

\bibitem[Porter(2015)]{porter15cyclotomic}
B.~Porter.
\newblock Cyclotomic polynomials.
\newblock 2015.

\bibitem[Regev(2005)]{regev05}
O.~Regev.
\newblock On lattices, learning with errors, random linear codes, and
  cryptography.
\newblock In H.~N. Gabow and R.~Fagin, editors, \emph{Proceedings of the 37th
  Annual {ACM} Symposium on Theory of Computing}, pages 84--93. {ACM}, 2005.

\bibitem[Regev(2009)]{regev2009lattices}
O.~Regev.
\newblock On lattices, learning with errors, random linear codes, and
  cryptography.
\newblock \emph{Journal of the ACM (JACM)}, 56\penalty0 (6):\penalty0 1--40,
  2009.

\bibitem[Regev(2010)]{regev2010learning}
O.~Regev.
\newblock The learning with errors problem.
\newblock \emph{Invited survey in CCC}, 7\penalty0 (30):\penalty0 11, 2010.

\bibitem[Rivest et~al.(1978{\natexlab{a}})Rivest, Adleman, and
  Dertouzos]{rivest1978data}
R.~L. Rivest, L.~Adleman, and M.~L. Dertouzos.
\newblock On data banks and privacy homomorphisms.
\newblock \emph{Foundations of Secure Computation, Academic Press}, pages
  169--179, 1978{\natexlab{a}}.

\bibitem[Rivest et~al.(1978{\natexlab{b}})Rivest, Shamir, and
  Adleman]{rivest1978method}
R.~L. Rivest, A.~Shamir, and L.~M. Adleman.
\newblock A method for obtaining digital signatures and public-key
  cryptosystems.
\newblock \emph{Commun. ACM}, 21\penalty0 (2):\penalty0 120--126,
  1978{\natexlab{b}}.

\bibitem[SEAL()]{sealcrypto2022}
SEAL.
\newblock {M}icrosoft {SEAL} (release 4.0).
\newblock \url{https://github.com/Microsoft/SEAL}, Mar. 2022.
\newblock Microsoft Research, Redmond, WA.

\bibitem[Sipser(2013)]{sipser2013introduction}
M.~Sipser.
\newblock \emph{Introduction to the Theory of Computation}.
\newblock Course Technology, third edition, 2013.

\bibitem[Stein(2012)]{stein2012algebraic}
W.~Stein.
\newblock Algebraic number theory, a computational approach.
\newblock \emph{Harvard, Massachusetts}, 2012.

\bibitem[Veugen(2014)]{veugen14}
T.~Veugen.
\newblock Encrypted integer division and secure comparison.
\newblock \emph{Int. J. Appl. Cryptogr.}, 3\penalty0 (2):\penalty0 166--180,
  2014.

\bibitem[Wood et~al.(2020)Wood, Najarian, and Kahrobaei]{wood2020homomorphic}
A.~Wood, K.~Najarian, and D.~Kahrobaei.
\newblock Homomorphic encryption for machine learning in medicine and
  bioinformatics.
\newblock \emph{ACM Computing Surveys (CSUR)}, 53\penalty0 (4):\penalty0 1--35,
  2020.

\end{thebibliography}
\addcontentsline{toc}{section}{References}
\bibliographystyle{abbrvnat}

\newpage
\printindex

\end{document}